\documentclass[a4paper,twoside,11pt,useAMS,usergraphicx]{Classes/CUEDthesisPSnPDF1}%, presently using

\usepackage{ulem}
\usepackage{hyperref}
\usepackage{bm}
\usepackage{StyleFiles/watermark}
\usepackage{lscape}
\usepackage{graphicx}
\usepackage{longtable}
\usepackage{lscape}
\usepackage{amssymb}
\usepackage{endnotes}
\usepackage{pifont}
\usepackage{titlesec}
\usepackage{lipsum}
\usepackage[sort]{natbib}
\usepackage{enumitem}

\newcounter{daggerfootnote}

\usepackage{multicol}
\usepackage{lipsum}
\usepackage{float}
\usepackage{adjustbox}
\newcommand\blfootnote[1]{%
  \begingroup
  \renewcommand\thefootnote{}\footnote{#1}%
  \addtocounter{footnote}{-1}%
  \endgroup
}
\usepackage{latexsym}
\usepackage{mathrsfs}
\bibpunct{(}{)}{;}{a}{,}{,}
\usepackage{subfig}
\usepackage{setspace}
\usepackage{booktabs, caption, makecell}

\usepackage{threeparttable}
\definecolor{dark-gray}{gray}{0.1}
\begin{document}
%\rom{2}
%\rom{20}
%\rom{200}
\renewcommand\baselinestretch{1.2}
\baselineskip=18pt plus1pt

\setcounter{secnumdepth}{3}
\setcounter{tocdepth}{3}

\setcounter{secnumdepth}{4}
\setcounter{tocdepth}{4}
\frontmatter % book mode only

\pagenumbering{empty}
\cleardoublepage
\thispagestyle{empty}
\begin{titlepage}
\centering
%\vspace*{0.2cm}

\begin{doublespacing}

{\bf \huge Unveiling diverse nature of core collapse supernovae }\\

\vspace*{1.4cm}
\vspace{0.6cm}
\includegraphics[width = 40mm]{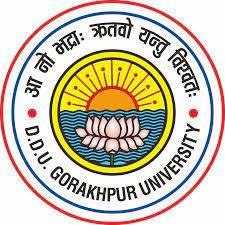} \\
\vspace{0.30cm}
{\large \bf THESIS} \\
{\large \sc Submitted For the Degree of } \\
\vspace{0.30cm}
{\LARGE \bf DOCTOR OF PHILOSOPHY}\\
\vspace{0.15cm}
{\small \sc \bf IN } \\
\vspace{0.25cm}
{\LARGE \bf PHYSICS} \\
\vspace{0.30cm}
{\large \sc by} \\
\vspace{0.5cm}
{\large \bf Amar Aryan} \\
{\large \bf DEPARTMENT OF PHYSICS}\\
{\large \bf  DEEN DAYAL UPADHYAYA GORAKHPUR UNIVERSITY}\\
{\large \bf  GORAKHPUR-273009 (U.P.) INDIA}\\
\vspace{0.60cm}

{\large \sc Research Centre}\\
\vspace{0.1cm}
{\large \bf ARYABHATTA RESEARCH INSTITUTE OF OBSERVATIONAL SCIENCES (ARIES)} \\
{\large \bf MANORA PEAK, NAINITAL-263001, INDIA}\\
\end{doublespacing}

\vspace{1cm}
{\large \bf SEPTEMBER 2023}\\
\end{titlepage}

%\include{Start/title/Univ_Certificates_1}
%\include{Start/title/Univ_Certificates_2}
%\include{Start/title/Univ_Certificates_3}
%\include{Start/title/5_Self_Cert}

%To create an empty space
\cleardoublepage
\thispagestyle{empty}

\vspace*{8cm}

\hspace*{0.0cm}{\huge \it  Dedicated to,}

\vspace*{1.0cm}

\hspace*{4.0cm} {\huge \it My Dada ji ...}\\
\\
%\vspace*{1.0cm}
%\hspace*{9.5cm} {\huge \it Supervisor}

%\newpage
\pagenumbering{roman}
%To create an empty space
\cleardoublepage
\thispagestyle{empty}

\begin{center}
\LARGE \bf {Acknowledgements} \\
\bigskip
\normalsize
\end{center}

\noindent

The Journey of my academic life commenced many years ago at ``Middle School Kubri", a government school in our village named ``Kubri". I vividly recall the unwavering encouragement and motivation I received from my teachers (among others) {\bf Sita Ram sir, Devnandan sir, and Sadhu sir}. Subsequently, for my secondary education (10th), I got admission to ``Gurukul Shikshalay", a school near our village. I consider myself fortunate to have had the opportunity to learn from exceptional teachers, including {\bf Manoj sir, Deglal sir, Chandrika sir, Ravi sir, Akhilesh sir, Parvej sir, Pawan sir and others}. They instilled a strong moral compass and laid the foundation within me for my academic success ahead.

In the pursuit of higher education, I came to Delhi in 2009. From my senior secondary (12th) school days through my bachelor's and master's degrees, I met numerous teachers and professors who provided unwavering support and motivation. I express my heartfelt gratitude to each one of them. I thank {\bf Meena sir} from my senior secondary school for his support, motivation, and guidance. I am also grateful to {\bf Mukharjee sir} for his assistance during my B.Sc college years. Moreover, I consider myself extremely fortunate to have attended lectures delivered by esteemed individuals such as {\bf Prof. Daya Shankar Kulshreshtha, Prof. Patrick Das Gupta, Prof. Shobhit Mahajan, Prof. H. P. Singh, Prof. T. R. Seshadri, Prof. Supriya Kar, Dr. Ashutosh Bhardwaj, Dr. Saurabh Sur, and others} during my M.Sc program. I extend my heartfelt thanks to all the professors and lecturers who enriched my educational Journey during my M.Sc days.

Upon finishing my M.Sc., I commenced my Journey at Aryabhatta Research Institute of Observational Sciences (ARIES) on July 13th, 2018. I embarked as a Junior Research Fellow to pursue  PhD, under the esteemed guidance of Dr. Shashi Bhushan Pandey. I extend my heartfelt gratitude to {\bf my PhD supervisor, Dr. Shashi Bhushan Pandey}, as well as {\bf my co-supervisor, Prof. Sugriva Nath Tiwari}, for their unwavering support, motivation, and guidance throughout this remarkable Journey. Their invaluable suggestions and directions were pivotal in completing this thesis. I am profoundly thankful to them for allowing me to explore the intellectual curiosity within my research and establish a solid foundation. I will forever appreciate the encouragement and excellent opportunities provided by my supervisor, Dr. S. B. Pandey, which allowed me to collaborate with eminent experts in the field of supernova research. I am grateful to {\bf Prof. Alexei V. Filippenko, Dr. Weikang Zheng, Prof. Keiichi Maeda, Prof. Jozsef Vinko, Prof. Dipankar Bhattacharya, and Prof. David A. H. Buckley} for their constant support, suggestions and guidance throughout this beautiful Journey.  I also acknowledge {\bf Prof. Craig Wheeler} and {\bf Prof. Van Dyk Schuyler} for valuable discussions regarding supernova observations and science. It was really a great experience for me to work in active collaboration with these eminent researchers.

I am very thankful to {\bf Ryoma Ouchi, Isaac Shivvers, Heechan Yuk, Sahana Kumar, Samantha Stegman, Goni Halevi, Timothy W. Ross, Carolina Gould, Sameen Yunus, Raphael Baer-Way, Asia deGraw, Dr. Abhay Pratap Yadav, Dr. Brajesh Kumar, Thomas G. Brink, Andrew Halle, Jeffrey Molloy, Charles D. Kilpatrick, Sanyum Channa, Maxime de Kouchkovsky, Michael Hyland, Minkyu Kim, Kevin Hayakawa, Kyle McAllister, Andrew Rikhter, Benjamin Stahl, and Yinan Zhu} for their kind support and help at various stages during my PhD. I immensely thank {\bf Ryoma Ouchi} for his support in several projects. I am also very thankful to {\bf Dr. Abhay Pratap Yadav} for his motivation and support.

I thank all the faculty members ({\bf Prof. Dipankar Banerjee, Dr. Brijesh Kumar, Dr. Alok Chandra Gupta, Dr. Jeewan C. Pandey, Dr. Ramakant S. Yadav, Dr. Saurabh, Dr. Kuntal Misra, Dr. Neelam Panwar, Dr. Indranil Chattopadhyay, Dr. Snehlata, Dr. Santosh Joshi, Dr. Yogesh Chandra Joshi, Dr. Manish K. Naja, Dr. Narendra Singh, Prof. Shantanu Rastogi, Prof. Ravi Shankar Singh, and others}), engineers ({\bf Mohit Joshi, Ashish Kumar, Krishna Reddy B., Sanjit Sahu and others}), post-doctoral fellows, research scholars ({\bf Prayag Bhaiya, Rahul Bhaiya, Teekendra, Atul, Vishnu, Vaibhav, Bakhtawar, Sumit and others}), and staff members {\bf (Abhishek Sharma, Himanshu Vidyarthi, Ram Dayal Bhatt, Amar Singh Meena, Hemant Kumar, and others)} of ARIES Nainital and the Department of Physics of DDU Gorakhpur University Gorakhpur profusely for their kind support throughout my study. In addition, I acknowledge the financial support of the {\bf BRICS grant DST/IMRCD/BRICS/Pilotcall/ProFCheap/2017(G)}. I also acknowledge the financial support from {\bf the Council of Scientific \& Industrial Research (CSIR), India, under file no. 09/948(0003)/2020-EMR-I}.

I thank my group members ({\bf Rahul Gupta, Amit Kumar, and Amit Kumar Ror}). I also thank my other batch-mates ({\bf Raj, Mahendar, Nikita, and Akanksha}), seniors ({\bf Mukesh bhaiya, Mridweeka di, Anjasha di, Ashwani bhaiya, Vineet bhaiya, Pankaj bhaiya, Amit, Ankur, Arpan, Krishan, Prajjwal, Aditya, Vinit, Jaydeep, and others)}, Juniors ({\bf Devanand, Mrinmoy, Bhavya, Kiran, Rahul, Shubham, Tushar, Nitin, Gurpreet, Naveen, Aayushi, Upasna, Shivangi, Dibya, Karan, Tarak, Srinivas, Vikrant, Monalisa, and others}) and canteen members (specially, Darshan, Jagdish dajyu, and Ravi dajyu) at ARIES and Devasthal campus, who have made this Journey even more memorable.

My family is my most valuable asset. Without their unwavering support, I can hardly imagine my fate. The words of wisdom by my {\bf Dadaji, Late   Jagdish Ram}, always consistently steered me on the path of life. A promise made by me to my Papa, {\bf Late Birendra Kumar Kashyap}, encouraged me to achieve several milestones, with many more yet to come. The unwavering support, motivation, and encouragement from {\bf my mummy, Smt. Yashoda Devi}, have been a constant presence in my life. The support from {\bf my bhaiya, Shri Hira Lal} and {\bf bhabhi, Smt. Soni Kashyap} can not be expressed in mere words. My brother went above and beyond, doing nothing less than a father would do for his son. My two nephews {\bf Astitva and Manas} kept me happy and entertained all the time. I also thank {\bf my two elder sisters, Paro di and Kanchan di}, for their unwavering support and unconditional care, treating me as if I were their own child. I am thankful to {\bf Deepak Jija Ji and Anil Jija Ji} for all their love and support. Their kids {\bf Aditya, Krishna, Saurya Sankar, and Alok Ranjan} have also been my constant entertainment source. Their presence filled my life with joy. In 2014, I felt blessed when I met my love {\bf Urvashi Rathore}. She is an inexhaustible source of motivation and has continuously provided me with selfless support in every situation. Her presence in my life has been a source of great comfort and strength.

 I can never forget the support of my dear {\bf uncles, Late Chander dev Kashayp, Shri Surendra Kashyap, Shri Rohit Kashyap, and Shri Bhuneshwar Kashyap.} I am really in debt to their love and affection, especially {\bf Sanhjhle Papa (Shri Surendra Kashyap)}. I extend a special thank you to {\bf Chhote chacha, Shri Bhuneshwar Kashyap} for standing on my side on many difficult occasions. I am also thankful to {\bf my dadi ji, Smt. Koyali Devi} and {\bf aunties, Kanta Devi, Geeta Devi, Kiran Devi and Sarita Devi} for all their love and blessings. I am glad to mention the names of my cousins, {\bf Bablu bhaiya, Baby di, Chhoti bhaiya, Moti bhaiya, Pankaj bhaiya, Priya di, Kajri di, Akash, Asha, Prabha, Paras, Puja, Guddu, Mansi and Golu} for their everlasting love and support. I extend a special thank you to {\bf Moti bhaiya} for standing by my side in hard times. I thank {\bf Pinky di} for her support. She always inspired me to achieve my goals. Thanks to {\bf Rajesh, Phupha Ji and Phuwa Ji} for their love. I am very thankful to {\bf my maternal uncles, Shri Vasudeo Pandit, Shri Mathura Pandit, and Shri Sita Ram Pandit}. I am also very grateful to Badi Mami, Manjhli Mami, and Chhoti Mami for their unconditional love and affection. I am also overwhelmed to thank my {\bf cousins from the maternal side, Sunil bhaiya, Deepak bhaiya, Pankaj, Seemu, Dinesh, Jitu, and Nikhil}, for their support. I am also very thankful to Dr. {\bf Shashi Bhushan Pandey's wife (ma'am) and their children} for giving me a family environment at ARIES and serving us tasty food on numerous occasions.

Friends play a significant role in our lives as they profoundly impact our emotional, social, and personal well-being. I thank my {\bf village friends, Rahul (Chacha), Vikram, Vijay, Rakesh Rana, Rahul Rana, Raja Rana, Dara, Panga, Ajit, Ayodhya, Brabha and others}. They always cheered each of my movements ahead. I also thank my Delhi friends, especially {\bf Lokesh}, for their support. I spent a lot of good times with {\bf my friends Nitesh, Ashish, and Mukesh} during my M.Sc program. During my PhD journey, whenever I visited Delhi, I had a delightful time staying with {\bf Kapildev, Avnish, and Nitesh}. Exploring Vijay Nagar with them added to the joy of my stay.

My research Journey has been an exhilarating adventure, marked by high phases and setbacks too. Maintaining my sanity throughout this entire endeavour proved to be a formidable task. However, I consistently relied on my abilities and confidently tackled each new challenge. I am immensely grateful to everyone who supported and assisted me along this Journey. {\bf This thesis is a result of blessings from several individuals}. Last but not least, I thank the almighty God for everything.
I look forward to reaching many more remarkable milestones in my life.

% \cleardoublepage

\smallskip
\vspace{0.5cm}
\hfill{--- Amar}
\thispagestyle{empty}

% Thesis Abstract -----------------------------------------------------
%\frontmatter
%\lhead{{ABSTRACT}}
%\section*{\Large \bf \sc {Abstract}}
%\addtotoc{Abstract}  

\cleardoublepage

\begin{center}
\LARGE \bf {\sc PREFACE} \\
\bigskip
\normalsize
\end{center}

\noindent 
~~~~Core-collapse supernovae (CCSNe) are catastrophic astrophysical phenomena that occur during the last evolutionary stages of massive stars having initial masses $\gtrsim$\,8\,M$_{\odot}$. These catastrophic events play a pivotal role in enriching our Universe with heavy elements and are also responsible for the birth of Neutron stars and stellar mass Black holes. Knowledge of the possible progenitors of CCSNe is fundamental to understanding these transient events. Additionally, the underlying circumstellar environment around possible progenitors and the physical mechanism powering the light curves of these catastrophic CCSN events also require careful investigations to unveil their nature. Based on the spectroscopic observational features, the CCSNe are primarily divided into H-rich and H-deficient categories. The H-rich CCSNe display unambiguous H-features in their spectra, while H-deficient CCSNe don't. Type Ib SNe are a subclass of H-deficient CCSNe that lack prominent H-features in their spectra but display distinct He-feature. 

The research work within the context of the present thesis is an attempt to investigate the possible progenitors, ambient media around the progenitors, and powering mechanisms behind the light curve of CCSNe.
Particularly, fractional contributions of different elements, including hydrogen, the key element discriminating Type Ib and Type IIb SNe, are studied in detail. We also employed several powering mechanisms to decipher underlying physical mechanisms behind Type Ib and Type IIb SNe. We have employed observational data from several telescopes, state-of-the-art simulation modules, and 1-dimensional hydrodynamic codes for such investigations. 

In this thesis, we have investigated the photometric and spectroscopic properties of two Type Ib CCSNe, namely, SN~2015ap and SN~2016bau. We aim to gain insight into their possible progenitors, the circumstellar environment surrounding them, and the powering mechanism for their light curve. We have analysed the photometric characteristics of the CCSNe mentioned above, encompassing their colour evolution, bolometric luminosity, photospheric radius, temperature, and velocity evolution. By analysing their light curves, we have computed the ejecta mass, synthesised nickel mass, and ejecta kinetic energy. Thus, the time domain astronomy of CCSNe is crucial to get insight into their several physical properties.
Furthermore, we have modelled the spectra of SN~2015ap and SN~2016bau at different stages of their development and also compared their spectra with several other similar SNe. The P~Cygni profiles of different lines present in the spectra are utilised to understand the velocity evolution of several line-emitting regions. The 1-dimensional stellar modelling of the possible progenitors and the comparison of the results of synthetic hydrodynamic explosions with actual observations indicate a 12\,M$_{\odot}$ progenitor exploding in a solar metallicity region as the potential progenitor for SN~2015ap. In contrast, a slightly less massive star exploding in a solar metallicity environment is the expected progenitor for SN~2016bau. At the pre-SN stage, the mass of the progenitor of SN~2016bau lies close to the boundary between the SN and a non-SN phase.

Type IIb SNe are another subclass of CCSNe and are thought to bridge the gap between H-rich and H-deficient CCSNe. At first, their spectra reveal noticeable H-features, but after a few weeks, the H-features diminish while prominent He-features begin to emerge. Type Ib/IIb CCSNe progenitors retain no to very small amounts of hydrogen during their explosions. However, the correct estimation of the amount of hydrogen retained before explosion by the underlying CCSN progenitor is subjected to contamination by the uncertainties associated with determining the extinction and distance of the CCSN. The photometric and spectroscopic investigations of Type IIb CCSNe are necessary to understand their progenitors, ambient medium, and powering mechanisms. Such analyses also decipher their link with the H-rich and H-deficient CCSNe.

In the present research work, we have performed the photometric and spectroscopic investigation of a Type IIb SN~2016iyc. Our findings indicate that SN~2016iyc lies towards the lower end of the distribution compared to similar CCSNe in terms of inherent brightness. The light curve analysis indicates that SN~2016iyc produces a relatively smaller amount of ejecta mass and suffers low nickel mass production. Based on the photometric and spectroscopic behaviour of SN~2016iyc, we performed the stellar evolution of models having initial zero-age main-sequence (ZAMS) masses in the range of 9--14\,M$_{\odot}$. The synthetic explosions of ZAMS star models with mass in the range of 12--13\,M$_{\odot}$ having the pre-SN radius, $R_{\mathrm{0}}$ within (240--300)\,R$_{\odot}$, produce bolometric luminosity light curves and photospheric velocities that match well with actual observations.  Additionally, ejecta mass $M_{\rm ej} =$ (1.89--1.93)\,M$_{\odot}$, explosion energy $E_{\rm exp} = $ (0.28--0.35) $\times 10^{51}$\,erg, and $M_{\rm Ni} < 0.09$\,M$_{\odot}$, are in good agreement with observed estimations; thus, SN~2016iyc probably exploded from a progenitor lying towards the lower mass limits for SNe~IIb. Additional hydrodynamic simulations have also been conducted to investigate the explosions of SN~2016gkg and SN~2011fu, aiming to compare intermediate- and high-luminosity examples among the extensively studied SNe Type IIb. The results obtained from modelling the potential progenitors and simulating the explosions of SN~2016iyc, SN~2016gkg, and SN~2011fu reveal a range of progenitor masses for SNe~IIb. The range of progenitor masses for Type IIb SNe identified under the present research work lies well within the established range of progenitor masses for CCSNe.

After discussing the properties of H-deficient SNe and the behaviour of a Type IIb SN retaining an intermediate amount of H-envelope, we provide interesting properties of H-rich and H-deficient SNe together that originate from progenitors, each having a mass of 25\,M$_{\odot}$ at ZAMS and zero metallicity. CCSNe from massive Population  III (Pop III) stars are thought to have had an enormous impact on the early Universe. The SNe from Pop III stars were responsible for the initial enrichment of the early Universe with heavy elements. Pop III stars played a key role in cosmic re-ionization. Thus, the investigations of the stellar evolution of Pop III stars and resulting SNe are essential. 
This thesis presents the results of 1-dimensional stellar evolution simulations of a rotating Pop III star having an initial mass of 25 M$_{\odot}$. Starting from ZAMS, the models are evolved until the onset of core collapse. The rapidly rotating models exhibit violent and intermittent mass loss episodes following the main sequence stage. Notably, the Pop III models exhibit smaller pre-SN radii compared to the model with solar metallicity.  Further, with models at the stage of the onset of core collapse, we perform the hydrodynamic simulations of the resulting SNe. As a consequence of the mass losses due to corresponding rotations and stellar winds, the resulting SNe span a class from weak H-rich to H-deficient CCSNe. This analysis demonstrates the substantial influence of initial stellar rotation on the evolution of massive stars and their resulting transients. Additionally, we observe that the absolute magnitudes of CCSNe originating from Pop III stars are much fainter compared to the ones originating from the star with solar metallicity. Based on the outcomes of our simulations, we conclude that within the range of explosion energies and nickel masses considered, these transient events exhibit very low luminosities. Consequently, detecting them at high redshifts would be a significant challenge.

Beyond discussing the CCSNe resulting from progenitors having ZAMS masses of 25\,M$_{\odot}$ or less, we have also studied the stellar evolution of a massive 100\,M$_{\odot}$ ZAMS star up to the onset of core collapse.
Based on initial mass, mass loss rate, rotation and metallicity, the resulting transient could fall into any category, PISN, PPISN, Type IIP-like SNe, and several H-rich/H-deficient SNe showing ejecta-CSM interaction signatures. However, in the presented thesis, we have investigated the consequences of a non-rotating 100\,M$_{\odot}$ ZAMS progenitor exploding into Type IIP-like CCSNe. We also have explored the effect of the variation of explosion energy and nickel mass on the light curves of resulting CCSNe.

The research work presented here has paved the way for new avenues of exploration within astronomy and astrophysics. Ultimately, we summarise our significant findings and discuss the potential prospects. We attempt to highlight the role of observations and simulations in synergistic investigations of transients. The increasing number of progenitor detections in high-resolution pre-explosion images and further refinement of available state-of-the-art stellar evolution codes would certainly protrude the knowledge of CCSNe.

%To create an empty space
\cleardoublepage
%\thispagestyle{empty}

%\documentstyle{article}
%\begin{document}
%\topmargin -1.5cm
%\textheight 10in
%\thispagestyle{empty}
\label{NOTATIONS AND ABBREVIATIONS}
%\addtotoc{Notations and abbreviations}  

\begin{center}
\Large{\bf Abbreviations, Notations and symbols}
\end{center}

\begin{tabbing}
cccccccccccccccccccccccccccccccc\=ccccccccccccccccccccccc\=cccccccccccccccccccccccccccc
cccccccccccccccccccccccccccc \kill

\textbf{$\alpha$} \> \textbf{R}ight \textbf{A}scension \> \\ 

\textbf{\AA} \> \textbf{Angstrom} \> \\ 

\textbf{ARIES} \> \textbf{A}ryabhatta \textbf{R}esearch \textbf{I}nstitute of observational Scienc\textbf{ES} \> \\    

\textbf{Ca~II} \> \textbf{Singly ionized Calcium} \> \\

%\textbf{BH} \>\textbf{B}lack \textbf{H}ole \> \\
\textbf{CCD}(s) \> \textbf{C}harge \textbf{C}ouple \textbf{D}evice(s) \> \\    
\textbf{CCSN}(e) \> \textbf{C}ore-\textbf{C}ollapse \textbf{S}uper\textbf{N}ova(e) \> \\  
\textbf{CSM} \> \textbf{C}ircum \textbf{S}tellar \textbf{M}aterial \> \\  
 
\textbf{$\delta$} \> \textbf{Dec}lination \> \\    
\textbf{\tt DAOPHOT} \> \textbf{D}ominion \textbf{A}strophysical \textbf{O}bservatory \textbf{Phot}ometry \> \\    

\textbf{DOT} \> \textbf{D}evasthal \textbf{O}ptical \textbf{T}elescope \> \\

\textbf{DFOT} \> \textbf{D}evasthal \textbf{F}ast \textbf{O}ptical \textbf{T}elescope \> \\    
   
\textbf{Eq} \>\textbf{Eq}uation\> \\

\textbf{$\epsilon_{nuc}$}  \> \textbf{Specific Luminosity due to nuclear reactions} \> \\

\textbf{Fe~II} \> \textbf{Singly ionized Iron} \> \\

\textbf{Fe-core} \> \textbf{Iron-core} \> \\
\textbf{Fig} \>\textbf{Fig}ure\> \\
\textbf{FITS} \> \textbf{F}lexible \textbf{I}mage \textbf{T}ransport \textbf{S}ystem \> \\    
\textbf{FoV} \> \textbf{F}ield \textbf{o}f \textbf{V}iew \> \\    
\textbf{FWHM} \> \textbf{F}ull \textbf{W}idth at \textbf{H}alf \textbf{M}axima \> \\    
\textbf{GRBs} \>\textbf{G}amma-\textbf{R}ay \textbf{B}urst\textbf{s}\> \\
%\textbf{GCN} \> \textbf{G}RB \textbf{C}oordinates \textbf{N}etwork \> \\    

\textbf{HN}(e) \> \textbf{H}yper\textbf{N}ova(e) \> \\     

\textbf{H~I} \> \textbf{Neutral Hydrogen} \> \\

\textbf{H-envelope} \> \textbf{Hydrogen-envelope} \> \\

\textbf{H-rich} \> \textbf{Hydrogen-rich} \> \\

\textbf{He~I} \> \textbf{Neutral Helium} \> \\

\textbf{He-core} \> \textbf{He-core} \> \\

\textbf{He-envelope} \> \textbf{Helium-envelope} \> \\

\textbf{He-rich} \> \textbf{Helium-rich} \> \\

\textbf{HST} \> \textbf{H}ubble \textbf{S}pace \textbf{T}elescope \> \\    
\textbf{\tt IRAF} \> \textbf{I}mage \textbf{R}eduction \textbf{A}nalysis \textbf{F}acilities \> \\    
\textbf{K} \> \textbf{Kelvin} \> \\    

\textbf{KN}(e) \> \textbf{K}ilo\textbf{N}ova(e) \> \\ 

\textbf{\tt MESA} \> \textbf{M}odules for \textbf{E}xperiments in \textbf{S}tellar \textbf{A}strophysics \> \\

\textbf{Mg~II} \> \textbf{Singly ionized Magnesium} \> \\

\textbf{MJD} \> \textbf{M}odified \textbf{J}ulian \textbf{D}ate \> \\    
\textbf{Mpc} \> \textbf{M}ega-\textbf{p}arse\textbf{c} \> \\    

\textbf{Myr} \> \textbf{M}illion\textbf{y}ea\textbf{r} \> \\

\textbf{M$_{\odot}$} \> \textbf{Solar Mass} \> \\

%\textbf{NASA} \> \textbf{N}ational \textbf{A}eronautics and \textbf{S}pace \textbf{A}dministration \> \\

\textbf{Na~I~D} \> \textbf{Neutral Sodium Doublet}\> \\

\textbf{NIR} \> \textbf{N}ear-\textbf{i}nfra\textbf{r}ed \> \\    
\textbf{NS}(s) \>\textbf{N}eutron \textbf{S}tar(s) \> \\

\textbf{O~I} \> \textbf{Neutral Oxygen} \> \\

\textbf{PSF} \> \textbf{P}oint \textbf{S}pread \textbf{F}unction \> \\    

\textbf{PISN}(e) \> \textbf{P}air-\textbf{I}nstability \bf{SN}(e) \> \\

\textbf{PPISN}(e) \> \textbf{P}ulsational \textbf{P}air-\textbf{I}nstability \bf{SN}(e)  \> \\

\textbf{R$_{\odot}$} \> \textbf{Solar Radius} \> \\

\textbf{S~II} \> \textbf{Singly ionized Sulpher} \> \\  

\textbf{SED}(s) \> \textbf{S}pectral \textbf{E}nergy \textbf{D}istribution(s) \> \\    

\textbf{Si~II} \> \textbf{Singly ionized Silicon} \> \\ 

\textbf{SN}(e) \> \textbf{S}uper\textbf{N}ova(e) \> \\        

\textbf{\tt SNEC} \> \textbf{S}uper\textbf{N}ova \textbf{E}xplosion \textbf{C}ode \> \\ 
   
\textbf{ST} \> \textbf{S}ampurnanand \textbf{T}elescope \> \\    
   
\textbf{UT} \> \textbf{U}niversal \textbf{T}ime \> \\     

\textbf{UV} \> \textbf{U}ltra\textbf{v}iolet \> \\ 
    
\textbf{WD}(s) \>\textbf{W}hite \textbf{D}warf(s) \> \\

\textbf{ZAMS} \>\textbf{Z}ero-\textbf{A}ge \textbf{M}ain-\textbf{S}equence \> \\

\end{tabbing}
%\end{document}
%\include{Start/title/Publication}
\cleardoublepage
\thispagestyle{empty}

\vspace*{5cm}

%~~~~~~~~~~~~~~~~~~~~~~~~~~~~~~~~~~~~~~~~~~~~~~~~~~~~~~~{\huge \it To Sai Baba}\\
%~~~~~~~~~~~~~~~~~~~~~~~~~~~~~~~~~~~~~~~~~~~~~~~~~~~~~~~{\huge \it  \& my parents}

\hspace*{1.0cm}

%\hspace*{3.0cm} {\large \rm My grand parents}

%\vspace*{4.0cm}

%\hspace*{3.0cm} {\huge \it Sai Baba}
%\vspace*{1.0cm}
%
%\hspace*{7.0cm} {\huge \it \& }
%\vspace*{1.0cm}

\hspace*{0.7cm} {{{\Large ``Life is like riding a bicycle. To keep your balance, you must keep moving."}\\}

\smallskip
\vspace{0.05cm}
\hfill{----- Albert Einstein}\\

\hspace*{8.0cm}
\vspace{3cm}
%\hspace*{3.0cm} {\large \rm My grand parents}

%\vspace*{4.0cm}

%\hspace*{3.0cm} {\huge \it Sai Baba}
%\vspace*{1.0cm}
%
%\hspace*{7.0cm} {\huge \it \& }
%\vspace*{1.0cm}

%\vspace{5cm}
%\hspace*{.0cm}
%{\Large ``Success is not final; failure is not fatal: It is the courage to continue that counts"}
%\smallskip
%\vspace{0.05cm}
%\hfill{----- Jayson DeMers}%Anthony T. Hincks: An author of life, Volume 1}

\tableofcontents
\listoffigures
\addcontentsline{toc}{chapter}{\sc List~of~figures} %adds References to contents page
\listoftables
\addcontentsline{toc}{chapter}{\sc List~of~tables} %adds References to contents page

\mainmatter % book mode only
\newpage
\pagenumbering{arabic}
\chapter{Introduction}\label{Ch:1}
\ifpdf
    \graphicspath{{Chapter1/Chapter1Figs/PNG/}{Chapter1/Chapter1Figs/PDF/}{Chapter1/Chapter1Figs/}}
\else
    \graphicspath{{Chapter1/Chapter1Figs/EPS/}{Chapter1/Chapter1Figs/}}
\fi

\ifpdf
    \graphicspath{{Chapter1/Chapter1Figs/JPG/}{Chapter1/Chapter1Figs/PDF/}{Chapter1/Chapter1Figs/}}
\else
    \graphicspath{{Chapter1/Chapter1Figs/EPS/}{Chapter1/Chapter2Figs/}}
\fi
%\section{Astrophysical Explosions and Their Importance}%\label{sn} 

Supernovae (SNe) are extreme astrophysical transient events occurring either when a White Dwarf (WD) is triggered into a runaway thermonuclear burning \citep[][]{1971ApJ...163..221C,1979ApJ...230L..37A,1997ApJ...478..678K} or during the last evolutionary stages of massive stars \citep[][]{1986NYASA.470..267W,1993ApJ...414L.105H, 1996ApJ...460..408T,1999ApJ...510..379M}. These catastrophic explosions represent the dynamical disruption of an entire star \citep{2012RSPTA.370..774W}. SNe are among the brightest astrophysical explosions; thus, they play a crucial role in determining cosmological distances and verifying cosmological models \citep[][]{2013JETPL..98..432B}. An astrophysical explosion occurs when magnetic, gravitational, or thermonuclear energy is released on dynamical timescales. The dynamical timescale is typically the sound-crossing time for the underlying system \citep[][]{2012RSPTA.370..774W}. Besides SNe, numerous explosive phenomena are continuously occurring in the Universe, including solar and stellar flares, eruptive phenomena in accretion disks, thermonuclear combustion on the surfaces of WDs and neutron stars (NSs), violent magnetic reconnection in NSs, cosmic gamma-ray bursts (GRBs) and kilonovae (KNe).  Each of these explosions represents a different type and a different amount of energy released.

Several transient events are associated with compact object binary systems involving NSs and/or Black holes (BHs). These transient events include X-ray novae \citep[][]{1997ApJ...491..312C}, X-ray bursters \citep[][]{1993SSRv...62..223L}, and soft gamma-ray repeaters \citep[][]{2008A&ARv..15..225M}. Short-GRBs originating from the merger of NS--NS or NS--BH are another example of transient events involving a binary system of NS or/and BH \citep{2005ApJ...630L.165L,2017ApJ...848L..34M,2017ApJ...848L..13A,2017ApJ...850L...1L}. Additionally, KNe are a class of transient events that are also associated with such binary merger events \citep[][]{2017LRR....20....3M,2019LRR....23....1M}. 

In contrast to the compact object binary systems mentioned above, the binary systems composed of a WD orbiting a companion star form a class of variable objects known as cataclysmic variables (CVs). There are two mechanisms through which the outbursts of radiant energy in CVs occur. The mechanism involving an accretion disk results in a ``dwarf nova" \citep[][]{1971AcA....21...15S,1974PASJ...26..429O} while the other mechanism involving thermonuclear burning on the surface of WD results in ``classical novae and recurrent novae" \citep[][]{1974ApJS...28..247S,2010ApJS..187..275S, 2012BASI...40..393K}. The outbursts in the accretion disk resulting in a ``dwarf nova" is rather a dramatic event than an explosion as the energy released is not on a dynamic scale. SNe are much more powerful events than novae. A detailed review of SNe is presented in the next section.
   
\section{Supernovae}%\label{sn}

SNe are extreme catastrophic stellar explosions whose effects pervade the whole of astronomy \citep[][]{1985supe.book.....M}. SNe explosions are so bright that they can outshine their host galaxy. These cataclysmic transient events enrich the Universe by creating and spreading heavier chemical elements \citep[][]{1986A&A...154..279M,2003MNRAS.340..908K, 2007ApJ...670....1G}, trigger the star formations \citep[][]{1977ApJ...217..473H,1978prpl.conf..368H, 1992AJ....103.1788V, 2013ApJ...762...50C}, and are also responsible for the birth of several compact objects including, NSs and BHs \citep[][]{1975Mercu...4...16S,1995ApJ...443..717B,1996NuPhA.606..137W,1999Natur.401..142I,2001PhRvD..63g3011B}.

\subsection{History of supernovae observations in ancient and modern astronomy}

~~~~~$\bullet$~{\bf Ancient SNe observations:} According to \citet[][]{Joglekar2011OldestSW}, the oldest possible supernova (SN) record dates back to around 4600\,$\pm$\,2000 BC, observed by unknown Indian observers. The name of the possible recorded SN is ``HB9" \citep[][]{Joglekar2011OldestSW}. Following \citet[][]{1985supe.book.....M}, the SN in the year AD\,185 occurred in the Centaurus constellation and is the oldest ever recorded SN by humankind. Evidence indicates that two Roman chronicles also reference the SN that the Chinese recorded in AD\,185. Further, in the year AD\,386, there are recorded pieces of evidence of another SN observation by Chinese observers in the Southern Dipper of the Sagittarius constellation \citep[][]{1982ASIC...90..355C}.
In the year AD\,393, the Chinese observers recorded the appearance of another SN in the constellation of Scorpio \citep[][]{1982ASIC...90..355C,2020MNRAS.497.1419H}. The SN\,1006 appeared in AD\,1006 in the constellation of Lupus and is considered probably the brightest one. This SN was so far south and below the horizon to the northern European observers; thus, records of SN\,1006 are mainly found in Arabic, Japanese, and Chinese texts. Another very widely observed SN in ancient times is the SN\,1054. It appeared in the constellation of Taurus, and the crab nebula is the remnant of this event. This SN was so bright that it could cast shadows and was visible even in the daylight for a few days \citep[][]{1983Obs...103..106B,1999PASP..111..871C}. Only a few observational records are available for another SN\,1186 occurring in the constellation of Cassiopeia \citep[][]{1999PASP..111..871C,2010ASPC..438..347K}. The pieces of evidence of its observations are found in Japanese and Chinese chronicles \citep[][]{1985supe.book.....M}. Later, the Danish astronomer Tycho Brahe performed careful observations of a very bright SN that occurred in AD\,1572 in the constellation of Cassiopeia. The corresponding SN is popularly known as ``Tycho Brahe's supernova" \citep[][]{1952Natur.170..364H}. The most recent SN to be seen through the naked eye in our Galaxy is the ``Kepler's supernova" \citep[][]{1977ApJ...218..617V} occurred in the year AD\,1604 in the constellation of Ophiuchus.

$\bullet$~ {\bf Modern-day SN astronomy:} The word ``super-novae" was used by Fritz Zwicky for the first time in a CalTech lecture course in 1931. However, in the year 1933, the word ``super-novae" was used in public during the ``December American Physical Society" meeting. The modern name ``Supernova" first came into existence in the year of 1938 \citep[][]{1985supe.book.....M}. The first SNe detection survey was started by Fritz Zwicky in 1933. With the help of a 45-cm Schmidt telescope at Palomar observatory, their group could discover twelve new SNe within three years. To discover the new SNe, they compared the new photographic plate images with the reference images of extragalactic space \citep[][]{2005oghp.book.....H}. Since then, humankind has made enormous progress in space- and ground-based astronomy for SNe and other transients. Currently, we are living in the era of Hubble Space Telescope (HST)\footnote{https://www.nasa.gov/mission\_pages/hubble/main/index.html} and James Webb Space Telescope (JWST)\footnote{https://www.nasa.gov/mission\_pages/webb/main/index.html} \citep[][]{2023arXiv230404869G}.

\subsection{SN Classification based on the explosion mechanisms}%\label{sec:triggering_mechanism}
Based on their explosion mechanism, the SNe are broadly classified into two main categories:

{\bf 1) Thermonuclear SNe:} These are the class of SNe resulting from the explosive thermonuclear burning causing the explosion of WDs \citep[][]{1990RPPh...53.1467W, 1986ARA&A..24..205W,1997Sci...276.1378N,2000ARA&A..38..191H}. However, the actual progenitor that explodes as a thermonuclear SN is still debated. Currently, two progenitor scenarios have been proposed; first, the single degenerate system \citep[][]{Whelan1973} and second, the double degenerate system \citep[][]{1984ApJS...54..335I}. The single degenerate scenario consists of a Carbon-Oxygen WD with a non-degenerate WD. The non-degenerate companion star transfers mass to the WD through Roche-lobe overflow. As soon as the WD approaches the Chandrasekhar mass limit, the fusion of Carbon/Oxygen develops that quickly consumes the entire WD, resulting in a thermonuclear SN \citep[][]{1985supe.book.....M,2020ApJ...892..121K,2006MNRAS.368.1095H,2012ApJ...756L...4H,2018SSRv..214...67N}. The double degenerate scenario consists of either two WDs merging or colliding to result in a subsequent explosion \citep[][]{vanRossum2016,2018MNRAS.473.5352L,2018ApJ...868...90T}.
  
\begin{figure*}[!t]
  \includegraphics[]{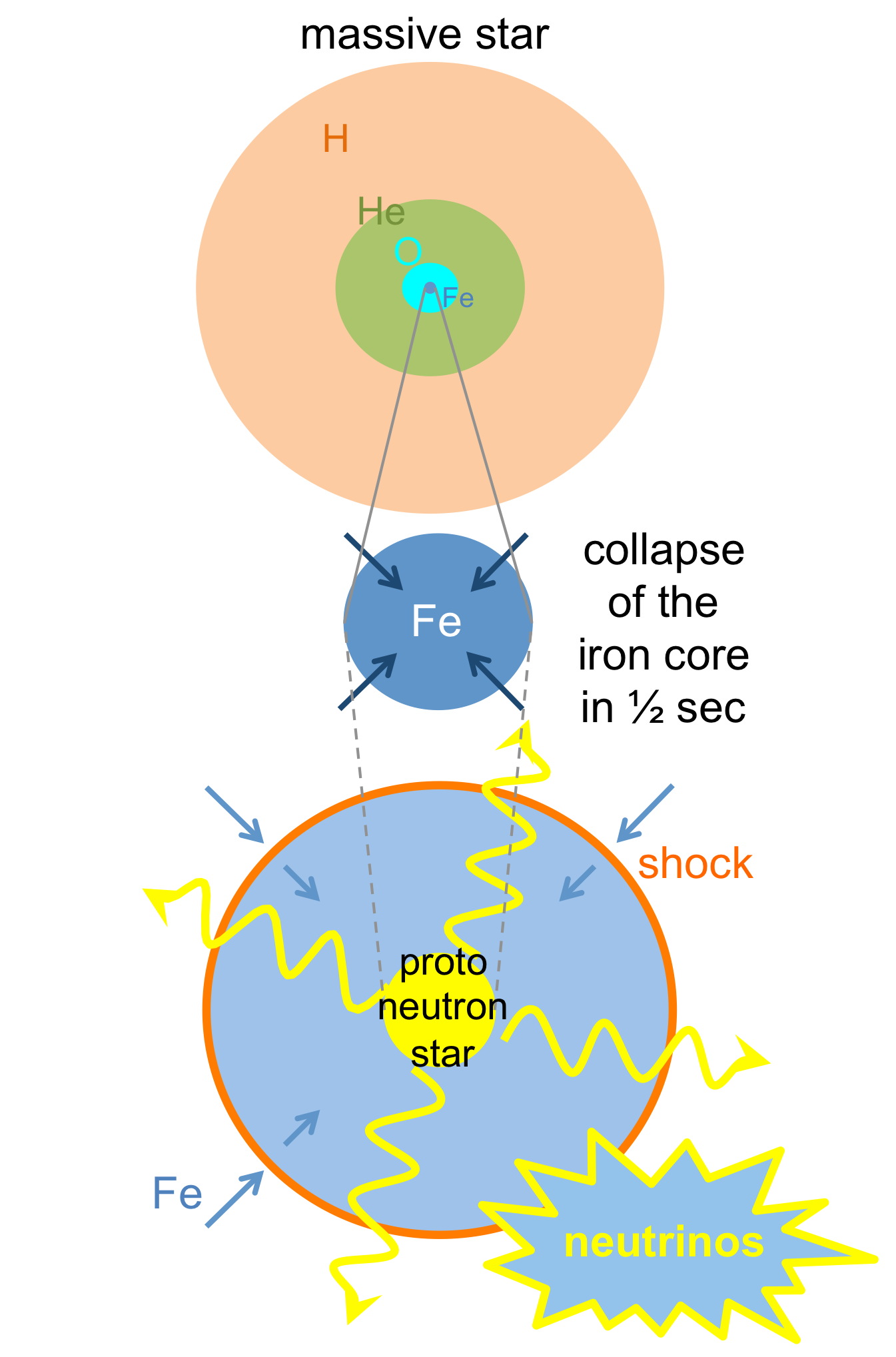}
  \caption{A visual representation of neutrino-driven explosion mechanism in CCSNe. The explosion mechanism relies on the absorption of neutrinos by the high-density post-shock gas. Figure credit: \citet[][]{2015PASA...32....9F}.}
  \label{fig:fig1}
\end{figure*}

{\bf 2) Core-collapse SNe (CCSNe):} Massive stars having the ZAMS mass of $\geqslant$8\,M$_{\odot}$ result into CCSNe as their terminating evolutionary stage \citep[][]{Garry2004,Woosley2005,2021Natur.589...29B,Smartt2015,2015PASA...32....9F}. Understanding the actual mechanism of the explosion for CCSNe is a long-standing problem. Theoretical studies spanning a period of more than half a century have been performed to understand these catastrophic phenomena, while much longer observational investigations are there. However, only recently has the mechanism of their explosion become a sharp focus. One of the most successful mechanisms in explaining the actual CCSNe process is the neutrino-driven explosion mechanism \citep[][]{2012ARNPS..62..407J,2006ApJS..163..335F, 2021Natur.589...29B,2017hsn..book.1095J}. Massive stars evolve for a few million years (10 - 40 million years, depending upon initial mass) of age. During a massive star's evolution, the star passes through successive burning phases of different elements. The massive star starts with burning Hydrogen in its core which continues for a major fraction of its entire evolution, and then it ignites Helium burning in the core. After that, successive burning of nuclear fusion reactions produces heavier nuclei, and the star develops an onion shell-like structure with heavier elements deeper in the star. During the last evolutionary stages, the massive star's core primarily comprises inert Iron. The inert Iron-core of the massive star keeps growing until it reaches the Chandrasekhar mass limit of nearly 1.5\,M$_{\odot}$, and becomes gravitationally unstable. Thus, after a few million years of evolution, the star's dense core implodes as there is hardly any radiation pressure support to the self-gravity of the core. Within less than a second, the core obtains nearly nucleon densities while the stellar material surrounding the core also follows the infall (Figure~\ref{fig:fig1}). The central temperature is so high that it dissociates the Iron nuclei into protons and neutrons. Also, the densities are so high that the protons transform into neutrons and produce neutrinos. At this stage, an NS is born.

The infalling material encounters sudden and abrupt deceleration due to the stiff NS surface, which launches a shock wave. The shock propagates outwards until it stalls at a certain distance from the core. The stalled shock causes the wobble in the NS. This wobbling motion causes the asymmetric distribution of matter, modulating the flux of neutrinos leaking from the NS. This asymmetric wobbling motion of dense matter is capable of deforming the space-time fabric, which could be detected in the form of Gravitational-waves (GW) \citep[][]{2020PhRvD.102b3027M,2019ApJ...876L...9R,2018ApJ...861...10M,2017MNRAS.468.2032A}.

 The wobbling shock starts rotating at the expense of the NS, which rotates in the opposite direction. The abundant densities are high enough to capture a few of the leaking neutrinos. The shock begins to expand in the direction where the matter below the shock catches the highest number of neutrinos. This is the decisive moment that marks the onset of the explosion. However, the shock will take some time to pass through the concentric envelopes of different elements and finally come out of the surface to mark an explosion. Meanwhile, the GWs and neutrinos have already propagated way ahead of the shock wave, and thus they are the first signals to come out of the surface of the collapsing star. After neutrino and GW signals, the shock break-out from the star's surface provides the first observational electromagnetic signature of the collapse.

\begin{figure*}[!t]
  \includegraphics[height=10cm,width=\columnwidth,]{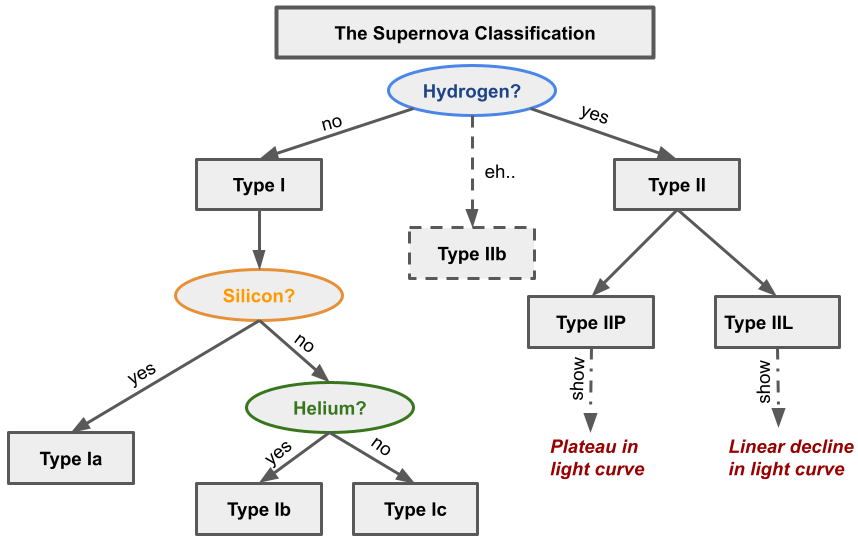}
  \caption{Classification of SNe primarily based on the observational features in their spectra and light curves using the results of (among others) \citet[][]{Schlegel1990}, \citet[][]{Filippenko1997}, and \citet[][]{Gal-yam16}.}
  \label{fig:fig2}
\end{figure*}

\subsection{SN Classification on the basis of observational properties}%\label{sec:taxonomy}

Based on observational behaviour, SNe are primarily classified into two categories; First, H-rich Type II SNe, and Second, H-deficient Type I SNe. Figure~\ref{fig:fig2} shows the various subclasses of Type II and Type I SNe. Figure~\ref{fig:fig_spec} shows the classification of several types of SNe explicitly based on their spectroscopic observational features. Extensive reviews are provided in (among others) \citet[][]{Filippenko1997} and \citet[][]{Gal-yam16} on SNe classification. A brief of Type I and Type II SNe is provided below:

\begin{figure*}[!t]
  \includegraphics[height=10cm,width=\columnwidth,]{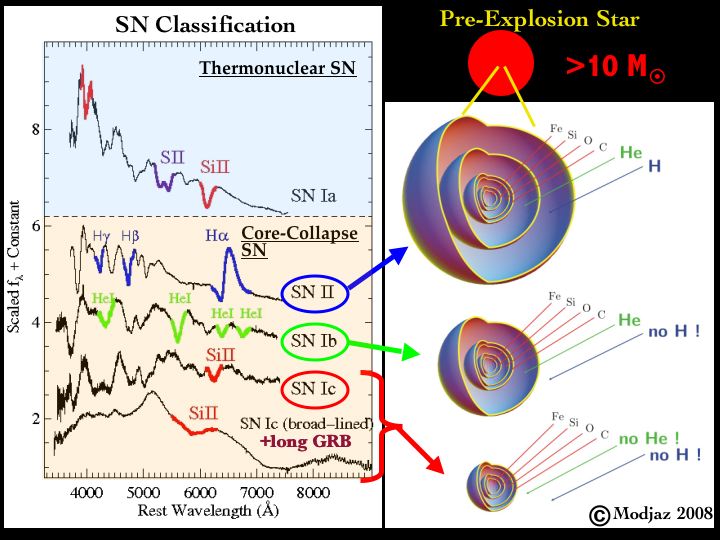}
  \caption{Spectra of several types of SNe (Left-hand panel). The thermonuclear SN Type Ia spectrum displays the strong features of S~II and Si~II and lacks any prominent H- or He-features. The Type II SN spectrum displays prominent features of hydrogen. Type Ib SN spectrum is well decorated with unambiguous He-features. Type Ic SNe spectra lack any prominent H- or He-features. Among CCSNe, the presence or absence of H- or He-features in the spectra depends upon the level of stripping their progenitor stars have gone through (Right-hand panel). The progenitors of Type II SNe suffer minimal stripping and retain most of their H-envelope; thus, they display prominent H-features in their spectra. Unlike Type II SNe, the progenitors of Type Ib SNe lose almost all of their H-envelopes through strong stellar winds or binary interaction; thus, their spectra lack any prominent H-feature but display strong He-features. Progenitors of Type Ic SNe suffer tremendous mass losses, and they cannot retain their outer H- as well as He-envelopes. Thus the spectra of Type Ic SNe are devoid of any prominent H- or He-features. Figure credit: \href{http://user.astro.columbia.edu/~mmodjaz/research.html}{Prof. Maryam Modjaz}.}
  \label{fig:fig_spec}
\end{figure*}

\begin{figure*}[!t]
  \includegraphics[width=\columnwidth]{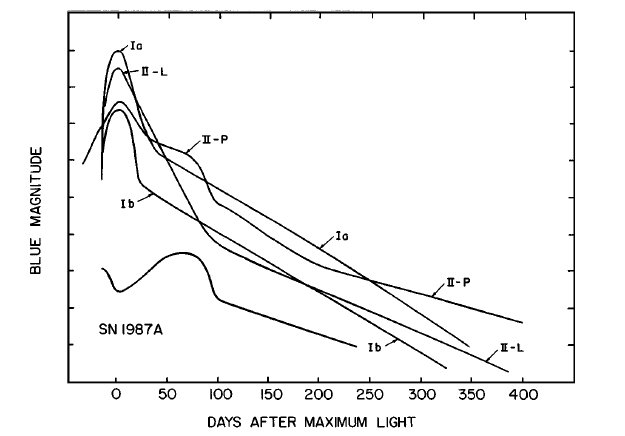}
  \caption{The light curves of various classes of SNe. The light curve of Type Ib SN represents the average of the light curves of Type Ib and Type Ic SNe. Figure credit: \citet[][]{Filippenko1997}.}
  \label{fig:fig3}
\end{figure*}
{\bf 1) Type I SNe:} These SNe lack prominent H-features in their spectra. Type I SNe display a wide range of spectroscopic features that are used to further sub-divide them into various categories; Type Ia SNe display strong Silicon features in their spectra, Type Ib SNe lack prominent Silicon features, but strong He-features dominate their spectra, Type Ic SNe neither display strong Silicon features nor prominent He-features. Rather the spectra are dominated by the prominent features of heavier elements/ions, including Calcium~II near-IR triplet, Oxygen~I $\lambda$\,7777 absorptions, and Ca~II H\&K absorption.

There exist a few more classes of interacting SNe among Type I. Type Ibn \citep[][]{2009MNRAS.400..866C,2015MNRAS.449.1921P, 2022ApJ...927...25M,2016ApJ...824..100M,2016MNRAS.461.3057S,2022ApJ...932...84M} and Type Icn \citep[][]{2022ApJ...932...84M,2022ApJ...927..180P} are the rare classes of Type Ib and Type Ic SNe respectively, which show narrow emission interaction features of Carbon and/or Oxygen in their spectra. Type Ic-BL is another class of SNe belonging to Type Ic SNe displaying broad absorption features in the spectra \citep[][]{2008MNRAS.383.1485V,2013MNRAS.432.2463M,2017ApJ...839...85C}. Type Ic-BL SNe are thought to be associated with Long-GRBs \citep[among many others,][]{2013MNRAS.432.2463M}. Intense investigations have been performed to understand the possible connections of GRBs and CCSNe (e.g.,\citealt[][]{1998ApJ...504L..87W,1999Natur.401..453B,2001BSAO...51...38S,2017AdAst2017E...5C,2021A&A...646A..50H,2022NewA...9701889K}).

{\bf 2) Type II SNe:} These SNe are identified by the strong presence of H-features in their spectra. They exhibit a wide range of spectroscopic and photometric behaviour. Type II SNe are further subdivided into two main categories, Type IIP and Type IIL, depending upon the shape of their light curves \citep[][]{Barbon1979,1985AJ.....90.2303D}. Type IIP SNe display a plateau in their light curves, and Type IIL SNe show a linear decline in their light curves after maximum brightness (Figure~\ref{fig:fig3}).

Besides these two main subcategories of Type II SNe, a few more subclasses also exist; Type IIn SNe are the subclass of Type II SNe which display narrow emission lines in their spectra \citep[][]{Schlegel1990,1996AJ....111.1271V,2000ApJ...536..239L,2014ApJ...790L..16M,2017hsn..book..403S,2021MNRAS.506.4715R}, Type IIb SNe initially display strong H-feature and in a few weeks their spectra are gradually dominated by He-features. Thus, Type IIb SNe are thought to be the link between H-deficient Type I and H-rich Type II SNe \citep[][]{1993Natur.365..232S}. Beyond Type I and Type II SNe, there exists a class of superluminous SNe (SLSNe) having luminosities about 10--100 times greater than above mentioned canonical SNe \citep[][]{2021A&G....62.5.34N}.

\section{Powering Mechanisms of Core-Collapse Supernovae}%\label{sec:intro_slsne}
Several models have been proposed as the possible powering mechanism for CCSNe. In this section, we provide short explanations of different powering mechanisms.
\begin{figure*}[!t]
  \includegraphics[width=\columnwidth]{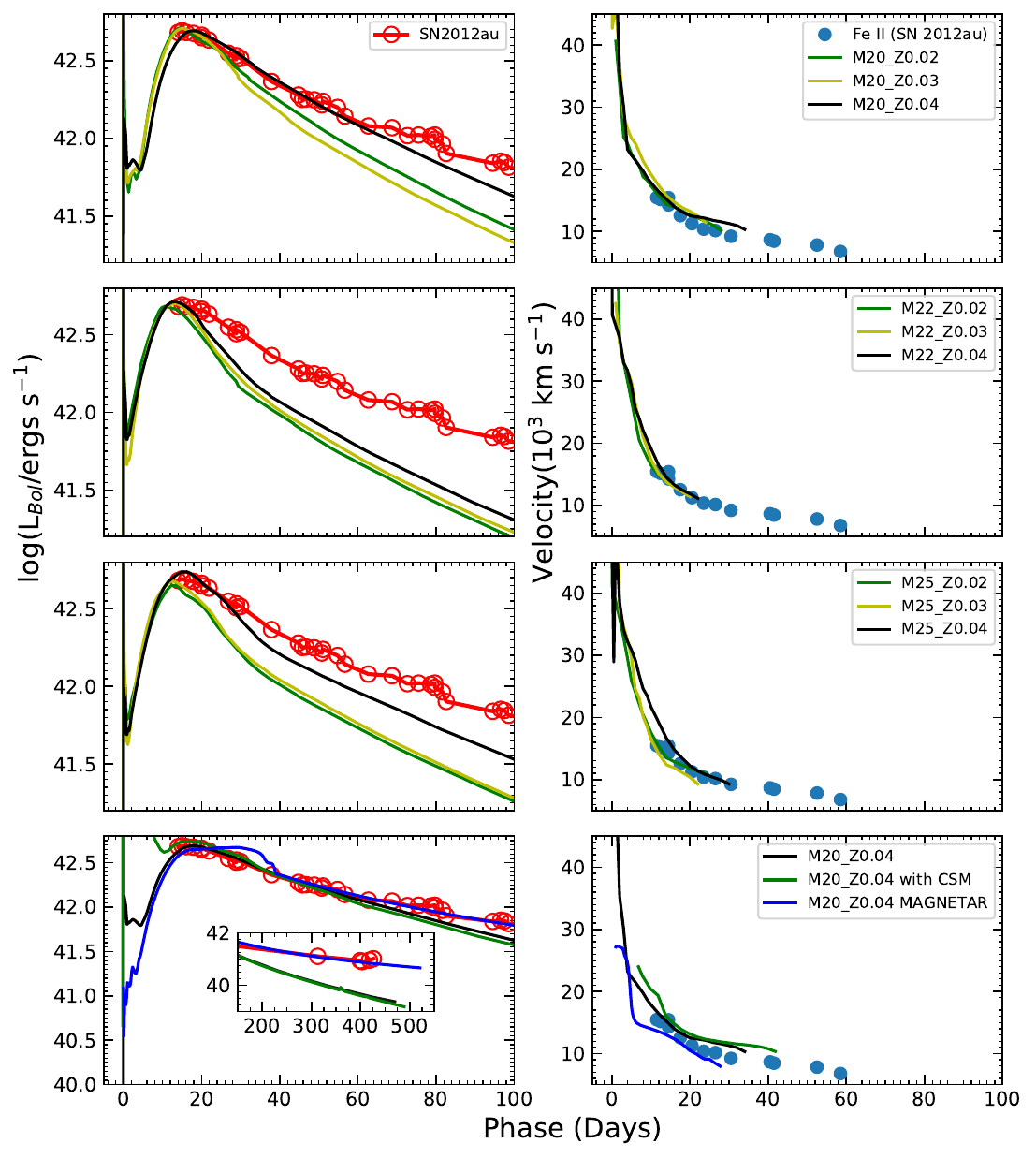}
  \caption{The results of the hydrodynamic modelling of SN~2012au, assuming the ``Radioactive decay model", ``Circumstellar interaction model", and the ``Magnetar-driven model". The ``Magnetar-driven model" seems best to match the observed quasi-bolometric light curve of SN~2012au. Figure is adapted from \citet[][]{Pandey2021}. }
  \label{fig:fig4}
\end{figure*}
\subsection{Radioactive decay model}
The radioactive decay of $^{56}$Ni and $^{56}$Co has been used to explain the observed light curves of SNe \citep[][]{1979ApJ...230L..37A,Arnett1980,Arnett1982,1997MNRAS.284..151M,2003A&A...404.1077E,2006A&A...450..241S}. In the process of the radioactive decay, $^{56}$Ni decays to $^{56}$Co which finally decays to stable $^{56}$Fe \citep[][]{Nadyozhin1994}. The deposition of gamma-rays resulting from the radioactive decay of $^{56}$Ni and $^{56}$Co are expected to thermalise in the homologously expanding SN ejecta and, after that, radiatively released to explain the light curves of several types of SNe including, Type Ia, Type Ib/c and Type II also \citep[e.g., as found by (among many others),][]{2003MNRAS.340..375P,Aryan2021,2022MNRAS.517.1750A,2003A&A...397..115V,Zheng2022}. 
The generalised mathematical expression governing the form of the output luminosity light curve is provided in \citet[][]{2008MNRAS.383.1485V,2009ApJ...704.1251C,2012ApJ...746..121C}.

\subsection{Magnetar-driven model}

The ``Magnetar-driven model" has been very successful in explaining the light curves of SLSNe \citep[][]{2012MNRAS.426L..76D,2021A&G....62.5.34N}. In this model, the energy input by the spin-down of a magnetar sitting in the centre of the SN ejecta governs the output luminosity light curve of the corresponding SN \citep[][]{Ostriker1971,Arnett1989,Maeda2007,Metzger2015,Kasen2010,Woosley2010}.  
The mathematical expression governing the output luminosity light curve is given in \citet[][]{2012ApJ...746..121C} and \citet[][]{Chatzopoulous2013}. Beyond SLSNe, a few other Type I and Type II SNe have also been thought to be powered by a ``Magnetar-driven" model. In \citet[][]{Maeda2007}, the authors have suggested the ``Magnetar-driven" model as the possible powering mechanism for a peculiar Type Ib SN~2005bf. In another recent work by \citep[][]{Pandey2021}, the photometric and spectroscopic properties combined with the hydrodynamic modelling of the exceptionally bright Type Ib SN~2012au indicated a "Magnetar-driven model" for the powering of the light curves as shown in the Figure~\ref{fig:fig4}. Many studies demonstrate the ``Magnetar-driven model" as the possible powering mechanism for several classes of CCSNe \citep[e.g.,][]{2019A&A...621A..64T,2017MNRAS.472..224S,2017ApJ...839...85C,2016ApJ...831...41W}. 

\subsection{Circumstellar interaction model}%\label{intro_grbs}

The ``Circumstellar interaction model" is accepted as one of the dominant powering mechanisms for some SLSNe and other interacting SNe which display circumstellar interaction features. Several Type Ibn SNe have shown the interaction features of their ejecta with the circumstellar material (CSM) \citep[e.g., among others,][]{2017A&A...602A..93K,2017MNRAS.471.4381S,2018MNRAS.475.2344V,2020MNRAS.491.6000S}. Unambiguous features of SN ejecta interacting with the CSM have also been identified in recent (among few other) Type Icn SN~2021csp \citep[][]{2022ApJ...927..180P}, SN~2022ann \citep[][]{2022arXiv221105134D}, and SN~2021ckj \citep[][]{2023A&A...673A..27N}. In some cases, the SNe progenitors are surrounded by the dense CSM. The CSM is produced due to the continuous or intermittent pre-explosion mass losses from the progenitor. Upon the occurrence of the explosion, the SN ejecta may vigorously interact with the surrounding CSM \citep[][]{Chevalier1982,Chevalier1994} resulting in the creation of a dual shock configuration; a forward shock moving in the CSM and a reverse shock that moves back into the SN ejecta. The kinetic energy from these shocks is transferred to the material, which is then released as radiation to power the light curves. The mathematical expression for the output luminosity light curve is given in \citet[][]{2012ApJ...746..121C} and \citet[][]{Chatzopoulous2013}.

\subsection{Hybrid models}%\label{sec:Anatomy_GRBs}
In some cases, the individual models mentioned above fail to reproduce the observed light curves of SNe, and one has to consider the contributions from more than one powering mechanism. Such a model, constituted out of two or more powering mechanisms to explain the light curves, is known as the ``hybrid model" \citep[please see,][]{2012ApJ...746..121C,Chatzopoulous2013,2019ApJ...874...68C,2018SSRv..214...59M}. Several combinations of two or more models are possible:

$\bullet$~ \citet[][]{2012ApJ...746..121C,Chatzopoulous2013} presented a ``hybrid model" where the final luminosity light curve has the combined contribution from the ``Radioactive decay model" and ``Circumstellar interaction model". Using this ``hybrid model", they attempted to explain the observed luminosity light curves of several SLSNe and Type IIn SNe.

$\bullet$~ In another scenario, the authors of \citet[][]{2017A&A...602A...9C} and \citet[][]{2017MNRAS.468.4642I} have constituted the ``hybrid model" by taking into account the collective contribution from ``Magnetar-driven model" and the ``Circumstellar interaction model" to explain the luminosity light curves of several SLSNe.

$\bullet$~ A collective contribution from the ``Radioactive decay model" and the ``Magnetar-driven model" has been utilised to explain the light curves of several SLSNe \citep[][]{2016ApJ...817L...8B,2019ApJ...872...90B}. 

$\bullet$~ To explain the light curve of a peculiar SLSN iPTF13ehe, the authors of \citet[][]{2016ApJ...828...87W} formulated a ``hybrid model" by taking collective contributions from three mechanisms, namely, ``Magnetar-driven model", ``Radioactive decay model", and the ``Circumstellar interaction model".

\begin{figure*}
\centering
  \begin{tabular}{cc}
  
    {\includegraphics[height=6.7cm,width=0.49\columnwidth]{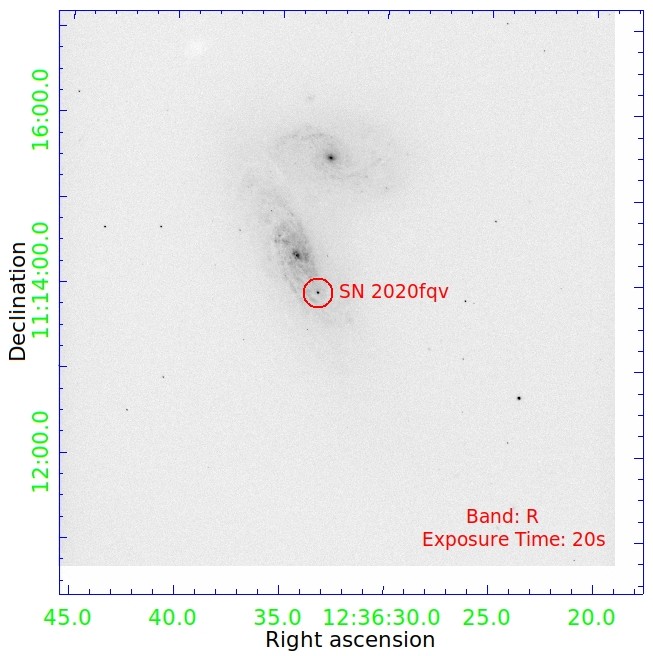}}
    {\includegraphics[height=6.7cm,width=0.49\columnwidth]{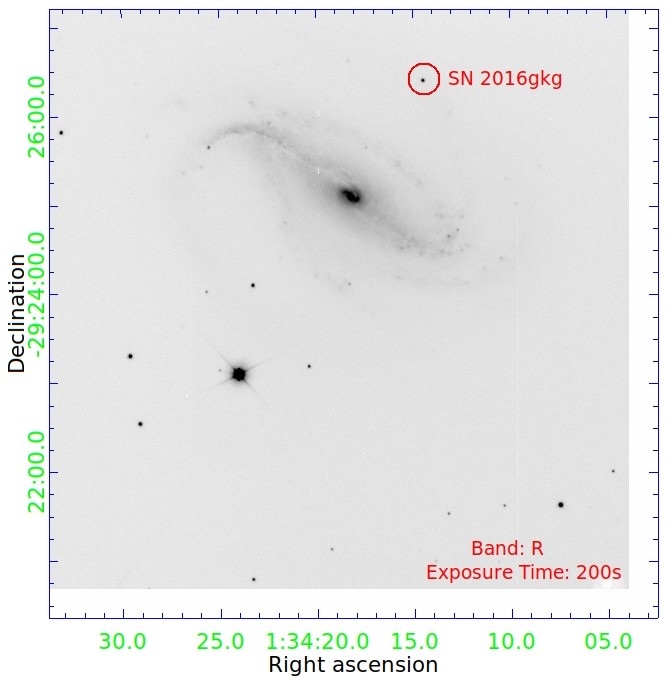}}\\
    {\includegraphics[height=6.7cm,width=0.49\columnwidth]{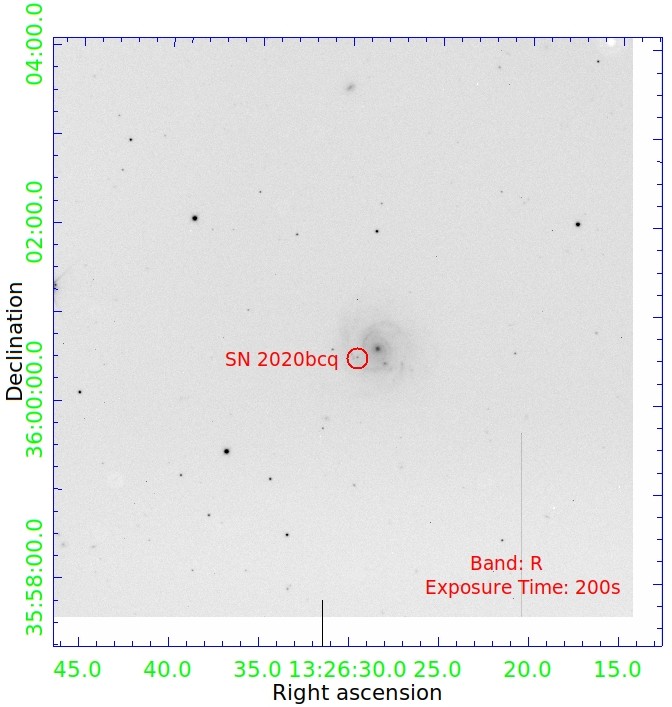}}
    {\includegraphics[height=6.7cm,width=0.49\columnwidth]{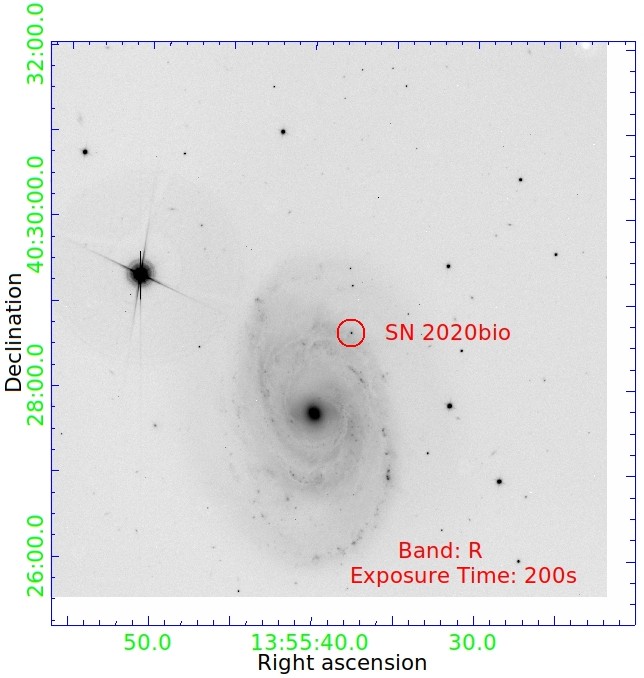}}\\
    {\includegraphics[height=6.7cm,width=0.49\columnwidth]{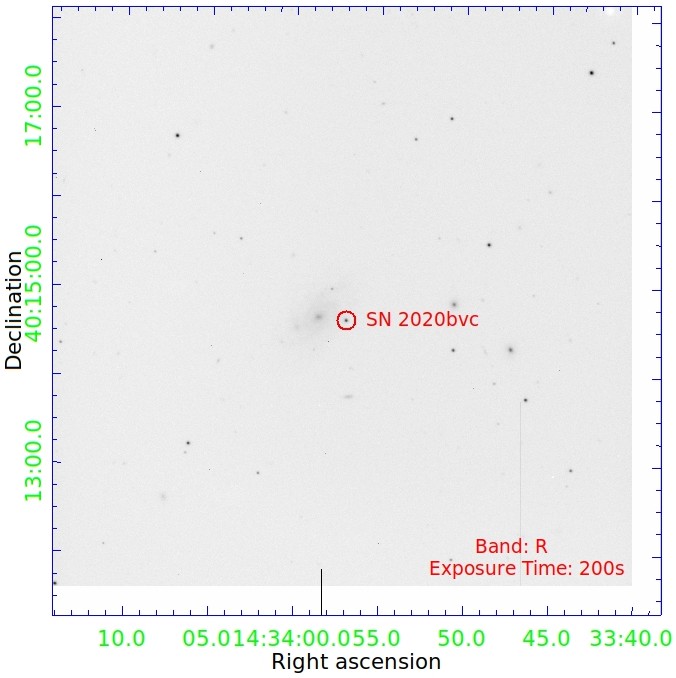}}
    {\includegraphics[height=6.7cm,width=0.49\columnwidth]{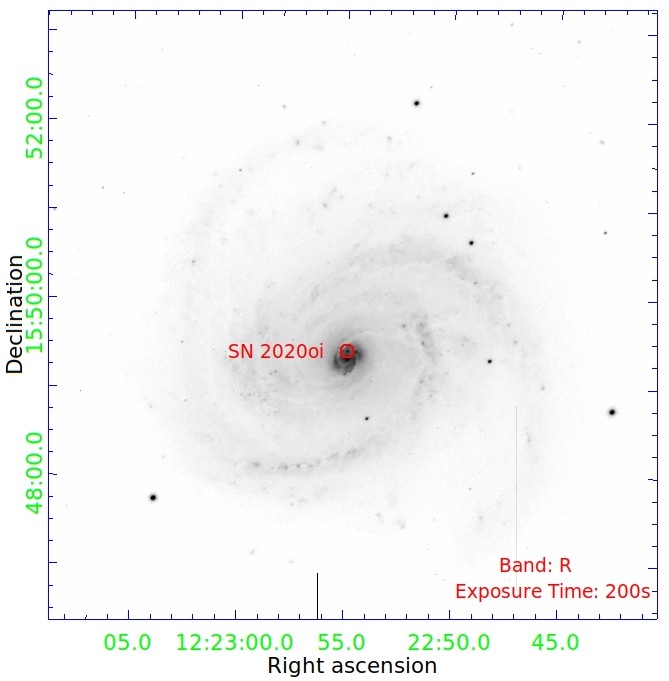}}
  \end{tabular}
  \caption{The first row shows the $R$-band images of SN~2020fqv and SN~2016gkg obtained using the 4K$\times$4K CCD imager \citep[][]{Pandey2018,2022JApA...43...27K} mounted on the axial port of 3.6 m DOT (details of 3.6 m DOT are provided in Chapter~\ref{Ch:2}). The remaining rows show the $R$-band images of SN~2020bcq, SN2020bio, SN2020bvc, and SN~2020oi using the 4K$\times$4K CCD imager.}
  \label{fig:figB}
\end{figure*}

\begin{figure*}
\centering
  \begin{tabular}{cc}
    {\includegraphics[height=8cm,width=\columnwidth]{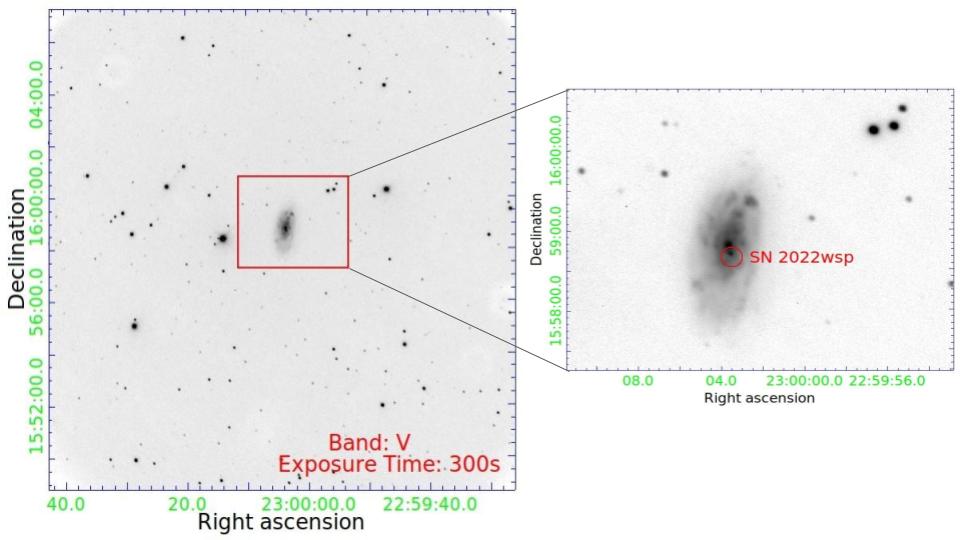}}\\
%    {\includegraphics[height=7cm,width=0.49\columnwidth]{SN2020bio_DOT.jpg}}\\
%    {\includegraphics[height=7cm,width=0.49\columnwidth]{SN2020bvc_DOT.jpg}}
    {\includegraphics[height=8cm,width=\columnwidth]{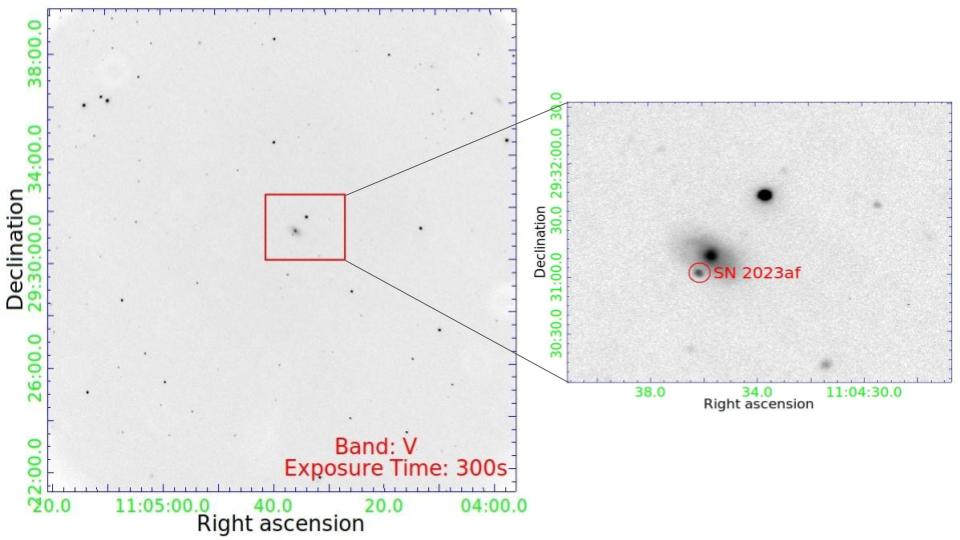}}
  \end{tabular}
  \caption{$V$-band images of SN~2020wsp and SN~2023af using the 2K$\times$2K CCD camera mounted at the axial port of 1.3 m DFOT (details of 1.3 m\,DFOT are provided in Chapter~\ref{Ch:2}).}
  \label{fig:figC}
\end{figure*}

\begin{figure*}
\centering
  \begin{tabular}{cc}
    {\includegraphics[height=8cm,width=0.5\columnwidth]{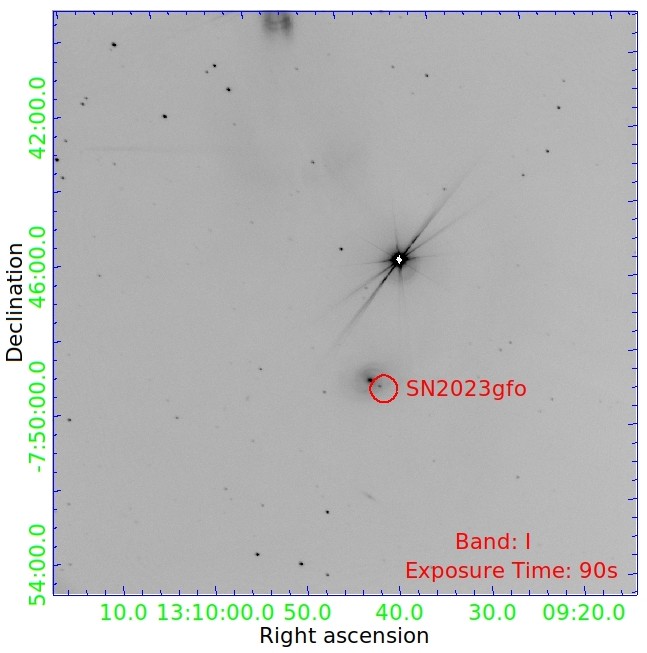}}
%    {\includegraphics[height=7cm,width=0.49\columnwidth]{SN2020bio_DOT.jpg}}\\
%    {\includegraphics[height=7cm,width=0.49\columnwidth]{SN2020bvc_DOT.jpg}}
    {\includegraphics[height=8cm,width=0.5\columnwidth]{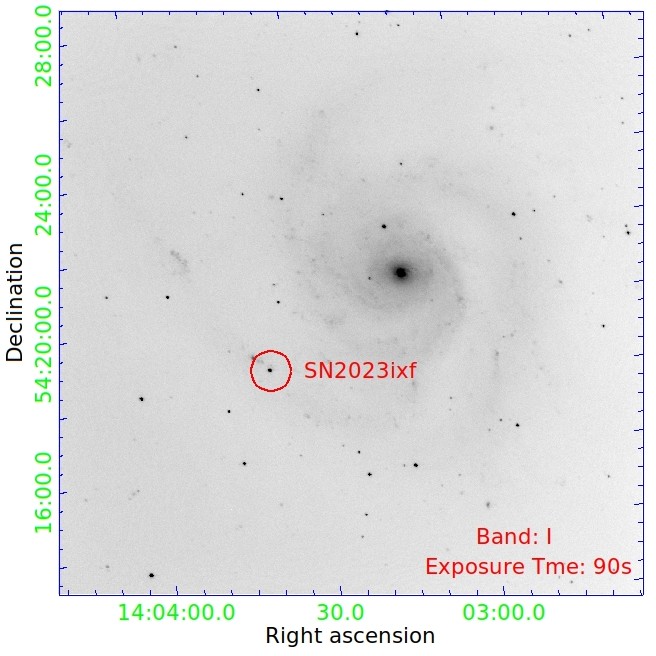}}
  \end{tabular}
  \caption{$I$-band images of SN~2023gfo and SN~2023ixf obtained using the 4K$\times$4K CCD camera mounted on the axial port of 1.04 m ST (details of 1.04 m\,ST are provided in Chapter~\ref{Ch:2}).}
  \label{fig:figD}
\end{figure*}

\section{Possible Progenitors and Ambient Environments of Core-Collapse Supernovae}

Understanding the physical properties of possible progenitors of CCSNe and their ambient surroundings is the prime focus of current research among the SN community. Researchers and scientists primarily depend on observations and observation-complemented simulations to unveil the nature of the possible progenitors and the surrounding media. Figures~\ref{fig:figB}, \ref{fig:figC}, and \ref{fig:figD} show the diversity in host galaxies for the occurrence of several CCSNe.

\subsection{Observational constraints on the   progenitors and their surroundings}   
Knowledge of the possible progenitors of CCSNe is among the most fundamental aspects of understanding these catastrophic explosions. The most efficient way to investigate the likely progenitors of CCSNe and their physical properties is by directly detecting progenitors in high-resolution pre-explosion images from space- and ground-based telescopes. Such a detection would provide direct evidence of the mass, luminosity, temperature, and other physical properties of the underlying progenitor \citep[e.g.,][]{2003PASP..115.1289V,2004Sci...303..499S,2006ApJ...641.1060L}.
 
$\bullet$~ {\bf Type IIP and Type IIL CCSNe}

With the help of the direct detections of red supergiant (RSG) progenitors of Type IIP SNe and the most massive WD progenitors in pre-explosion images, a star requires to have a mass of at least 8$\pm$1\,M$_{\odot}$ to finally terminate its life as a CCSN \citep[][]{Smartt2009}. Although the direct detection of progenitors in the pre-explosion images is the most efficient way to understand them, only a few such detections have been obtained due to the uncertainties associated with the temporal and spatial position of the occurrence of an SN. It is impossible to predict the actual location and timing of the occurrence of an SN. As per the review provided in \citep[][]{Smartt2015}, there are only 18 detections of precursor objects and 27 upper limits in archival images from the ground- and space-based telescopes. Among these, most of the detections of progenitor stars are for Type IIP, IIL, or IIb SNe. Additionally, only one detection for Type Ib SN progenitor is available. Beyond these detections, 14 upper limits are also there for Type Ibc CCSNe.

\begin{figure*}[!t]
  \includegraphics[width=\columnwidth]{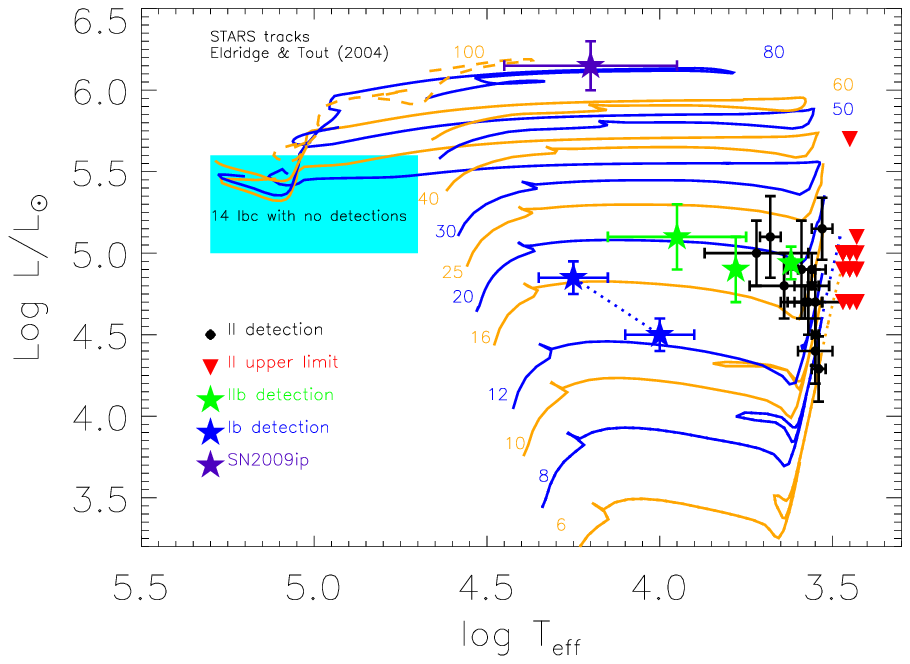}
  \caption{The positions of all the detected Type II and Type Ibc SNe progenitors. Upper limits to the detections are also shown. Two positions connected by dashed lines correspond to two progenitors presented by \citet[][]{Bersten2014} and \citet[][]{2015MNRAS.446.2689E}.  There is a debate over SN~2009ip to be a genuine CCSN, although it had a very massive progenitor \citep[][]{2010AJ....139.1451S,2011ApJ...732...32F,2011MNRAS.415.2020S,2013MNRAS.430.1801M,2013ApJ...763L..27P,2013MNRAS.433.1312F}. The stellar evolutionary tracks in this figure are from \citet[][]{2004MNRAS.353...87E}. Figure credit: \citet[][]{Smartt2015}.}
  \label{fig:fig5}
\end{figure*} 

In the volume-limited calculations, the frequency of the occurrence of Type IIP SNe is highest among several Types of SNe \citep[][]{Smartt2009b,2007ApJ...661.1013L,2008ApJ...673..999P,1999A&A...351..459C}; therefore, Type IIP SNe progenitor population is currently the best understood observationally with the help of available direct detections and upper limits. A few Type IIP SNe having clear detections of their progenitors are discussed here. The first example is SN~2003gd from a nearby galaxy, M74. Through the archival images of M74 imaged some 6--9 months before the explosion utilising the Hubble Space Telescope and the Gemini North telescope, researchers could identify the possible progenitor of SN~2003gd \citep[][]{2003PASP..115.1289V}. The identified progenitor has a $V$-band magnitude of 25.8$\pm$0.15 \citep[][]{Smartt2009}. Further investigations by \citep[][]{2004Sci...303..499S} reveal that the possible progenitor of SN~2003gd is an RSG. The initial mass of the identified progenitor likely lies in the range of  8$^{+4}_{-2}$\,M$_{\odot}$. The metallicity at the SN~2003gd explosion site was probably near solar \citep[][]{Smartt2009}. 
       
The following Type IIP SN with progenitor detection in pre-explosion images is SN~2005cs, which occurred in the whirlpool galaxy, M51. The investigations of HST images by \citet[][]{2005MNRAS.364L..33M} and \citet[][]{2006ApJ...641.1060L} constrain the progenitor to be an RSG again having a mass of 8$\pm$2\,M$_{\odot}$. SN~2008bk is another Type IIP SN with its progenitor detection in pre-explosion images having a significant confidence level. \citet[][]{2008ApJ...688L..91M} could constrain the progenitor to be an RSG with a mass of 8.5$\pm$1.0\,M$_{\odot}$. The metallicity of the host galaxy at the SN~2008bk explosion site lies between the metallicity of Small Magellanic Cloud (SMC) and Large Magellanic Cloud (LMC) \citep []{Smartt2009}.

Unlike the vast majority of CCSNe in the local Universe exploding in the star-forming regions \citep[][]{2003PASP..115....1V}, SN~2004dj and SN~2004am are the examples of CCSNe originating in star clusters \citep[][]{Smartt2009b}. The SN~2004dj occurred in a well-studied star cluster Sandage 96 \citep[][]{2004ApJ...615L.113M}. The authors of \citet[][]{2004ApJ...615L.113M} estimated the age of the cluster to be nearly 14\,Myr and the corresponding initial mass of the identified progenitor to be around 15\,M$_{\odot}$. In another work, \citet[][]{2005ApJ...626L..89W} estimated an initial progenitor mass of 12\,M$_{\odot}$. A main-sequence mass in the range of 12--20\,M$_{\odot}$ is inferred for the identified progenitor by the calculations of \citet[][]{2009ApJ...695..619V}. 
Like SN~2004dj, the SN~2004am exploded in another super star cluster L in M82. The identified progenitor is estimated to have an initial mass of 12$^{+7}_{-3}$\,M$_{\odot}$ \citep[][]{Smartt2009} derived from the star cluster age of 18$^{+17}_{-8}$\,Myr \citep[][]{2008A&A...486..165L}. A list of Type II CCSNe with secure progenitor detections and upper limits on detection are given in  \citet[][]{Smartt2015} in Table 1 and Table 2 there.

There are detections of progenitors for three more Type IIP SNe, namely, SN~1999ev, SN~2004A, and SN~2004et, but the significance level of their detections could be better. Following \citet[][]{2005MNRAS.360..288M}, if the detected progenitor for SN~1999ev is an RSG then the corresponding mass of the progenitor is around 15--18\,M$_{\odot}$. For SN~2004A, \citet[][]{2006MNRAS.369.1303H} suggests the initial mass of an RSG progenitor to be 9$^{+3}_{-2}$\,M$_{\odot}$. Further, a yellow supergiant star of initial mass around 15\,M$_{\odot}$ is claimed to be the possible progenitor of SN~2004et by \citet[][]{2005PASP..117..121L}, but later \citet[][]{Smartt2009b} questioned its identification as the detected object was visible at the same luminosity even after around four years later since the SN explosion. Finally, \citet[][]{Smartt2009b} suggested a supergiant star having an initial mass of 9$^{+5}_{-1}$\,M$_{\odot}$ as the detected progenitor of SN~2004et. A few more detections of Type IIP SNe progenitors were claimed in pre-explosion images, but those detections are debatable \citep[please see,][]{2007ApJ...661.1013L,2008PASP..120.1259L,Smartt2009b}.

Following \citet[][]{Smartt2009b}, the occurrence frequency of Type IIL SNe is lowest. These SNe have very short or no plateau in their light curves, probably due to low-mass H-envelope not being capable of sustaining a longer duration of recombination \citep[][]{Smartt2009}. A Type IIL SN progenitor could lose mass through strong stellar winds or binary interaction, resulting in a corresponding low-mass H-envelope. A higher mass progenitor than the Type IIP SNe progenitor is expected if the mass loss occurs through stellar winds \citep[][]{Smartt2009}. There are two incidences of progenitor detections for Type IIL SNe as tabulated by \citet[][]{Smartt2015}. The authors of \citet[][]{2011ApJ...742....6E} could constrain the properties of the progenitor of SN~2009hd using the pre-explosion HST images. The estimated magnitude and colour limits indicate a luminous RSG star with the possibility that the progenitor star could have been rather yellow than red. The investigations by \citet[][]{2010ApJ...714L.280F} and \citet[][]{2010ApJ...714L.254E} utilising the archival HST images claim the progenitor detection for SN~2009kr. The authors of \citet[][]{2010ApJ...714L.280F} claim a yellow supergiant (YSG) star having a mass of 15$^{+5}_{-4}$\,M$_{\odot}$ while the estimated initial mass of the YSG progenitor by \citet[][]{2010ApJ...714L.254E} is 18--24\,M$_{\odot}$. 

Last but not least, the recent discovery of one of the nearest H-rich Type II SN~2023ixf in the pinwheel galaxy (M101) has opened new avenues to understanding the possible progenitors of H-rich CCSNe. The analysis by \citet[][]{2023arXiv230604722K} indicate towards an RSG progenitor of 11\,M$_{\odot}$ for SN~2023ixf.

$\bullet$~ {\bf Type IIb CCSNe}

Type IIb SNe begin by displaying unambiguous signatures of H-features in their spectra and later evolve to exhibit strong He-features along with weaker H-features \citep[][]{Filippenko1997}. For Type IIb SNe also, a few  cases have been reported \citep[][]{Smartt2015} where the progenitor stars are detected in pre-explosion images:

(a) \citet[][]{Crockett2008} identified the progenitor star for SN~2008ax in the pre-explosion images from HST archival images. Based on their detections, they proposed two possible scenarios for the progenitor; First, a single massive star that has stripped off most of its H-envelope utilising radiatively driven mass-loss mechanisms. Finally, the exploding progenitor is an He-rich Wolf-Rayet (WR) star retaining only a tenuous H-envelope. Second, an interacting binary progenitor system where the mass loss primarily due to binary interaction is responsible for the production of stripped progenitor.

(b) \citet[][]{Maund2011} and \citet[][]{Van2011} have claimed to identify the progenitor of SN~2011dh in pre-explosion archival images obtained through HST. By comparing the position of the detected progenitor on the HR diagram with several stellar evolution tracks, \citet[][]{Maund2011} propose a single YSG star at the end of core Carbon-burning having an initial mass of 13$\pm$3\,M$_{\odot}$ as the possible progenitor for SN~2011dh. While comparing the position of the detected object on HR diagram with several stellar evolution tracks, \citet[][]{Van2011} propose an initial mass in the range of 17--19\,M$_{\odot}$.

(c) SN~2013df is another Type IIb SN with its progenitor detected in the pre-explosion images with a significant confidence level. \citet[][]{Van2014} have claimed to confirm the detection of progenitor star of SN~2013df in pre-explosion archival images from HST obtained around 14 years before the actual SN explosion. The identified progenitor is a YSG star with an initial mass of 13--17\,M$_{\odot}$. The positions of detected progenitors of Type II CCSNe on the HR diagram are shown in Figure~\ref{fig:fig5}. The upper limits are also shown there.

$\bullet$~ {\bf Type Ibc CCSNe}

Following \citet[][]{Smartt2015}, only one case of progenitor identification has been reported for a Type Ibc SN. For the site of SN iPTF13bvn, the HST pre-explosion images indicate the presence of a blue star. But, \citet[][]{Cao2013} indicated that the detected progenitor was not within the 1$\sigma$ error circle in their alignment of a ground-based image having significantly high resolution. Still, the detection could not be strongly rejected as the detection was well within 3$\sigma$. Later, utilising the alignment with HST images, \citet[][]{2015MNRAS.446.2689E} supported the argument mentioned above and suggested it to be the first-ever detection of a progenitor for Type Ibc SN (particularly Type Ib). Further analysis by \citet[][]{Cao2013} and \citet[][]{Groh2013} proposed that the identified progenitor probably was a single massive with an initial mass of around 30\,M$_{\odot}$ that later evolved into a WN star. Alternatively, binary progenitor scenarios with initial masses of 20 + 19 M$_{\odot}$ or 10  + 8 M$_{\odot}$ were also proposed \citep[][]{Bersten2014,2015MNRAS.446.2689E}. Figure~\ref{fig:fig5} shows the corresponding positions of proposed progenitors. 

\subsection{Simulation-based investigations on the possible progenitors and surroundings}

As mentioned in the previous subsection, direct detection of progenitors in pre-explosion images is the most successful method to constrain their physical properties and the ambient environment around them. With the help of observationally derived parameters, reliable progenitor models can be built for several classes of SNe. With the help of many state-of-the-art simulation tools, those models can be evolved up to their late evolutionary stages. Such simulation-based studies are essential to explain and understand the physics behind observed phenomena. 

\begin{figure*}[!t]
  \includegraphics[width=1.05\columnwidth,height=10cm]{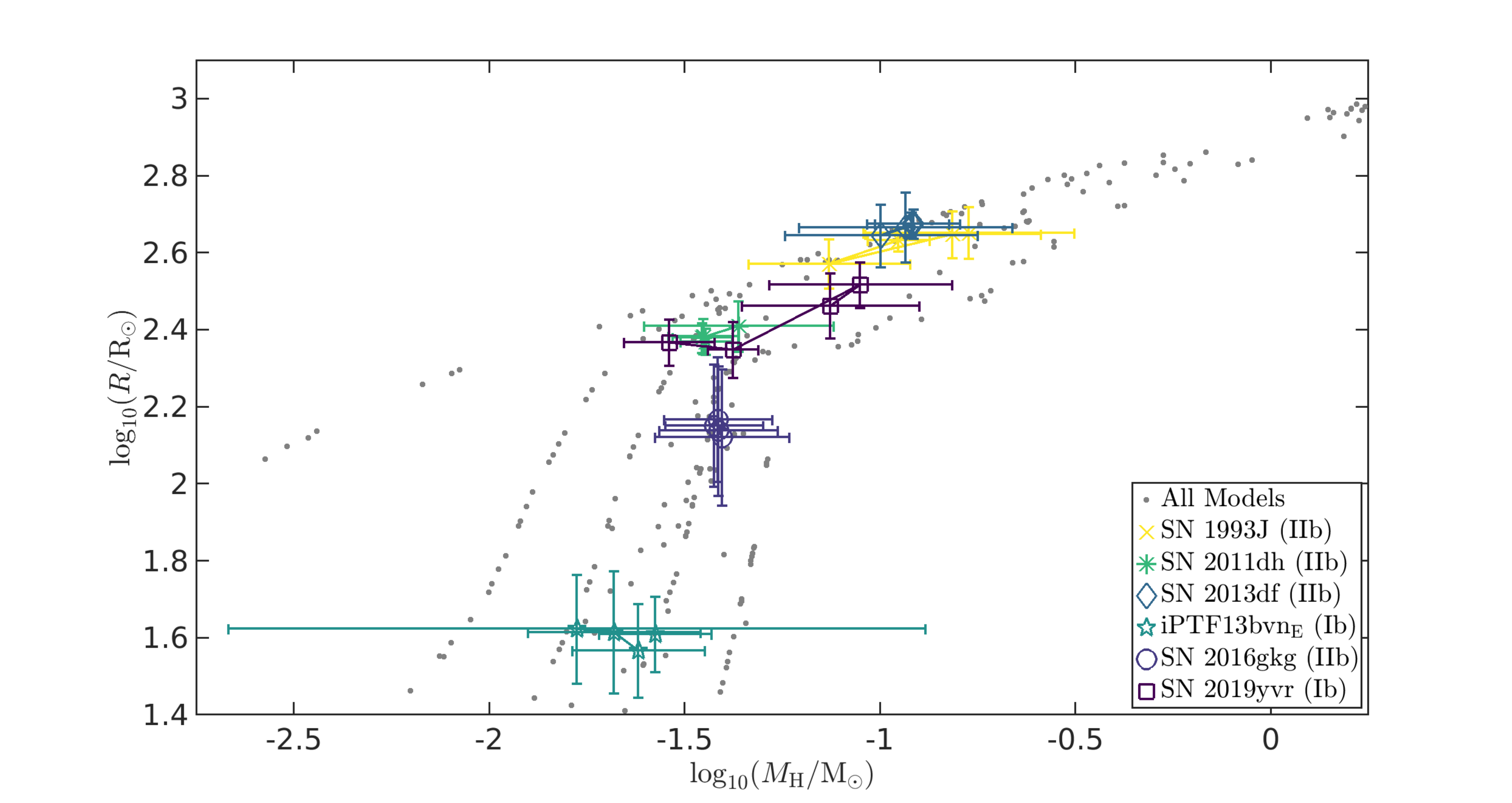}
  \caption{Mean hydrogen mass as a function of mean progenitor stellar radius along with their standard deviation. Figure credit: \citet[][]{2022MNRAS.511..691G}.}
  \label{fig:radius_mass}
\end{figure*}

Utilising the 1-dimensional hydrodynamic simulations, \citet[][]{2017MNRAS.470.1642F} attempts to explain the pre-explosion outbursts activities in RSG progenitors via wave heating. The author also finds that the wave heating in massive stars could explain some flash-ionised SNe and a few diversity observed in Type IIP and IIL CCSNe. As the surface abundances play an essential role in decorating the spectrum with the features of several elements/ions, authors of \citet[][]{2019MNRAS.483..887D} attempt to investigate the surface abundances of RSG stars utilising 1-dimensional simulations. 
In another work, \citet[][]{2018MNRAS.476.1853F} examine the effect of wave heating in massive H-poor stars serving as the progenitors of Type IIb/Ib CCSNe. They find that only a subset of these progenitors is expected to experience pre-explosion outbursts due to wave heating. Type Ib/IIb CCSNe are stripped SNe that arise from progenitors that have retained very little to no hydrogen at the time of their explosions. However, the uncertainties associated with determining the extinction and distance of the underlying CCSN affect the correct estimation of the amount of hydrogen retained by the progenitor at the pre-explosion stage. While studying the progenitor channels for Type IIb CCSNe, \citet[][]{2022MNRAS.511..691G} find that the post-interaction mass-loss rate plays a vital role on the amount of hydrogen retained in the envelope by the underlying progenitor at the time of the explosion. Figure~\ref{fig:radius_mass} shows the mean of the amount of hydrogen as a function of the mean stellar progenitor radius for all the models in their study, including several Type Ib/IIb CCSNe. Earlier, the authors of \citet[][]{2019ApJ...885..130S} also explored the impact of single- vs binary-progenitor systems for Type IIb CCSNe. They found that both the progenitor systems contributed to roughly the same number of Type IIb SNe at solar metallicity. However, at lower metallicities, the binary-progenitor channel dominated over single. At the time of investigating the evolution of He-rich progenitor stars through 1-dimensional simulations, \citet[][]{2018MNRAS.481L.141K} find that the amount of $^{56}$Ni synthesised plays a key role in separating rapidly fading Type I SNe with normal Type Ibc SNe.

1-dimensional stellar evolutions are helpful in investigating the pre-explosion activities, while post-explosion properties, including bolometric or multi-band light curves, are explored with the help of synthetic explosions of stellar models on the verge of core collapse. Utilising several state-of-the-art 1-dimensional stellar evolution codes, artificial models are computed based on the observational behaviour of the underlying SN. The stellar-model is evolved through various stages of its life until it reached the onset of core collapse. Further, the model on the stage of the beginning of core collapse is exploded synthetically to produce simulated bolometric light curves, multi-band light curves, and many other photospheric properties. These simulated properties are compared with actual observations. Such a comparison of simulated results with actual results further helps to put important constraints on the physical properties of the underlying SN progenitor.

Exploring the lack of consensus over the progenitor masses of Type IIP CCSNe, \citet[][]{2007A&A...461..233U} could put important constraints over pre-explosion characteristics of the exploding star and several SN properties with the help of hydrodynamic and time-dependent atmosphere models. They estimate that the exploding progenitor of the Type IIP SN~1999em had a pre-SN radius of 500$\pm$200\,R$_{\odot}$. It ejected an enormous amount of matter as indicated by an ejecta mass of 19.0$\pm$1.2\,M$_{\odot}$. The estimated explosion energy and the amount of $^{56}$Ni synthesised were (1.3$\pm$0.1)$\times$10$^{51}$\,erg and 0.036$\pm$0.009\,M$_{\odot}$, respectively. They attempted to derive approximate connections between basic physical and observationally-obtained parameters and concluded that the hydrodynamic and atmospheric models needed more consistency. They also concluded that the hydrogen recombination event in all the Type IIP SNe is time-dependent at their photospheric epochs.

The hydrodynamic simulations are essential to investigate the shortcomings of the proposed model and the effect of multi-dimensions. In \citet[][]{2009A&A...506..829U}, the authors find out that the inferred progenitor mass of SN~2004et through hydrodynamic simulations is much higher than obtained through pre-explosion images earlier. The mismatch between the progenitor mass estimates through pre-explosion images and hydrodynamic simulations was speculated to be the multi-dimensional effect. 
 
With time, attempts to incorporate the improvements suggested in previous studies to make more realistic models were made. \citet[][]{2011ApJ...729...61B} presented bolometric light curves of Type IIP SNe obtained using a 1-dimensional Lagrangian hydrodynamic code that utilises flux-limited radiation diffusion approximation \citep[][]{1981ApJ...248..321L}. Utilising their code, the authors obtained remarkable agreement between the observationally-obtained parameters and simulated results for SN~1999em. In another work by \citet[][]{2014ApJ...787..139D}, the authors presented hydrodynamic modelling of a Type IIP SN~2012aw utilising accurate spectroscopic and photospheric observations. They estimated that the underlying progenitor had an envelope mass of around 20\,M$_{\odot}$, progenitor radius of  $\sim$430\,R$_{\odot}$, explosion energy of $\sim$1.5$\times$10$^{51}$\,erg and initial $^{56}$Ni mass of around 0.06\,M$_{\odot}$. In \citet[][]{2017MNRAS.464.3013P}, the authors attempted to improve the current understanding of the possible progenitors of under luminous CCSNe Type IIP through radiation hydrodynamics modelling. 
Their investigations indicated that the low luminosity Type IIP CCSNe originated from relatively lower mass progenitors, and the catastrophic explosions were less energetic than intermediate luminosity Type IIP CCSNe. In a recent work, the authors of \citet[][]{2020A&A...642A.143M} attempted to estimate the physical parameters of Type II SNe. Using statistical interference techniques, they simultaneously fitted the bolometric light curve and the corresponding photospheric velocity evolution to hydrodynamic models. They found that the progenitor mass constraints were in very well agreement with the progenitor mass limits derived through pre-explosion images. Finally, they displayed that hydrodynamic modelling of progenitors can be used to put robust constraints over the physical properties of Type II SNe.
In relatively recent work, \citet[][]{2022A&A...660A..41M} performed a similar study and estimated the physical parameter distribution of an ensemble containing 53 Type II SNe through hydrodynamic modelling. Their analysis indicated a range of physical parameters for Type II SNe considered in their study with ejecta mass lying in the range of (7.9--14.8)\,M$_{\odot}$, explosion energies lying between 0.15$\times$10$^{51}$\,erg and 1.40$\times$10$^{51}$\,erg. Additionally, the inferred $^{56}$Ni mass for the SNe in the sample lies in the range of (0.006--0.069)\,M$_{\odot}$. A few studies have also attempted to perform the hydrodynamic simulations of Type IIn \citep[][]{2022MNRAS.515.3597L,2020A&A...635A.127K,2015MNRAS.449.4304D} and Type IIL SNe \citep[][]{2019MNRAS.485.5120B,Sukhbold2016} also.
   
Compared to other Type CCSNe, the literature is significantly populated with hydrodynamic modelling of Type IIP SNe. Specifically, the hydrodynamic modelling of Type IIP SNe is relatively more straightforward than other CCSNe since the former occurs in an environment with very low density \citep[][]{2000ApJ...545..444B,2006ApJ...641.1029C}. Additionally, the progenitors of Type IIP SNe have extended and nearly spherically symmetric outer H-envelopes, which suppress any inhomogeneities arising from diverse explosions \citep[][]{2005ASPC..342..330L}. Unlike Type IIP SNe, stripped-envelope CCSNe of Type Ibc and Type IIb have relatively compact progenitors. The complicated stages of envelope burning in them make it very difficult to perform their stellar evolution and then compute the hydrodynamic simulations of their explosions. Only a handful of such studies have been performed for stripped-envelope CCSNe. 

\begin{landscape}
\begin{figure*}
\centering
    \includegraphics[height=11.0cm,width=9.5cm,angle=-90]{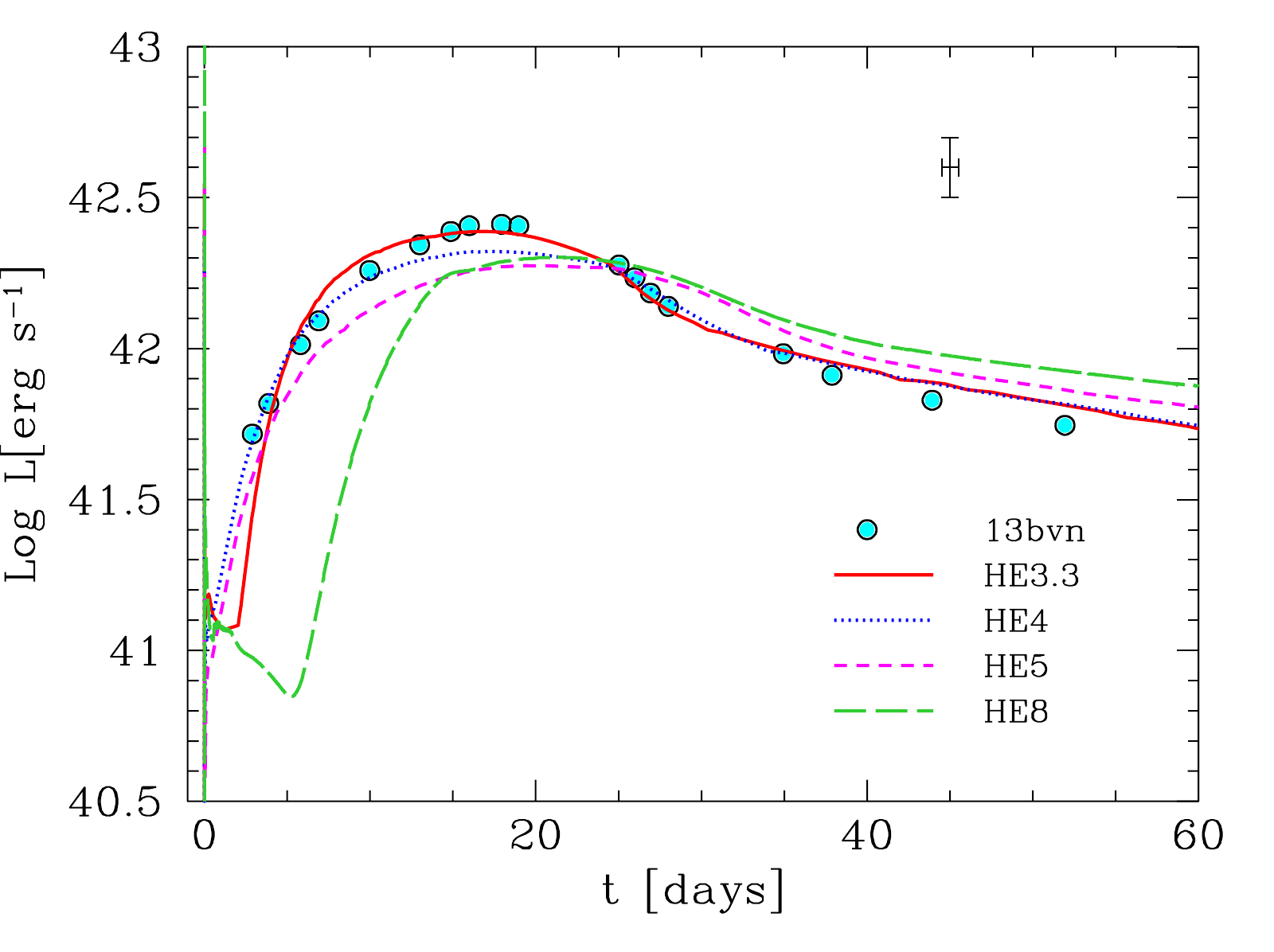}
    \includegraphics[height=11.0cm,width=9.5cm,angle=-90]{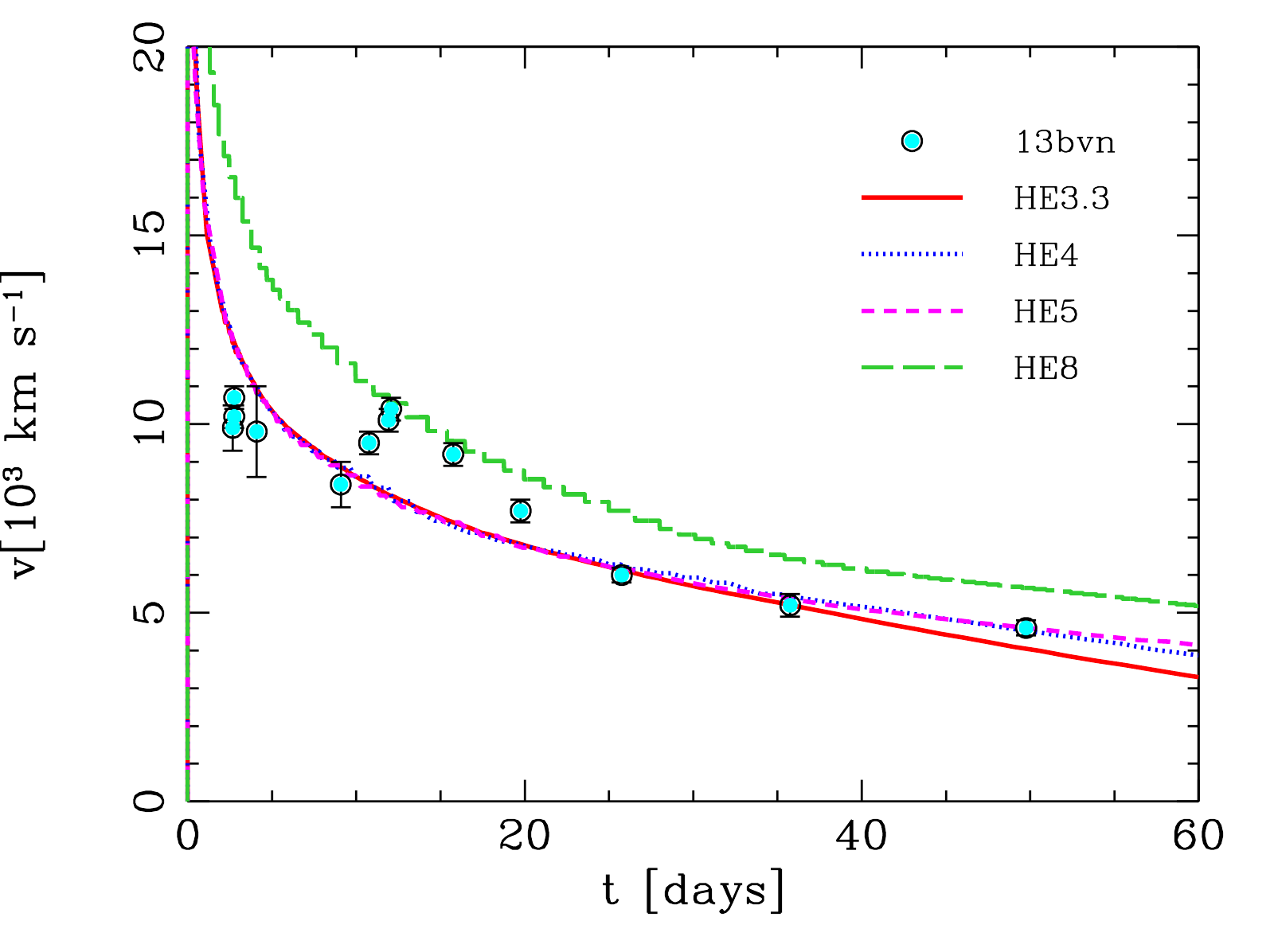}
   \caption {Hydrodynamic modelling outcomes for iPTF13bvn. The model names correspond to the progenitor star's pre-SN mass (e.g., for {\tt HE3.3}, the pre-SN mass is 3.3,M$_{\odot}$). A comparison was made between the model bolometric light curves (displayed in the left panel) and the evolution of photospheric velocity (shown in the right panel) with the actual observations. The model featuring a 3.3,M$_{\odot}$ pre-SN mass exhibited the closest agreement with the observed data. Figure credit: \citet[][]{Bersten2014}.}
    \label{fig:fig6}
\end{figure*}
\end{landscape}
        
After confirming the location of the progenitor candidate of the SN iPTF13bvn, the authors of \citet[][]{2014A&A...565A.114F} perform hydrodynamic modelling of the bolometric light curve to constrain the amount of $^{56}$Ni synthesised and the ejecta mass. They find that iPTF13bvn synthesised an amount 0.05\,M$_{\odot}$ of $^{56}$Ni and the corresponding ejecta mass was 1.9\,M$_{\odot}$. They also find that the bolometric light curve does not follow the single massive WR-star possible progenitor scenario as predicted in earlier studies. In another work by \citet[][]{Bersten2014} also, the authors predict a binary interacting progenitor system for iPTF13bvn and perform the evolutionary calculations. Their models could explain the light curve shape (left panel in Figure~\ref{fig:fig6}), photospheric velocity evolution (right panel in Figure~\ref{fig:fig6}), the absence of Hydrogen, and also the pre-SN photometry. The results from their hydrodynamic modelling suggest that the pre-SN stage mass of the progenitor was $\approx$\,3.3\,M$_{\odot}$ and the initial mass of the progenitor could not be larger than $\approx$\,8\,M$_{\odot}$. Later, \citet[][]{2015MNRAS.446.2689E} also incorporated a set of several binary progenitor models to put constraints over the probable binary system of iPTF13bvn. According to their investigations, the two companions in the binary progenitor system of iPTF13bvn would have masses of 10\,M$_{\odot}$ and 20\,M$_{\odot}$.

A few studies have presented the stellar evolution of the possible progenitors of Type IIb SNe also and have performed the hydrodynamic simulations of their explosions. \citet[][]{Bersten2012} utilised a set of several hydrodynamic models to explore the properties of underlying progenitor. According to their estimates, SN~2011dh probably exploded from a large progenitor star with a radius of $\sim$\,200\,R$_{\odot}$. The exploding star probably had an He-core of 3--4\,M$_{\odot}$ and retained a thin H-envelope at the pre-SN stage. The corresponding initial mass estimate for the progenitor was in the range of 12--15\,M$_{\odot}$. They estimated an ejecta mass of around 2\,M$_{\odot}$, explosion energy of (6--10)$\times$10$^{50}$\,erg, and about 0.06\,M$_{\odot}$ of $^{56}$Ni synthesised. The small initial progenitor mass constraints ruled out the possibility of a single-star evolution scenario as the progenitor of SN~2011dh.

\begin{figure*}[!t]
  \includegraphics[width=\columnwidth]{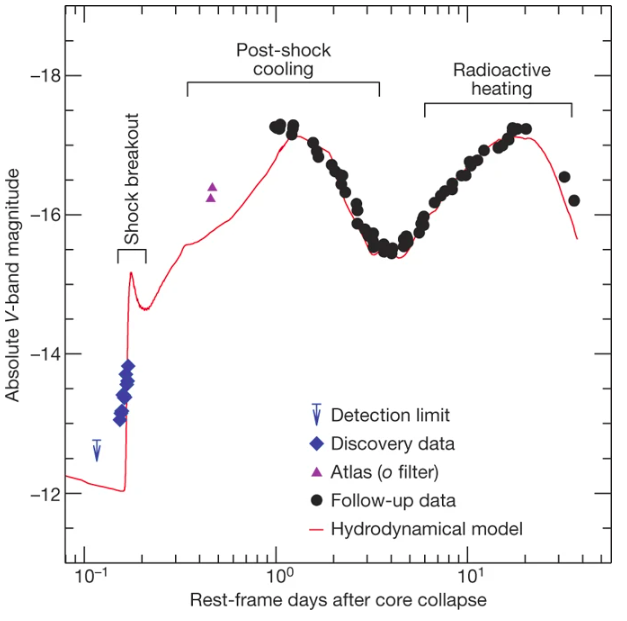}
  \caption{The outcome of hydrodynamic modelling for SN~2016gkg successfully replicated its distinct phases: the initial shock-breakout phase, the subsequent peak during shock cooling, and the later peak arising from the radioactive decay of $^{56}$Ni and $^{56}$Co. Figure credit: \citet[][]{Bersten2018}.}
  \label{fig:fig7}
\end{figure*}

The authors of \citet[][]{Morales2015} performed hydrodynamic simulations to model the pseudo-bolometric light curve and the photospheric velocity evolution of SN~2011fu. Their analysis indicates that the SN~2011fu exploded from a progenitor star having an initial mass of 13--18\,M$_{\odot}$. Further, they found that the exploding star released a kinetic energy of 1.3$\times$10$^{51}$\,erg, and the amount of $^{56}$Ni synthesised was 0.15\,M$_{\odot}$.

The serendipitous identification of Type IIb SN~2016gkg in its extremely early stages served the purpose of performing unprecedented explorations on CCSNe. It revealed a rapid brightening in optical wavelengths of nearly 40\,magnitudes per day during the very early phases of detection \citep[][]{Bersten2018}. In \citet[][]{Bersten2018}, authors performed the hydrodynamic modelling of the explosion to account for the different phases of the evolving SN (the preferred model is shown in Figure~\ref{fig:fig7}). The preferred model in their work had an initial mass of 18\,M$_{\odot}$ with solar abundances and a pre-SN mass of 5\,M$_{\odot}$. The object in the preferred model had an H-rich envelope of radius 320\,R$_{\odot}$ and the corresponding mass of the H-rich envelope was found to be 0.01\,M$_{\odot}$.

\section{Core-Collapse Supernovae from Population III stars }

The birth of the very first generation of stars, known as the Population III (Pop III) stars, marked the end of cosmic dark ages as they were responsible for generating fundamental modifications to the early Universe through the production of ionizing photons and enrichment with heavy elements \citep[][]{2013RPPh...76k2901B}. There are multiple investigations which indicate that the Pop III stars were intrinsically very massive \citep[e.g.,][]{1984ApJ...277..445C,1986MNRAS.223..763M}; however, a few studies have also investigated the possibilities of the existence of relatively low mass Pop III stars \citep[e.g.,][]{2009Sci...325..601T,2016ApJ...826....9I,2020MNRAS.494.1871W}. Pop III stars have been postulated to have multiple cosmological consequences, including production of 3\,K microwave background radiation \citep[][]{1978Natur.275...35R}, chemical enrichment \citep[e.g.,][]{2020MNRAS.491.4387K}, dust formation \citep[e.g.,][]{2001MNRAS.325..726T}, cosmic reionization \citep[e.g.,][]{1979MNRAS.188..781H}, etc.

Due to their cosmological importance, Pop III stars are fascinating objects of research despite their non-observance to date. It has been established through numerical studies that sufficiently large, very massive stars directly collapse to form BHs, while there is a tendency to explode in smaller ones \citep[][]{1984ApJ...277..445C}. The authors of \citet[][]{2004MmSAI..75..312N} mentions that massive stars with masses greater than around 20 -- 25\,M$_{\odot}$ form BHs to terminate their life. The stars with non-rotating BHs collapse to result in faint SNe, while the rotating BH case can produce highly energetic SNe popularly known as hypernovae (HNe) \citep[][]{2004MmSAI..75..312N}. Several studies are there to understand the evolution of massive Pop III stars. The authors of \citet[][]{2003A&A...399..617M} investigate the evolution of massive Pop III stars in the mass range of [120 -- 1000]\,M$_{\odot}$ and the consequences of mass losses. The evolutions of two single Pop III stars and one binary system of Pop III stars are studied in \citet[][]{2008MNRAS.384.1533L}. Based on the sample of available models in their work, they find that Pop III SNe have a fainter peak but long plateau phase in their light curves. In another work, \citet[][]{2010ApJ...724..341H} have studied the evolution and explosion of Pop III stars in the mass range of [10--100]\,M$_{\odot}$. Their studies indicate that most of the non-rotating Pop III stars become blue supergiants (BSGs) at the end of their life to result in SNe resembling the light curve of SN~1987A finally; however, a few are found to become RSGs, and the light curves of resulting SNe resemble Type IIP SNe.

\begin{figure*}[!t]
  \includegraphics[width=\columnwidth]{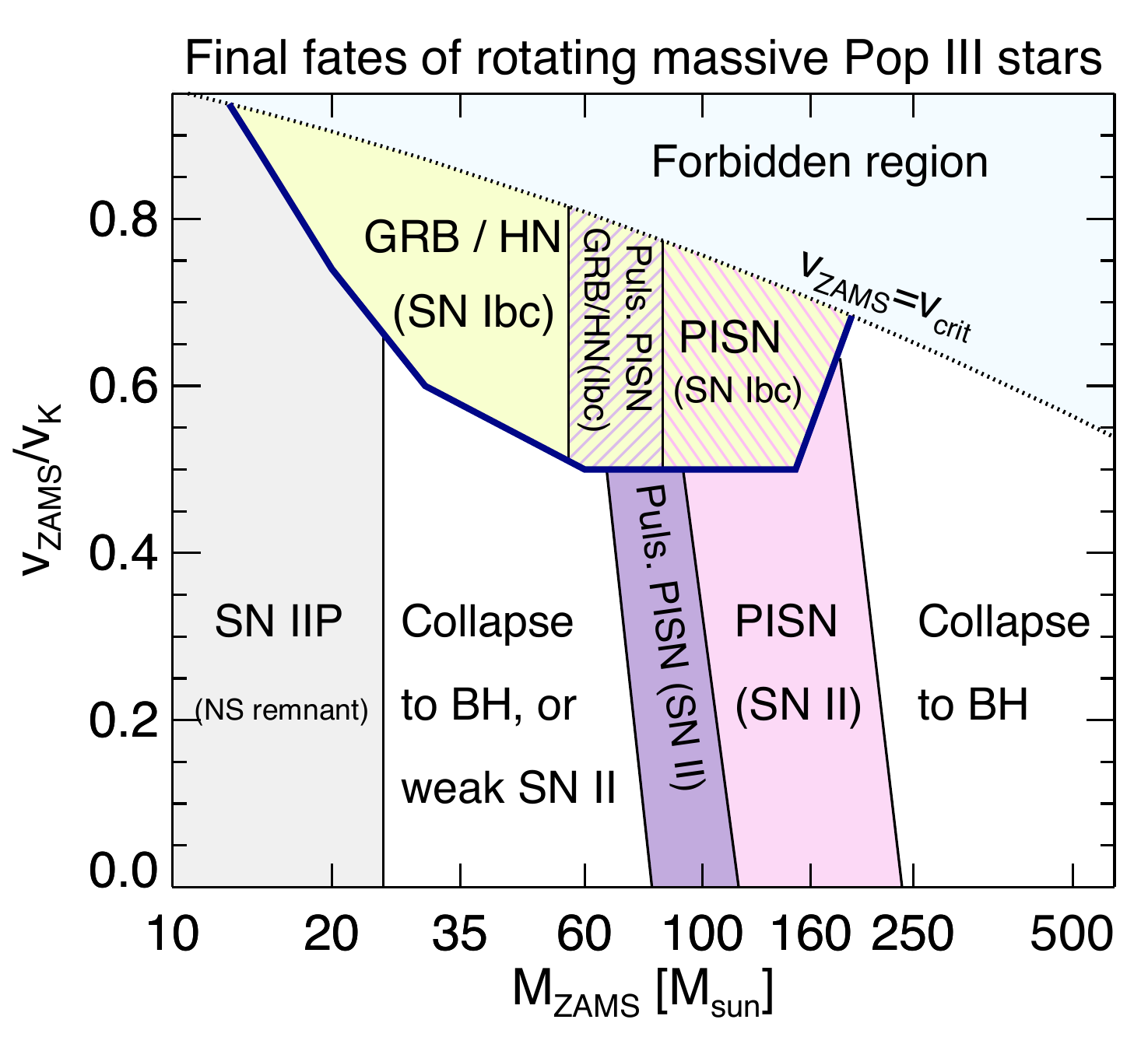}
  \caption{Phase diagram depicting the terminating fates of massive Pop III stars. In the regions above the dotted line, surface rotational velocity surpasses the critical threshold, thus that is the forbidden region. Further, the chemically homogeneous evolution is effective in the mass range of around [13 -- 190]\,M$_{\odot}$, indicated by the thick (blue) solid line. Figure credit: \citet[][]{2012A&A...542A.113Y}.}
  \label{fig:fig8}
\end{figure*}

In \citet[][]{2012A&A...542A.113Y}, the authors investigated the effect of including rotation and magnetic fields on the evolution of massive Pop III stars in the mass grid covering a range of [10--1000]\,M$_{\odot}$. In the considered mass grid, the authors identified the effect of chemically homogeneous evolution for a limited mass and rotational velocity range. They find that the effect of chemically homogeneous mixing is not seen beyond a ZAMS mass of 190\,M$_{\odot}$. Depending upon the rotational velocity and ZAMS mass, the final fates of the Pop III stars could span a wide range of transient events, including several classes of SNe, GRBs, HNe, PISNs, PPISNs, or collapse directly to form a BH. A corresponding phase diagram in the plane of ZAMS mass and rotational velocity at ZAMS is depicted in Figure~\ref{fig:fig8}.
To determine the signatures of detection and identification of Pop III SNe, the authors of \citet[][]{2016ApJ...821..124T} performed a multicolour light curve simulation of several Pop III CCSN model resulting from Pop III stars having mass in the range of [25 -- 100]\,M$_{\odot}$. They find that the multicolour light curves could be significantly helpful in identifying the Pop III SNe through current and future transient surveys. For the past few decades, researchers have been trying hard to find observational signatures of Pop III SNe \citep[e.g.,][]{2009ApJ...698L..68F,2014ApJ...792L..32I,2019PASJ...71...59M,2022ApJ...937...61Y,2022GReGr..54...24P,2019ApJ...876...97E,2016ApJ...823...83M}.

\section{Thesis Layout}

There are seven Chapters in this thesis. The research work under the presented thesis investigates the photometric and spectroscopic behaviour of a number of CCSNe of different types. An attempt to understand the underlying progenitor, possible powering mechanisms, and surrounding environments has been made by utilising the photometric and spectroscopic data from several national and international telescopes. The research work under this thesis also uses several state-of-the-art simulation tools to perform the 1-dimensional stellar evolution of the possible progenitors and simulate their explosions to replicate the actual SN explosions. Part of the research work in this thesis contains observational data-based research work complemented by simulation-based analysis for a subset of new CCSNe. In contrast, some parts involve simulation-based investigations of the evolution and explosion of proposed SN models from the literature. A brief overview of each Chapter is provided below:

{\bf Chapter 1:}  In this Chapter, we present a comprehensive perspective on the existing knowledge within the research field, focusing on the powering mechanisms, potential progenitors, and surrounding environments of catastrophic CCSNe. This introductory Chapter also provides an overview of SNe's historical observations, current progress in the field, and persisting issues of active research to set the context of present research work.

{\bf Chapter 2:} Data from several telescopes and numerous state-of-the-art simulation tools are required to address issues outlined in Chapter~\ref{Ch:1}, concerning the possible progenitors, powering mechanisms, and ambient media around CCSNe. The details of observations, data reduction, and modelling tools employed to carry out the research work in the presented thesis are discussed in Chapter~\ref{Ch:2}. We also provide the details of several national and international telescopes utilised to obtain the photometric and spectroscopic data.

{\bf Chapter 3:} Understanding the possible progenitors of CCSNe is one of the most fundamental tasks in understanding these catastrophic explosions. Therefore, within Chapter~\ref{Ch:3}, we aim to illuminate the origins of the underlying progenitors of two Type Ib CCSNe, SN~2015ap and SN~2016bau, by performing their photometric and spectroscopic analyses. Our analyses show that SN~2015ap has intermediate luminosity among a subset of similar SNe, while SN~2016bau suffers very high host galaxy extinction and appears relatively less luminous. With the help of synergistic investigations utilising optical data from several telescopes and state-of-the-art simulation tools, we find that SN~2015ap originated from a 12\,M$_{\odot}$ ZAMS progenitor, while a slightly less massive ZAMS progenitor is expected for SN~2016bau. Both the SNe seem to originate at the solar metallicity abundance sites of their corresponding host galaxies. The results of this work are published in \citet[][]{Aryan2021} and \citet[][]{2022JApA...43...87A}.

{\bf Chapter 4:} Having discussed the properties of two H-deficient Type Ib SNe in Chapter~\ref{Ch:3}, we perform the photometric and spectroscopic investigations of a Type IIb SN~2016iyc in this Chapter. Type IIb SNe are thought to be the interconnecting link between H-rich and H-deficient CCSNe. 
Thus, the strategic investigations of Type IIb SNe are essential to enlighten the possible connections between H-rich and H-deficient CCSNe. The progenitors of Type Ib/IIb CCSNe retain very little to no hydrogen at the time of their explosions. However, the correct estimation of the amount of hydrogen retained before explosion by the underlying CCSN progenitor is subjected to contamination by the uncertainties associated with determining the extinction and distance of the CCSN.

In this Chapter, we investigate the photometric and spectroscopic analysis of a Type IIb SN~2016iyc. Our studies indicate SN~2016iyc lies towards the fainter luminosity regions among a subset of similar types of SNe. Based on the photometric and spectroscopic behaviours, the stellar evolution of progenitor models with ZAMS mass in the range of (9--14)\,M$_{\odot}$ is performed. Finally, the comparison of simulation-produced light curves and photospheric velocity evolution indicate that SN~2016iyc probably originated from a star having ZAMS mass in the range of (12--13)\,M$_{\odot}$. Our analysis also suggests that the host galaxy metallicity at the site of the occurrence of SN~2016iyc is probably solar. This Chapter also presents the hydrodynamic modelling of two other Type IIb SNe, SN~2011fu and SN~2016gkg. Based on the analysis conducted in this Chapter, we also find that Type IIb SNe progenitors preserve some trace amount of hydrogen at their pre-SN phase. This hydrogen mass appears to fall somewhere between the amount retained by the progenitors of H-rich and H-deficient SNe during their pre-SN stages. The results are published in \citet[][]{2022MNRAS.517.1750A}.

{\bf Chapter 5:} After discussing the properties of H-deficient SNe in Chapter~\ref{Ch:3}, the behaviour of a Type IIb SN retaining an intermediate amount of H-envelope in Chapter~\ref{Ch:4}, this Chapter discusses H-rich and H-deficient SNe together that originate from progenitors each starting with a mass of 25\,M$_{\odot}$ at ZAMS and zero metallicity. CCSNe from massive Pop III stars
are postulated to have had an enormous impact on the early Universe. The SNe from Pop III stars were responsible for the initial enrichment of the early Universe with heavy elements, and they also played a pivotal role in cosmic reionization. This Chapter comprises pure simulation-based results of the stellar evolution of a 25\,M$_{\odot}$ ZAMS progenitor with zero metallicity and investigates the effects of rotation on the resulting SNe. The hydrodynamic simulation of the resulting CCSNe is also presented in this Chapter. The rapidly rotating models display violent mass losses after their main-sequence phases compared to non-rotating and slowly rotating models. Further, the light curves of rapidly rotating models resemble Type Ib/c SNe, while slowly rotating models mimic the light curves of Type IIP SNe. These results are published in \citet[][]{2023MNRAS.521L..17A}.

{\bf Chapter 6:} Chapter~\ref{Ch:3}, Chapter~\ref{Ch:4}, and Chapter~\ref{Ch:5} discuss the properties of canonical CCSNe arising from progenitors having ZAMS masses of 25\,M$_{\odot}$ or less. However, in this Chapter, we present the pure simulation-based results of the stellar evolution of a 100\,M$_{\odot}$ ZAMS progenitor star and discuss the resulting transient. Depending upon the initial mass, mass loss rate, rotation and metallicity, the resulting transient could fall into any category, PISN, PPISN, Type IIP-like SNe, and several H-rich/H-deficient SNe showing ejecta-CSM interaction signatures. Starting from ZAMS, the 100\,M$_{\odot}$ ZAMS model is evolved up to the onset of core collapse. The evolution of various physical parameters, including radius, temperature, and density, are discussed as the model star passes through various stages of its life. Finally, the consequences of the model exploding as Type IIP SNe are explored. The results are published in \citet[][]{2022JApA...43....2A}.

{\bf Chapter 7:} This Chapter provides a comprehensive summary of the presented thesis along with its potential future prospects.

 %Introduction
\chapter{\sc Observations, Data analysis, and Modelling tools}\label{Ch:2}      
\ifpdf
    \graphicspath{{Chapter2/Chapter2Figs/JPG/}{Chapter2/Chapter2Figs/PDF/}{Chapter2/Chapter2Figs/}}
\else
    \graphicspath{{Chapter2/Chapter2Figs/EPS/}{Chapter2/Chapter2Figs/}}
\fi

%\newpage
%\thispagestyle{empty}
%\mbox{}

\section{Introduction}\label{Ch2_introduction}

As outlined in Chapter~\ref{Ch:1}, the research work under the presented thesis focuses on unveiling the diverse nature of CCSNe, including their progenitors, powering mechanisms, and surrounding media, with the help of numerous telescopes and state-of-the-art simulation tools. Telescopes help us to capture the electromagnetic signals emitted during and after the SN explosion. Observations from various ground- and space-based telescopes are essential to understand the fundamental aspects of these catastrophic transients. In the current era, telescopes equipped with delicate back-end instruments and detectors allow scientists to capture and analyse the photons across the electromagnetic spectrum, including radio, optical, ultraviolet (UV), X-ray, and gamma-ray. With the steady progress of telescope technology, our understanding of SNe has significantly expanded, shedding light on their progenitors, the processes that drive these calamitous cosmic explosions, and their impact on the persistent surrounding.

As mentioned above, the role of telescope-based observations in SN astronomy is essential. Similarly, state-of-the-art simulation tools and techniques are essential for understanding these catastrophic explosions. Simulations help scientists understand and synthetically recreate the complex mechanisms resulting in SNe by incorporating nuclear reactions, hydrodynamics, radiative transfer processes, and gravity. Also, with the help of multi-dimensional simulations, we can explore the effects of initial stellar mass, metallicity (Z), rotations, etc., on the resulting SNe. Utilising simulation tools, we can model a massive star and study detailed evolution through various stages of its life to eventually analyse its explosion into a synthetic SN.

Furthermore, simulations can complement the interpretation of observational data by providing theoretical frameworks for comparison. Utilising observational data and simulation-based investigations together, people have refined their understanding of SNe, validated theoretical models, and made predictions for future observations. In short, simulation tools play a pivotal role in advancing our knowledge of SNe and contribute towards enhancing the understanding of stellar evolution, nucleosynthesis, and the dynamics of the universe.

In this Chapter, we provide details of the telescopes and back-end instruments utilised to obtain the photometric and spectroscopic data to carry out the research meeting the goals specified in the synopsis. We also present a detailed explanation of photometric and spectroscopic data reduction processes. Ultimately, we discuss the simulation tools to perform progenitor modelling and simulate the synthetic explosions.

\section{Details of Telescopes and Instrumental Setups} \label{Ch2_Telescopes and detectors}

To meet the goals specified in the synopsis, we required a significant amount of photometric and spectroscopic data from several national and international telescopes equipped with modern back-end instruments. We primarily utilised data from 3.6\,m Devasthal optical telescope, 3.0\,m Shane telescope, 1.3\,m Devasthal fast optical telescope, 1.04\,m Sampurnanand telescope, 1.0\,m Nickel Telescope, 0.76\,m Katzmann Automatic Imaging Telescope. Additionally, for the case of SN~2015ap, we could obtain the host galaxy images from the Hubble space telescope. Furthermore, we also utilised the archival $U$-band data for SN~2015ap from the UV/optical telescope\footnote{https://swift.gsfc.nasa.gov/about\_swift/uvot\_desc.html} (UVOT) onboard Swift\footnote{https://swift.gsfc.nasa.gov}. Below are the details of each telescope, along with a summary of available back-end instruments.

\subsection{3.6 m Devasthal Optical Telescope (DOT)} \label{sec:telescopes_3.6}
The 3.6\,m Devasthal Optical Telescope\footnote{https://www.aries.res.in/facilities/astronomical-telescopes/360cm-telescope} is a remarkable, custom-built instrument that stands tall in the awe-inspiring Devasthal region of India. It is a national facility commissioned in the year 2016. The 3.6\,m DOT is maintained and operated by the Aryabhatta research institute of observational sciences (ARIES) in the Nainital district of Uttarakhand, India. With its sizeable primary mirror spanning 3.6 meters in diameter, this telescope is the country's largest steerable optical telescope, offering an exceptional capacity to follow and study numerous celestial objects with remarkable clarity and precision. Due to the longitudinal advantage of the telescope's strategic location in the Devasthal region coupled with its favourable weather conditions and low light pollution, the telescope is a heaven for astronomers and researchers willing to work on unravelling the mysteries of the cosmos. The 3.6\,m DOT is an almost perfect fusion of engineering, scientific expertise, and natural scenic beauty, making it a beacon of astronomical follow-up and discovery.

\begin{figure}
\includegraphics[width=\columnwidth]{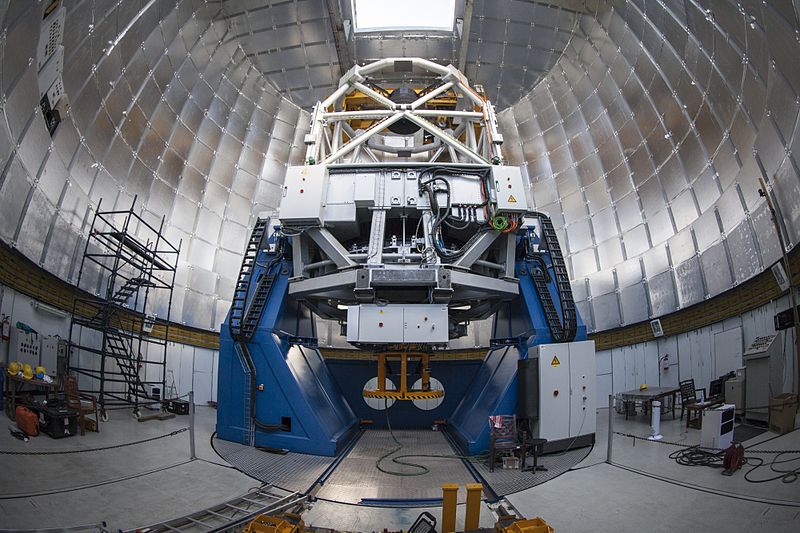}
\caption{The 3.6\,m\,DOT situated at Devasthal site of ARIES, Nainital, Uttarakhand, India. Image credit: \href{https://en.wikipedia.org/wiki/3.6m_Devasthal_Optical_Telescope}{3.6 m DOT}}
\label{fig:3.6m_DOT}
\end{figure}

Equipped with advanced optics and modern technology, the 3.6\,m DOT provides astronomers and researchers with an exceptional view of the cosmos. Its high-resolution imaging capabilities enable detailed observations of transient events, distant galaxies, stars, and planetary bodies, to unveil the secrets of the Universe. The 3.6\,m DOT offers three Cassegrain ports for instrument mounting; one main-axial port and two side ports, namely, side-port1 and side-port2. Currently, the 3.6\,m DOT is utilised with four available back-end instruments depending upon the observing time requested from the accepted proposals.

The 4K$\times$4K CCD imager is an optical imaging instrument mounted on the main axial port of the 3.6\,m DOT that covers a wavelength range of (400--900)\,nm. It has a plate scale of nearly 0.1$''$\,pixel$^{-1}$ and offers a field of view (FoV) of 6.5$'$ $\times$ 6.5$'$. The Bessell $UBVRI$ photometric filters \citep[][]{1990PASP..102.1181B} and SDSS $ugriz$ photometric filters \citep[][]{1996AJ....111.1748F} are available with the 4K$\times$4K CCD imager. Extensive details on the calibration and characterisation of 4K$\times$4K CCD imager are given in \citet[][]{Pandey2018} and \citet[][]{2022JApA...43...27K}.

TIRCAM2 is another imaging instrument mounted on the side-port2 of 3.6\,m DOT. It can provide near-infrared photometric observations in the wavelength range of (1--3.7)\,$\mu$m. Utilising the InSb array, it has a plate scale of around 0.17$''$\,pixel$^{-1}$  and offers an FoV of 86.5$''$ $\times$ 86.5$''$. The broadband $J$, $H$, $K$ and narrow-band BrG, $K$-cont, nbL filters are available with the TIRCAM2. The details of TIRCAM2 are described in the instrument paper \citep[][]{2018JAI.....750003B}.

ADFOSC is a low-resolution optical spectrograph and camera available with 3.6\,m DOT. The details about ADFOSC are provided in \citet[][]{2019arXiv190205857O}. Further, TANSPEC is a medium-resolution spectrograph and camera effective in the wavelength range of (550--2540)\,nm. The details of TANSPEC can be found in \citet[][]{2022PASP..134h5002S}.

\subsection{3.0 m Shane Telescope at Lick Observatory}

The Shane Telescope\footnote{https://www.lickobservatory.org/explore/research-telescopes/shane-telescope/} is one of the largest telescopes at the Lick observatory with the primary mirror spanning a diameter of around 3\,m ($\sim$120 inches). It was built with an equatorial fork mount to follow celestial objects moving across the sky. This telescope has an adaptable 3-foci design (prime focus, Cassegrain focus, and coud\'e focus). The adaptable 3-foci design allows several kinds of instruments to be used that favour a variety of observations and research.  

\begin{figure}
\includegraphics[height=12cm,width=0.49\columnwidth]{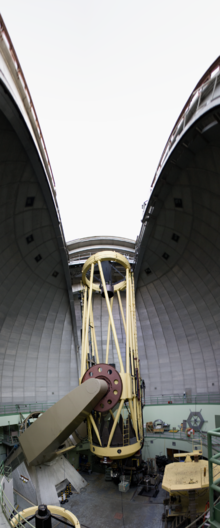}
\includegraphics[height=12cm,width=0.49\columnwidth]{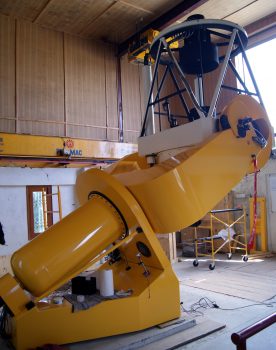}
\caption{{\em Left:} The 3\,m\,Shane telescope at the Lick observatory. Image credit: \href{https://en.wikipedia.org/wiki/C._Donald_Shane_telescope}{3 m Shane Telescope}; {\em Right:} The 1.3\,m DFOT at the Devasthal site of ARIES. Image credit: \href{https://www.aries.res.in/facilities/astronomical-telescopes/130cm-telescope}{1.3 m DFOT}}
\label{fig:Shane_n_DFOT}
\end{figure}

The Kast double spectrograph mounted at the Cassegrain focus is among the most important instruments available with the Shane Telescope. As the name suggests, it combines two separate spectrographs, one optimised for the red and the other for the blue wavelengths. Both spectrographs are housed in a single structure and share some components. The approximate range of good performance for the blue lens is in the wavelength range of (300--700)\,nm with an approximate range of resolution within (700--1700), while the red lens is optimised to show good performance in the wavelength range of (400--1100)\,nm offering an approximate resolution within the range of (500--3000). one can visit its homepage\footnote{https://mthamilton.ucolick.org/techdocs/instruments/kast/kast\_quickReference.html} comprising a complete list of the available instruments with the 3.0\,m Shane telescope.

\subsection{1.3 m Devasthal Fast Optical Telescope (DFOT)}

The 1.3\,m Devasthal fast optical telescope\footnote{https://www.aries.res.in/facilities/astronomical-telescopes/130cm-telescope} is a modern Ritchey-Chretien Cassegrain telescope with a 1.3\,m diameter primary mirror and a focal ratio of f/4 offering an unvignetted FoV of diameter 66$'$ at its axial port. It has an equatorial-fork mounting requiring the rotation of only one axis while tracking the celestial objects in the sky. An in-house developed motorised filter changer is equipped with the telescope, capable of hosting eight filters out of the available filters at a time. The available filters are Bessel $UBVRI$, SDSS $ugriz$, and narrow-band H$_{\alpha}$, O[III], Si[III] interference filters. This telescope has provided extensive support to researchers by providing optical photometric data for research works involving transient astronomy, star clusters, AGNs, stellar variability, etc. Owing to the darkness and very good seeing conditions at the Devasthal site, the telescope has significantly contributed to the follow-up of faint objects despite its relatively smaller size of the primary mirror in the international context.

The 1.3\,m DFOT can be equipped with two available CCDs depending upon the requirement of the observers. The primary instrument that serves the purpose for most of the observers is the 2K$\times$2K CCD camera. It is a conventional back-illuminated and thermoelectrically cooled CCD with a pixel size of 13.5\,$\mu$m. It provides a plate scale of around 0.54$''$\,pixel$^{-1}$. The other 512$\times$512\,pixels CCD is also available for mounting on the axial port of 1.3\,m DFOT depending upon the requirement of the observer. This is also a back-illuminated and thermoelectrically cooled CCD with a pixel size of 16\,$\mu$m. Extensive details on the telescope and the available instruments can be found in \citet[][]{2012SPIE.8444E..1TS,2022JAI....1140004J}.

\subsection{1.04 m Sampurnanand Telescope (ST)}

The 1.04\,m Sampurnanand telescope\footnote{https://www.aries.res.in/facilities/astronomical-telescopes/104cm-telescope} is among one of the oldest telescopes in the country.
Installed in 1972, it is an RC reflector with a Cassegrain focus and possesses a primary parabolic mirror spanning a diameter of 1.04\,m. It is mounted on an equatorial two-pier English mount. With f/13 focal ratio, it produces an FoV of 45$'$ with corrector at the Cassegrain end. Additionally, there are three finder telescopes available alongside the 1.04\,m ST with diameters of 264\,mm (reflector telescope with focal ratio of f/14), 200\,mm (refractor telescope with focal ratio of f/15), and 110\,mm (refractor having focal ratio of f/7), respectively. The 200\,mm refractor telescope covering an FoV of around 20$'$ is usually utilised for guiding the main 1.04\,m telescope.
The telescope has already displayed its scientific importance with its involvement in research on transients, star clusters, young star-forming regions, HII regions, AGNs, etc. As per the information on the telescope's homepage, the data from 1.04\,m ST have been utilised in nearly 364 scientific research papers and 45 PhD theses.

Currently, the observations at the 1.04\,m ST are primarily carried out through two major instruments, namely, the 4K$\times$4K CCD imager and ARIES Imaging Polarimeter (AIMPOL). The 4K$\times$4K CCD imager is similar to the one equipped at the main axial port of the 3.6\,m DOT. The broadband $UBVRI$ filters can be incorporated with the back-end instruments. The plate scale with the 4K$\times$4K CCD imager is around 0.23$''$\,pixel$^{-1}$. The 4K$\times$4K CCD is used to obtain multi-band photometry, while the AIMPOL offering an FoV of around 8$'$ is utilised to measure the liner polarisation in the specified broadband filters.
\begin{figure*}
\includegraphics[width=\columnwidth]{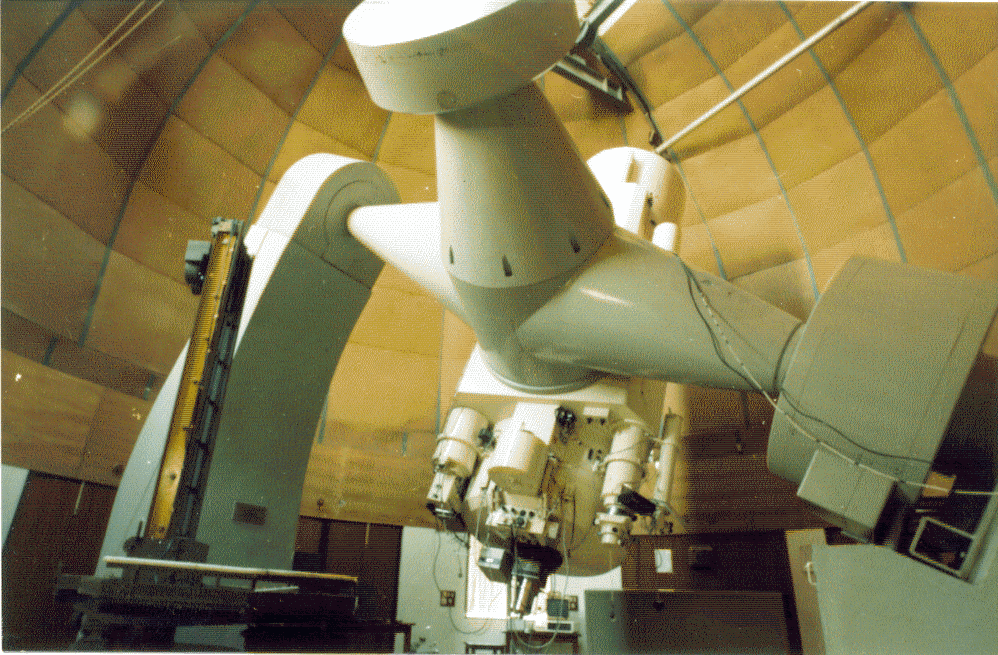}
\caption{ The 1.04\,m ST at Nainital campus of ARIES. Image credit: \href{https://www.aries.res.in/facilities/astronomical-telescopes/104cm-telescope}{1.04 m ST}}
\label{fig:ST}
\end{figure*}

\subsection{1.0 m Nickel Telescope at Lick Observatory}

The 1.0\,m Nickel telescope\footnote{https://mthamilton.ucolick.org/techdocs/telescopes/Nickel/intro/} is one of the newer telescopes at the Lick observatory. It was built within the Lick observatory workshops at the University of California, Santa Cruz. It has the same optical features as the Cassegrain focus of the 3.0\,m Shane telescope mentioned above so that instrument can be shared between these two telescopes. Similar optical features allow instruments designed for the 3\,m Shane telescope to be tested at the 1\,m Nickel telescope so that the earlier one could be free for research purposes.

The 1\,m Nickel telescope has a focal ratio of f/17. Unlike the 3-foci Shane telescope, the f/17 cassegrain focus is the only port available for instrument mounting. This telescope has a modern direct imaging CCD camera of 2K$\times$2K pixels. The CCD has 15\,$\mu$m pixels, a plate scale of 0.184$''$\,pixel$^{-1}$ providing a square FoV of 6.3$'$. The available filter wheel with the detector system offers five filter-holding positions. Out of five positions, one is always empty, and the remaining four positions are used to hold the available filters. The $UBVRI$ filters are most commonly employed with this 1\,m the available CCD camera. Extensive details on the CCD camera are available at the detector's homepage\footnote{https://mthamilton.ucolick.org/techdocs/instruments/nickel\_direct/intro/}.

\begin{figure}
\includegraphics[height=12cm,width=0.49\columnwidth]{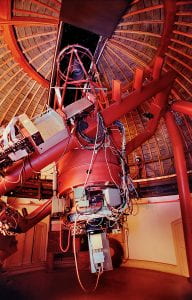}
\includegraphics[height=12cm,width=0.49\columnwidth]{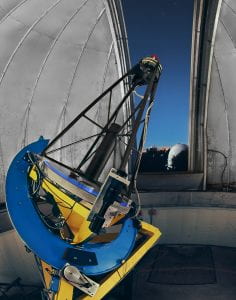}
\caption{{\em Left:} The 1\,m\,Nickel telescope at the Lick observatory. Image credit:\href{https://www.lickobservatory.org/explore/research-telescopes/nickel-telescope/}{1 m Nickel}; {\em Right:} The 0.76\,m KAIT at the Lick observatory. Image credit: \href{https://www.lickobservatory.org/explore/research-telescopes/katzman-automatic-imaging-telescope/}{0.76 m KAIT}.}
\label{fig:Nickel_n_KAIT}
\end{figure}

\subsection{0.76 m Katzmann Automatic Imaging  Telescope (KAIT) at Lick Observatory}

The Katzmann Automatic Imaging Telescope\footnote{https://www.ucolick.org/public/telescopes/kait.html} is a robotic telescope with a primary mirror spanning a diameter of around 76\,cm equipped with a CCD imaging camera. It has a focal ratio of f/8.2 with a plate scale of around 33.2$''$\,mm$^{-1}$ at its focal plane. It requires minimum human support to look for SNe in the sky throughout each day of the year. This telescope has been optimised to facilitate the early discovery of SNe. 

The operation of the 0.76\,m KAIT is well organised. It first checks the weather and analyses if the weather conditions are satisfactory for the observations to begin. After that, KAIT opens its dome and starts to observe a preprogrammed list of galaxies to look for new SNe. The newly observed data are compared with previous data from the surveyed area to look for changes. If there are changes that indicate any possible new SN, KAIT automatically emails the astronomers at the University of California, Berkeley. The astronomers at the University of California, Berkeley, follow up by making their own observations to verify the probable SN event. After finishing the observations on the arrival of morning lights, KAIT closes its dome and suspends the activities until the upcoming evening. Extensive details on the KAIT system are provided in \citet[][]{Filippenko2001}. 

\subsection{Hubble Space Telescope (HST)}

\begin{figure*}
\includegraphics[width=\columnwidth]{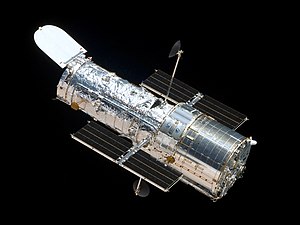}
\caption{ The Hubble Space Telescope, figure credit: \href{https://en.wikipedia.org/wiki/Hubble_Space_Telescope}{Hubble Space Telescope}. }
\label{fig:HST}
\end{figure*}

The Hubble space telescope\footnote{https://www.nasa.gov/mission\_pages/hubble/main/index.html} is the first ever astronomical observatory to be placed in space to orbit around Earth. It is a 2.4\,m, f/24 Ritchey-Chretien telescope that can observe near-ultraviolet to optical to near-infrared wavelength range. It utilises two primary camera systems, namely, Advance Camera for Surveys (ACS; installed on HST in 2002) and Wide Field Camera 3 (WFC3; installed on HST in 2009), to obtain photometric observations. Beyond the photometric cameras, the HST is also equipped with several spectrographs. The Space Telescope Imaging Spectrograph (STIS) and the Cosmic Origin Spectrograph (COS) are two complementary instruments mounted on HST that provide detailed spectra of several astronomical sources. Extensive details on the available photometric and spectroscopic instruments systems are provided on the HST instruments homepage\footnote{https://www.nasa.gov/content/goddard/hubble-space-telescope-science-instruments}.

\section{Photometric Data Reduction } \label{Data reduction (Photometry)}

Photometric data reduction refers to the process of extracting useful information for the sources of interest from photospheric observations stored in raw Flexible Image Transport System (FITS) files. The FITS files store information on the intensity or flux of the astronomical sources. A careful process is required to extract the information by eliminating the spurious signals from the actual signals of interest originating from the desired sources. In this section, we discuss the details of the process of photometric data reduction for the SNe considered in this thesis.

\subsection{Pre-processing} \label{Pre_processing}

The raw FITS files from photometric observations are contaminated due to several unwanted signals. The sources of the contaminating signals may be of electric origin or extragalactic. One needs to account for and eliminate these spurious signals from science FITS files to obtain a precise measure of the flux or intensity. The process of removing unwanted signals of electric or extragalactic origin from science frames is known as pre-processing. There are several elements in pre-processing, discussed in detail below:

$\bullet$~{\bf Bias correction}

The bias is referred to an offset signal present in each pixel of the CCD, even without any incident light. The bias signal act as a source of contamination or noise to the original signal obtained by exposing the CCD for a desired duration and must be subtracted from each frame of interest before any further cleaning.

The process of bias correction starts by obtaining multiple bias frames throughout the course of entire observations in an observing night. A bias frame is acquired with zero-exposure time, keeping the other settings of CCD the same as for obtaining the science images but with the shutter closed. Minor fluctuations in the bias signal level can be eliminated by median combining the multiple bias frames and forming a master-bias frame. To median-combine several bias frames and obtain a master-bias frame, we use the $ZEROCOMBINE$ task from {\tt IRAF}\footnote{The distribution of {\tt IRAF} is managed by the National Optical Astronomy Observatory, operated by AURA, Inc., in collaboration with the U.S. NSF through a cooperative agreement.} \citep[][]{1986SPIE..627..733T,1993ASPC...52..173T}. 

The master-bias frame is then subtracted from each frame obtained through non-zero exposure time. The bias correction process involves the pixel-wise subtraction of the bias value from the corresponding pixel in all other frames with non-zero exposure time. To obtain the bias-corrected frames by subtracting the master-bias from all the non-zero exposure frames, we utilise the $IMARITH$ or $CCDPROC$ task from {\tt IRAF}. After bias correction, images are free from bias signal and left with astronomical signals containing other sources of noise. Bias correction is an important step as it eliminates the instrumental signature from the data that allows precise measurements of the astronomical signal. It also helps to improve the signal-to-noise ratio (SNR).      

$\bullet$~{\bf Flat-field correction}

Flat correction is another critical step in pre-processing of photometric observational images. 
When a CCD is exposed to uniform light, all the pixels of the CCD are exposed to nearly the same number of photons. In an ideal case, all the pixels of the CCD should return the same counts, provided they all have similar pixel sensitivity. But, in real scenarios, pixel sensitivity varies. The sources of pixel sensitivity variation are associated with pixel-to-pixel size disparity, deposition of dust grains over the pixels, vignetting effects, etc. Thus, the process of flat correction aims to correct for the uneven illumination and pixel-to-pixel sensitivity variations across the CCD, which can lead to systematic errors in the data.

To correct the pixel-to-pixel sensitivity variation, we must perform the flat-field correction. We obtain several flat frames in different filters by exposing the CCD to a uniform source of light, e.g., the twilight sky during evening and morning. In some cases, dome flat frames are also preferred. The flat frames are obtained in as many filters as the science frames because the responses of the telescope optics system and dust grains are wavelength dependent. Due to this reason, we can not use the flat frames in one filter for other frames with different filters. The flat and science frames are obtained with similar CCD settings. However, exposure durations can be different.   

All the flat frames in different filters are first subjected to bias correction. As mentioned earlier, the bias correction is performed through the {\tt IRAF} tasks $IMARITH$ or $CCDPROC$. Once the bias correction is applied to all the flat frames, we generate the normalised master flat frame for each filter incorporating the $FLATCOMBINE$ task in the {\tt IRAF}. As a next step, the bias-corrected science frames in each filter are divided by the normalised master flat frame of the corresponding filter. To perform this step, we utilise the $IMARITH$ and $CCDPROC$ tasks in the {\tt IRAF}. Similar to bias correction, flat correction is exceptionally crucial to perform accurate measurements of astronomical sources.

$\bullet$~{\bf Cosmic rays removal}

Cosmic rays are highly energetic and charged particles (e.g., positively charged protons, muons, negatively charged electrons, etc.) originating beyond our solar system. These charged particles continuously shower onto our Earth's atmosphere. The cosmic rays can easily penetrate the dome and telescope system and fall on several CCD pixels producing huge counts in suffering pixels. As the cosmic rays are continuously falling on Earth from several directions all the time, it is very unlikely that any CCD exposure would not suffer from the cosmic ray hits. Cosmic ray removal is an essential step in the pre-processing process for science images, as cosmic rays can create spurious signals affecting astronomical measurements' accuracy.

The number of cosmic rays hitting a CCD frame depends on the exposure time, although bias frames having zero exposure durations also display some cosmic ray hit features. As mentioned above, cosmic ray hits cause a large count on a single pixel, resulting in a bright spot with different profiles, unlike stellar ones. We utilise the {\tt IRAF} tasks $COSMICRAYS$ or $CRMEDIAN$ to remove the cosmic rays from all the observed frames. Another option to remove the cosmic rays from photometric frames is by median combining multiple frames as cosmic rays from random directions hit the CCD chip in random positions.

$\bullet$~{\bf Image stacking} \label{Image_stacking}

In several cases, the sources of our interest are not visible in single frames, and we need to stack multiple images of the same filter to detect the source. To detect the source of interest with significant SNR in the science frames, we first need to align and then combine the images of individual filters. To align the science images of a particular band, we utilise the $IMALIGN$ task while the $IMCOMBINE$ task from {\tt IRAF} is used to stack the aligned images.

$\bullet$~{\bf Template subtraction}

There are several incidences when an SN occurs close to nuclei or explodes in the bright spiral arm of the host galaxy. In such cases, the SN flux can be significantly contaminated with the flux of the corresponding host galaxy. This situation becomes even more problematic as the SN fades with time. In such situations, magnitude estimations will be erratic if the host galaxy contribution is not eliminated; hence, template subtraction becomes essential to remove the host galaxy contribution. To perform the template subtraction, we need to either obtain the pre-explosion template FITS files of the SN field from the sky surveys or, if the archival template FITS files from sk surveys are unavailable, the template can also be obtained by observing the same supernova field under good observing conditions, once the SN has gone beyond the detection limit of the employed telescope.

With the template image available, we need to do a few checks before subtracting the science images from the template image. First, we need to subtract the median sky background for each science and template image. The sky background counts for the template and science frames are determined at the locations in the images free from any sources. After that, the mean Full-Width-at-Half-Maximum (FWHM) values are constrained for the template and science frames using several bright enough isolated field stars. Between template and science frames, the frame with better FWHM is convolved to match the FWHM of the other so that the two images have similar point spread functions (PSFs). Once the FWHM matching is done, the next task is to match the flux of several sources in the template and science frames near the SN field. Then, a flux scaling factor between the science and template frames is determined. If the flux scaling factor is $>$1, we multiply the scaling factor with the frame with the lower flux or divide it from the frame with the higher flux. As a final step, we subtract the science frame from the template frame, as the science and template frames have achieved similar FWHM and flux values. The final image, resulting from template subtraction, is now ready for photometry.

\subsection{Photometry} \label{Photometry}

Once pre-processing of the raw data is performed, the science images are ready for photometry to estimate the brightness of the source SN of our interest. Photometry is a fundamental technique in astronomy involving the measurement of the flux of electromagnetic radiation emitted by astronomical sources. It plays an essential role in understanding the behaviours of stars, galaxies, transient events, and other astronomical phenomena. The estimated flux values can be expressed in terms of magnitude units that can be further employed to extract valuable information about the brightness, colour, variability, and other properties of the astronomical sources of interest. Two techniques are there to perform the photometry on a pre-processed science image depending upon the crowding in the considered field:

$\bullet$~{\bf Aperture photometry}

Aperture photometry is a beneficial and handy technique to estimate the total flux of an astronomical source within a defined aperture. Aperture photometry is preferred to estimate the magnitudes of sources in sparse astronomical fields consisting of uncrowded astronomical sources. In this photometry technique, we select an aperture of a suitable radius around the source of our interest to estimate the associated flux. The selected aperture could be a circular or annular region centred on the source of interest. However, a circular aperture is preferred over an annular one for aperture photometry because the field stars exhibit PSF profiles with functional Moffat, modified Lorentzian, or Gaussian.
To find the coordinates of the centre of the selected aperture, we utilise the $CENTERPARS$ task using {\tt DAOPHOT} available in {\tt IRAF}. This task fits one of the three functions mentioned above to determine the centre of the selected aperture.
The total flux in the defined aperture also consists of sky fluxes; thus, one has to remove the sky fluxes to measure the source flux precisely. To remove the sky background, we use the {\tt DAOPHOT} $FITSKYPARS$ task in {\tt IRAF}.

Selecting an appropriate size of the aperture is essential to perform precise photometry. The chosen aperture should be big enough to collect almost all the flux from the source of interest, but at the same time, it should not be very big to introduce additional noises from the sky background and nearby sources. For example, assuming a Gaussian function for the stellar profile, an aperture with a radius of around 3 to 4 times the FWHM of PSF is sufficient to collect more than 99\% of the star flux. When estimating the total flux, one can apply an aperture correction to account for the uncertainties associated with determining the appropriate aperture size. One can use the $PHOT$ task of {\tt DAOPHOT} in {\tt IRAF} to perform the aperture photometry.

We can summarise the aperture photometry as follow. We identify our source of interest in the field and select an appropriate aperture size. After that, we estimate the sky contributions and subtract that from the total flux in the defined aperture. We can also apply aperture corrections to account for the uncertainties of selecting an appropriate aperture radius. Finally, we perform the aperture photometry with the help of $PHOT$ task of {\tt DAOPHOT} available with {\tt IRAF}.

$\bullet$~{\bf PSF photometry}

Using aperture photometry in a crowded field is unsuitable since the nearby stars can contaminate the photometry of the source of our interest. Additionally, aperture photometry's linear sky background assumption is invalid in a crowded field. The sources in a crowded field can be so close that their PSFs may overlap, leading to erratic aperture photometry. With these limitations of aperture photometry, PSF photometry is the preferred technique to obtain reliable photometry in a crowded field. 
The stellar profile extends to several pixels due to atmospheric extinction. The Moffat, modified Lorentzian, or Gaussian functions are used to fit the stellar profiles in the PSF photometry. One of these three functions generates a standard fit to the PSF image by choosing a few bright and isolated field stars. One should be careful that the selected bright star is not saturated. This process of choosing appropriate stars to form the PSF is done using the {\tt DAOPHOT} task $PSTSELECT$, and the formation of the standard PSF is performed using the {\tt DAOPHOT} task $PSF$ in {\tt IRAF}. The FWHM of the stellar profile is utilised to determine the fitting radius, while the fitting parameters are adjusted to match the observed PSF. Additionally, the standard PSF image is scaled to match the PSF in the observed image and subsequently subtracted to isolate the source flux. The PSF magnitude of the object of interest is then estimated using the {\tt DAOPHOT} $ALLSTAR$ task in the {\tt IRAF}.

\subsection{Atmospheric extinction}
Atmospheric extinction refers to the phenomena of absorption or scattering of light from astronomical objects by the Earth's atmosphere. This causes significant attenuation in the observed brightness of those celestial sources. The effect of atmospheric extinction is particularly significant for ground-based observatories. The amount of decrement in the brightness of incoming light passing through Earth's atmosphere is primarily a function of the wavelength of incoming light, the altitude of the site, and the constituents of the atmosphere. There are two prime mechanisms in Earth's atmosphere contributing to atmospheric extinction:

$\bullet$~{\bf Scattering:} Earth's atmosphere consists of several molecules, aerosols, and other particles capable of scattering the incoming light from celestial objects in all directions. Two types of scattering phenomena occur in Earth's atmosphere that significantly decrease the intensity of incoming light. The first one is Rayleigh scattering. The coefficient of Rayleigh scattering varies inversely to the fourth power of the incoming light. As a result, blue and violet suffer more scattering than other longer wavelength lights, e.g., red and infrared light. Another type is the aerosol scattering arising from aerosols in Earth's atmosphere.

$\bullet$~{\bf Absorption:} A few atmospheric constituents, including ozone, water vapour, and a few other gasses, can selectively absorb specific wavelengths of light. For example, the ozone layer at a height of around 15 to 40\,km in the stratosphere significantly absorbs the incoming ultraviolet radiation.

Thus, atmospheric extinction has three main contributors; two from scattering (i.e., Rayleigh and aerosol scattering) and one from absorption. The total extinction coefficient is the sum of these three extinction coefficients. Extensive detail on atmospheric extinction is provided in \citet[][]{1975ApJ...197..593H}.

In addition to the atmospheric extinction, the local atmospheric conditions of any observatory on several seasons also affect the observations. The relation between the observed magnitude, m($\lambda$,$z$) of any celestial object on the Earth and its magnitude, m$_{0}$($\lambda$), outside the Earth's atmosphere is given by the famous Bouguer's linear formula as follow:

\begin{equation}\label{eq:mag8}
m(\lambda, z) = m_{o}(\lambda) + 1.086 \times A_\lambda \times M(z)
\end{equation}

Here, $A_\lambda$ denotes the wavelength-dependent extinction coefficient, $z$ is the zenith angle, and $M(z)$ represents the airmass at a zenith distance of $z$. Airmass refers to the extent of the Earth's atmosphere that the light from a celestial object has to traverse when observed from the surface of the Earth along the line of sight. The airmass for a specific astronomical source, assuming a plane-parallel atmosphere, can be obtained using the object's declination ($\delta$), the observer's latitude ($\phi$), the zenith angle ($z$), and the hour angle ($h$) at the time of observation. The formula to estimate the airmass is as follows:

\begin{equation}\label{eq:airmass}
M(z) = sec(z) = (sin\phi ~ sin\delta ~ + ~ cos\phi ~ cos\delta ~ cosh)^{-1}
\end{equation}

The assumption of a plane-parallel geometry for Earth's atmosphere is not true. Thus a correction term $\Delta$X is applied in the above Eq. to account for the curvature of the atmosphere. The correction term is given by:

\begin{equation}\label{eq:airmass_del}
\Delta X = 0.00186[sec(z) - 1] + 0.002875[sec(z) - 1]^2 + 0.0008083([sec(Z) - 1]^3
\end{equation}

With the correction term $\Delta$X, the final airmass equation takes the form as: 

\begin{equation}\label{eq:airmass_tot}
M(z) = sec(z) - \Delta X
\end{equation}

Eq.~\ref{eq:airmass_tot} indicates that light from any celestial object close to the zenith has to traverse a shorter path than when the same source is near the horizon. As a result, the source at the later position suffers from a higher extinction. In general, the extinction of a celestial source is directly related to the airmass and is also influenced by the wavelength of the incoming light. As mentioned earlier, it is necessary to account for atmospheric extinction and airmass when calculating instrumental magnitudes using aperture or PSF photometry techniques. This correction is applied to transform the measured magnitudes to what they would be if the observations were conducted outside the atmosphere. The formula for this correction is as follows:

\begin{equation}\label{eq:mag_instrumental}
m_\lambda^{0} = m_\lambda - A_\lambda \times M(z)
\end{equation}

In this Eq., $m_\lambda^{0}$ represents the instrumental magnitude at zero airmass (corresponding to the value if the observations were conducted outside the Earth's atmosphere), $m_\lambda$ represents the instrumental magnitude at a specific airmass.

\subsection{Photometric Calibration}

The instrumental magnitude of a source obtained through different photometric systems would be different. A photometric system comprises a telescope equipped with a filter and detector system at any observatory. We need to transform or calibrate the instrumental magnitudes to some standard system to use those magnitudes from different photometric systems together. The process of converting photometric system-dependent instrumental magnitudes to an independent standard system is known as photometric calibration. Once calibrated to a standard system, the magnitudes from several photometric systems can be used together for further analysis.

Along with the science frames, several photometric standard stars with a range of colours and brightnesses should also be observed under good observing conditions. The transformation from instrumental to calibrated magnitudes can be done through colour-colour or colour-magnitude transformation equations. In the present thesis, we have mainly employed the transformation equations from \citet[][]{Stahl2019} and \citet[][]{2022JApA...43...27K}.

\section{Spectroscopic Data Reduction}\label{Data_reduction_Spectroscopy}  
Spectroscopy in astronomy refers to the study of the variation of the flux of a celestial object in small wavelength bins. When light passes through a medium (e.g., gas or a cloud of dust), it interacts with the atoms, ions, and molecules present in the medium. These interactions cause certain wavelengths of light to be absorbed or emitted, which are reflected as absorption or emission feature in the spectrum. This is the basic principle behind spectroscopy. To observe and record the spectra of a source, spectrographs are attached with the telescopes. Spectrographs are instruments with the ability to disperse light into its component wavelengths.
Spectroscopy is an excellent method for exploring the chemical composition of celestial sources and extracting crucial information such as surface gravity, radial velocities, redshift, metallicity, and many more. Under the context of the present thesis, we have conducted a comprehensive analysis utilising slit-based low-resolution spectroscopic data of several CCSNe. These observations were made possible through the utilisation of many telescopes and detectors, enabling a detailed investigation of the physical and chemical properties of these catastrophic explosions. 
 
The spectrograph receives light from the celestial object, which is then dispersed perpendicular to the slit axis, referred to as the ``dispersion axis". The slit serves the purpose of blocking unwanted light and selectively collecting light from the source of our interest. In a specific observation, the optimal width of the slit is selected to match the seeing-limited FWHM. The dispersion axis carries valuable information about the distribution of flux along the spatial axis of the CCD.

To ensure precise wavelength and flux calibrations, the spectra from arc lamps such as Fe-Ar, Fe-Ne, and Neon, as well as standard star spectra, are also obtained. Additionally, bias and flat frames are also acquired to facilitate preprocessing of the raw CCD data. The subsequent subsections will outline the primary steps followed during the process of spectroscopic data reduction.

\subsection{Pre-processing} 

As mentioned previously, while discussing the pre-processing of photometric data, it is also mandatory to preprocess the raw spectroscopic data to eliminate the CCD artefacts and cosmic ray hits. Thus, the spectroscopic data also deal with bias, flat, and cosmic ray corrections. Techniques similar to photometric data preprocessing are employed for bias and cosmic ray hit corrections. However, instead of using uniformly illuminated twilight sky flat frames, halogen lamp spectra are obtained for applying the flat correction.

To create the master flat frame, multiple bias-corrected halogen lamp frames are median combined using the $FLATCOMBINE$ task in IRAF. The next step involves generating a normalised flat frame. This is achieved by fitting a high-order spline function to the halogen spectrum extracted from the master flat frame, utilising the $RESPONSE$ task in IRAF. The fitted function is then divided to the master flat frame to produce the normalised master flat frame.

Applying the flat correction involves dividing the normalised master flat frame from the bias-corrected arc-lamp frames, standard star frames, and science frames. This is accomplished using the $IMARITH$ task in IRAF. Additionally, to account for the non-linear behaviour typically observed at the edges of the CCD, any necessary trimming of the 2-dimensional spectrum edges can be performed using the $IMCOPY$ task in IRAF.

\subsection{Spectrum~extraction} 

In this section, we discuss the process of extracting a 1-dimensional spectrum from a pre-processed 2-dimensional spectroscopic image. The first step for extracting the spectrum involves the identification of a dispersion line. Generally, the dispersion line is expected to be located in the middle of the spatial axis. However, to determine its precise location, a cut is made along the spatial axis, and the peak coordinates of the Gaussian profile are recorded. Once we identify the centre of the dispersion line, the next step involves extracting the aperture and subtracting the background.

The aperture, characterised by its base-to-base width and peak, can be easily traced as the cross-section of the dispersion line or the stellar profile has a Gaussian-like shape. Background windows are selected on either side of the Gaussian peak. A low-order ``Chebyshev function" is employed to estimate the background profile, which is then subtracted from the aperture along that column.

It is important to consider that the dispersion line is not exactly perpendicular to the spatial axis of the CCD due to geometric distortions caused by telescope optics, grating alignment, and the gradient of atmospheric refraction with respect to the wavelength. Consequently, we generate the 1-dimensional spectrum along the dispersion axis by extracting the peak counts along the dispersion axis and subsequently summing them across the spatial axis.

 All the above-mentioned tasks to extract the 1-dimensional spectrum can be performed using the $APALL$ task in {\tt IRAF}. 
Extracting the standard star spectra observed on the same night also follows a similar procedure. However, the extraction of arc lamp spectra for the wavelength calibration of the object and standard star spectra is performed with the same trace and centring position. To obtain this, the $APALL$ task is used again, with the object or standard star spectrum as the reference. The extracted spectrum of the source of interest consists of pixel numbers versus count values, which need to be converted into wavelength and flux. We perform the wavelength and flux calibration tasks to achieve these conversions.

\subsection{Wavelength and Flux calibration} 

~~~~~$\bullet$~{\bf Wavelength calibration}

This task aims to transform the pixel numbers on the x-axis of the extracted spectrum into wavelength units. The arc-lamp spectra have several characteristic emission features with precisely known corresponding wavelengths.
Utilising the $IDENTIFY$ task in {\tt IRAF}, these emission features can be easily identified.
The line identification output data are fitted using low-order polynomials to prepare the pixel-to-wavelength mapping solution. Further, we use the fitted polynomial onto other 1-dimensional spectra of the source and standard stars utilising the $DISPCOR$ task in {\tt IRAF}. As a result, we obtain the wavelength-calibrated spectra of the source of interest and standard stars.

$\bullet$~{\bf Flux~calibration} 

The flux calibration task aims to transform the y-axis instrumental counts of the wavelength-calibrated spectrum to flux units (i.e., in units of erg\,s$^{-1}$\,cm$^{-2}$\,\AA$^{-1}$).
To establish the relationship between CCD counts and flux at different wavelengths, we utilise the standard-star spectra and employ splines to fit the continua. This fitting process gives sensitivity functions that provide a mapping from CCD counts to flux for each specific wavelength. Finally, these sensitivity functions are utilised to calibrate and adjust the flux values in each individual spectrum of the source of our interest. The detailed processes of wavelength and flux calibration are achieved utilising {\tt IRAF} routines and custom Python and IDL-based scripts\footnote{https://github.com/ishivvers/TheKastShiv}. These processes are explicitly described in \citet[][]{2012MNRAS.425.1789S}.

\section{Supernova Progenitor Modelling Using the Modules for Experiments in Stellar Astrophysics (MESA)}

{\tt MESA}, the acronym for the Modules for Experiments in Stellar astrophysics \citep[][]{Paxton2011,Paxton2013,Paxton2015,Paxton2018, Paxton2019,2023ApJS..265...15J} represents a collection of powerful, open-source libraries designed to serve to a diverse range of computational needs within the realm of stellar astrophysics. These libraries are well known for their reliability, efficiency, and ability to operate seamlessly in multi-threaded environments. {\tt MESA} can tackle a diverse range of astrophysical situations. It is capable of performing the stellar evolution of very low mass stars to very old ages, can evolve massive stars from pre-main sequence up to the core-collapse stage, evolve very massive stars to the PISN or PPISN stages depending upon their masses, and also capable of simulating the evolution of accretion onto an NS. Thus, {\tt MESA} can mimic various astrophysical events ongoing in our Universe. However, within the context of the research work in the presented thesis, we have utilised {\tt MESA} to evolve massive stars (M$_{\rm ZAMS}$\,$>$\,8\,M$_{\odot}$) from pre-main sequence up to the onset of core collapse. These models at the stage of the beginning of core collapse serve the purpose of inputting pre-SN models to other simulation tools that simulate their synthetic, hydrodynamic explosions.

\begin{figure*}
\centering
  \begin{tabular}{cc}
    {\includegraphics[height=7cm,width=0.49\columnwidth]{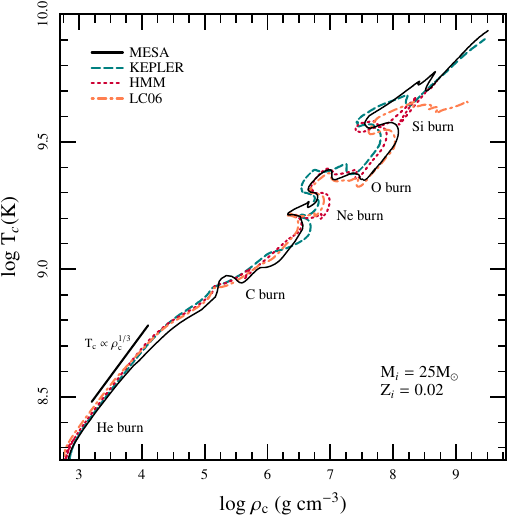}}
    {\includegraphics[height=7cm,width=0.49\columnwidth]{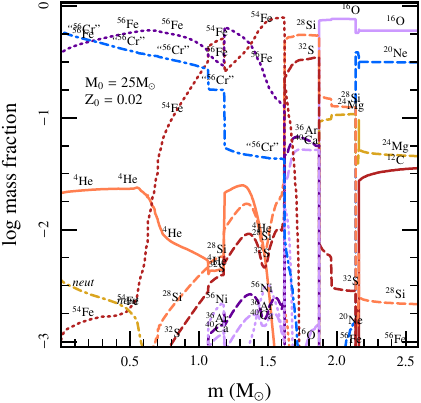}}
  \end{tabular}
  \caption{{\em Left:} The variation of the core temperature against core density as the 25\,M$_{\odot}$ model with Z $=$ 0.02 evolves through various stages of burning in its core is shown. The results of other available evolutionary codes are also shown for comparison.
 {\em Right:} The mass fractions of several species at the stage of the onset of core collapse are shown. The figures are adapted from \citet[][]{Paxton2011}.}
  \label{fig:fig_MESA_use}
\end{figure*}

The capability of {\tt MESA} to simulate the stellar evolution of a massive star starting from pre-main sequence up to the onset of core collapse is well documented in \citet[][]{Paxton2011}. The authors of \citet[][]{Paxton2011} have evolved a 25\,M$_{\odot}$ ZAMS mass star up to the onset core collapse stage. They have studied the evolution of the model on the HR diagram. The authors have also investigated the variation of the core-temperature versus core-density curve (as shown in the left-hand panel of Figure~\ref{fig:fig_MESA_use}) throughout the entire evolution of the model. They have also studied models' mass fractions (as shown in the right-hand panel of Figure~\ref{fig:fig2}) at the onset of the core collapse stage. With these capabilities, {\tt MESA} caters to excellent opportunities to investigate the entire evolution of massive stars.

\section{Tools to Simulate the Synthetic Explosions of pre-SN Models}
Once the models have arrived at the stage of the onset of core collapse, we can utilise the output of {\tt MESA} in appropriate form as input to other SN explosion codes capable of simulating the synthetic explosions of the {\tt MESA} pre-SN models. There are several codes available to simulate synthetic explosions from pre-SN models. However, we have utilised two codes for the research work performed in the present thesis mentioned below:

\subsection{SuperNova Explosion Code (SNEC)}

\begin{figure*}
\centering
  \begin{tabular}{cc}
    {\includegraphics[height=7cm,width=0.49\columnwidth]{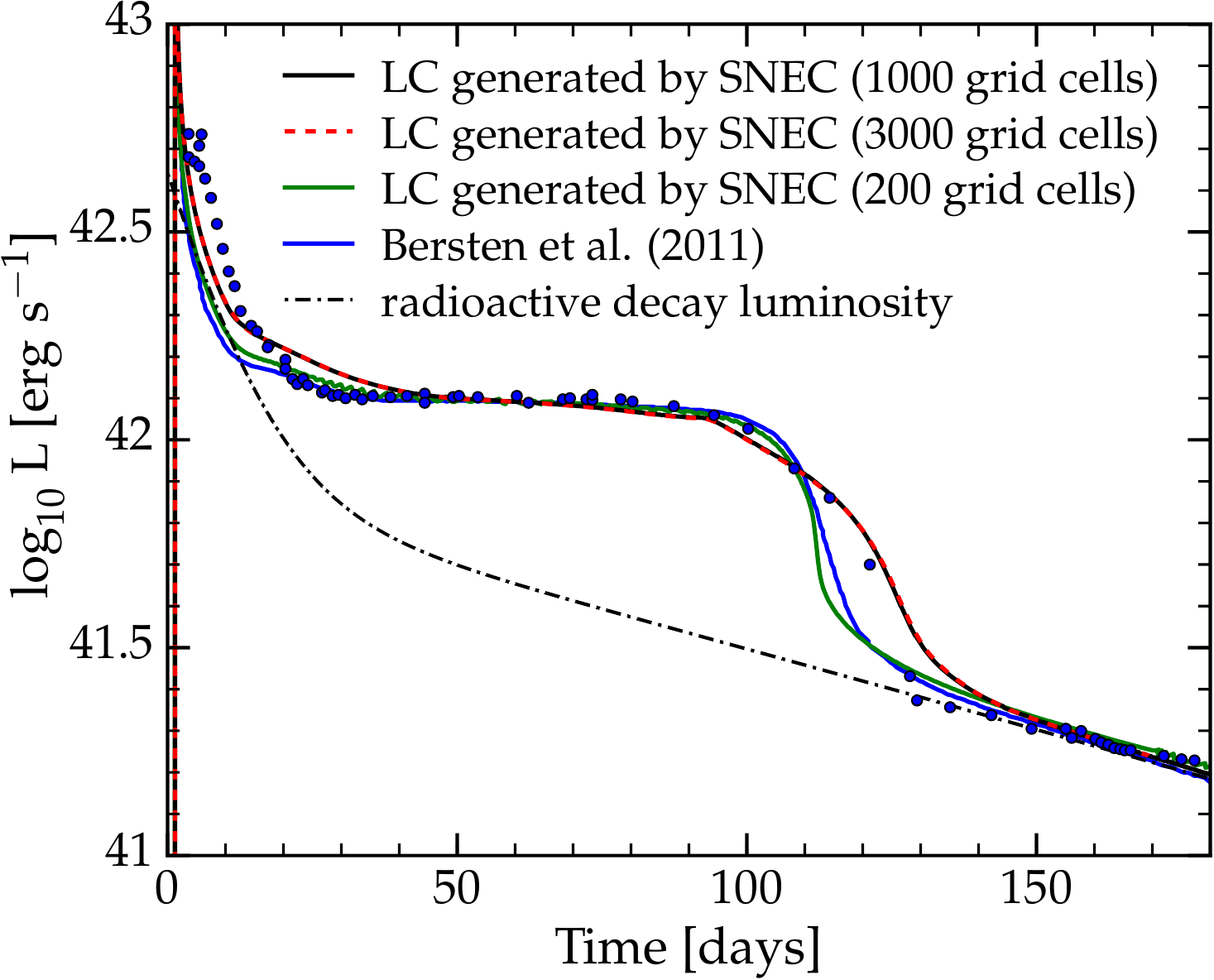}}
    {\includegraphics[height=7cm,width=0.49\columnwidth]{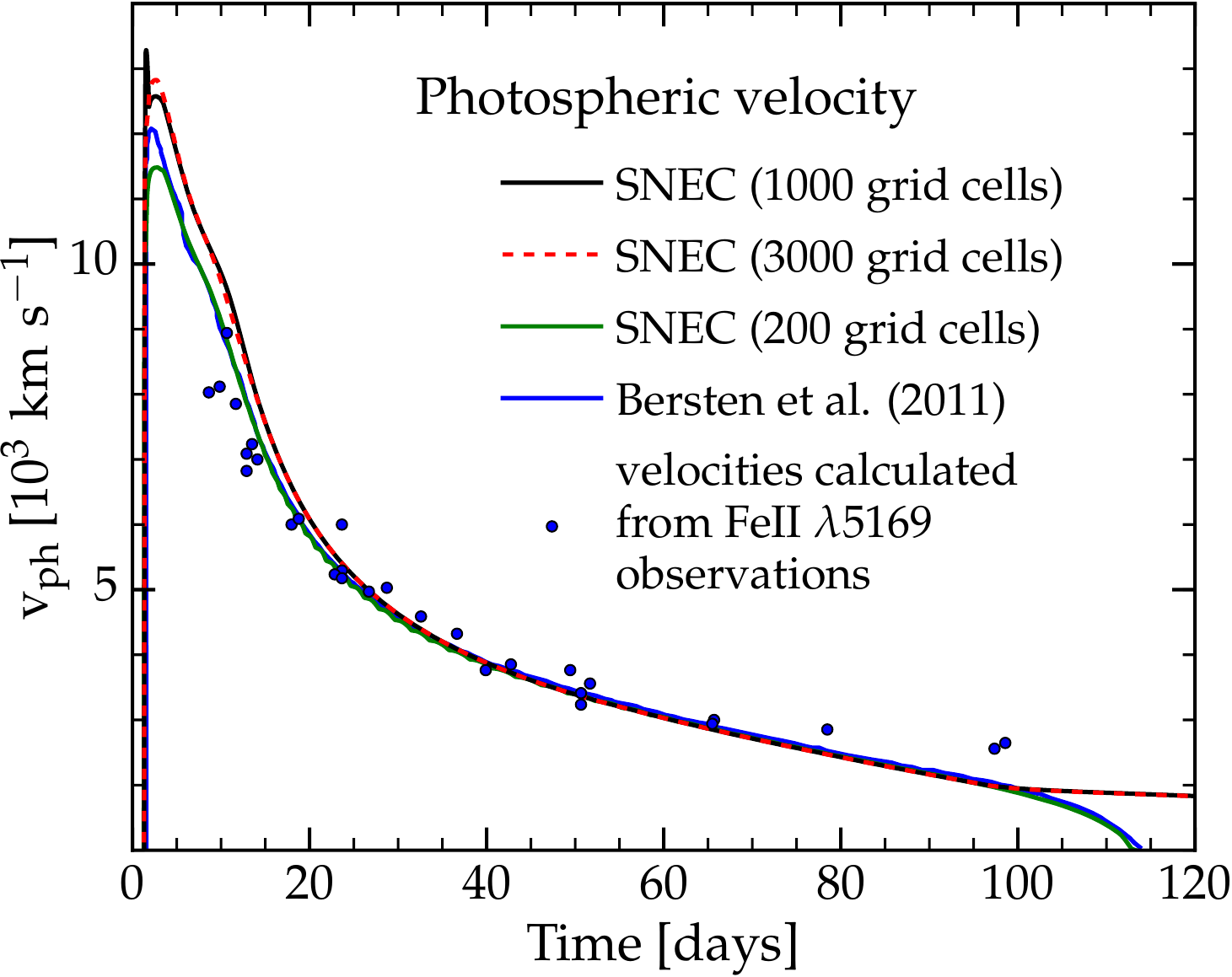}}
  \end{tabular}
  \caption{{\em Left:} The comparison of {\tt SNEC} generated bolometric light curves with actual observations of SN~1999em are shown. The effect of increasing grid cells is also explored.
 {\em Right:} The comparison of corresponding {\tt SNEC} generated velocity evolutions is compared with actual Fe~II line velocities of SN~1999em. The results of \citet[][]{2011ApJ...729...61B} are also shown for comparison purposes. Figure credit: \citet[][]{Morozova2015}.}
  \label{fig:fig_SNEC_use}
\end{figure*}

{\tt SNEC}, the acronym for SuperNova Explosion Code \citet[][]{Morozova2015} is a publicly available, 1-dimensional Lagrangian hydrodynamic code. {\tt SNEC} solves the radiation energy transport equation within the flux-limited diffusion approximation to simulate the synthetic explosion of pre-SN models. To synthetically explode the pre-SN model, {\tt SNEC} requires structure and chemical composition information. Thus, once a certain mass ZAMS model has reached the onset of the core collapse stage using {\tt MESA}, the pieces of information regarding the structure and chemical composition of the pre-SN model are mapped in the appropriate form, and those pieces of information serve as input to {\tt SNEC}, that explodes the pre-SN model.

The setups for the explosion in {\tt SNEC} are controlled by a file named {\tt parameters}. The type of explosion, explosion energy, $^{56}$Ni mass, $^{56}$Ni distribution within the model, the mass of the excised core, number of grid cells, duration upto which the light curves are generated etc., are controlled by the user in the {\tt parameters} file. For the explicit details on the working of {\tt SNEC} and underlying explosion parameters, one should consider \citet[][]{Morozova2015}. After finishing the calculation, the code generates the bolometric light curve, multi-band light curves, and the photospheric velocity evolution of the SN, along with a few more physical and chemical characteristics. The model-generated light curves and velocity evolution are compared with actual observations as shown in the left- and right-hand panels of Figure~\ref{fig:fig_SNEC_use}.

\begin{figure*}
\centering
  \begin{tabular}{cc}
    {\includegraphics[height=7cm,width=0.49\columnwidth]{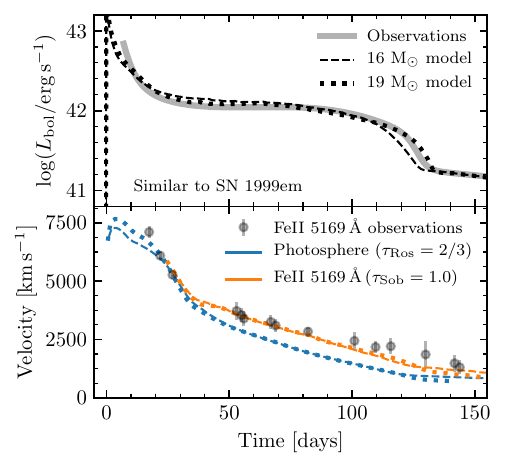}}
    {\includegraphics[height=7cm,width=0.49\columnwidth]{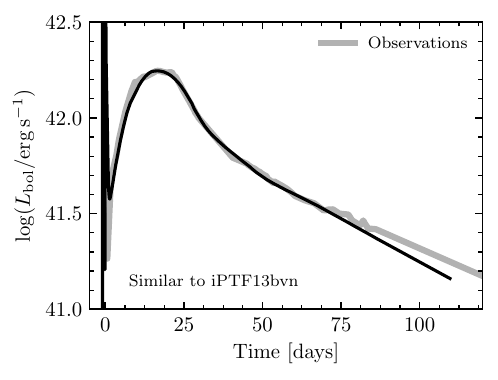}}
  \end{tabular}
  \caption{{\em Left:} The comparison of {\tt STELLA} generated bolometric light curves with actual observations (top panel) of SN~1999em are shown. The comparison of corresponding velocity evolution is also shown (bottom panel). {\em Right:} The capacity of {\tt STELLA} to model a few stripped envelope is displayed. This figure compares the {\tt STELLA} produced bolometric light curve with a Type Ib SN iPTF13bvn observations. Figure credit: \citet[][]{Paxton2018}.}
  \label{fig:fig_STELLA_use}
\end{figure*}

\subsection{STELLA}
{\tt STELLA} \citep[][]{Blinnikov1998, Blinnikov2000, Blinnikov2006} is another 1-dimensional multi-group radiation hydrodynamics code. A simplified version of {\tt STELLA} is available publicly in {\tt MESA}.
In contrast to {\tt SNEC}, it tackles the radiative transfer equations through the intensity momentum approximation in each frequency bin. Within {\tt MESA}, the utilisation of {\tt STELLA} commonly involves employing a maximum of 40 frequency groups. This selection of 40 frequency bins equips {\tt STELLA} with the ability to generate SEDs of exploding models; however, with 40 frequency bins, {\tt STELLA} fails to produce complete spectra. Extensive details on massive star explosions and utilisation of {\tt STELLA} to generate bolometric luminosity light curves, multi-band light curves, and velocity evolutions are well documented in \citet[][]{Paxton2018}. The utilisation of {\tt STELLA} to model the observations of H-rich and H-deficient CCSNe is shown in the left- and right-hand panels of Figure~\ref{fig:fig_STELLA_use}.

%*************************************************************************
 %Observation
\chapter{\sc Constraints on the progenitor masses of two Type Ib Supernovae, SN 2015ap and SN 2016bau}
\label{Ch:3}
%\blfootnote{ The results of this chapter are:}
\blfootnote{ The results of this Chapter are published in: \textbf{{Aryan}, Amar,} {Pandey}, S. B., {Zheng}, W. et al., 2020,  {\textit{MNRAS}, {\textbf{505}}, 2530} and \textbf{{Aryan}, Amar,} {Pandey}, S. B., {Kumar}, A. et al., 2022,  {\textit{JApA}, {\textbf{43}}, 87.}}
\ifpdf
    \graphicspath{{Chapter3/Chapter3Figs/JPG/}{Chapter3/Chapter3Figs/PDF/}{Chapter3/Chapter3Figs/}}
\else
    \graphicspath{{Chapter3/Chapter3Figs/EPS/}{Chapter3/Chapter3Figs/}}
\fi
    
\section{Current Understanding in the Research Area}
\label{sec:Introduction}
As detailed in earlier Chapters~\ref{Ch:1} and \ref{Ch:2}, CCSNe are among the most powerful astronomical explosions, occurring during the final stellar evolutionary stages of massive stars having initial mass $\geqslant$8\,M$_{\odot}$ \citep[][]{Garry2004, Woosley2005, Groh2017}). Knowledge of the possible progenitors of CCSNe is one of the most fundamental tasks in understanding these catastrophic explosions. Therefore, in this Chapter, we aim to explore the origins of the underlying progenitors of two Type Ib CCSNe by analysing their photometric and spectroscopic data along with simulations involving hydrodynamic modelling of their explosions. 

Due to the limited understanding of the potential progenitors of H-stripped CCSNe, people have started paying special attention to this research area. For SNe~Ib/c, broadly, two scenarios are proposed for their progenitor system. The first involves relatively low-mass progenitors ($>$\,11\,M$_{\odot}$) in binary systems  \citep[][]{Podsiadlowski1992, Nomoto1995, Smartt2009}, where the primary star lost its H-envelope through the transfer of mass to a companion star. The second considers massive WR stars ($>$\,20--25\,M$_{\odot}$) that lose mass via stellar winds \citep[e.g.,][]{Gaskell1986, Eldridge2011, Groh2013}. For a Type Ib SN iPTF13bvn, the authors of \citet[][]{Cao2013} reported the detection of a possible progenitor in pre-explosion images within a 2$\sigma$ error radius, consistent with a massive WR progenitor star. Stellar evolutionary models also support the massive WR progenitor scenario. Based on observational evidence from early- and nebular-phase spectroscopy of  SNe~Ib, both massive WR stars and interacting binary progenitors are proposed. The hydrodynamic modelling of the possible progenitors (identified either via direct imaging as in the case of iptf13bvn \citep[][]{Cao2013} or indirect methods which include nebular-phase spectral modelling \citealt[][]{Jerkstrand2015, Uomoto1986}) and simulating their synthetic explosions can be vital to understanding their nature, physical conditions, circumstellar environment, and chemical compositions. Unfortunately, only a handful of such studies have been performed in the cases of stripped-envelope SNe, including the Type Ib SN iptf13bvn \citep[][]{Cao2013, Bersten2014, Paxton2018}, the famous Type IIb SN~2016gkg \citep[][]{Bersten2018}, and the Type IIb SN~2011dh \citep[][]{Bersten2012}. Our work takes such studies one step further as we perform the stellar evolution of the possible progenitors of two new SNe~Ib and simulate their synthetic explosions.

Another aspect of CCSNe is understanding the underlying powering mechanisms behind their light curves. The main powering mechanism in normal SNe~Ib/c is radioactive decay of $^{56}$Ni and $^{56}$Co, leading to the deposition of energetic gamma rays that thermalise in the homologously expanding ejecta \citep[e.g.,][]{Arnett1980, Arnett1982, Arnett1996, Nadyozhin1994, Chatzopoulous2013, Nicholl2017}. Some SNe~Ib (e.g., SN~2005bf, \citealt[][]{Maeda2007}) have also shown evidence for the light curves being powered by the spin-down of a young magnetar  \citep[e.g.,][]{Ostriker1971, Arnett1989, Maeda2007, Kasen2010, Woosley2010, Chatzopoulous2013, Nicholl2017}. In the post-photospheric phase, when the SN ejecta become optically thin, the light curves of CCSNe are powered by energy deposition from the radioactive decay of $^{56}$Ni to $^{56}$Co  and finally to $^{56}$Fe. In many cases, however, the SN progenitor is embedded within dense circumstellar matter (CSM), so when the SN explosion occurs, the SN ejecta may violently interact with the CSM, resulting in the formation of forward and reverse shocks that deposit their kinetic energy into the material which is radiatively released and thus powers the light curve \citep[e.g.,][]{ Chevalier1982, Chevalier1994, Moriya2011, Ginzberg2012, Chatzopoulous2013, Nicholl2017}.

In this Chapter, we explore the photometric and spectroscopic behaviour of SN~2015ap and SN~2016bau, primarily using the data obtained using the KAIT, Nickel, and Shane telescopes at Lick Observatory. Based on the analysis, we model and attempt to place constraints on the properties of the progenitors of these two SNe. In Sec.~\ref{sec:Data_red}, details about various telescopes and reduction procedures are presented. 
Sec.~\ref{sec:Photometric} provides methods to correct for the Milky Way and the host-galaxy extinction. Photometric properties of the two SNe, such as their bolometric light curve, temperature, radius, and velocity evolution, are also discussed. Sec.~\ref{sec:Spectral} includes the analysis describing the spectral evolution of SN~2015ap and SN~2016bau, as well as comparisons with other similar and well-studied SNe; we also model the spectra of these SNe using {\tt SYN++}. In Sec.~\ref{mosfit}, we have modelled the multi-band light curves of SN~2015ap and SN~2016bau utilising {\tt MOSFiT}. Further, quasi-bolometric light-curve modelling is performed in Sec.~\ref{sec:lc_model}; light curves corresponding to various powering mechanisms of SNe are fitted to the observed quasi-bolometric light curves. The assumptions and methods for modelling the possible progenitors of the two SNe and their evolution until the onset of core-collapse using {\tt MESA} are presented in  Sec.~\ref{sec:mesa}. We discuss the assumptions and methods for producing the synthetic explosions using {\tt SNEC} and {\tt STELLA} in Sec.~\ref{sec:snec}; here, the comparisons between the parameters obtained through synthetic explosions and observed ones are presented. In Sec.~\ref{sec:Discussions}, we discuss our analysis's major results and findings. Finally, in Sec.~\ref{sec:Conclusions}, we summarise our work by briefly discussing the major outcomes and providing concluding remarks. 

\begin{figure*}
\centering
\includegraphics[width=\textwidth]{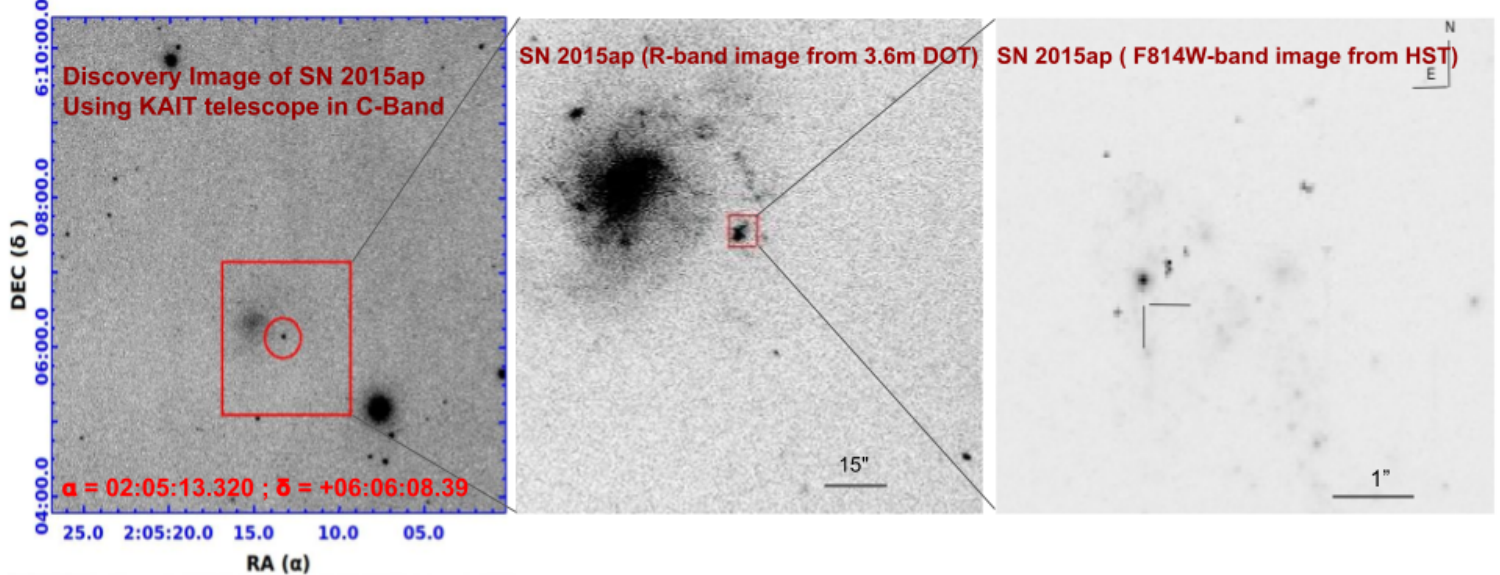}
\caption{Left panel: The discovery image of SN~2015ap using the KAIT telescope in $C$-filter. The SN is marked inside the red circle. Middle panel: A zoomed version of an $R$-band image with $2' \times 2'$ FoV obtained on 6 Oct. 2020 using the 4K$\times$4K CCD imager mounted on the 3.6\,m DOT. Right panel: A further zoomed, about $6'' \times 6''$ FoV of the location of SN~2015ap taken on 12 Aug. 2020 using the HST in the F814W filter. The SN seems to be fainter than the detection limit of the HST at this time. A log of HST observations is included in the Appendix.}
\label{fig:finding_chart15ap}
\end{figure*}

\section{Data acquisition and reduction }
\label{sec:Data_red}

SN~2015ap was discovered by KAIT as part of the Lick Observatory Supernova Search \citep[LOSS;][]{Filippenko2001}, in an 18\,s unfiltered image (close to the $R$ band; see \citealt{Li2003}) taken at  11:16:31 on 2015~Sep.~08 \citep[][]{Ross2015}, at 17.41$\pm$0.06\,mag. The object was also marginally detected one day earlier on Sep. 07.47 with 18.07$\pm$0.22\,mag. We measure its J2000.0 coordinates to be $\alpha=02^{\rm h}05^{\rm m}13.^{\rm s}32$, $\delta=+06^{\circ}06^{'}08.^{''}4$, with an uncertainty of $0.^{''}5$ in each coordinate. SN~2015ap is $28.^{''}3$ west and $16.^{''}2$ south of the nucleus of its host galaxy IC~1776, which has a redshift of $z$ = 0.011375$\pm$0.000017 \citep[][]{Chengalur1993} and a barred-spiral (SB(s)d) morphology as mentioned in \citealt[][]{deVaucouleurs1991}. SN~2016bau was discovered by Ron Arbour in an unfiltered image taken at 23:22:33 on 2016~Mar.~13  \citep[][]{2016TNSTR.215....1A}, at 17.8\,mag.  Its J2000.0 coordinates are given as $\alpha=11^{\rm h}20^{\rm m}59.^{\rm s}02$, $\delta=+53^{\circ}10^{'}25^{''}6$. SN~2016bau is $35.^{''}3$ west and $15.^{''}2$ north of the nucleus of its host galaxy NGC~3631, which has $z$ = 0.00384$\pm$0.00014 \citep[][]{Falco1999} and a  morphology of SAc~C \citep[][]{Ann2015}. Both of these SNe were classified as SN~Ib based on well-developed features of He~I, Fe~II (blended), and Ca~II, a few days after maximum brightness.

Figure~\ref{fig:finding_chart15ap} shows finder charts of SN~2015ap obtained with different telescopes. In the HST image, for the location of SN~2015ap and the cluster from which it came, astrometry was done between the ground-based image (picked up from the ESO\footnote{https://www.eso.org/public/} archive) and the HST image. A coordinate transformation was performed between the two images, which allowed the position of the SN to be found. Then, photometry was done using {\tt DOLPHOT} \citep[][]{2000PASP..112.1383D,2016ascl.soft08013D}, a photometry pipeline developed for HST data. While the SN was not found in this finder chart, we outlined only the star cluster close to the reported position of SN~2015ap.

The $B$, $V$, $R$, and $I$ follow-up images of SN~2015ap and SN~2016bau were obtained with both the 0.76\,m KAIT and the 1\,m Nickel telescope at the Lick Observatory. We reduced all the images using a custom pipeline\footnote{https://github.com/benstahl92/LOSSPhotPypeline} detailed by \citet[][]{Stahl2019}. Here, we briefly summarise the process of photometry. Image subtraction techniques were employed to eliminate the background light from the host galaxy, utilising supplementary images acquired subsequent to the SN's decline below our detection threshold. Point-spread-function (PSF) photometry was executed through the utilization of DAOPHOT \citep[][]{Stetson1987} from the IDL Astronomy User\textquotesingle s Library\footnote{http://idlastro.gsfc.nasa.gov/}.
Three nearby stars were chosen from the Pan-STARRS1\footnote{http://archive.stsci.edu/panstarrs/search.php} catalogue for calibration. The magnitudes of these objects were initially converted to \citet{Landolt1992} magnitudes using the empirical prescription presented by \citet[][Eq. 6]{Torny2012} and thereafter converted to the KAIT/Nickel natural system. We measured all the apparent magnitudes in the KAIT4/Nickel2 natural system. The ultimate outcomes were then transformed to the standard system using local reference sources and accounting for colour terms for KAIT4 and Nickel2, as documented by \citet[][]{Stahl2019}. In addition, the $U$-band photometry was obtained from the {\it Swift} Optical/Ultraviolet Supernova Archive (SOUSA; https://archive.stsci.edu/prepds/sousa/; \citealt[][]{Brown2014}).

There are a total of 17 spectra of SN~2015ap; 15 of them were taken with the Kast spectrograph\footnote{https://mthamilton.ucolick.org/techdocs/instruments/kast} \citep{Miller&Stone1993} on the 3\,m Shane telescope at Lick Observatory and the other 2 were taken with LRIS\footnote{https://www2.keck.hawaii.edu/inst/lris/lrishome.html} \citep{Oke1995} on the Keck-I 10\,m telescope. Except for the earliest spectrum taken on 10.531 Sep. 2015 (UT dates are used throughout this Chapter) using the Lick Shane/Kast system, 16 spectra in this Chapter were published by \citet[][]{Shivvers2018}; refer to that paper for details of the observations. Also, there are 8 spectra of SN~2016bau, all obtained with the Lick Shane/Kast system. We used  the 600/4310 grism on the blue side, the 300/7500 grating on the red side, the d5500 dichroic, and a long slit $2''$ wide. For LRIS we used the 600/4000 grism on the blue side, the 400/8500 grating on the red side, and a $1''$ slit. The spectra were binned to  2\,\AA\ pixel$^{-1}$.

\section{Photometric Properties}
\label{sec:Photometric}
In this section, we discuss various photometric properties of SN~2015ap and SN~2016bau, including their colour evolution, extinction, quasi-bolometric light curves, and various blackbody parameters.

\begin{figure}
	\includegraphics[width=\columnwidth]{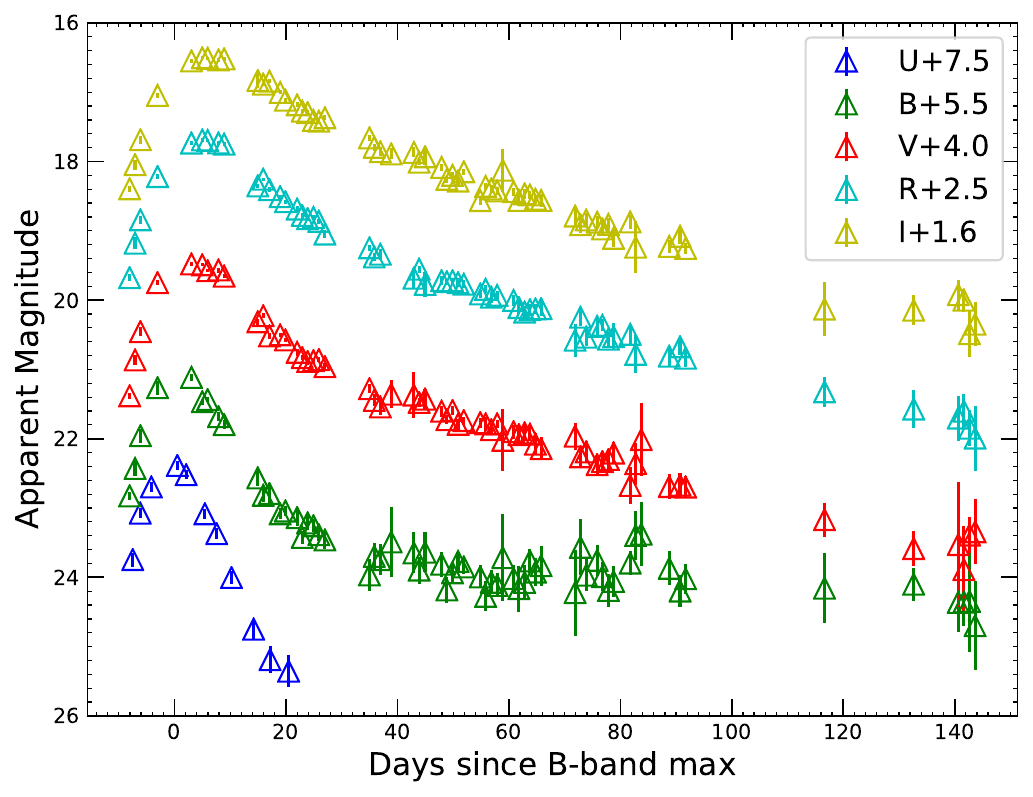}
    \caption{$UBVRI$ light curves of SN~2015ap, where $BVRI$ data were obtained with KAIT while the $U$-band data were taken from the UVOT mounted on {\it Swift} (https://swift.gsfc.nasa.gov).}
    \label{fig:UBVRI_LC}
\end{figure}

\begin{figure}	
    \includegraphics[width=\columnwidth]{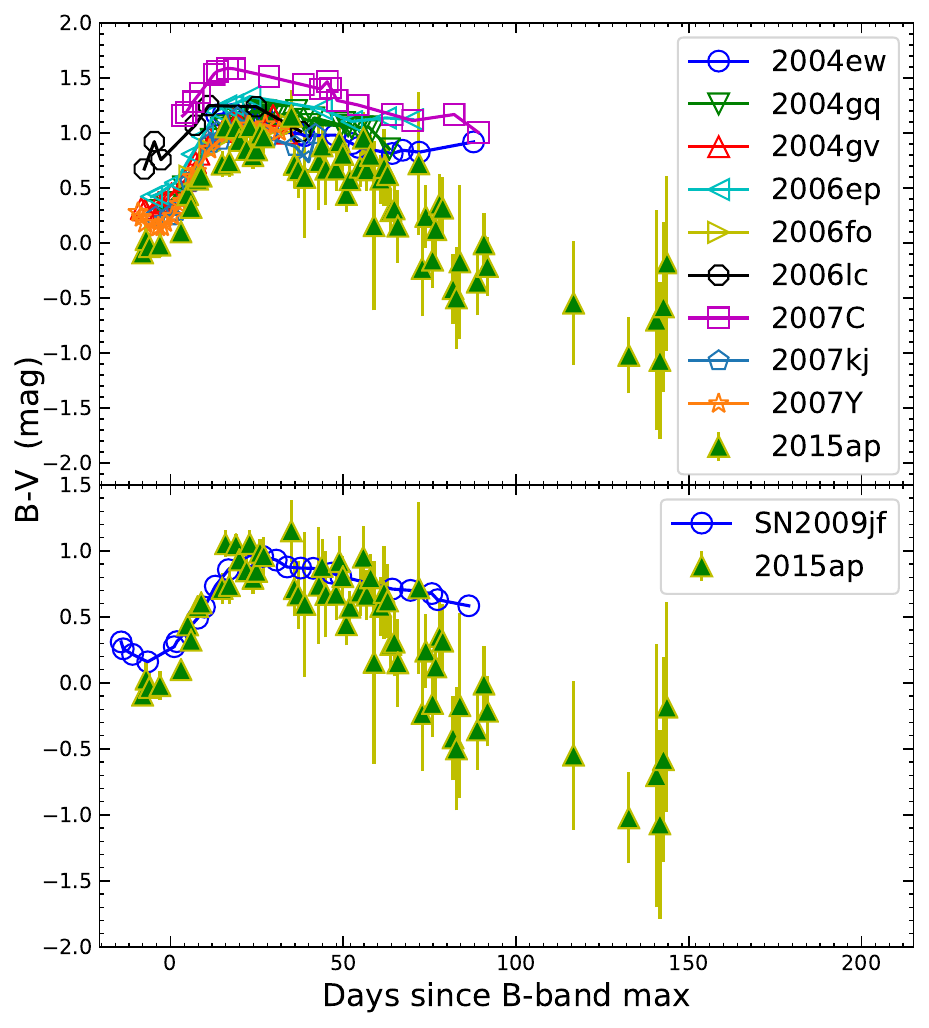}
    \caption{The top panel shows a comparison of the $(B-V)$ colour of SN~2015ap (corrected for Milky Way extinction) with that of other Type Ib SNe. The data for SNe other than SN~2015ap are taken from \citet[][]{Stritzinger2018}. The bottom panel shows the $(B-V)$ colour curves of SN~2015ap and SN~2009jf (both corrected for Milky Way extinction).}
    \label{fig:color_curve_2015ap}
\end{figure}

\subsection{Photometric behaviour of SN~2015ap}
\label{subsec:Photometric_SN2015ap}

Most of the analysis in this Chapter has been performed with respect to $B$-band maximum brightness. To find its date, we fit a sixth-order polynomial to the $B$ data sampling the photospheric phase. The resulting date of $B$-band maximum is MJD 57282.47$\pm$2.56. 
To determine the explosion epoch ($t_{\rm exp}$) of SN~2015ap, we use $R$-band data. The $R$ magnitudes are converted into fluxes, and a sixth-order polynomial is fitted. We extrapolate the polynomial, and the epoch corresponding to zero flux is taken as the explosion epoch, MJD 57272.72$\pm$1.49, which is in good agreement with \citet[][]{Prentice2019}.

Figure~\ref{fig:UBVRI_LC} shows the $UBVRI$ light curves of SN~2015ap. The rising rate for the $U$-band light curve is faster than that of other bands, and similarly, the $U$ decline rate is faster, making the $U$ light curve much narrower compared to other bands. As we go to longer wavelengths, we see that the light curves become broader, with the $I$-band light curve being the broadest.

$\bullet$~{\bf Colour evolution and extinction correction}

Distances are taken from the NASA Extragalctic Database\footnote{https://ned.ipac.caltech.edu/} (NED), and a cosmological model with H$_{0}$ = 73.8\,km\,s$^{-1}$\,Mpc$^{-1}$, $\Omega_{m}$ = 0.3, and $\Omega_{\Lambda}$ = 0.7 is assumed throughout.
For SN~2015ap, we corrected for Milky Way (MW) extinction using NED following \citet[][]{Schlafly2011}. In the direction of SN~2015ap, the Galactic extinctions for the $U$, $B$, $V$, $R$, and $I$ bands are 0.185, 0.154, 0.117, 0.092, and 0.064\,mag, respectively. 

The top panel of Figure~\ref{fig:color_curve_2015ap} shows a comparison of the $(B-V)$ colour of SN~2015ap with that of other Type Ib SNe (all corrected for MW extinction). SN~2015ap seems to be the least reddened and lies below nearly all of the other SNe~Ib.

Following \citet[][]{Prentice2019}, the host-galaxy contamination is negligible and hence ignored. To further support this assumption, the $(B-V)$ colour curve of SN~2009jf \citep[][]{Sahu2011} is shown in the bottom panel of Figure~\ref{fig:color_curve_2015ap}, corrected for an MW colour excess of $E(B-V)_{\rm MW} = $ 0.112\,mag (host extinction is negligible). In order to match the ($B-V$) colour curve of SN~2009jf, we need not apply any shift to the MW-corrected SN~2015ap $(B-V)$ colour curve.

Figure~\ref{fig:miller} illustrates the position of SN~2015ap in the Miller diagram \citep[][]{Richardson2014}. The distance modulus ($\mu$) of SN~2015ap is 33.269\,mag; thus, SN~2015ap appears to be a normal SN~Ib. For SN~2016bau, we calculate $\mu =$ 32.65\,mag; based on its position in Figure~\ref{fig:miller}, it seems to be a moderately luminous, normal SN~Ib.

$\bullet$~{\bf Quasi-bolometric and bolometric light curves}

To obtain the quasi-bolometric light curves, we used the {\tt superbol} code \citep{Nicholl2018}. We first provided the extinction-corrected $U$, $B$, $V$, $R$, and $I$ data to {\tt superbol}. Thereafter, it mapped the light curve in each filter to a common set of times through the processes of interpolation and extrapolation. It then fits blackbody curve to the spectral energy distribution (SED) at each epoch, up to the observed wavelength, to give the quasi-bolometric light curve by performing trapezoidal integration. The peak quasi-bolometric luminosity obtained through integrating the flux over a wavelength range of 4000--10,000\,\AA\, is $10^{(42.548\pm0.019)}$\,erg\,s$^{-1}$, in agreement with \citet[][]{Prentice2019}.

Figure~\ref{fig:bol_compare} shows a comparison of the quasi-bolometric light curve of SN~2015ap with that of other H-stripped CCSNe.
The code {\tt superbol} also provides the  bolometric light curve, including the additional blackbody corrections to the observed quasi-bolometric light curve, by fitting a single blackbody to observed fluxes at a particular epoch and integrating the fluxes trapezoidally for a wavelength range of 100--25,000\,\AA.
The top panel of Figure~\ref{fig:BB_param_SN2015ap} provides the resulting quasi-bolometric and bolometric light curves of SN~2015ap.

\begin{figure}
	\includegraphics[width=\columnwidth]{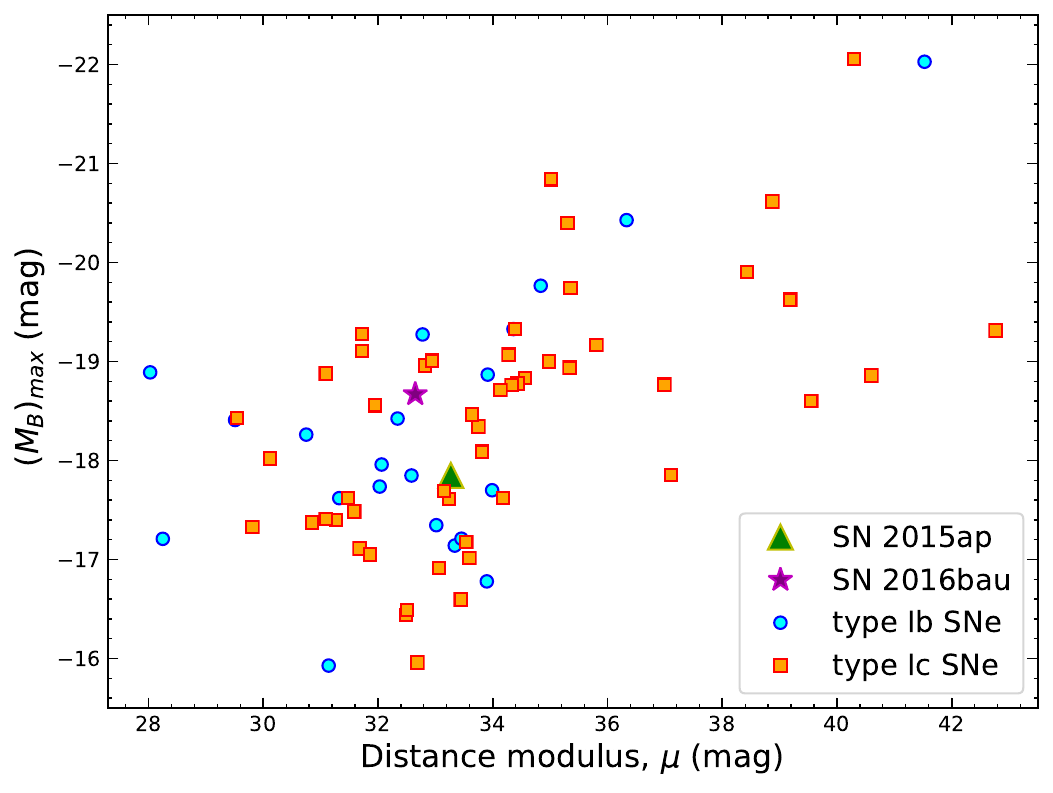}
    \caption{The position of SN~2015ap and SN~2016bau in the Miller diagram.}
    \label{fig:miller}
\end{figure}

\begin{figure}
	\includegraphics[width=\columnwidth]{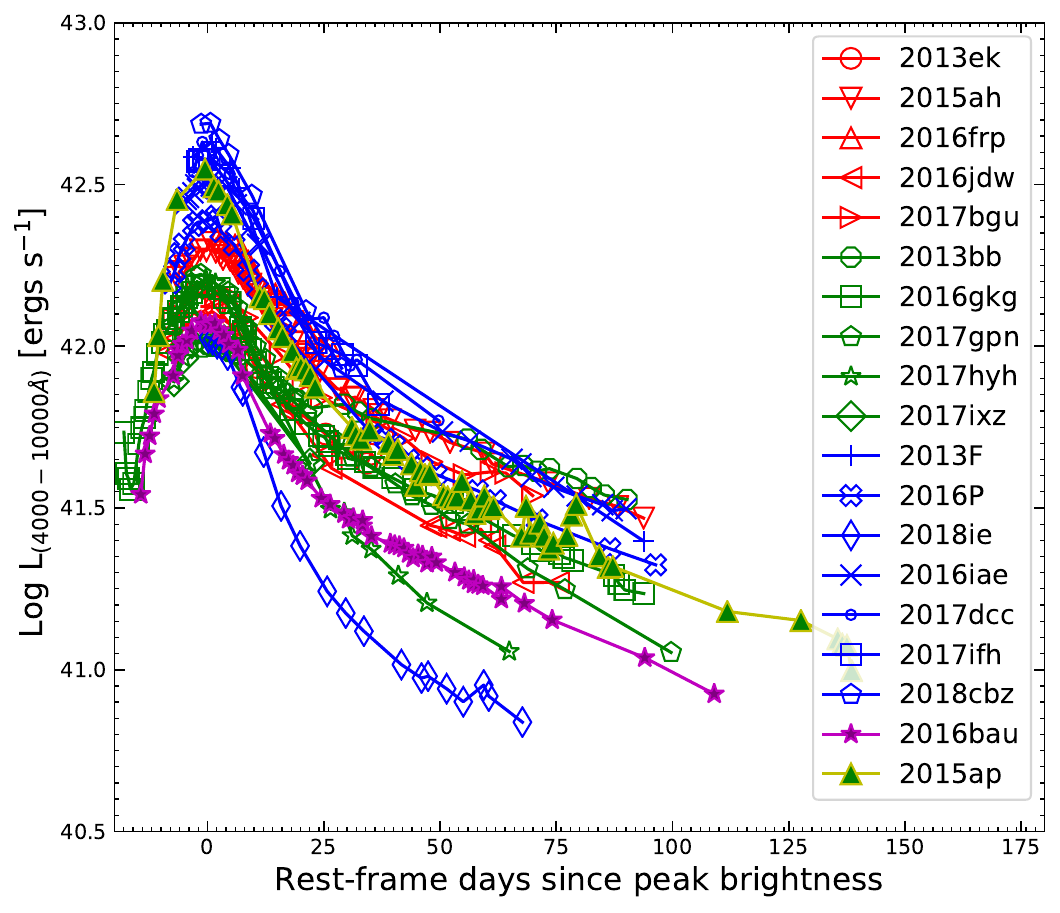}
    \caption{Comparison of quasi-bolometric light curves of SN~2015ap and SN~2016bau, obtained by fitting a blackbody to the SED and integrating the fluxes over the wavelength range of 4000--10,000\,\AA\, with other stripped-envelope CCSNe. Symbols are colour coded: red, green, and blue correspond to Type Ib, IIb, and Ic SNe, respectively.}
    \label{fig:bol_compare}
\end{figure}

$\bullet$~{\bf Temperature, radius, and velocity evolution}
\label{tempradvel_SN2015ap}

From {\tt superbol}, the photospheric temperature ($T_{\rm BB}$) and radius ($R_{\rm BB}$) evolution of SN~2015ap are also obtained. During the initial phases, the photospheric temperature is high, reaching about 11,400\,K at $-6.18$\,d. Further, as the SN ejecta expand, cooling occurs and the temperature tends to fall, dropping to 4490\,K on +21.66\,d, then remaining nearly constant (Fig.~\ref{fig:BB_param_SN2015ap}, second panel from top). A conventional evolution in radius is also seen. Initially, at an epoch of -8\,d, the photospheric radius is 4.14$\times$ 10$^{14}$\,cm. Thereafter, the SN expands and its radius increases, reaching a maximum radius of 3.27$\times$10$^{15}$\,cm, beyond which the photosphere seems to recede into the SN ejecta (Fig.~\ref{fig:BB_param_SN2015ap}, third panel from top).
From the  prior knowledge of the explosion epoch and radii at various epochs, we can estimate the photospheric velocity evolution of this SN, using $v_{\rm ph} = R_{\rm BB}/t$, where $t$ is the time since the explosion. The bottom panel of Figure~\ref{fig:BB_param_SN2015ap} shows the velocity evolution of SN~2015ap.

From the known value of $t_{\rm exp}$, a rise time ($t_{\rm rise}$) of 14.8$\pm$2.2\,d is obtained. The photospheric velocity near maximum light is 9000\,km\,s$^{-1}$ \citep[][]{Prentice2019}. With the known values of $t_{\rm rise}$, the photospheric velocity near maximum light, and a constant opacity ($\kappa$) of 0.07\,cm$^2$\,g$^{-1}$, we also obtain the ejecta mass ($M_{\rm ej}$) and kinetic energy ($E_{\rm ke}$) from the \citet{Arnett1982} model by following Equations (1) and (3) of \citet[][]{Wheeler2015}: 2.2$\pm$0.6\,M$_{\odot}$ and (1.05$\pm$0.31)$\times$10$^{51}$\,erg, respectively. Our derived ejecta mass is slightly higher than that of \citet[][]{Prentice2019}, but less than \citet[][]{Anjasha2020}.
Corresponding to a peak luminosity of (3.53$\pm$0.16)$\times$10$^{42}$\,erg\,s$^{-1}$, the amount of $^{56}$Ni synthesised is 0.14$\pm$0.02\,M$_{\odot}$, calculated following \citet[][]{Prentice2016}. Precise measurement of the amount of $^{56}$Ni generated in a CCSN is vital in deducing various progenitor scenarios and explosion processes associated with different CCSNe, as it heavily relies on both the explosion characteristics and the core structure of the underlying progenitor. Consequently, the measurement of nickel production plays an essential role in understanding the progenitors of CCSNe and their explosion mechanisms \citep[][]{2019MNRAS.483.3607S}.
 
\begin{figure}
\includegraphics[width=\columnwidth]{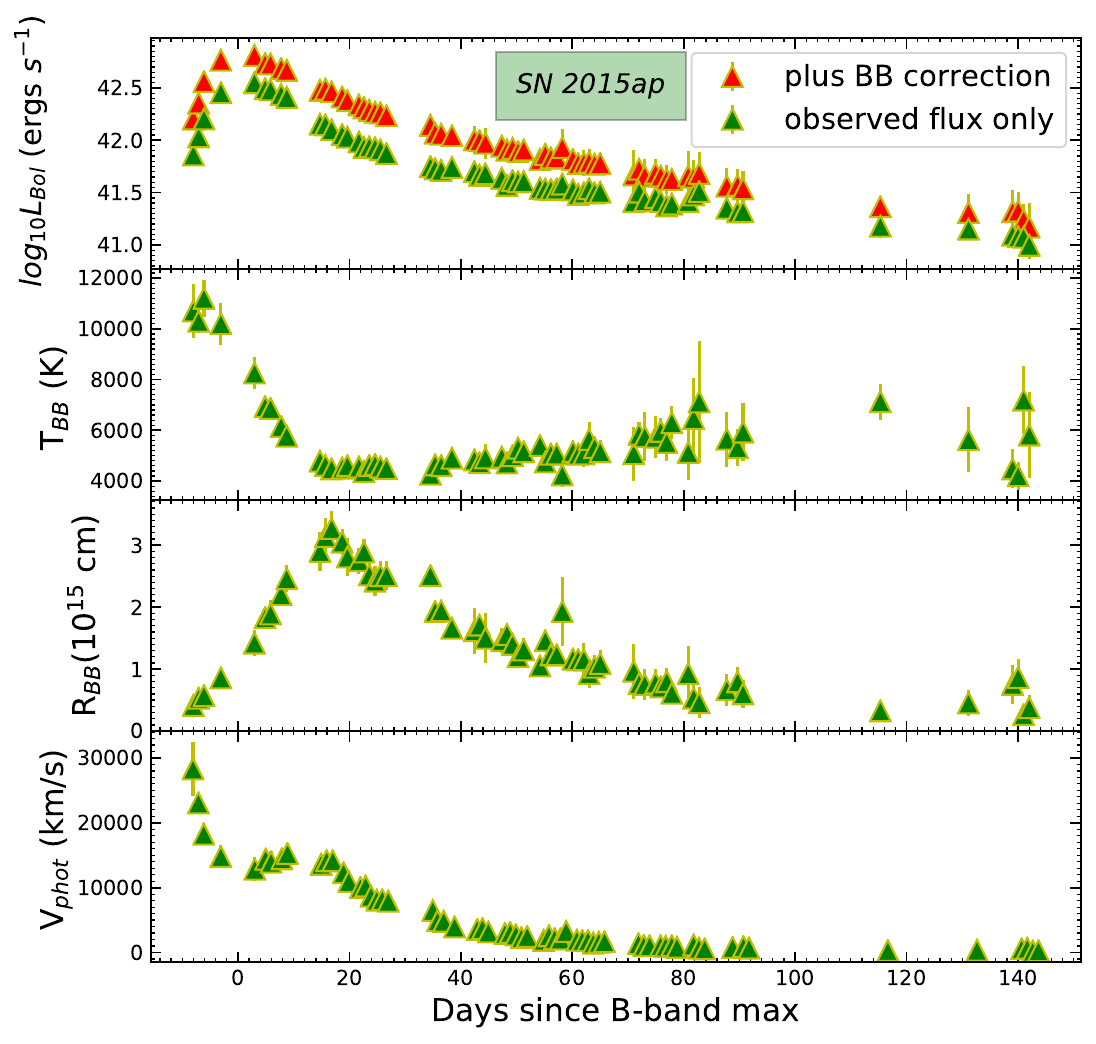}
\caption{The top panel shows the bolometric and quasi-bolometric light curves of SN~2015ap. Second panel: the temperature evolution of SN~2015ap. Third and fourth panels:  the radius and velocity evolutions (respectively) obtained using blackbody fits. }
\label{fig:BB_param_SN2015ap}
\end{figure}

\subsection{Photometric properties of SN~2016bau}
\label{subsec:Photometric_SN2016bau}

\begin{figure}
	\includegraphics[width=\columnwidth]{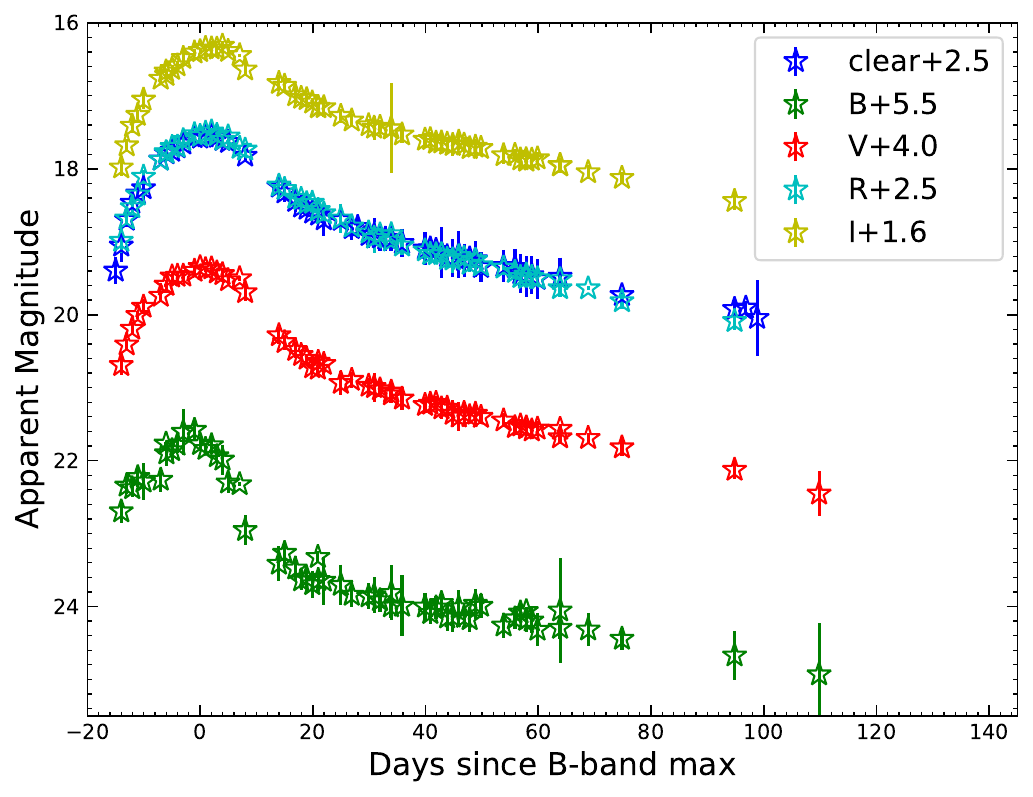}
    \caption{$BVRI$ and Clear filter light curves of SN~2016bau.}
    \label{fig:CBVRI_SN2016bau}
\end{figure}

Following a method similar to that for SN~2015ap, the $B$-band maximum and the explosion epochs of SN~2016bau were determined to be MJD 57477.37$\pm$1.99 and MJD 57462.54$\pm$0.97, respectively. Figure~\ref{fig:CBVRI_SN2016bau} shows the $BVRI$ and Clear ($C$) filter light curves of SN~2016bau. Light curves in the shorter-wavelength bands are narrower compared to those in the longer-wavelength bands, following a trend similar to that of SN~2015ap. Also, the $C$ filter almost exactly replicates the $R$-band light curve, as expected \citep{Li2003}.

$\bullet$~{\bf Colour evolution and extinction correction}

\begin{figure}	\includegraphics[width=\columnwidth]{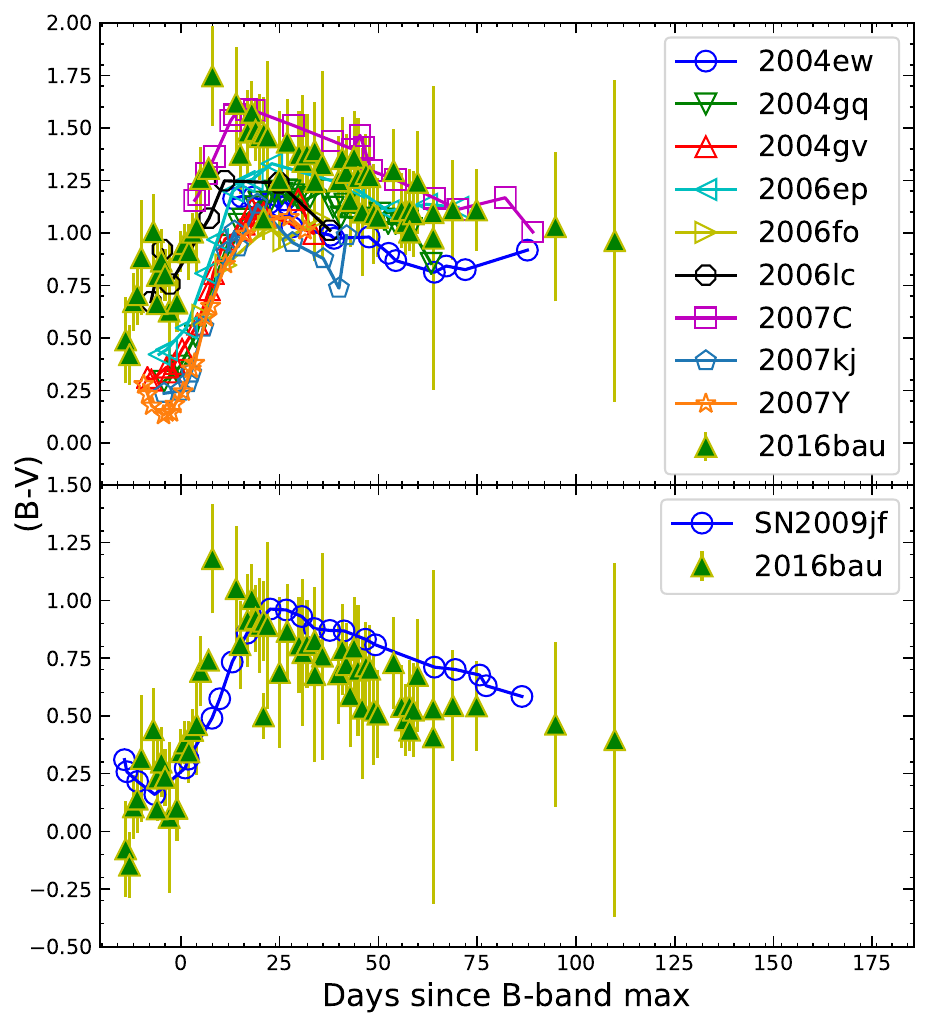}
    \caption{The top panel shows a comparison of the $(B-V)$ colour of SN~2016bau with that of other SNe~Ib (all corrected for MW extinction). The data for the other SNe are taken from \citet[][]{Stritzinger2018}. The bottom panel shows the $(B-V)$ colour curves of SN~2016bau and SN~2009jf, both corrected for MW extinction. To match the SN~2009jf colour curve, a shift of 0.566\,mag is required for the SN~2016bau colour curve.}
    \label{fig:color_curve_2016bau}
\end{figure}

Similar to SN~2015ap, we corrected for MW extinction using NED following \citet[][]{Schlafly2011}. In the direction of SN~2016bau, the Galactic extinction for the $B$, $V$, $R$, and $I$ bands is 0.060, 0.045, 0.036, and 0.025\,mag, respectively. The top panel of Figure~\ref{fig:color_curve_2016bau} shows a comparison of the $(B-V)$ colour of SN~2016bau with that of other SNe~Ib. SN~2016bau seems to be heavily reddened and lies above nearly all of the other SNe~Ib. To correct the host-galaxy extinction, we made use of the colour curve of SN~2009jf. We calculated the differences in the colours of the two SNe and took their weighted mean. In this calculation, we made use of data only in the range 0 to +20\,d for the reasons mentioned by \citet[][]{Stritzinger2018}. The resulting host-galaxy extinction is $E(B-V)_{\rm host} =$ 0.566$\pm$0.046\,mag. Thus, in order to match the SN~2009jf $(B-V)$ colour curve, we need to shift the MW-corrected $(B-V)$ colour curve of SN~2016bau downward by 0.566\,mag, as shown in  the bottom panel of Figure~\ref{fig:color_curve_2016bau}.

$\bullet$~{\bf Quasi-bolometric and bolometric light curves}

To obtain the quasi-bolometric and bolometric light curves of SN~2016bau, we used {\tt superbol}, as in the case of SN~2015ap. The extinction-corrected  $B$, $V$, $R$, and $I$ data were given as input to {\tt superbol}. The top panel of Figure~\ref{fig:BB_param_SN2016bau} shows the quasi-bolometric and bolometric light curves of SN~2016bau. Figure~\ref{fig:bol_compare} shows the comparison of the quasi-bolometric light curve of SN~2016bau obtained by integrating the flux over the wavelength range 4000--10,000\,\AA\, with other H-stripped CCSNe from \citet[][]{Prentice2019}. Here, SN~2016bau also seems to lie on the moderately bright end.   
  
$\bullet$~{\bf Temperature, radius, and velocity evolution}
\label{tempradvel_SN2016bau}

Figure~\ref{fig:BB_param_SN2016bau} also shows the evolution of the photospheric temperature ($T_{\rm BB}$) and radius ($R_{\rm BB}$) of SN~2016bau, obtained using {\tt superbol}. During the initial phases, the photospheric temperature is very high, reaching about 17,000\,K near 0\,d. Thereafter, as the SN ejecta expand, cooling occurs and the temperature falls, reaching 6000\,K at around +20\,d, then remaining nearly constant (Fig.~\ref{fig:BB_param_SN2016bau}, second panel from top). A conventional evolution in radius is also seen (Fig.~\ref{fig:BB_param_SN2016bau}, third panel from top). Initially, at an epoch of around $-6$\,d, the photospheric radius is 0.28$\times$ 10$^{15}$\,cm. Following this, the supernova expands and its radius increases, reaching a maximum radius of 0.96$\times$10$^{15}$\,cm, beyond which the photosphere seems to recede within the SN ejecta.

\begin{figure}
%\centering
\includegraphics[width=\columnwidth]{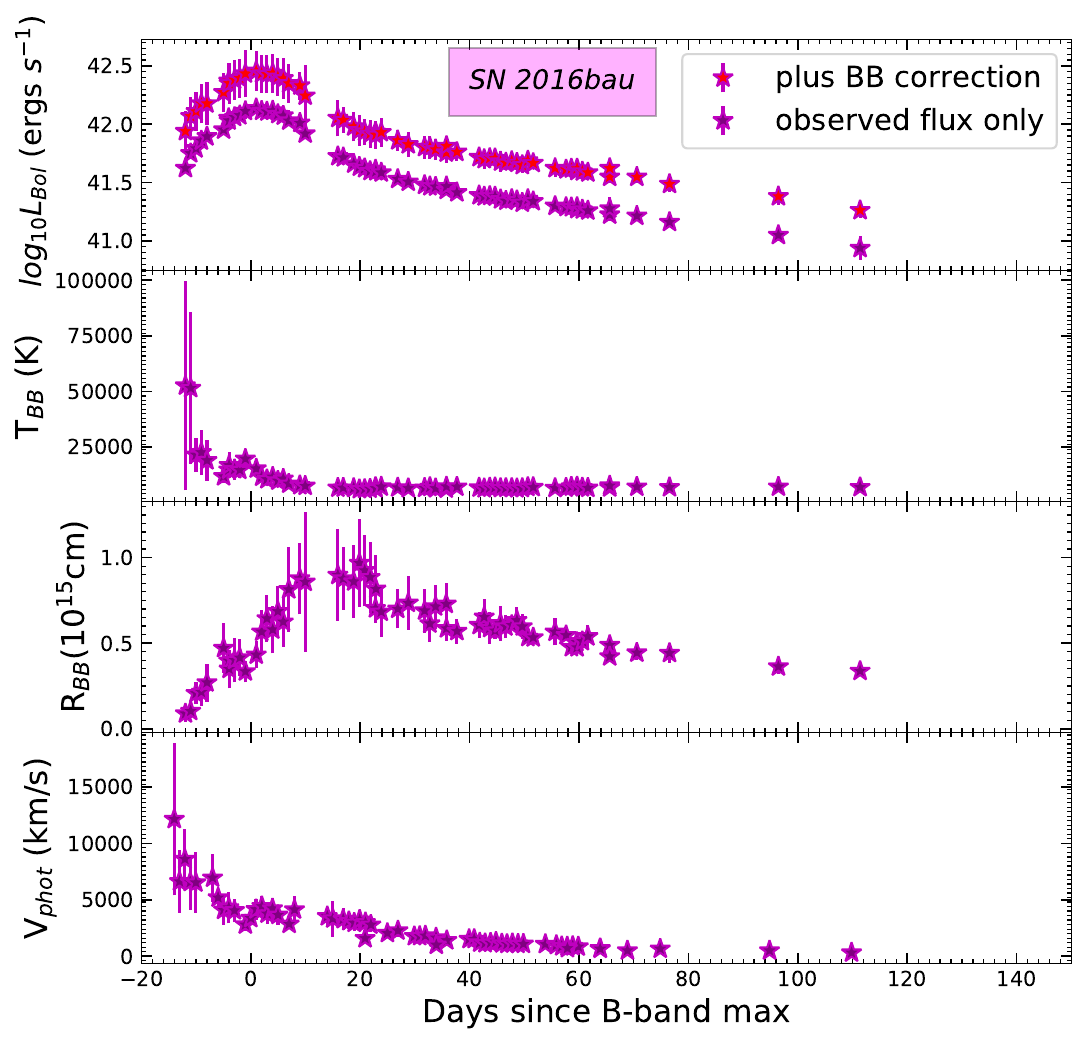}
\caption{The top panel shows the bolometric and quasi-bolometric light curves of SN~2016bau. Second panel: the temperature evolution of SN~2016bau. Third and fourth panels:  the radius and velocity evolution (respectively) obtained using blackbody fits.}
\label{fig:BB_param_SN2016bau}
\end{figure}
 
From the prior knowledge of explosion epoch and radii at various epochs, we estimate the photospheric velocity evolution of this SN in the same way as for SN~2015ap. The bottom panel of Figure~\ref{fig:BB_param_SN2016bau} shows the velocity evolution of SN~2016bau.
   
From the prior knowledge of $t_{\rm exp}$, a rise time ($t_{\rm rise}$) of 17.09$\pm$1.29\,d is obtained. The photospheric velocity near maximum light is $\sim$\,5000\,km\,s$^{-1}$, obtained from a blackbody fit. With the known values of $t_{\rm rise}$, the photospheric velocity near maximum light, and a constant opacity ($\kappa$) 0.07\,cm$^2$\,g$^{-1}$, we obtain the ejecta mass ($M_{\rm ej}$) and kinetic energy ($E_{\rm ke}$) following Eq. (1) and Eq. (3) of \citet[][]{Wheeler2015}; the results are 1.6$\pm$ 0.3\,M$_{\odot}$ and (0.24$\pm$0.04)$\times$ 10$^{51}$\,erg, respectively. Following \citet[][]{Prentice2016}, an amount of 0.055$\pm$0.006\,M$_{\odot}$ of $^{56}$Ni is synthesised, corresponding to a peak luminosity of (1.19$\pm$0.08)$\times$10$^{42}$\,erg\,s$^{-1}$.

\section{Spectral Properties}
\label{sec:Spectral}
In this section, we discuss the spectral features of SN~2015ap and SN~2016bau and further compare their properties with other similar SNe. We modelled the spectra of these two SNe at different epochs using {\tt SYN++} \citep[][]{Branch2007, Thomas2011} and performed the spectral matching of the 12, 13, and 17\,M$_{\odot}$ spectral models given by \citet[][]{Jerkstrand2015} with the spectra of SN~2015ap and SN~2016bau at phases around 100\,d past explosion. In this section, we also estimate the velocities of various lines present in the spectra, using their absorption troughs.

\subsection{Spectroscopic nature of SN~2015ap}
\label{subsec:Spec_evol_SN2015ap}

\begin{figure}
\includegraphics[width=\columnwidth]{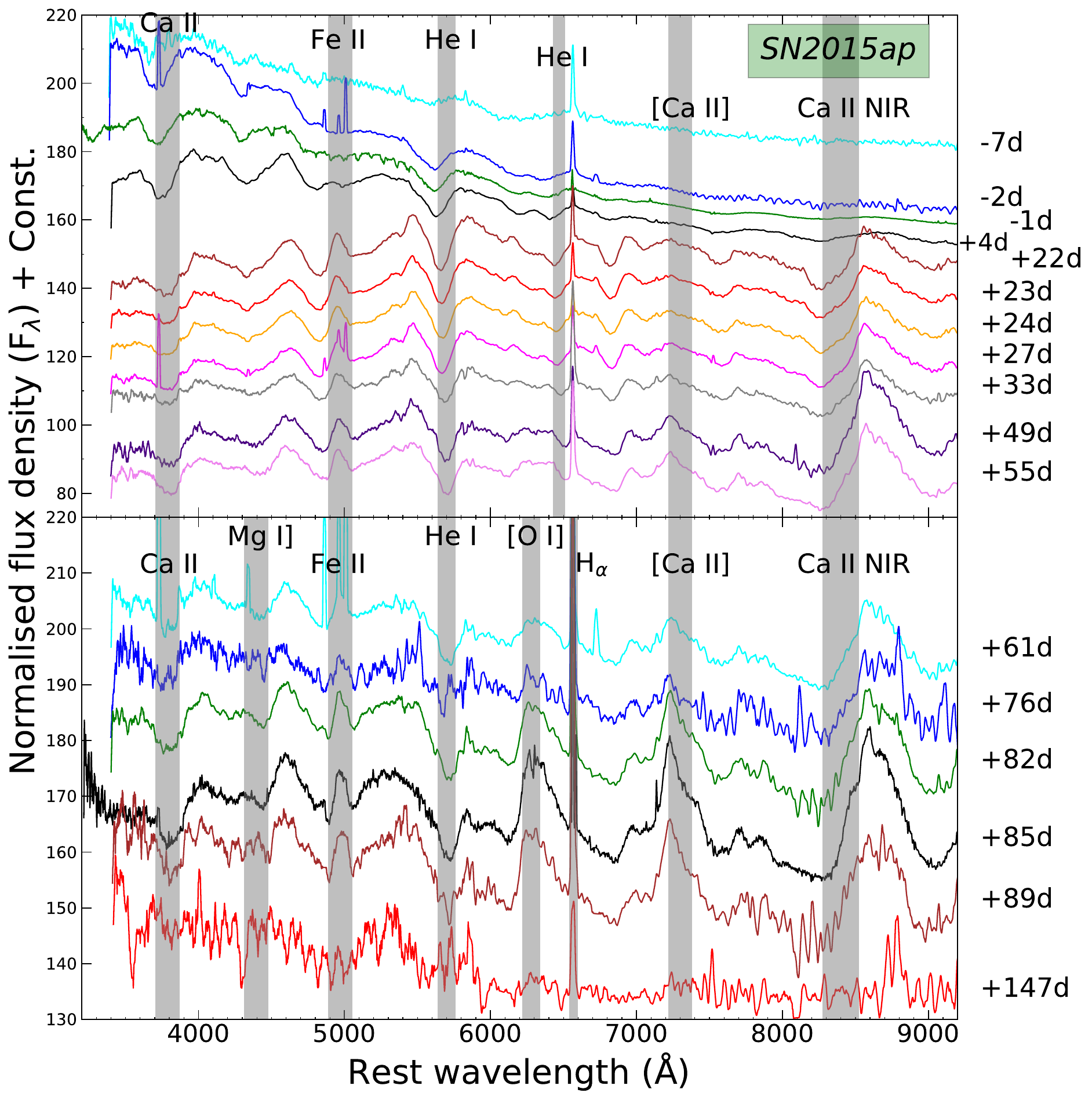}
\caption{Top panel: the early-phase spectra up to +55\,d. Bottom panel: the spectral evolution beyond +55\,d since $B_{\rm max}$ of SN~2015ap.}
\label{fig:spectral_evol_SN2015ap}
\end{figure}

The top and bottom panels of Figure~\ref{fig:spectral_evol_SN2015ap} show the early phases (-7\,d to +55\,d) and late phases (+61\,d to +147\,d) spectral evolution, respectively. It is quite evident from the top panel of  Figure~\ref{fig:spectral_evol_SN2015ap} that initially SN~2015ap shows broad-lined features (at -7\,d and -2\,d), which later evolve to spectra of a normal Type Ib SN. The characteristic He~I line at 5876\,\AA\, of a typical SN~Ib is clearly seen in the spectra of SN~2015ap. We also see unambiguous He~I features at 6678\,\AA\, and 7065\,\AA, but these two lines are not as prominent as the one at 5876\,\AA. In the very early phases (up to +4\,d), the He~I line at 7065\,\AA is hard to identify. We see that the He~I absorption at 5876\,\AA\, is stronger than any other absorption features, and it is present in every spectrum up to +147\,d. The Fe~II feature near 5169\,\AA\, is very hard to be identified, and it seems to be highly blended with He~I 5016\,\AA. 

The spectral evolution also shows the Ca~II near-infrared (NIR) feature, which is almost absent in the very early phases (-7\,d to +4\,d), but starts to develop very strongly from +22\,d onward. It is so strong that each spectrum on and after +22\,d shows it, even the lowest-SNR spectrum at +147\,d. The forbidden [Ca~II] feature near 7300\,\AA\, is almost absent at very early phases (-7\,d to +4\,d), then begins to develop very obviously in the spectra from +22\,d, and is present up to +89\,d. This [Ca~II] feature can also get blended with [O~II] emission at 7320\,\AA\, and 7330\,\AA. It almost disappears in the spectrum at +147\,d, though its absence may not be real owing to the very poor SNR of that spectrum. We also see the Ca~II~H\&K feature near 3934\,\AA, which is present in every spectrum.
From the bottom panel of Figure~\ref{fig:spectral_evol_SN2015ap}, we see that as the spectra evolve, the absorption features of different lines tend to disappear and the emission features become more prominent. We see a very weak semiforbidden Mg~I] line, at 4571\,\AA. However, we can clearly see the forbidden emission lines of [O~I] and [Ca~II], which indicates the onset of the nebular phase.

$\bullet$~{\bf Spectral comparison}
%\label{subsec:spec_com_SN2015ap}

To investigate the spectroscopic behaviour of SN~2015ap, we compare its spectral features with those of other well-studied SNe~Ib such as SN~2004gq \citep[][]{Modjaz2014}, SN~2008D \citep[][]{Modjaz2014}, SN~2012au \citep[][]{Pandey2021}, SN~2009jf \citep[][]{Sahu2011, Modjaz2014}, iPTF13bvn \citep[][]{Srivastav2014a}, SN~2007uy \citep[][]{Modjaz2014, Milisavljevic2010}, SN~2007gr \citep[][]{Valenti2008, Modjaz2014}, and SN~2005bf \citep[][]{Modjaz2014}.

The top panel of Figure~\ref{fig:spectral_comparison} shows the early phase (-7\,d) spectral comparison of SN~2015ap with these well-studied SNe~Ib. The spectral features of SN~2015ap look similar to those of SN~2008D and SN~2012au, compared to other SNe~Ib. We can see that the Ca~II~H\&K feature, the Mg~II feature, and the He~I P~Cygni profile match very well with those of SN~2008D, while in other SNe in the comparison sample, these features are much more developed. The blue end of the spectrum matches nicely with SN~2008D, while the redder part is featureless and much closer to SN~2012au and SN~2005bf. The He~I 5876\,\AA\, feature of SN~2015ap is completely different from that of SN~2007uy, but resembles that of SN~2008D and seems to be less evolved compared to that of SN~2004gq, SN~2012au, SN~2009jf, and iptf13bvn. The Fe~II profile in SN~2015ap is hard to detect, which may be due to a very high initial optical opacity. We try to estimate the velocities using the various absorption features. As the spectrum at this epoch is continuum dominated, only a few absorption features were visible. The velocity estimated using He~I absorption lines  is $\sim$\,14,100\,km\,s$^{-1}$.

\begin{figure}
%\centering
\includegraphics[height=20cm]{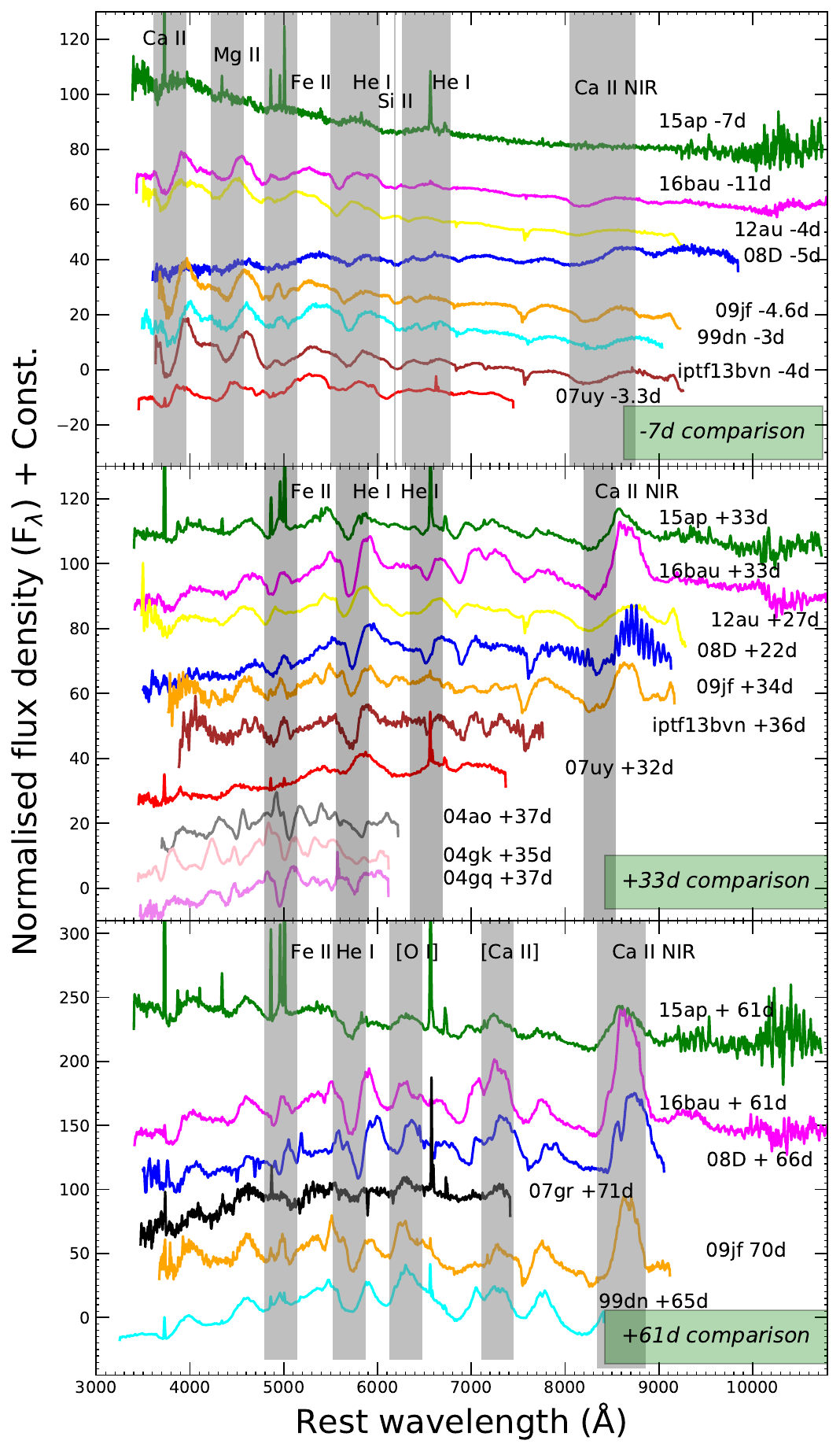}
\caption{Spectral comparison of the spectra of SN~2015ap and SN~2016bau at epochs -7\,d (top panel), +33\,d (middle panel), and +61\,d (bottom panel). Various important SN~Ib features have been compared with those of other similar-type SNe.}
\label{fig:spectral_comparison}
\end{figure}

Further, we compare the +33\,d spectrum of SN~2015ap with our comparison sample (middle panel of Figure~\ref{fig:spectral_comparison}). At this epoch, the spectrum of SN~2015ap shows various P~Cygni profiles. We see that it almost exactly replicates the +22\,d spectrum of SN~2008D. The blended He~I and Fe~II profiles near 5016\,\AA\, well match those of other SNe~Ib, except for SN~2007uy, where the peak is almost absent, and for SN~2005bf, where it is double-peaked. The He~I feature at 5876\,\AA\, matches well with other SNe~Ib except for the cases of SN~2007uy and SN~2005bf, where the absorption features seem to be much broader. Here, once again, SN~2007uy seems to match much less and SN~2008D seems to best match the +22\,d spectrum of SN~2015ap. The velocities estimated using the Ca~II NIR triplet and He~I absorption features are $\sim$\,7800\,km\,s$^{-1}$ and 10000\,km\,s$^{-1}$, respectively.

For a much clear comparison, we compared the +61\,d  spectrum of SN~2015ap (bottom panel of Fig.~\ref{fig:spectral_comparison}) with spectra of other well-studied SNe~Ib. At this epoch also, the spectrum is still much closer to that of SN~2008D compared to other SNe~Ib. The blended Fe~II and He~I  features near 5016\,\AA\, show an almost similar profile to that of SN~2008D. However, the 5876\,\AA\, He~I profile is narrower in SN~2015ap compared to SN~2008D. We can see the onset of the appearance of the [O~I] line in SN~2015ap, which differs from SN~2008D. The He~I profile is well matched in the cases of SN~1999dn and SN~2009jf, but the Ca~II NIR triplet of these SNe differs from that of SN~2015ap. Here the velocities estimated using the Ca~II NIR triplet and He~I absorption features are $\sim$\,6900\,km\,s$^{-1}$ and 8100\, km\,s$^{-1}$, respectively.

$\bullet$~{\bf Spectral modelling}
%\label{subsec:syn_SN2015ap}

After confidently identifying the features present in the spectra of SN~2015ap and comparing them with those of other well-studied SNe, we tried to model a few spectra at various epochs using {\tt SYN++}. The top panel in Figure~\ref{fig:syn_SN2015ap} shows the early-phase (-7\,d) spectrum of SN~2015ap. This spectrum displays weak and broad P~Cygni profiles of He~I, Ca~II~H\&K, and the Ca~II~NIR triplet, and also some blended features of Fe~II. We also show the best-matching synthetic spectrum, generated by {\tt SYN++}. The absorption features due to Ca~II~H\&K, Ca~II~NIR triplet, He~I, and Fe~II multiplets are easily reproduced. The photospheric velocity and blackbody temperature associated with the best-fit spectrum are 12,000\,km\,s$^{-1}$  and 12,000\,K, respectively. We perform {\tt SYN++} matching to four additional spectra from epochs -2\,d to +33\,d (Fig.~\ref{fig:syn_SN2015ap}). With the passage of time, the SN expands, cools gradually, and its expansion velocity decreases slowly, so we see a gradual decrease in the values of these fitted parameters. The photospheric velocities during the phase of -7\,d to +33\,d vary from 13,000\,km\,s$^{-1}$ to 6800\,km\,s$^{-1}$, and the blackbody temperature varies from 12,000\,K to 4500\,K, which are in good agreement with those obtained photometrically from blackbody fits. Owing to the local thermodynamic equilibrium (LTE) approximation, {\tt SYN++} does not work well for the later epochs and cannot be used to fit the spectra.

To get an estimate of the progenitor mass, we used the +98.75\,d post-explosion spectrum of SN~2015ap and plotted it along with the 12, 13, and 17\,M$_{\odot}$ model spectra from \citet[][]{Jerkstrand2015} at 100\,d, after scaling by a factor of exp$(-2 \times \Delta t /111.4)$ \citep[][]{Jerkstrand2015}, where $\Delta t$ = 1.25, is the time difference between the epoch of the model spectrum and the epoch of the observed spectrum. We can see that the 12\,M$_{\odot}$ and 17\,M$_{\odot}$ model spectra seem to reproduce the observed spectrum of SN~2015ap. The 12\,M$_{\odot}$ spectrum very nicely matches the observed spectrum throughout the entire wavelength range, while the 17\,M$_{\odot}$ spectrum slightly overproduces the flux near the Ca~II~NIR triplet close to 8500\,\AA. However, the 13\,M$_{\odot}$ model spectrum fails to explain the observed fluxes throughout the entire wavelength range. Thus, based on this analysis, a range of 12--17\,M$_{\odot}$ is expected for the possible progenitor mass of SN~2015ap, in agreement with that described by \citet[][]{Anjasha2020}.

\begin{figure}
	\includegraphics[width=\columnwidth]{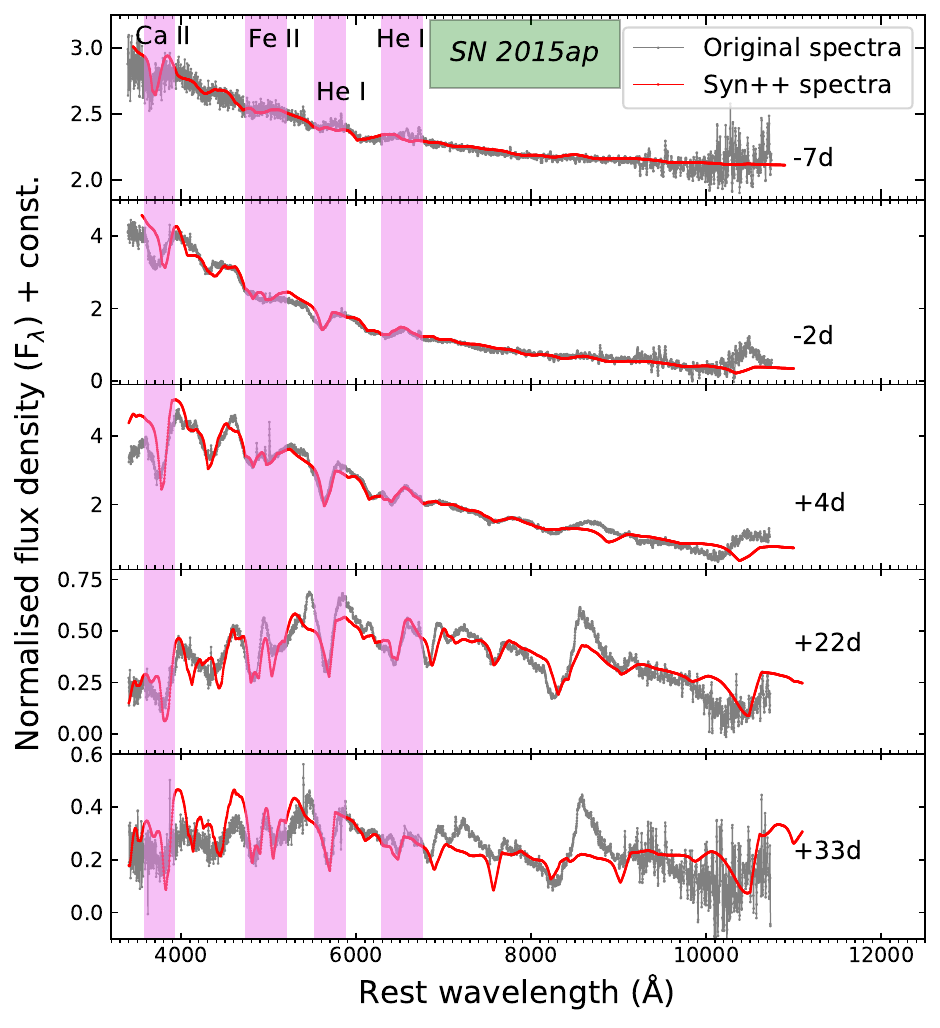}
    \caption{{\tt SYN++} modelling of the spectra of SN~2015ap at epochs -7\,d, -2\,d, +4\,d, +22\,d, and +33\,d. Prominent He~I features could be produced nicely.}
    \label{fig:syn_SN2015ap}
\end{figure}

\begin{figure}
%\centering
\includegraphics[width=\columnwidth]{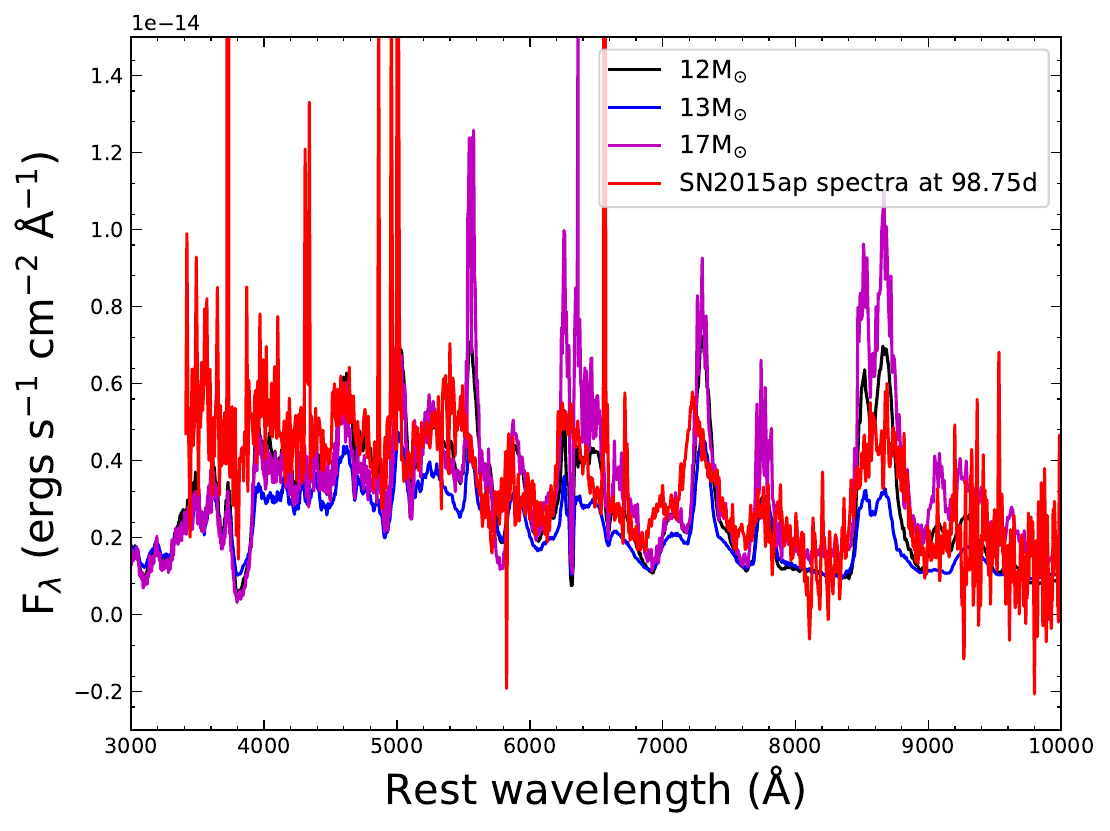}
\caption{The $t = $+98.75\,d  spectrum of SN~2015ap plotted along with the 12, 13, and 17\,M$_{\odot}$ models from \citet[][]{Jerkstrand2015} at 100\,d scaled with an exponential factor exp$(-2 \times 1.25/111.4)$. The 12\,M$_{\odot}$ and 17\,M$_{\odot}$ models seem to best match the observed spectrum of SN~2015ap.}
\label{fig:jerkstrand_SN2015ap}
\end{figure} 

\begin{figure*}
\centering
\includegraphics[width=\textwidth]{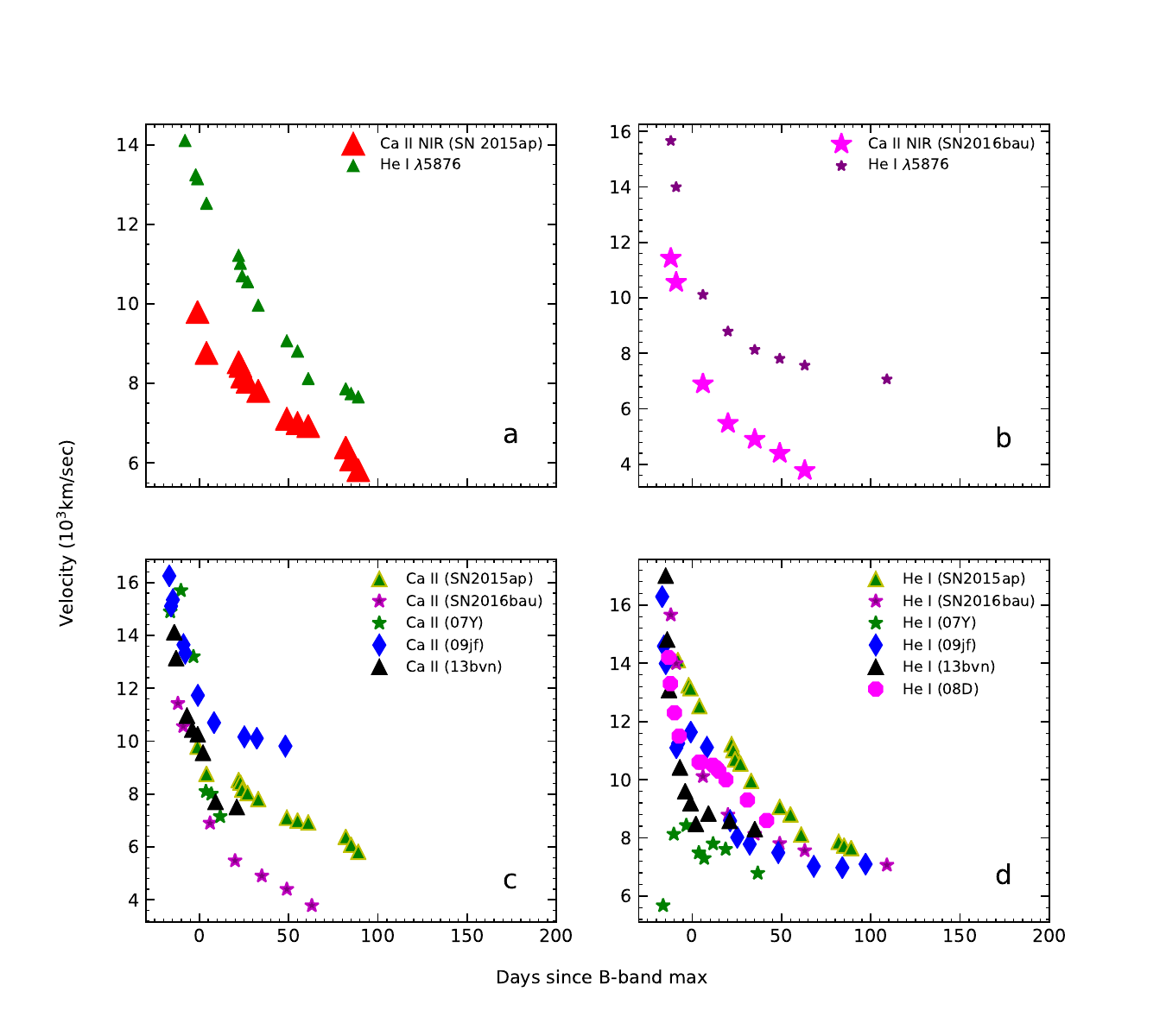}
\caption{(a) The temporal evolution of velocities of He~I 5876\,\AA\, and the Ca~II NIR triplet for SN~2015ap. (b) Same as (a) but for SN 2016bau. The bottom two panels show the comparison of these line velocities of SN~2015ap and SN~2016bau with SNe 2007Y \citep[][]{Stritzinger2009}, 2009jf \citep[][]{Sahu2011}, iPTF13bvn \citep[][]{Srivastav2014a}, and 2008D \citep[][]{Modjaz2014}. }
\label{fig:vel_evolve}
\end{figure*}

$\bullet$~{\bf Velocity evolution of various lines of SN~2015ap}
%\label{subsec:Vel_evol_SN2015ap}

We used the blue-shifted absorption minima of P~Cygni profiles and the special-relativistic Doppler formula \citep[][]{Sher1968} to obtain the velocities of several lines. Figure~\ref{fig:vel_evolve}$a$ shows the He~I and the Ca~II NIR triplet velocity evolution of SN~2015ap. In the initial few days, the line velocities tend to decrease rapidly.
At an epoch of +24\,d, the velocities estimated using He~I 5876\,\AA\, and the Ca~II NIR triplet are $\sim$\,10,600\,km\,s$^{-1}$ and 8200\,km\,s$^{-1}$, respectively. The velocities estimated using these two lines drop to $\sim$\,9100\,km\,s$^{-1}$ and 7100\,km\,s$^{-1}$ (respectively) at an epoch of +49\,d. In the late phases, the velocities decline gradually.

Figure~\ref{fig:vel_evolve}$c,d$, show comparisons of velocities obtained using the Ca~II NIR triplet and He~I 5876\,\AA\, features with other well-studied SNe. Initially the Ca~II NIR line velocity of SN~2015ap evolves in a manner similar to iptf13bvn, thereafter it decays slowly, attaining a velocity of $\sim$\,7600\,km\,s$^{-1}$ at +89\,d. We see that the ejecta velocity of SN~2015ap obtained using He~I 5876\,\AA\, is higher than that of other SNe~Ib but closer to SN~2008D.   

\subsection{Spectral properties of SN~2016bau}
\label{subsec:Spec_evol_SN2016bau}

\begin{figure}
%\centering
\includegraphics[width=\columnwidth]{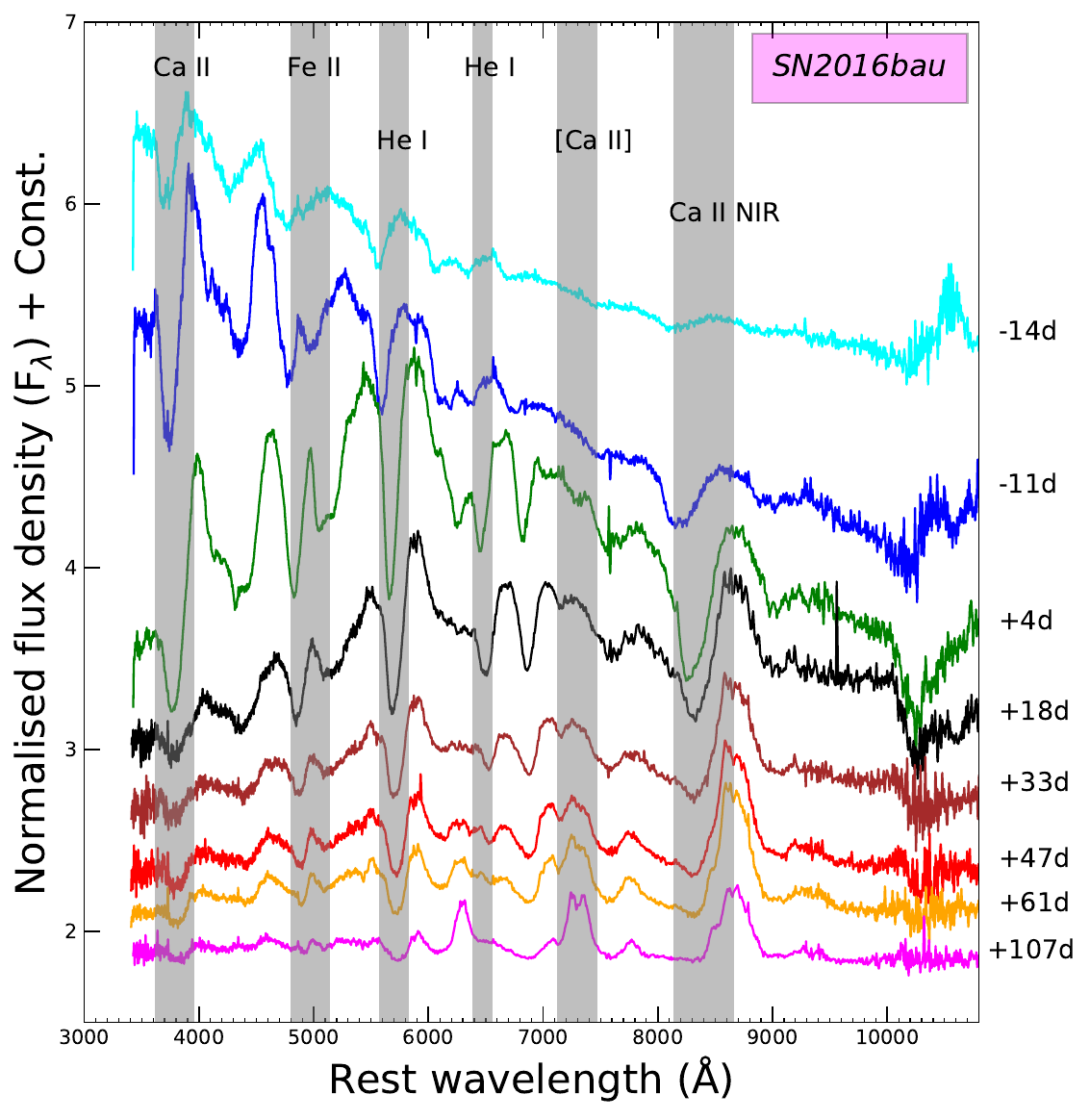}
\caption{Various line identifications in the overall spectral evolution of SN~2016bau. Strong He~I features along with other important lines are indicated.}
\label{fig:spec_evol_SN2016bau}
\end{figure}

Figure~\ref{fig:spec_evol_SN2016bau} shows the spectral evolution of SN~2016bau for a period of -14\,d to +107\,d.  In the first spectrum (-14\,d), we see well-developed He~I features (5876, 6678, and 7065\,\AA). Also, the spectrum on this particular epoch displays a weak and broad Ca~II~NIR feature, but with the passage of time (beyond $\sim$+33\,d), prominent Ca~II NIR features start to appear. We also see that in the initial phases, there is a strong Ca~II~H $\&$ K feature, but it becomes progressively weaker in later phases. As the phase approaches the date of $B$-band maximum brightness, the He~I features start to appear much more strongly, as seen in the spectra at -11\,d, +4\,d, and +18\,d. Beyond $-11$\,d, the longer wavelengths exhibit strong Ca~II~NIR features. The spectra also display very strong P~Cygni profiles of He~I at 5016\,\AA\, after $t =  $ -11\,d. The emergence of very strong features of He~I in the early-time spectra confirms this SN to be of Type Ib. %The log of the spectroscopic observations for SN~2016bau is provided in Table~\ref{tab:SN2016bau_spec_obs}.

$\bullet$~{\bf Spectral comparison}
%\label{subsec:spec_com_SN2016bau}

To investigate the spectroscopic behaviour of SN~2016bau, we have compared its spectral features with those of other well-studied SNe~Ib. We used a sample similar to that for SN~2015ap.
The top panel of Figure~\ref{fig:spectral_comparison} shows the early-phase (-11\,d) spectral comparison of SN~2016bau with other well-studied H-stripped CCSNe. The spectral features of SN~2016bau look much more similar  to those of SN~2012au, SN~2009jf, and SN~1999dn compared to other SNe. we can see that the Ca~II~ H\&K feature, the Mg~II feature, and the He~I P~Cygni profile of SN~2016bau match very well those of SN~2012au, SN~2009jf, and SN~1999dn. The He~I 5876\,\AA\, feature of SN~2016bau is completely different from that of SN~2007uy, SN~2015ap, and SN~2008D. As with SN~2015ap, the blended Fe~II profile in SN~2016bau is hard to detect, which may be due to a very high initial optical opacity. The spectrum at this epoch has nicely developed He~I features. The velocity estimated using the He~I 5876\,\AA\, absorption line is $\sim$\,15,600\,km\,s$^{-1}$.

The middle panel of Figure~\ref{fig:spectral_comparison} shows the +33\,d spectral comparison of SN~2016bau with other well-studied SNe~Ib. The spectrum at this epoch contains many P~Cygni profiles of various lines. These spectral features look most similar to those of SN~2008D and SN~2015ap, compared to other SNe~Ib. We see that the Ca~II NIR feature, the Fe~II feature, and the He~I P~Cygni profile match very well those of SN~2008D and SN~2015ap, compared to other SNe. SN~2009jf also seems to match nicely in the redder part of the spectrum. The He~I 5876\,\AA\, feature of SN~2016bau is completely different from those of SN~2007uy, SN~2005bf, SN~2004gq, and SN~2007gr, but resembles those of SN~2008D, SN~2015ap, and SN~2009jf. The He~I 5876\,\AA\, profile of SN~2016bau seems to be more asymmetric and narrower than that of iptf13bvn. At this epoch, the velocities estimated using the absorption features of He~I and the Ca~II NIR triplet are $\sim$\,8100\,km\,s$^{-1}$ and 4900\,km\,s$^{-1}$, respectively.

For much clearer comparisons, we also compared the +61\,d spectrum of SN~2016bau (bottom panel of Figure.~\ref{fig:spectral_comparison}) with other well-studied H-stripped CCSNe spectra. At this epoch, the bluer part of the spectrum is much closer to SN~2008D compared to other SNe~Ib. Here the blended Fe~II and He~I feature near 5016\,\AA\, shows a nearly similar profile to those of SN~2008D and SN~2009jf. However, the 5876\,\AA\, He~I profile is broader in SN~2016bau compared with SN~2008D. We can see that the onset of the appearance of [O~I] in SN~2016bau is slightly different from that of SN~2008D, while it matches nicely that of SN~2009jf. The He~I profile is well matched in the case of SN~2008D. Here the velocities estimated using the Ca~II NIR and He~I absorption features are $\sim$\,3700\,km\,s$^{-1}$ and 7600\,km\,s$^{-1}$, respectively.

$\bullet$~{\bf Spectral modelling}
%\label{subsec:syn_SN2016bau}

After confidently identifying various spectral features, we tried to model the spectra of SN~2016bau at different epochs using {\tt SYN++}. The top panel of Figure~\ref{fig:syn_SN2016bau} shows the early-phase (-14\,d) spectrum of SN~2016bau. It contains many broad P~Cygni profiles of He~I, Ca~II~H\&K, the Ca~II~NIR triplet, and also some blended features of Fe~II. In this figure, we have also presented the best-matching synthetic spectrum generated by {\tt SYN++}. The modelled spectrum easily reproduces the absorption features of Ca~II~H\&K, Ca~II~NIR, He~I, and the Fe~II multiplet. The photospheric velocity and blackbody temperature associated with the best-fit spectrum are 16,000\,km\,s$^{-1}$ and 9000\,K respectively. We performed {\tt SYN++} matching for four additional spectra which covers a period of -11\,d to +33\,d (subsequent panels of Fig.~\ref{fig:syn_SN2016bau}). With the passage of time, the SN expands and cools gradually, and its expansion velocity decreases slowly, so we see a gradual decrease in the fit parameters such as velocity and temperature in the later phases. The photospheric velocity during the phase of -14\,d to +33\,d varies from 16,000\,km\,s$^{-1}$ to 8000\,km\,s$^{-1}$ and the blackbody temperature ranges from 9000\,K to 4000\,K, in good agreement with values obtained photometrically from blackbody fits.

\begin{figure}
	\includegraphics[]{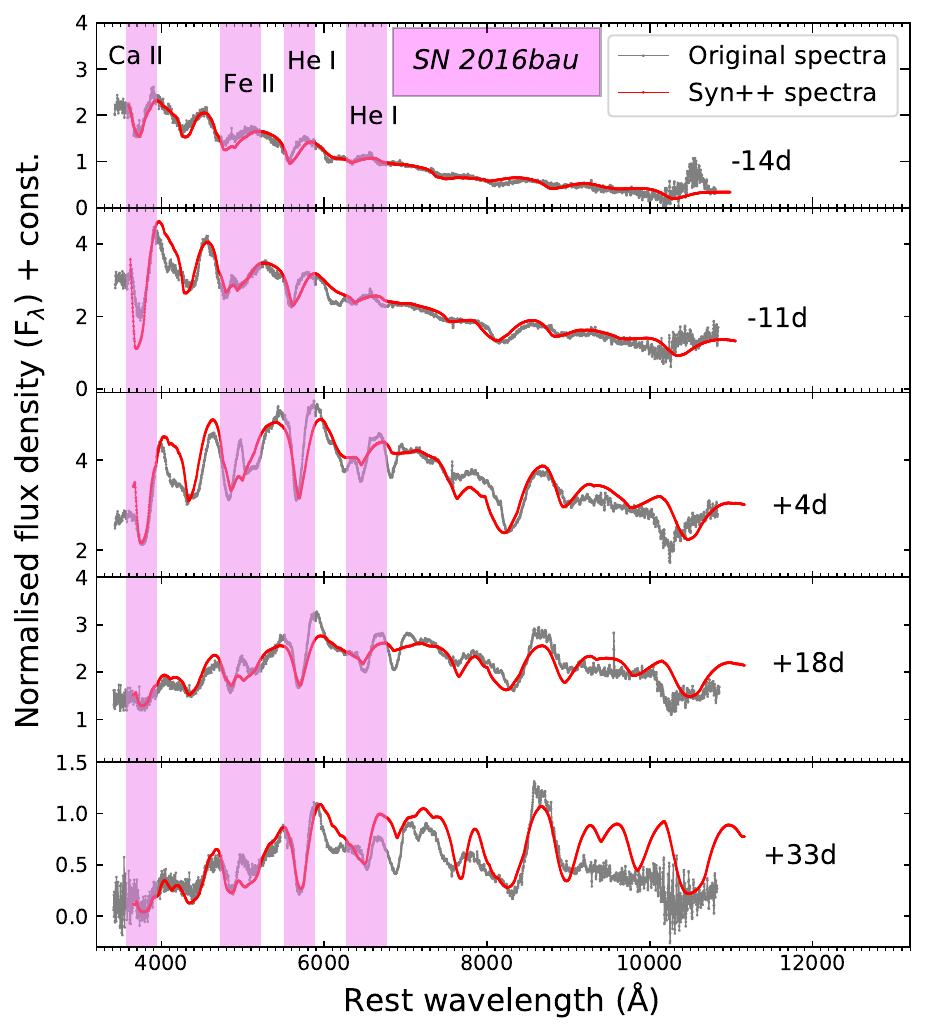}
    \caption{{\tt SYN++} modelling of the spectra of SN~2016bau at epochs -14\,d, -11\,d, +4\,d, +18\,d, and +33\,d.}
    \label{fig:syn_SN2016bau}
\end{figure}

We also try to match the +121\,d spectrum of SN~2016bau with the 12\,M$_{\odot}$, 13\,M$_{\odot}$, and 17\,M$_{\odot}$ model spectra at +100\,d  from \citet[][]{Jerkstrand2015}, scaled with a factor of exp$(-2 \times \Delta t /111.4)$ \citep[][]{Jerkstrand2015}, where $\Delta t$ = 21, is the time difference of the epoch of model spectrum and the epoch of observed spectrum. We can see that all three models over--predict the observed fluxes from 3000\,\AA\, to around 6000\,\AA, beyond which the 12\,M$_{\odot}$ model spectrum seems to best describe the observed spectrum. It could nicely explain the [Ca~II] emission near 7300\,\AA\, and the Ca~II~NIR feature near 8500\,\AA. The 13\,M$_{\odot}$ model spectrum also produces the [Ca~II] emission but fails to explain the observed fluxes near the Ca~II-NIR triplet. The 17\,M$_{\odot}$ model spectrum overpredicts the flux throughout the entire wavelength range and thus fails to explain the spectrum of SN~2016bau. Hence, based on our analysis, a slightly low-mass progenitor ($\leq $12\,M$_{\odot}$) is expected.

\begin{figure}
\includegraphics[width=\columnwidth]{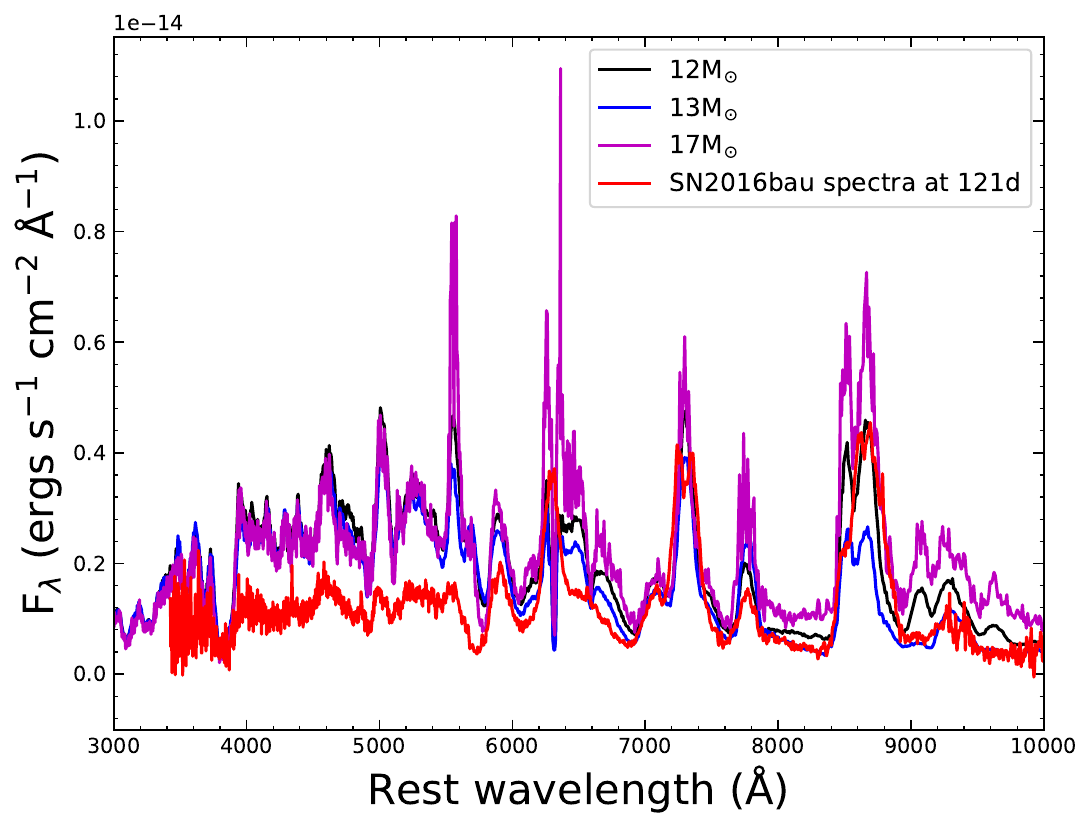}
\caption{The $t =  $+121\,d spectrum of SN~2016bau plotted along with the 12, 13, and 17\,M$_{\odot}$ models from \citet[][]{Jerkstrand2015} at 100\,d scaled with an exponential factor exp$(-2 \times 21/111.4)$. The 12\,M$_{\odot}$ model seems to best match the spectrum of SN~2016bau.}
\label{fig:jerkstrand_SN2016bau}
\end{figure}

$\bullet$~{\bf Velocity evolution of Various lines of SN~2016bau}
%\label{subsec:Vel_evol_SN2016bau}

We used the blue-shifted absorption minima of P~Cygni profiles to obtain the velocities of He~I and Ca~II NIR lines. Figure~\ref{fig:vel_evolve}$b$ shows the He~I and Ca~II NIR velocity evolution of SN~2016bau. In the initial few days, the line velocities tend to decrease rapidly, but in later phases, the velocities decline gradually. At an early epoch of -14\,d, the velocities estimated using He~I 5876\,\AA\, and the Ca~II NIR triplet are $\sim$\,15,600\,km\,s$^{-1}$ and $\sim$\,11,400\,km\,s$^{-1}$, respectively. The velocities estimated using these two lines drop to $\sim$\,7800\,km\,s$^{-1}$ and $\sim$\,4400\,km\,s$^{-1}$ (respectively) at an epoch of +47\,d. Beyond +47\,d, the velocities continue to decline at a slower rate as compared to the initial decline rate. 
Figure~\ref{fig:vel_evolve}$c,d$ show the comparison of these two line velocities with other well-studied SNe. Initially, the Ca~II NIR velocity declines very fast but in the later epochs, only a gradual decline is seen. It evolves in a manner very similar to  iPTF13bvn, but in the late phases, the velocities are slower than other SNe. The velocities obtained using He~I 5876\,\AA\,of SN~2016bau evolve in a manner similar to that of other SNe~Ib, but much closer to SN~2009jf and iPTF13bvn. The He~I line velocity of SN~2016bau at +33\,d reaches $\sim$\,8200\,km\,s$^{-1}$, nearly equal to those of SN~2009jf and iPTF13bvn.  

\section{Fitting the Multi-band Light Curves using {\tt MOSFiT}} 
\label{mosfit}
{\tt MOSFiT}\citep[][]{Guillochon2018} is a Python-based package that downloads data from openly available online catalogues, generates the Monte Carlo ensembles of semi-analytical light curve fits to the downloaded data sets along with their associated Bayesian parameter posteriors and provides the fitting results back. Besides fitting data downloaded from openly available online catalogues, one can also perform a similar analysis to the private data sets. {\tt MOSFiT} employs various powering mechanisms for the light curves of different types of SNe. Some of them are; a) {\tt default} model incorporating the Nickel-Cobalt decay \citep[][]{Nadyozhin1994}, b) {\tt magnetar} model that takes a magnetar engine with simple spectral energy distribution \citep[][]{Nicholl2017}, and c) {\tt csm} model which is interacting CSM--SNe \citep[][]{Chatzopoulous2013, Villar2017}. A detailed description of all the models available through {\tt MOSFiT} is provided in \citet[][]{Guillochon2018}.

\begin{figure}
	{\includegraphics[width=\columnwidth]{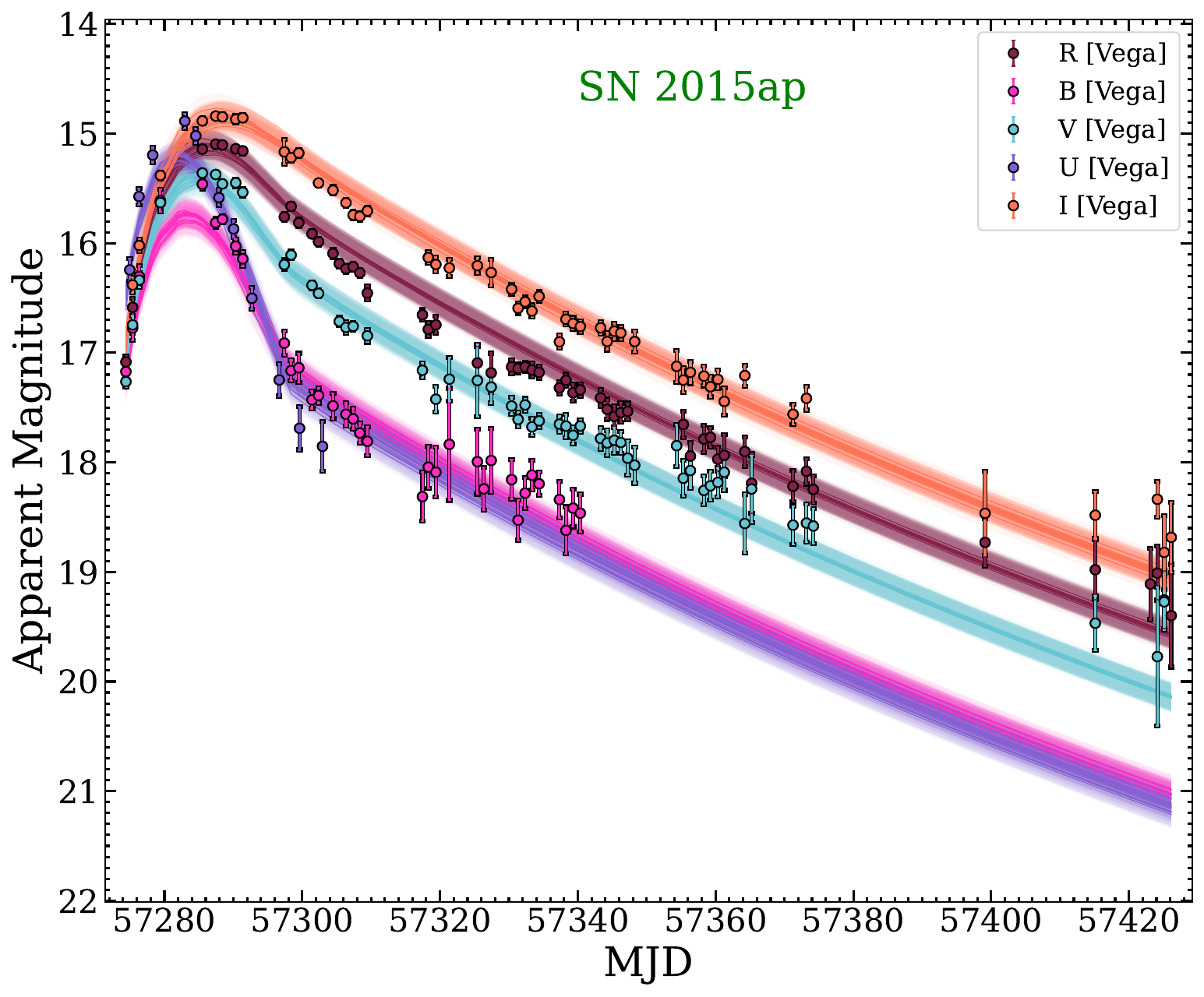}}
	\caption{The results of {\tt MOSFiT} fittings to the multi band light curves of SN~2015ap. For Type Ib SNe, the radioactive decay of nickel and cobalt is considered to be the prominent powering mechanism for their light curves. Thus, the {\tt default} model from {\tt MOSFiT} has been employed to fit the light curves.}
  \label{fig:mosfit_SN2015ap}
\end{figure} 

\begin{figure*}[!t]
  \includegraphics[width=\columnwidth,height=20cm]{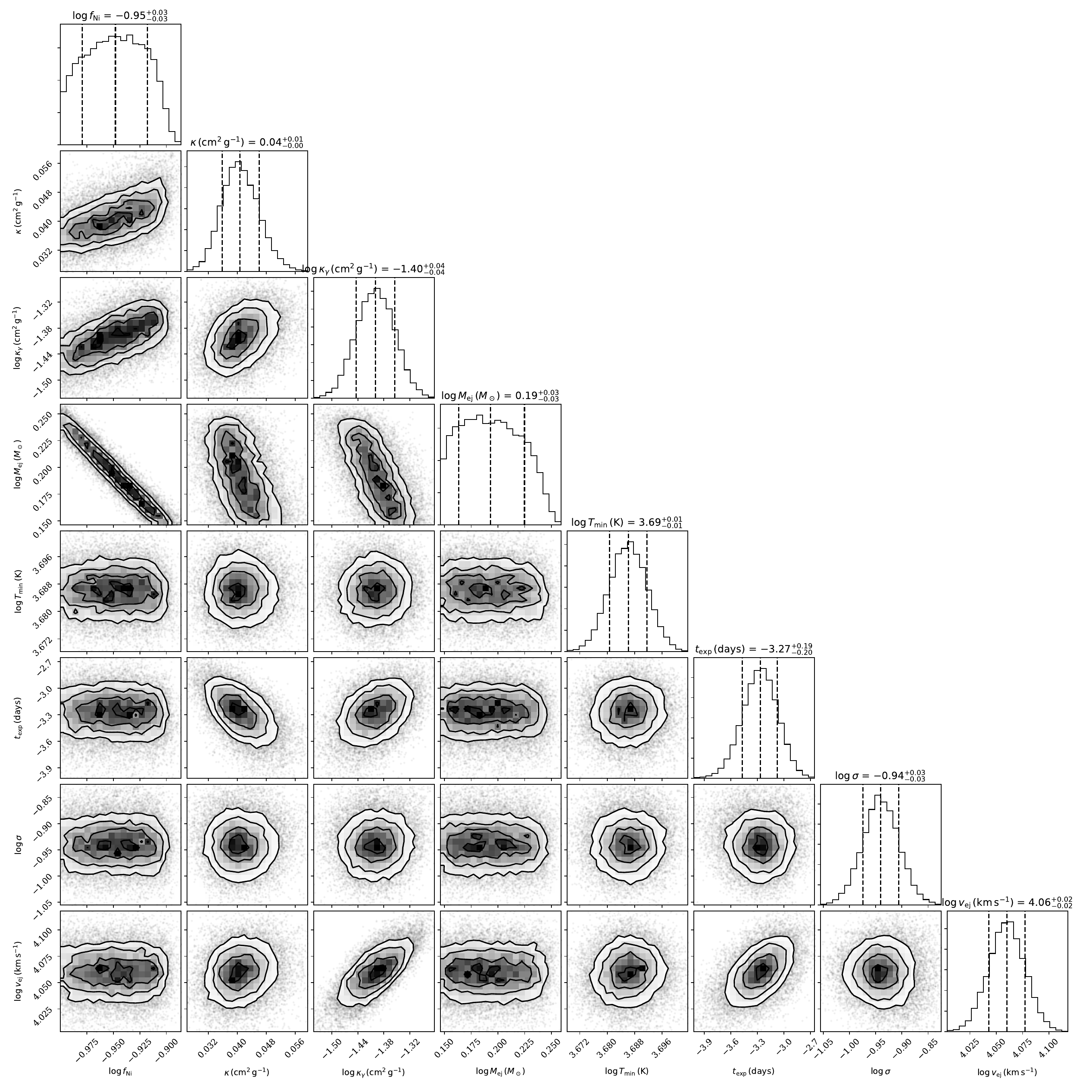}
  \caption{The corner plot of the fitting parameters from the {\tt default} model for SN~2015ap using {\tt MOSFiT}.}
  \label{fig:corner_SN2015ap}
\end{figure*}  

$\bullet$~ {\bf For SN~2015ap}:
The {\tt default} model incorporating the Nickel-Cobalt decay is considered the prominent powering mechanism for the Type Ib SNe. Thus, we tried to fit the multi-band light curves of SN~2015ap utilising the {\tt default} model. The multi-band light curves were corrected for total extinction (host galaxy plus Milky Way). The results of {\tt MOSFiT} fittings utilising the {\tt default} models are displayed in Figure~\ref{fig:mosfit_SN2015ap}. A corresponding corner plot is also shown in Figure~\ref{fig:corner_SN2015ap}. The ejecta mass obtained through {\tt MOSFiT} fitting is $\sim$1.5\,M$_{\odot}$ which is close to the value obtained by \citet[][]{Prentice2019}. The detailed fitting parameters are tabulated in \ref{tab:mosfit}.

\begin{figure}
	{\includegraphics[width=0.99\columnwidth]{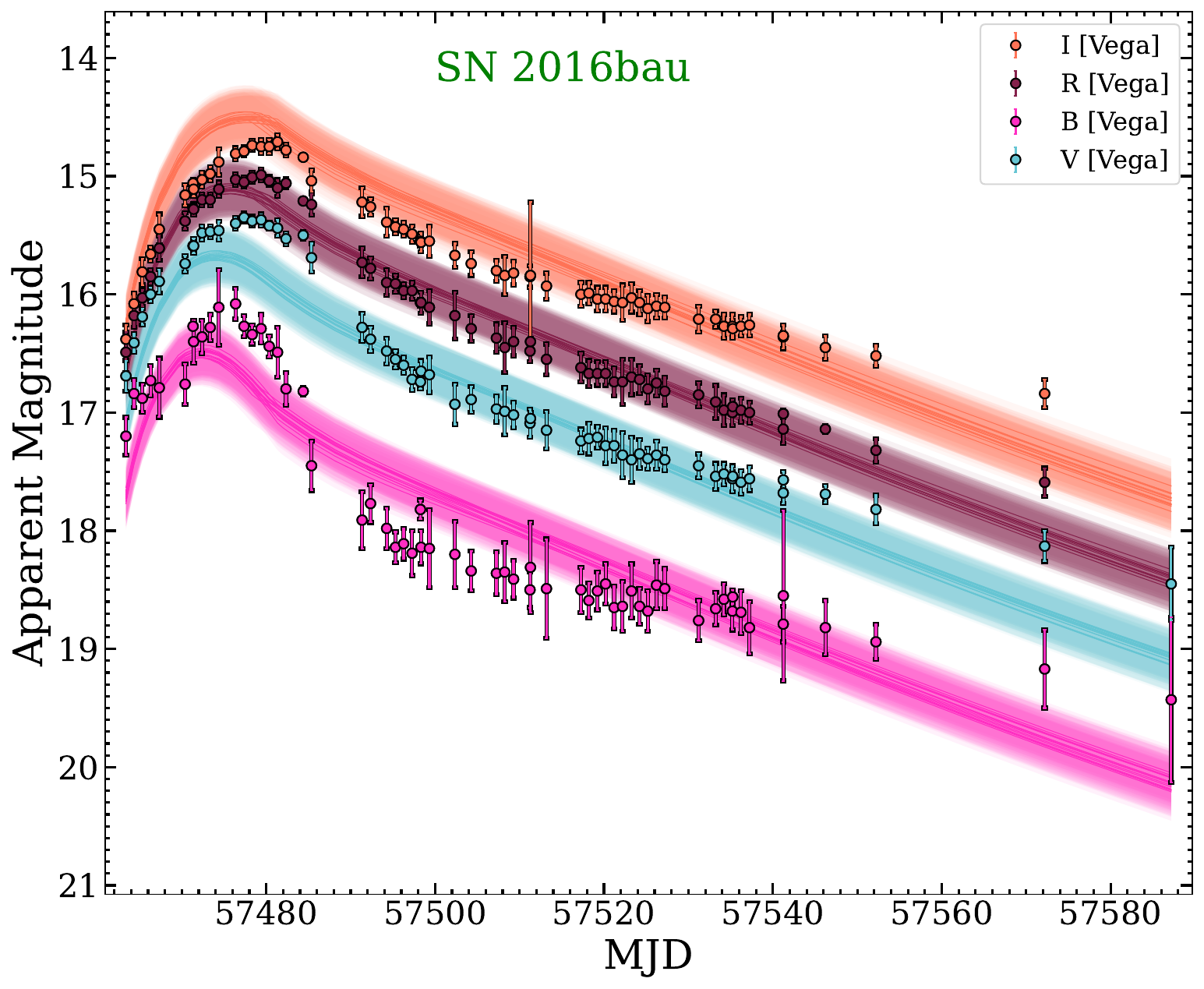}}
	\caption{The results of {\tt MOSFiT} fittings to the multi band light curves of SN~2015ap. For Type Ib SNe, the radioactive decay of nickel and cobalt is considered to be the prominent powering mechanism for their light curves. Thus, the {\tt default} model from {\tt MOSFiT} has been employed to fit the light curves.}
  \label{fig:mosfit_SN2016bau}
\end{figure} 

\begin{landscape}
\begin{table*}
\centering
\begin{scriptsize}
\begin{tabular}{l|ccccccccc}
\hline
\hline
Source name & $\log\, M_{\rm ej}$\,(M$_{\odot})$ & $\log\, f_{\rm Ni}$ & $\kappa\,({\rm cm}^{2}\,{\rm g}^{-1})$ & $\log\, \kappa_\gamma\,({\rm cm}^{2}\,{\rm g}^{-1})$ & $\log\, v_{\rm ej}\,({\rm km\,s}^{-1})$ & $\log\, T_{\min}\,{\rm (K)}$ & $\log\, \sigma$ & $t_{\rm exp}\,{\rm (days)}$ \\
\hline
SN2015ap & 0.19$_{-0.03}^{+0.03}$ & -0.95$_{+0.03}^{-0.03}$ & 0.04$_{-0.00}^{+0.01}$ & -1.40$_{-0.04}^{+0.04}$ & 4.06$_{-0.02}^{+0.02}$ & 3.69$_{-0.01}^{+0.01}$ & -0.94$_{+0.03}^{-0.03}$ & -3.27$_{+0.19}^{-0.20}$   \\\\
SN2015ap$^{*}$ & 0.25$_{-0.12}^{+0.16}$ & -0.24$_{+0.15}^{-0.22}$ & 0.10$_{-0.03}^{+0.05}$ & 1.73$_{-1.58}^{+1.48}$ & 4.45$_{-0.03}^{+0.03}$ & 3.41$_{-0.25}^{+0.24}$ & -0.80$_{+0.07}^{-0.05}$ & -3.64$_{+0.29}^{-0.40}$\\\\
\hline
SN2016bau & 0.19$_{+0.02}^{-0.01}$ & -0.94$_{+0.02}^{-0.02}$ & 0.05$_{-0.00}^{+0.00}$ & -1.46$_{-0.06}^{+0.06}$ & 3.95$_{-0.02}^{+0.02}$ & 3.82$_{-0.01}^{+0.01}$ & -0.65$_{+0.02}^{-0.02}$ & -4.88$_{+0.14}^{-0.06}$   \\\\
SN2016bau$^{*}$ & -0.51$_{+0.23}^{-0.18}$ & -1.01$_{+0.21}^{-0.25}$ & 0.10$_{-0.03}^{+0.04}$ & 2.45$_{-3.41}^{-2.48}$ & 3.58$_{-0.02}^{+0.14}$ & 3.55$_{-0.38}^{+0.31}$ & -1.72$_{+0.23}^{-0.18}$ & -4.41$_{+0.29}^{-0.36}$\\\\
\hline
\hline
\end{tabular}
\end{scriptsize}
\caption{Best-fit parameters and 68\,\% uncertainties for the {\tt default} model. The parameters for the source with * are the results from \citet[][]{Meyer2020} presented here for comparison with our studies. In this table, $M_{ej}$ is the ejecta mass, $f_{Ni}$ is the nickel mass fraction, $\kappa$ is the Thomson electron scattering opacity, $\kappa_{\gamma}$ is gamma-ray opacity of the SN ejecta, $v_{ej}$ represents the ejecta velocity, $T_{min}$ is an additional parameter for temperature floor [ please see \citet[][]{Nicholl2017}, for further details ], $\sigma$ is an additional variance parameter which is added to each uncertainty of the measured magnitude so that the reduced $\chi^{2}$ approaches 1, and $t_{\rm exp}$ is the epoch of explosion since first detection.  
\label{tab:mosfit}}
\end{table*}
\end{landscape}

$\bullet$~ {\bf For SN~2016bau}: 
Similar to SN~2015ap, the {\tt default} model is employed for the multi-band light curves of SN~2016bau after correcting for the total extinction. The fitting results are displayed in Figure~\ref{fig:mosfit_SN2016bau}. The detailed fitting parameters are tabulated in Table~\ref{tab:mosfit}.

\section{MINIM Modelling of the Quasi-bolometric Light Curve}
\label{sec:lc_model}

In this section, we fit the radioactive decay (RD) and the magnetar (MAG)  powering mechanisms, as discussed by \citealt[][]{Chatzopoulous2013} (see also \citealt{Wheeler2017, Kumar2020, Kumar2021}), by employing the {\tt MINIM} \citep[][]{Chatzopoulous2013} code. {\tt MINIM} is a $\chi^{2}$-minimisation fitting code that utilises the Price algorithm \citep[][]{Brachetti1997}.  In the RD model, the radioactive decay of $^{56}$Ni and $^{56}$Co leads to the deposition of energetic gamma-rays that are assumed to thermalise in the homologously expanding SN ejecta and thus powering the light curve. In the MAG model, the light curves are powered by the energy released by the spin-down of a young magnetar located in the centre of the SN ejecta. Following \citet[][]{Prentice2019}, we have adopted a constant opacity, $\kappa$ = 0.07\,cm$^2$\,g$^{-1}$, for both SN~2015ap and SN~2016bau.

\subsection{SN~2015ap}

Figure~\ref{fig:minim_SN2015ap} shows the results of RD and MAG model fittings to the quasi-bolometric light curve of SN~2015ap. The fitted and calculated parameters are listed in Tables \ref{tab:parameter_RD} and \ref{tab:parameter_MAG}. The ejecta mass ($M_{\rm ej}$) in the RD and MAG models was calculated using Eq. 1 from \citet[][]{Wheeler2015}. Although the RD model seems to reasonably fit the observed quasi-bolometric light curve, it seems unable to reproduce not only the observed peak luminosity but also the light curve at late phases, after +50\,d. The nickel mass and ejecta mass obtained from these models are somewhat smaller than the ones inferred directly from the observed rise time and peak luminosity in Sec.\ref{tempradvel_SN2015ap}. As the RD model is the prominent powering mechanism for normal SNe~Ib, we tried to fit the RD model only up to a relatively early phase ($\sim$\,+20\,d). In the inset of Figure~\ref{fig:minim_SN2015ap}, we see that the early phase is nicely fitted in this case. A nickel mass ($M_{\rm Ni}$) of 0.181$\pm$ 0.006\,M$_{\odot}$ and an ejecta mass ($M_{\rm ej}$) of 2.1$\pm$0.09\,M$_{\odot}$ obtained through this fitting are close to the observed values listed in Sec.\ref{tempradvel_SN2015ap}. The fitted and calculated values from the photospheric phase are collected in Table~\ref{tab:parameter_RD}. 

The MAG model fits the whole observed light curve, both around peak brightness as well as during the late phase, better than the RD model. However, the fitted parameters seem to be unphysical, particularly the very slow initial rotation ($P_{i} \approx $40\,ms) and the very low initial rotational energy ($E_{p} \approx $10$^{49}$\,erg). This is not surprising given that the magnetar model contains {\it two} timescales, one for the rising part and another for the declining part of the light curve \citep{Chatzopoulous2013}.
Thus, the possibility of SN~2015ap powered by spin-down of a magnetar is less likely.

\begin{table*}
\caption {Minimum $\chi^2$/dof parameters for SN~2015ap and SN~2016bau for the RD model.}
\label{tab:parameter_RD}
\begin{center}
{\scriptsize
\begin{tabular}{ccccccccccccc}
\hline\hline
	&	$M_{\mathrm{Ni}}$$^{a}$	&	$t_\mathrm{d}$$^{b}$	&	$A_{\mathrm{\gamma}}$$^{c}$	&	$M_{\mathrm{ej}}$$^{b}$ 	&	$\chi^2/\mathrm{dof}$	\\
	&	(M$_{\odot}$)	&	(days)	&	&	(M$_{\odot}$) 		\\
\hline
\hline
SN~2015ap\\
\hline
 whole light curve    &	0.094   $\pm$ 0.004	&	8.0  $\pm$  2.0	&	30.05   $\pm$   1.05 & 0.64 $\pm$ 0.3 & 6.2		\\
 +20\,d data    &	0.181   $\pm$ 0.006	&	14.5  $\pm$  0.3	&	5.3   $\pm$   0.3 & 2.1 $\pm$ 0.09 & 1.2		\\
\hline
SN~2016bau\\
\hline
 whole light curve    &	0.08   $\pm$ 0.01	&	20.5  $\pm$  0.7	&	7.8   $\pm$   0.5 & 2.8 $\pm$ 0.2 & 22.7		\\
  +20\,d data    &	0.065   $\pm$ 0.001	&	18.09  $\pm$  0.3	&	9.95   $\pm$   0.4 & 1.81 $\pm$ 0.07 & 1.2		\\

\hline\hline
\end{tabular}}
\end{center}

{$a$, mass of $^{56}$Ni synthesised;
$b$, effective diffusion timescale;
$c$, optical depth for the $\gamma$-rays measured 10\,d after the explosion;
$d$, ejecta mass, with $\kappa = $0.07\,cm$^2$\,g$^{-1}$.}

\end{table*}

\subsection{SN~2016bau}

Figure~\ref{fig:minim_SN2016bau} shows the results of the RD and MAG model fittings to the quasi-bolometric light curve of SN~2016bau. The fitted and calculated parameters are listed in Tables \ref{tab:parameter_RD} and \ref{tab:parameter_MAG}. The RD model can fit the observed peak luminosity, but huge deviations from the observed light curve are seen in the later phases. The model also fails to match the observed stretch factor of the light curve. Like SN~2015ap, we also tried to fit only the early part, before +20\,d post-peak (see the inset in Fig.~\ref{fig:minim_SN2016bau}). 
In this case, the fitted parameters, such as the nickel mass of 0.065$\pm$0.002\,M$_{\odot}$ and the ejecta mass of 1.81$\pm$0.07\,M$_{\odot}$ (see Table~\ref{tab:parameter_RD}), are very close to the observed values in Section~\ref{tempradvel_SN2016bau}. Similar to SN~2015ap, the slow rotation ($P_i \approx $70\,ms), low magnetar rotational energy ($E_p \approx $10$^{48}$\,erg), and very high progenitor radius ($\sim$\,1200\,R$_{\odot}$) make the MAG model physically unrealistic for SN~2016bau, despite the better fit to the whole light curve in Figure~\ref{fig:minim_SN2016bau}.

\begin{landscape}
\begin{table*}
\caption{Minimum $\chi^2$/dof parameters for SN~2015ap and SN~2016bau for the MAG model.}
\label{tab:parameter_MAG}
\begin{center}
{\scriptsize
\begin{tabular}{ccccccccccccc}
\hline\hline
	&	$R_0$$^{a}$	&	$E_p$$^{b}$	  & 	$t_d$$^{c}$	&	$t_p$$^{d}$	 & $v_{\rm exp}$$^{e}$      & $M_{\rm ej}$$^{f}$  	& $P_i$$^{g}$   &    $B$$^{h}$    &   $\chi^2/\mathrm{dof}$\\
	&($10^{13}$\,cm) & ($10^{51}$\,erg) & (days)          & (days)         & ($10^3$\,km\,s$^{-1}$) & (M${_\odot}$)    & (ms)          & ($10^{14}$\,G)   &		\\
\hline
\hline
SN~2015ap\\
\hline
      & 0.594   $\pm$  0.003 & 0.01210 $\pm$  0.00007 & 8.7  $\pm$  0.1 & 12.06   $\pm$  0.08 & 6.03  $\pm$  0.02  & 0.75$\pm$ 0.02 & 40.6 $\pm$0.1 & 25.5$\pm$ 0.2 & 1.54 \\
\hline
SN~2016bau\\
\hline
      & 7.9   $\pm$  0.7 & 0.00403 $\pm$  0.00007 & 8.5   $\pm$  0.1 & 15.1  $\pm$  0.1  & 6.2  $\pm$  0.5  & 1.26$\pm$ 0.02 & 70.4$\pm$ 0.6 & 52.6$\pm$ 0.8 & 1.4 \\
\hline\hline
\end{tabular}}
\end{center}

{$a$, progenitor radius; $b$, magnetar rotational energy; $c$, effective diffusion timescale (in days); $d$, magnetar spin-down timescale;
$e$, SN expansion velocity; $f$, $\kappa = $0.07\,cm$^2$\,g$^{-1}$ is used; $g$, initial period of the magnetar; $h$, magnetic field of the magnetar.}
\end{table*}
\end{landscape}

\begin{figure}
	\includegraphics[width=\columnwidth]{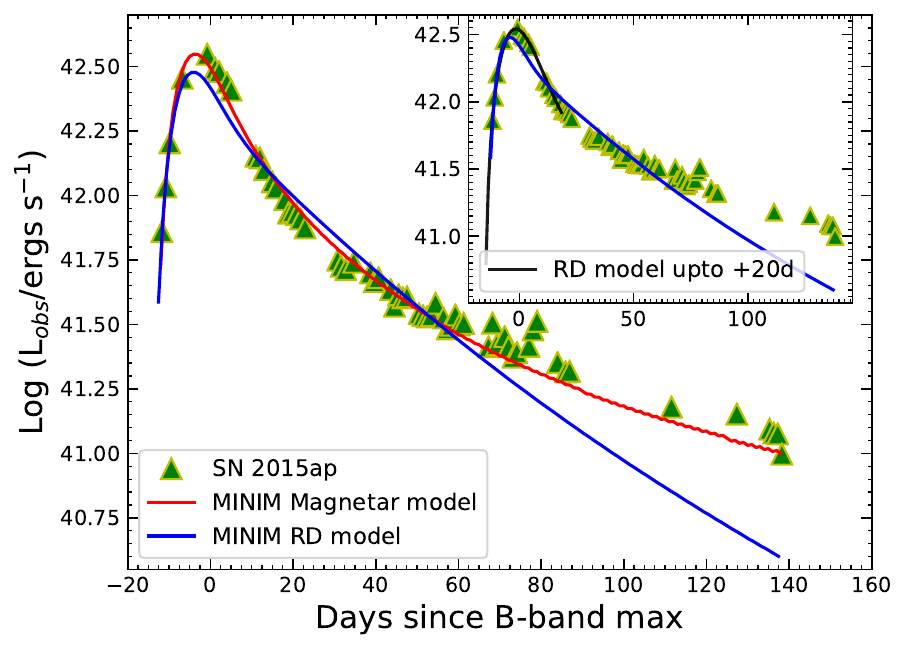}
 	\caption{{\tt MINIM} modelling of the quasi-bolometric light curve of SN~2015ap. The inset shows the RD model fitting only for the early phase (up to +20\,d).}
   	\label{fig:minim_SN2015ap}
\end{figure}

\begin{figure}
	\includegraphics[width=\columnwidth]{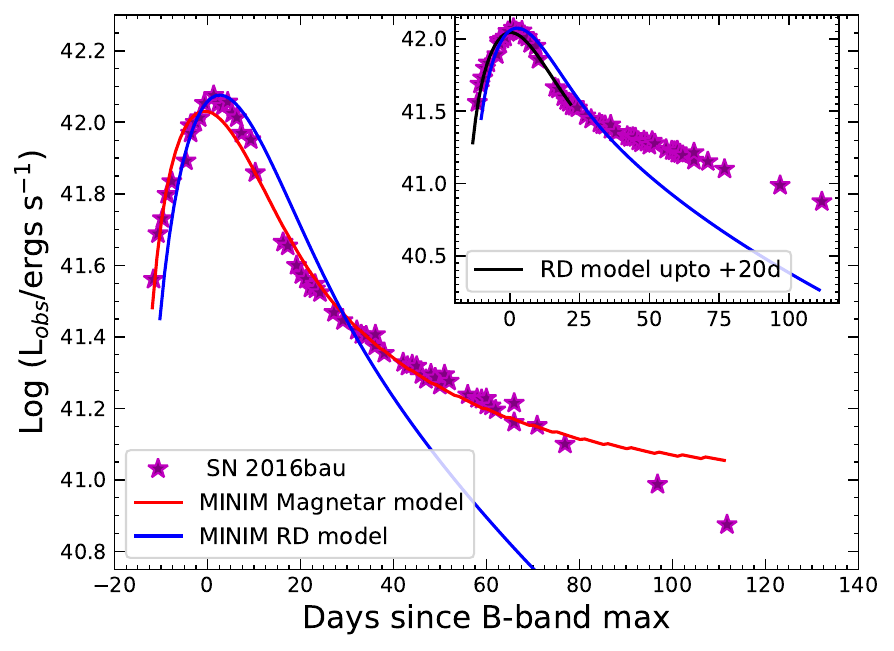}
 	\caption{{\tt MINIM} modelling of the quasi-bolometric light curve of SN~2016bau. The figure in the inset shows the RD model fitting only for the early phase (up to +20\,d).}
   	\label{fig:minim_SN2016bau}
\end{figure}

\subsection{Summary of {\tt MINIM} modelling}
The failure of the semi-analytical RD models to simultaneously fit the early and late parts of the light curves of SN~2015ap and SN~2016bau highlights the issues related to the validity of such kinds of models in the case of stripped-envelope SNe. The difficulty in explaining the late-phase decline rate with the assumed diffusion model has already been explored by \citet{Wheeler2015}. 

Figures~\ref{fig:minim_SN2015ap} and \ref{fig:minim_SN2016bau} reveal another issue: the model light curve is steeper after +30--40\,d than the observed one. This indicates that the early- and late-phase data cannot be fitted simultaneously with the same model parameters. 
Even though the nickel mass of the second model fits (insets in Figs.~\ref{fig:minim_SN2015ap} and \ref{fig:minim_SN2016bau}) the observed peak luminosity, the decline rate, which is related to the ejecta mass (as in Eq. 1 of \citealt[][]{Wheeler2015}) is clearly too fast, suggesting an underestimated $M_{\rm ej}$. 

This issue is very probably related to the assumption of the constant ejecta density profile (and also the constant opacity). The less steep late part of the light curve needs more ejecta mass to trap the heating gamma-rays originating from the Ni and Co decay. The early part, however, suggests ejecta that dilute much faster than what can fit the late part. Within the context of the constant-density model, this dichotomy means that fitting only the early part results in a lower ejecta mass compared to fitting only the late part (Figure~\ref{fig:minim_SN2015ap_tot}). 

\begin{figure}
	\includegraphics[width=\columnwidth]{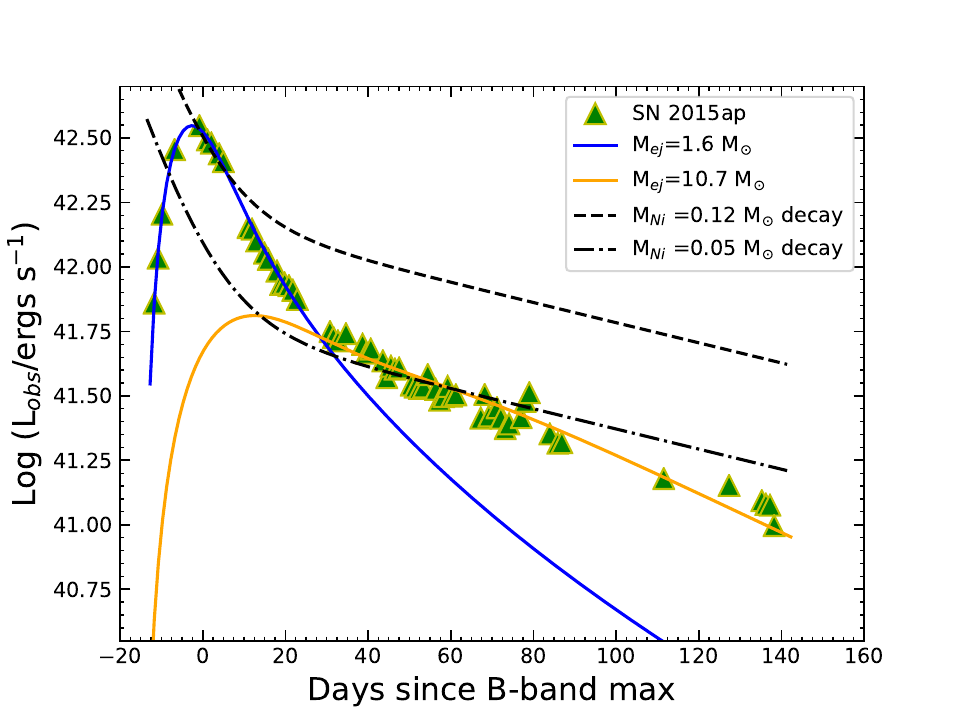}
 	\caption{Detailed {\tt MINIM} modelling to the quasi-bolometric light curve of SN~2015ap. All of the models with different $M_{\rm ej}$ and $M_{\rm Ni}$ are shown together. The models with $M_{\rm ej} = $1.6\,M$_{\odot}$ and 10.7\,M$_{\odot}$ show the RD model fittings considering only the early- and late-phase data, respectively. The other two models show the effect of the amount of nickel present. }
   	\label{fig:minim_SN2015ap_tot}
\end{figure}

Since this issue cannot be solved self-consistently with the constant-density model, and it can plague the mass estimates that consider only the early or late part of the light curve, we investigate it further in the following sections by applying more realistic models for the progenitors and the SN light curves. 

\section{MESA Modelling of a 12 M$_{\odot}$ ZAMS Progenitor Star}
%as possible progenitor}
\label{sec:mesa}

Adoption of a 12\,M$_{\odot}$ ZAMS progenitor star for SN~2015ap is primarily based on the results of comparison of the three models from \citet[][]{Jerkstrand2015} and also in the literature; in addition, it supports the observed amount of ejecta mass for SN~2015ap. The ejecta mass for SN~2016bau was calculated to be $\sim$\,1.6\,M$_{\odot}$. On account of such an ejecta mass and the result of \citet[][]{Jerkstrand2015} spectral model matching, we also choose a 12\,M$_{\odot}$ ZAMS star as a possible progenitor for SN~2016bau.

We first evolve the 12\,M$_{\odot}$ ZAMS star until the onset of core-collapse, using the 1-dimensional stellar evolution code {\tt MESA}, version 11701\citep[][] {Paxton2011,Paxton2013,Paxton2015,Paxton2018}. We do not consider rotation and assume an initial metallicity of Z = 0.02. Convection is modelled using the mixing theory of \citet[][]{Henyey1965}, adopting the Ledoux criterion. We set the mixing-length parameters to $\alpha$ = 3.0 in the region where the mass fraction of hydrogen is greater than 0.5, and set it to 1.5 in the other regions. Semi-convection is modelled following \citet[][]{Langer1985} with an efficiency parameter of $\alpha_{\mathrm{sc}}$ = 0.01. For the thermohaline mixing, we follow \citet[][]{Kippenhahn1980}, and set the efficiency parameter as $\alpha_{\mathrm{th}}$ = 2.0. We model the convective overshooting with the diffusive approach of \citet[][]{Herwig2000}, with $f$= 0.01 and $f_0$ = 0.004 for all the convective core and shells. We use the `Dutch' scheme for the stellar wind, with a scaling factor of 1.0. The `Dutch' wind scheme in MESA combines results from several papers. Specifically, when $T_{\mathrm{eff}} > $10$^4$\,K and the surface mass fraction of hydrogen is greater than 0.4, the results of \citet[][]{Vink2001} are used, and when $T_{\mathrm{eff}} > $10$^4$\,K and the surface mass fraction of hydrogen is less than 0.4, the results of \citet[][]{Nugis2000} are used. In the case when $T_{\mathrm{eff}} < $10$^4$\,K, the \citet[][]{dejager1988} wind scheme is used. 

SNe~Ib have been considered to originate from  massive stars which lose almost all of their H-envelope, most probably due to binary interaction \citep[e.g.,][]{Yoon2010, Dessart2012, Eldridge2016, Ouchi2017}. Here, to produce such a stripped model, we artificially strip the H-envelope, mimicking the binary interaction. Specifically, after evolving the model until the exhaustion of helium, we impose an artificial mass-loss rate of $\dot{M} \gtrsim $10$^{-4}$\,M$_{\odot}$\,yr$^{-1}$ until the total hydrogen mass of the star goes down to 0.01\,M$_{\odot}$. After the hydrogen mass reaches the specified limit, we switch off the artificial mass loss and evolve the model until the onset of core collapse. At the time of the core collapse, our model has a total mass of 3.42\,M$_{\odot}$.
 
\section{Synthetic Explosions of Modelled Progenitors Using SNEC and STELLA}
\label{sec:snec}

In this section we briefly discuss the assumptions and setups to produce  artificial explosions using {\tt SNEC} and {\tt STELLA} for SN~2015ap and SN~2016bau.

\subsection{Results of synthetic explosions for SN~2015ap}
Using the progenitor model on the verge of core-collapse obtained through {\tt MESA}, we have conducted the radiation hydrodynamic simulations. For this purpose, we use the publicly available codes {\tt SNEC} and {\tt STELLA}.

{\tt SNEC} is a 1-dimensional Lagrangian hydrodynamic code, which also solves radiation energy transport with the flux-limited diffusion approximation. The code generates the bolometric light curve and the photospheric velocity evolution of the SN, along with a few other observed parameters.
The setup for the calculation using {\tt SNEC} closely follows \citet[][]{Ouchi2019}. Here, we briefly summarise the important parameters and modifications made to \citet[][]{Ouchi2019}. First, we excise the innermost 1.4\,M$_{\odot}$ before the explosion, assuming it collapses to form an NS. The number of cells is set to be 70. Although this number is relatively small, we have confirmed that the SN's light curve and photospheric velocity are well converged in the time domain of interest. We tried following two possible powering mechanisms for SN~2015ap using {\tt SNEC}, as follows.

\begin{figure}
\includegraphics[width=\columnwidth]{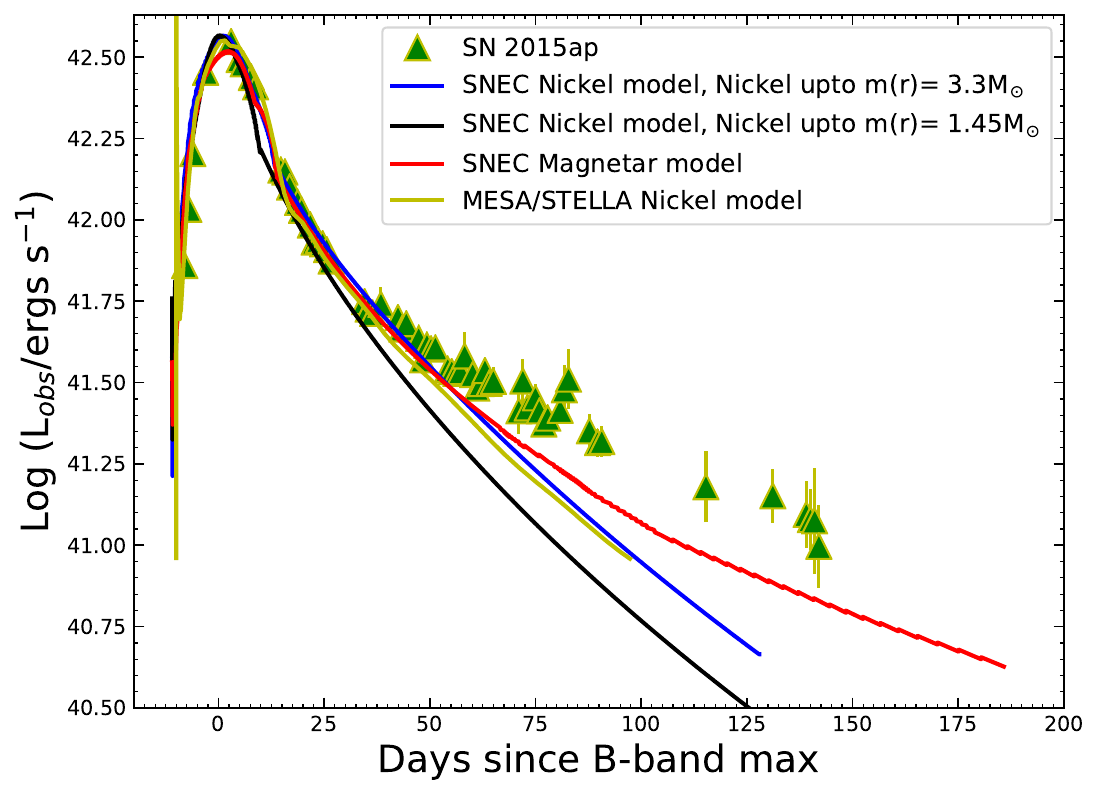}
\caption{Comparison of the quasi-bolometric light curve of SN~2015ap with those obtained using {\tt SNEC} by considering the Ni--Co decay and the magnetar model. This figure also depicts the result of the Ni--Co decay model obtained using {\tt STELLA}.}
\label{fig:lum}
\end{figure}

\begin{figure}
\includegraphics[width=\columnwidth]{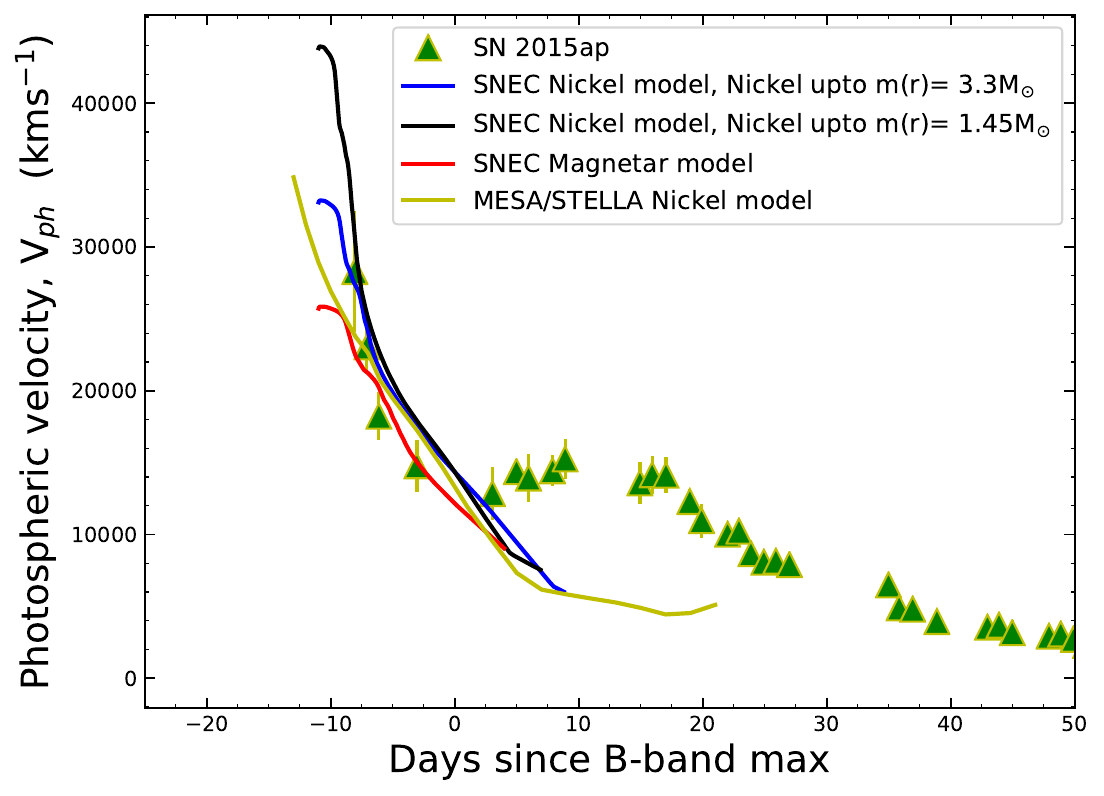}
\caption{Comparison of the observed velocity evolution of SN~2015ap with that produced by {\tt SNEC} and {\tt STELLA} for the two models employed.}
\label{fig:velocity}
\end{figure}

$\bullet$~{\bf Ni--Co decay}

The radioactive decay of Ni and Co is considered to be the most prominent mechanism for the powering of light curves of SNe~Ib \citep[e.g.,][]{Karamehmetoglu2017}. {\tt SNEC} incorporates this model by default. Here, we provide the setup of the explosion parameters to incorporate the Ni--Co decay model. 
The code does not include a nuclear-reaction network, and $^{56}$Ni is given by hand. We considered two scenarios of nickel distribution. In one case, the mass of Ni is set to be $M_{\mathrm{Ni}}$ = 0.135\,M$_{\odot}$ and distributed from the inner boundary up to the mass coordinate $m (r)$ = 3.3\,M$_{\odot}$. For this model, the explosion is simulated as a {\tt piston}, with the first two computational cells of the profile boosted outward with a velocity of 4.0$\times$10$^{9}$\,cm\,s$^{-1}$ for a time interval of 0.01\,s. The total energy ($E_{\rm tot}$) of the model is 3.7$\times$10$^{51}$\,erg.  For the second case, the mass of Ni is set to be $M_{\mathrm{Ni}}$ = 0.1\,M$_{\odot}$ and distributed near the centre, from the inner boundary up to the mass coordinate of $m (r)$ = 1.45\,M$_{\odot}$. The explosion is simulated as a {\tt piston}, with the first two computational cells of the profile boosted outward with a velocity of 5.0$\times$10$^{9}$\,cm\,s$^{-1}$ for a time interval of 0.01\,s. The total energy of the model, in this case, is 6.5$\times$10$^{51}$\,erg. Thus, our results provide a range of $M_{\mathrm{Ni}}$ and total energy, depending on the distribution of nickel mass.

$\bullet$~{\bf Spin-down of a magnetar}

We also tried a magnetar-powering mechanism for the light curve as SN~2015ap; it shows some resemblance to SN~2008D, which has broader features in its early-time spectra and also some X-ray emission in the later phases. The magnetar models also explain a few SNe~Ib, one such example being SN~2005bf \citep[][]{Maeda2007}. For this model, the explosion is the {\tt piston} type, with the first two computational cells of the profile boosted outward with a velocity of 3.3$\times$10$^{9}$\,cm\,s$^{-1}$ for a time interval of 0.01\,s. The total energy of the model, in this case, is 2.2$\times$10$^{51}$\,erg. The most important change made to \citet[][]{Ouchi2019} is that we add the magnetar heat to the ejecta.

Following \citet[][]{Metzger2015}, the magnetar spin-down luminosity is given by
\[
L_{\mathrm{sd}} = L_{\mathrm{sd_i}} (1 + t/t_{\mathrm{sd}})^{-2}.
\]
\noindent
Here, $L_{\mathrm{sd_i}}$ is the spin-down luminosity at $t$ = 0, and $t_{\mathrm{sd}}$ is the initial spin-down time. We inject this luminosity into the whole ejecta above the mass cut uniformly in mass.
For the initial spin-down luminosity, we assume $L_{\mathrm{sd{_{i}}}}$ = 1.3$\times$10$^{43}$\,erg\,s$^{-1}$, while for the initial spin-down time, we assume $t_{\mathrm{sd}}$ = 12\,d. In this model, we do not include the effect of Ni heating. Following \citet[][]{Metzger2015} (Equations 2 and 3), corresponding to $L_{\mathrm{sd{_{i}}}}$ = 1.3$\times$10$^{43}$\,erg\,s$^{-1}$ and $t_{\mathrm{sd}}$ = 12\,d, we obtain a magnetic field ($B$) of 5.1$\times$10 $^{14}$\,G and an initial period ($P_i$) of 43.06\,ms for the modelled magnetar. These values of $B$ and $P_i$ are very close to those obtained using {\tt MINIM}.

We also use the public version of {\tt STELLA}, available with {\tt MESA}. The default radioactive decay of $^{56}$Ni and $^{56}$Co as a powering mechanism is used for SN~2015ap. Nearly similar parameters as in the case of {\tt SNEC} are used for {\tt STELLA}. The {\tt MESA} setups are unchanged for {\tt STELLA} calculations. We use a total energy after explosion of 3.6$\times$10$^{51}$\,erg and a $^{56}$Ni mass of $0.193$\,M$_{\odot}$, which is slightly more than the $^{56}$Ni mass used in {\tt SNEC}. To avoid the numerical problem caused by the high-velocity material, we removed the outer layer of the progenitor where the density is less than 10$^{-5}$\,g\,cm$^{-3}$ \citep[see also][]{Moriya2020}.

$\bullet$~{\bf Summary of the modelling}

Figure~\ref{fig:lum} shows the comparison of the observed quasi-bolometric luminosity with that produced by {\tt SNEC}. We find that the radioactive decay models with $^{56}$Ni mass in the range 0.1--0.135\,M$_{\odot}$ and total energy in the range (3.7--6.5)$\times$10$^{51}$\,erg could nicely explain the light curve. Our results also signify that the distribution of $^{56}$Ni mass plays an important role for explaining the observed light curves.
The magnetar model could also explain the quasi-bolometric light curve, but we do not see very strong evidence that SN~2015ap is powered by a magnetar. This figure also shows the results of {\tt STELLA} calculations. We see that, although the explosion energy is similar, we need a slightly higher amount of nickel to properly match the observed light curve of SN~2015ap.
  
Figure~\ref{fig:velocity} illustrates a comparison of observed photospheric velocity evolution produced by blackbody fitting with that produced by {\tt SNEC}. We see that the Ni--Co decay and magnetar models initially show higher velocities, but in the later epochs, they replicate the observed photospheric velocities well.  This figure also shows the Fe~II 5169\,\AA\, velocity evolution produced by {\tt STELLA}. Owing to the unambiguous absence of Fe~II 5169\,\AA\, features in SN~2015ap, we used the photospheric velocity obtained through a blackbody fit for our comparison of Fe~II 5169\,\AA\, line velocities. We can see a very good match between the {\tt STELLA} velocities and observed ones. The modelling parameters, along with the observed values, are listed in Table \ref{tab:parameter_model}.

\subsection{Comparison of synthetic explosions with observations of SN~2016bau}
Using a similar progenitor model on the verge of core-collapse, obtained through {\tt MESA}, as in the case of SN~2015ap, we carried out the radiation hydrodynamic simulations using {\tt SNEC} and {\tt STELLA}. For calculations using {\tt SNEC}, the setup closely follows that of \citet[][]{Ouchi2019}.  First, we excise the innermost 1.53\,M$_{\odot}$ before the explosion, assuming it collapses to form an NS. The number of cells is set to 70. We try two possible powering mechanisms for SN~2016bau using {\tt SNEC}, as follows.

\begin{table*}
\caption{Observed and modelled parameters for SN~2015ap and SN~2016bau}
\label{tab:parameter_model}
\begin{center}
{\scriptsize
\begin{tabular}{ccccccccccccc}
\hline\hline
	&	$M_{\mathrm{Ni}}$	&	$M_{\mathrm{ej}}$	&	$E_{\mathrm{tot}}$	\\
	&	(M$_{\odot}$)	&	(M$_{\odot}$)	& $10^{51}$\,erg		\\
\hline
\hline
SN~2015ap\\
\hline
 Arnett's model$^{a}$    &	0.14   $\pm$ 0.02	&	2.2  $\pm$ 0.6	&	* \\

 {\tt SNEC} (Ni distributed up to $M(r) = 3.3\,M_{\odot}$)    &	    0.135	&	  2.02	&	   3.7 	\\
 
 {\tt SNEC} (Ni distributed up to $M(r) = 1.45\,M_{\odot}$)    &	    0.1	&	  2.02	&	   6.5 	\\

 {\tt SNEC}  Magnetar model   &	0.0	&	  2.02	&	   2.2 	\\

 From {\tt STELLA}    &	0.193	& 1.92		&	3.6	\\
\hline
SN~2016bau\\
\hline
 Arnett's model$^{a}$   &	0.055   $\pm$ 0.006	&	1.6  $\pm$ 0.3	&	** \\

%SNEC Ni model    &	0.05	&	1.89	&	1.6 	\\

{\tt SNEC} (Ni distributed up to $M(r) = 3.3\,M_{\odot}$)    &	    0.045	&	  1.89	&	   1.23 	\\
 
{\tt SNEC} (Ni distributed up to $M(r) = 1.57\,M_{\odot}$)    &	    0.03	&	  1.89	&	   1.93 	\\

{\tt SNEC}  Magnetar model   &	0.0	&	1.89	&	0.8 	\\

 From {\tt STELLA}    &	0.065  	&	1.89  	&	1.6	\\
\hline\hline
\end{tabular}}
\end{center}
%\par
{$a$: Calculated using $t_{\rm rise}$, $\kappa = 0.07$\,cm$^2$\,g$^{-1}$, and $L_{\rm peak}$.}\\
{$*$: Instead, a kinetic energy of the ejecta $(E_{k})$ = 1.05$\times$ 10$^{51}$\,erg is obtained.}\\
{$**$: Instead, a kinetic energy of the ejecta $(E_{k})$ = 0.24$\times$10$^{51}$\,erg is obtained.}\\
\end{table*} 

$\bullet$~{\bf Ni--Co decay}

The setup for the calculation using {\tt SNEC} is similar to that of SN~2015ap.
Considering the radioactive decay of Ni--Co the most prominent mechanism for powering light curves of SNe~Ib, we employed this model as the powering mechanism for SN~2016bau. Here we briefly describe the setup of the explosion parameters incorporated in the Ni--Co decay model.

We considered two cases of $^{56}$Ni mass distribution. In the first case, the mass of $^{56}$Ni synthesised is set to be $M_{\mathrm{Ni}}$ = 0.045\,M$_{\odot}$. Then, it is distributed from the inner boundary up to the mass coordinate of $M(r)$ = 3.3\,M$_{\odot}$. Thereafter, we simulate the explosion as a {\tt piston}, with the first two computational cells of the profile boosted outward with a velocity of 4.2$\times$10$^{9}$\,cm\,s$^{-1}$ for a time interval of 0.01\,s. The model has a total energy of 1.3$\times$10$^{51}$\,erg. For the second case, the mass of $^{56}$Ni synthesised is set to be $M_{\mathrm{Ni}}$ = 0.03\,M$_{\odot}$ and distributed from the inner boundary up to the mass coordinate of $M(r)$ = 1.57\,M$_{\odot}$. Thereafter, we simulate the explosion as a {\tt piston}, with the first two computational cells of the profile boosted outward with a velocity of 4.9$\times$10$^{9}$\,cm\,s$^{-1}$ for a time interval of 0.01\,s, and the model has a total energy of 1.93$\times$10$^{51}$\,erg. We find that, depending on the $M_{\mathrm{Ni}}$ distribution, the radioactive decay models with $^{56}$Ni mass in the range 0.03--0.045\,M$_{\odot}$ and total energy in the range (1.23--1.93)$\times$10$^{51}$\,erg can nicely explain the light curve.

\begin{figure}
%\centering
\includegraphics[width=\columnwidth]{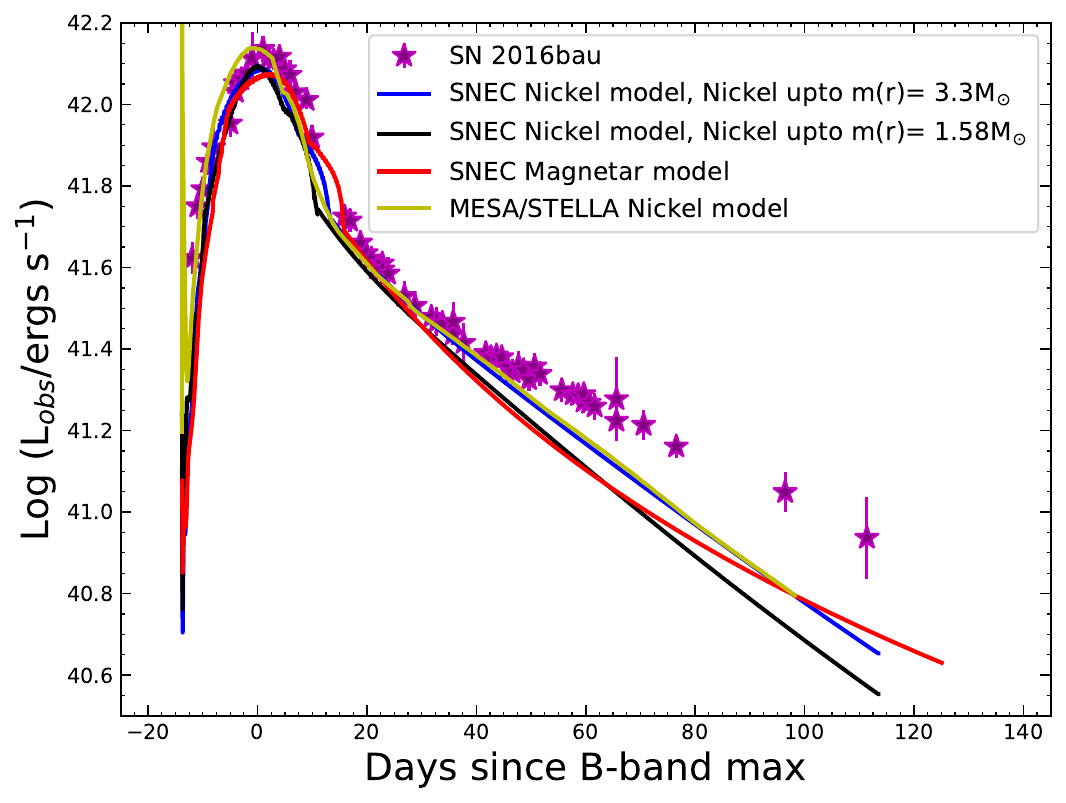}
\caption{Comparison of the quasi-bolometric light curve of SN~2016bau and light curves produced by {\tt SNEC} and {\tt STELLA}, considering the radioactive decay model.}
\label{fig:lum_SN2016bau}
\end{figure}

\begin{figure}
%\centering
\includegraphics[width=\columnwidth]{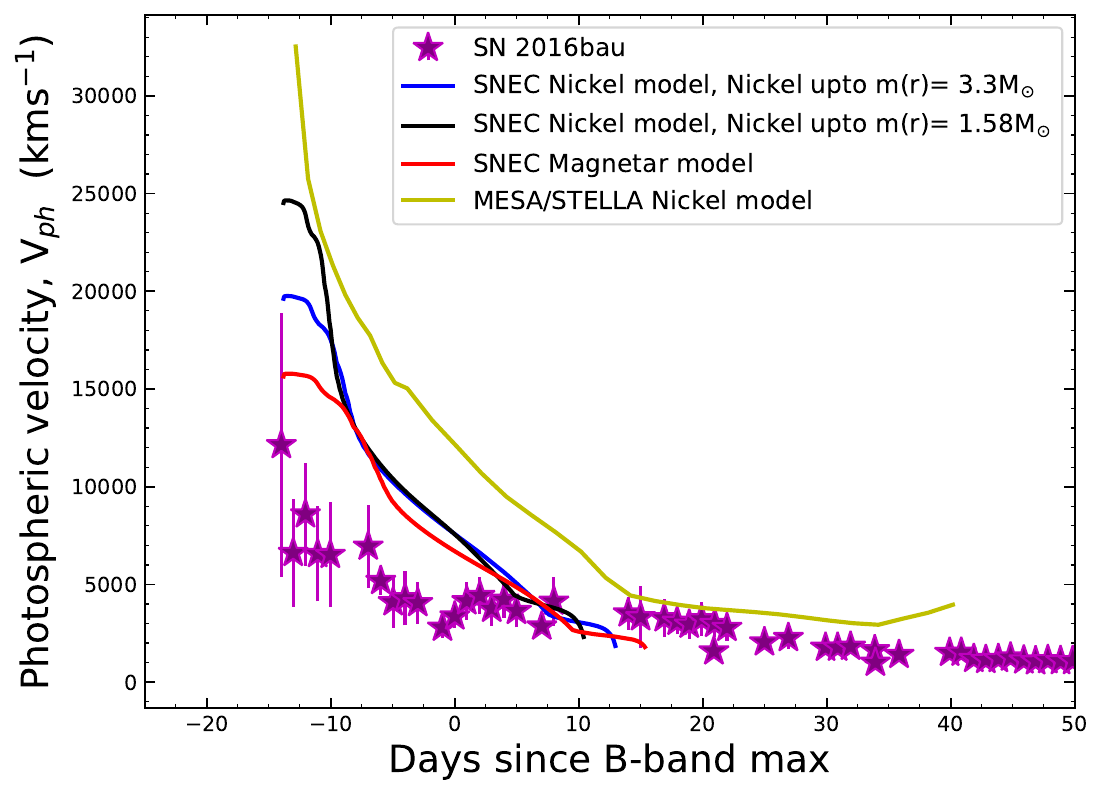}
\caption{Comparison between the observed velocity evolution of SN~2016bau and the photometric velocity evolution produced by {\tt SNEC} and {\tt STELLA}, considering the radioactive decay model.}
\label{fig:velocity_SN2016bau}
\end{figure}

$\bullet$~{\bf Spin-down of a magnetar}

We also tried the magnetar powering mechanism for SN~2016bau. The explosion is simulated as a {\tt piston}, with the first two computational cells of the profile boosted outward with a velocity  of 3.6$\times$10$^{9}$\,cm\,s$^{-1}$ for a duration of 0.01\,s and a total model energy of 0.8$\times$10$^{51}$\,erg. Like SN~2015ap, we inject $L_{\mathrm{sd_i}}$ = 0.4$\times$10$^{43}$\,erg\,s$^{-1}$ to the ejecta above the mass cut uniformly in mass. We assume $t_{\mathrm{sd}}$ = 16.0\,d and do not include the effect of Ni heating. Corresponding to $L_{\mathrm{sd{_{i}}}}$ = 0.4$\times$10$^{43}$\,erg\,s$^{-1}$ and $t_{\mathrm{sd}}$ = 16\,d, we obtain a magnetic field ($B$) of 7.9$\times$10$^{14}$\,G and an initial period ($P_i$) of 71.8\,ms for the modelled magnetar. These values of $B$ and $P_i$ are very close to those obtained using {\tt MINIM}.

We also perform {\tt STELLA} calculations for SN~2016bau by employing the default radioactive decay of $^{56}$Ni and $^{56}$Co powering mechanism. We used similar parameters as in the case of {\tt SNEC} and {\tt STELLA}. The {\tt MESA} setups are unchanged for {\tt STELLA} calculations. We used a total energy after explosion of 1.6$\times$10$^{51}$\,erg and a $^{56}$Ni mass of 0.065\,M$_{\odot}$, slightly above the $^{56}$Ni mass used in {\tt SNEC}.

$\bullet$~{\bf Summary of the modelling}

Figure~\ref{fig:lum_SN2016bau} shows the comparison of the observed quasi-bolometric luminosity with that produced by {\tt SNEC} for the radioactive decay and magnetar spin-down models. We find that the radioactive decay model from {\tt SNEC} could successfully explain the light curve. The magnetar model also shows a good match, but we do not see any significant signs of SN~2016bau being powered by a magnetar. This figure also shows the results of {\tt STELLA} calculations; they also match the observed light curve of SN~2016bau nicely, with parameters similar to those of {\tt SNEC}.

Figure~\ref{fig:velocity_SN2016bau} shows a comparison of the observed photospheric velocity evolution produced by blackbody fitting with that produced by the radioactive decay model and the magnetar spin-down model using {\tt SNEC}. Here, the models show high initial velocities, which drop at later epochs. We see that the velocities produced by the models initially deviate from the observed photospheric velocities but tend to follow velocities similar to the observed ones at later epochs. The figure also shows a comparison of the Fe~II 5169\,\AA line velocity obtained using {\tt STELLA}  with the photospheric velocity obtained through blackbody fits. Similar to the case of SN~2015ap, this SN also lacks unambiguous features of the Fe~II 5169\,\AA line.

From Figure~\ref{fig:velocity_SN2016bau}, it is evident that our model overestimates the velocity by a factor of nearly two. The $M_{\rm ej}$ from Arnett's model is $\sim$\,1.6\,M$_{\odot}$. The diffusion time is given by $(M^3/E)^{1/4}$. Since this combination is fixed, $M^3/E$ is roughly constant. But $M^3/E \approx M^3/(Mv^2) = M^2/v^2 \approx$ constant. Thus, we have $M \approx v$ to meet the observational constraints. So, if we need to decrease the velocity by a factor of two, the ejecta mass would also decrease by the same factor. Then the He star mass is likely $\sim$\,2--2.5\,M$_{\odot}$, which is at the boundary between an SN and a non-SN. Since $E \approx Mv^2 \approx v^3$, the energy may go down quite substantially in this case.
 
\section{Discussion}
\label{sec:Discussions}
We present a detailed photometric and spectroscopic analysis of two Type Ib SNe, namely SN~2015ap and SN~2016bau. From our analysis, SN~2015ap is an intermediate-luminosity normal SN~Ib, while SN~2016bau is highly extinguished by host-galaxy dust. In this section, we discuss the major outcomes of our present analysis.

The photometric properties of both the SNe were analysed by determining the bolometric luminosity of their light curves. For SN~2015ap, we calculate a $^{56}$Ni mass of 0.14$\pm$0.02\,M$_{\odot}$ and an ejecta mass of 2.2$\pm$0.6\,M$_{\odot}$, while for SN~2016bau, the $^{56}$Ni mass and ejecta mass were 0.055$\pm$0.006\,M$_{\odot}$ and 1.6$\pm$0.3\,M$_{\odot}$, respectively. The photospheric temperature, radius, and velocity evolution for both SNe were also explored. Based on the derived physical quantities from the temporal evolution of these two SNe, we tried to constrain possible powering mechanisms using a semi-analytical model called {\tt MINIM}. We found that the semi-analytical RD model failed to simultaneously fit both SNe's early and late phases, raising issues related to the validity of such models in the case of stripped-envelope SNe. One solution to such a situation is to fit the early-phase and late-phase data with different sets of model parameters. Another cause for the failure of the RD model can be attributed to the assumption of the constant ejecta density profile. For the MAG model, the fitted parameters seemed to be unphysical in both SNe, especially the very slow initial rotation (for SN~2015ap, $P_{i} \approx$ 40\,ms, and for SN~2016bau, $P_{i} \approx$ 70\,ms) and the very low initial rotational energy (for both SNe, $E_{p} \approx$10$^{49}$\,erg). Also, no signs of these SNe powered by a magnetar mechanism were evident either from photometry or spectroscopy, so this possibility was discarded. 

The spectroscopic behaviour of both SNe was also studied using the present and archival data. These SNe showed unambiguous He~I features from very early to late phases, confirming them to be SNe~Ib. Their spectral features match those of other well-studied SNe~Ib. Additionally, SN~2015ap closely resembled SN~2008D, which had shown X-ray emission. The spectra of our two SNe at various epochs were modelled using SYN++. For SN~2015ap, the spectral modelling indicated a range of photospheric temperatures and velocities from 13,000\,km\,s$^{-1}$ to 6800\,km\,s$^{-1}$ and from 12,000\,K to 4500\,K (respectively) during the time interval -7\,d to +33\,d. For SN~2016bau, these two parameters ranged from 16,000\,km\,s$^{-1}$ to 8000\,km\,s$^{-1}$ and from 9000\,K to 4000\,K (respectively) during the time interval -14\,d to +33\,d. The spectra of these two SNe at particular epochs were compared with model spectra of  12, 13, and 17\,M$_{\odot}$ progenitor stars. For SN~2015ap, the 12\,M$_{\odot}$ and 17\,M$_{\odot}$ model spectra showed reasonable matches, indicating a progenitor in the mass range 12--17\,M$_{\odot}$; for SN~2016bau, the 12\,M$_{\odot}$ model spectrum could explain the observed spectrum to some extent better than other models, indicating a $\leq$ 12\,M$_{\odot}$ progenitor.

Based on the photometric and spectroscopic properties described above, a 12\,M$_{\odot}$ ZAMS star was chosen as the possible progenitor for both SNe. This 12\,M$_{\odot}$ ZAMS progenitor was evolved up to the onset of core collapse using {\tt MESA}. The {\tt MESA} outputs on the onset of core collapse were fed as input to {\tt SNEC} and {\tt STELLA}, which simulate the synthetic explosions. The RD and MAG models were employed using {\tt SNEC} while only the RD model was employed in {\tt STELLA}. Here also, the models failed to fit the late part of the light curve simultaneously. The cause can be attributed to the various assumptions, including the spherically symmetrical explosions and the use of constant ejecta density profiles throughout. Similar to the case of {\tt MINIM}, the MAG model provided unphysical parameters during the {\tt SNEC} analysis.

Our analysis favours a 12\,M$_{\odot}$ ZAMS star as a possible progenitor for SN~2015ap based on outputs of the models reasonably explaining the bolometric luminosity light curve and the photospheric velocity. However, in the case of SN~2016bau, the model velocities are higher by a factor of almost 2, demanding that the ejecta mass be lower by a similar factor. This would imply a helium star of mass $\sim$\,2--2.5\,M$_{\odot}$, which is at the boundary for exploding as an SN. Thus, a slightly lower mass ZAMS star could also be the possible progenitor of SN~2016bau. Hydrodynamic evolution models of such low-mass stars to reach the stage of core collapse and then undergo synthetic explosions are extremely difficult to perform, but this can be taken as a challenge for the future.   

\section{Conclusions}
\label{sec:Conclusions}
Photometric and spectroscopic analyses of the Lick/KAIT-discovered Type Ib SN~2015ap and another Type Ib SN~2016bau, both having extensive follow-up observations made with various telescopes, are discussed. For both SNe, photometric data corrected for the Milky Way and host-galaxy extinction were used to estimate the quasi-bolometric luminosity light curves and study the photospheric radius, temperature, and velocity evolution. Spectral properties of SN~2015ap and SN~2016bau were then explored in detail. We modelled the spectra, studied the spectral evolution, and compared the spectra at various epochs with those of other well-studied SNe~Ib, which further confirmed that these two SNe are Type Ib SN. 

We attempted to determine the progenitor masses of SN~2015ap and SN~2016bau. Our results support 12\,M$_{\odot}$ progenitors for these two SNe. The 12\,M$_{\odot}$ ZAMS progenitor was evolved up to the onset of core-collapse using {\tt MESA}. The output of {\tt MESA} was incorporated as input to {\tt SNEC} and {\tt STELLA}, which produced the artificial explosions replicating the actual SN explosions. Considering the decay of Ni and Co to be the most prominent powering mechanism for SNe~Ib, we tried this powering mechanism for SN~2015ap and SN~2016bau, using {\tt SNEC} and {\tt STELLA}. We found that the quasi-bolometric luminosity could nicely be explained by our models, while the velocity evolution obtained from {\tt SNEC} and {\tt STELLA} satisfactorily agrees with the observed one. We also explored the effect of the distribution of nickel mass near the centre and up to near the surface. Lower amounts of nickel were required to match the light curve for the case of centrally distributed nickel in comparison to the case where the nickel was distributed up to near the surface. Based on the above conclusions, our analysis supports a star having $M_{\rm ZAMS}$ = 12\,M$_{\odot}$ as the possible progenitor for SN~2015ap. For SN~2016bau, a slightly lower ZAMS progenitor is expected.
 %SN2015ap n SN2016bau
\chapter{\sc Low-luminosity Type IIb SN~2016iyc arising from a low mass progenitor}
\label{Ch:4}
\blfootnote{ The results of this Chapter are published in: \textbf{{Aryan}, Amar,} {Pandey}, S. B., Zheng, W. et al., 2022,  {\textit{MNRAS}, {\textbf{517}}, 1750}.}
\ifpdf
    \graphicspath{{Chapter4/Chapter4Figs/JPG/}{Chapter4/Chapter4Figs/PDF/}{Chapter4/Chapter4Figs/}}
\else
    \graphicspath{{Chapter4/Chapter4Figs/EPS/}{Chapter4/Chapter4Figs/}}
\fi

\section{Introduction}
\label{sec:Introduction_ch4}
After studying the temporal and spectral properties of two H-deficient Type Ib SNe followed by the stellar evolution and synthetic explosions of their possible progenitors in Chapter~\ref{Ch:3}, we perform the photometric and spectroscopic analyses of a Type IIb SN in this Chapter. Type IIb SNe are a subclass of catastrophic CCSNe. These SNe form a transition class of objects that link H-rich Type II and H-deficient Type Ib SNe (\citealt[][]{Filippenko1988, Filippenko1993, Filippenko1997, Smartt2009}). Their early-phase spectra show strong H features and distinct He-features start to appear a few weeks later; thus, these SNe are thought to be partially stripped by retaining some trace H-envelope, and the He core is exposed once the envelope becomes optically thin. It is relevant to mention that SNe Type Ib/IIb are differentiated just on the basis of the presence of some trace amount of hydrogen in the exploding star. However, the correct estimation of the retained amount of hydrogen is difficult due to the uncertainties associated with determining the extinction and distance of underlying CCSNe \citep[][]{2022MNRAS.511..691G}.

The predominant powering mechanisms in SNe~IIb are the radioactive decay of $^{56}$Ni and the deposition of internal energy by the shock in the ejecta \citep[e.g.,][]{Arnett1980, Arnett1982, Arnett1996, Nadyozhin1994, Chatzopoulous2013, Nicholl2017}. In a few cases, the SN progenitors are also surrounded by dense CSM; thus, the SN ejecta may violently interact with the CSM. The interaction of CSM with the SN ejecta results in the formation of a two-component shock structure: a forward shock moving into the CSM and a reverse shock moving back into the SN ejecta. Both of these shocks deposit their kinetic energies into the material that is radiatively released, powering the light curves of the SNe \citep[e.g.,][]{ Chevalier1982, Chevalier1994, Moriya2011, Ginzberg2012, Chatzopoulous2013, Nicholl2017}.

Understanding the possible progenitors of stripped or partially-stripped CCSNe is a challenging task. Methods to investigate the SN progenitors and their properties include (a) direct detections of objects in pre-explosion images and (b) modelling of certain mass ZAMS stars as the possible progenitors based on the observed photometric and spectroscopic properties of the SNe. Direct detections of progenitors are rare owing to the uncertainty associated with the spatial positions and the infrequent occurrence of these transient phenomena. One has to be very lucky to get such pre-explosion images. However, for SNe~IIb, four cases of the direct detection of objects in pre-explosion images have been claimed. These include SN~1993J \citep[][]{Filippenko1993-IAUC,Aldering1994}, SN~2008ax \citep[][]{Crockett2008}, SN~2011dh \citep[][]{Maund2011,Van2011}, and SN~2013df \citep[][]{Van2014}, indicating either massive Wolf-Rayet (WR) stars ($M_{\rm ZAMS} \approx 10$--28\,M$_\odot$; \citealt{Crockett2008}) or more extended yellow supergiants (YSGs) with $M_{\rm ZAMS}$ = 12--17\,M$_\odot$ \citep[][]{Van2013, Folatelli2014, Smartt2015} as SN~IIb progenitors. Following \citet[][]{Smartt2009b} and \citet[][]{Van2017}, there have only been $\sim 34$ cases of direct CCSNe progenitor detections. With these direct detections, the progenitors of SNe~IIP are red supergiants (RSGs); SN~IIn progenitors are luminous blue variables; the progenitors of SNe~IIL are still debated, with only the case of SN~2009kr suggesting RSG or yellow supergiant progenitors; and the progenitors of SNe~Ib/c are either low-mass stars in a binary system \citep[][]{Podsiadlowski1992, Nomoto1995, Smartt2009} or a single massive WR star \citep[e.g.,][]{Gaskell1986, Eldridge2011, Groh2013}.  

The second method, progenitor modelling using stellar evolution codes to constrain the nature of the possible progenitors of stripped or partially-stripped CCSNe, identified either via direct imaging as in the case of iPTF13bvn \citep[][]{Cao2013} or indirect methods including nebular-phase spectral modelling \citep[][]{Jerkstrand2015, Uomoto1986}, and simulating the synthetic explosions of their pre-SN models, is also vital to understand their nature, physical conditions, circumstellar environment, and chemical compositions. But, the progenitor modelling of such objects using various stellar evolution codes is difficult due to the complicated shell-burning stages. Another problem associated with such modelling is the obscure nature of the mixing-length-theory parameter ($\alpha_{\rm MLT}$). The basis of $\alpha_{\rm MLT}$ has no physical origin \citep[][]{Joyce2018,Viani2018}. Furthermore, \citet[][]{Joyce2018} mentions that $\alpha_{\rm MLT}$ is neither a physical constant nor a computational one; it is rather a free parameter, so the value of $\alpha_{\rm MLT}$ must be determined individually in each stellar evolution code.

Owing to the above-mentioned difficulties, only a handful of such studies, including progenitor modelling followed by their synthetic explosions, have been performed in the case of stripped or partially-stripped CCSNe, including the Type Ib SN~iPTF13bvn \citep[][]{Cao2013, Bersten2014, Paxton2018}, the famous Type IIb SN~2016gkg \citep[][]{Bersten2018}, a few other Type IIb SNe including SN~2011dh \citep[][]{Bersten2012}, SN~2011fu \citep[][]{Morales2015}, two Type Ib SNe~2015ap and 2016bau (results presented in Chapter~\ref{Ch:3}), and another Type Ib SN~2012au \citep[][]{Pandey2021}.

Considering these limited studies, our work goes one step further, as we perform the 1-dimensional stellar evolution of the possible progenitor models for the low-luminosity Type IIb SN~2016iyc and simulate synthetic explosions. Our studies in this work point toward SN~2016iyc originating from the lower-mass end of the ZAMS progenitor systems observed for Type IIb CCSNe.

This Chapter is divided into eight sections, including an introduction in Sec.~\ref{sec:Introduction_ch4}. Sec.~\ref{sec:Data_red_ch4} provides details about various telescopes and reduction procedures, including the discovery of SN~2016iyc using the KAIT at Lick Observatory as well as recalibrated photometry of SN~2016gkg. 
In Sec.~\ref{sec:Photometric_ch4}, methods to correct for the extinction, photometric properties, including the bolometric light curve, blackbody temperature, and radius evolutions, are discussed. We present the analyses describing the spectral properties and comparisons with other similar and well-studied SNe in Sec.~\ref{sec:Spectral_ch4}; we also model the spectra of these SNe using {\tt SYN++}. The assumptions and methods for modelling the possible progenitor of SN~2016iyc and the evolution of the models until the onset of core collapse using {\tt MESA} are presented in  Sec.~\ref{sec:mesa_snec_ch4}. Further, in this section, we discuss the assumptions and methods for simulating the synthetic explosions using {\tt SNEC} and {\tt STELLA}. Here, comparisons between the parameters obtained through synthetic explosions and observations are presented. We also perform hydrodynamic modelling of the synthetic explosions of SN~2016gkg and SN~2011fu in Sec.~\ref{sec:SN2016gkg_model_ch4}. In Sec.~\ref{sec:Discussions_ch4}, we discuss our major results and findings. We summarise our work and provide concluding remarks in Sec.~\ref{sec:Conclusions_ch4}. 

\section{Data acquisition and reduction }
\label{sec:Data_red_ch4}
SN~2016iyc was discovered by \citet[][]{de2016} in an 18\,s unfiltered image taken at 03:28:00 on 2016~Dec.~18 (UT dates are used throughout this Chapter) by the 0.76\,m KAIT as part of the Lick Observatory Supernova Search (LOSS). Its brightness was $17.81\pm0.11$\,mag, and the object was not detected earlier on Dec. 04.14 with an upper limit of 19.0\,mag.
We measure its J2000.0 coordinates to be $\alpha=22^{\rm h}09^{\rm m}14.^{\rm s}29$,
$\delta=+21^{\circ}31^{'}17.^{''}3$, with an uncertainty of $0.^{''}5$ in each coordinate.
SN~2016iyc is $14.^{''}0$ west and $10.^{''}4$ north of the nucleus of the host galaxy UGC~11924, which has
redshift $z$ = 0.012685$\pm$0.000017 \citep[][]{1993AJ....105.1271G}, with a spiral morphology (Sd).

%according to the NASA/IPAC Extragalactic Database
%(NED\footnote{http://ned.ipac.caltech.edu/}),

$B$, $V$, $R$, and $I$ multiband follow-up images of SN~2016iyc were obtained with both KAIT and the 1\,m Nickel telescope at Lick Observatory; KAIT also obtained additional unfiltered ({\it Clear (C)}-band) images. Although unfiltered and thus nonstandard, $C$ is most similar to the $R$ band \citep[][]{Li2003}, and has been widely used for SN observations by KAIT.

All images were reduced using a custom pipeline\footnote{https://github.com/benstahl92/LOSSPhotPypeline}
detailed by \citet[][]{Stahl2019}. Here we briefly summarise the photometric procedure. Image subtraction was conducted to remove the host-galaxy contribution, using additional images obtained after the SN had faded below the detection limit. Point-spread-function (PSF) photometry was obtained using DAOPHOT \citep[][]{Stetson1987} from the IDL Astronomy User’s Library\footnote{http://idlastro.gsfc.nasa.gov/}.
Several nearby stars were chosen from the Pan-STARRS1\footnote{http://archive.stsci.edu/panstarrs/search.php} catalogue for calibration purpose;
their magnitudes were first transformed into the \citet{Landolt1992} system using the empirical prescription presented by \citet[][Eq.~6]{Torny2012}, and then into the KAIT/Nickel natural system.
All apparent magnitudes were measured in the KAIT4/Nickel2 natural system. The final results were transformed to the standard system using local calibrators and colour terms for KAIT4 and Nickel2 \citep[][]{Stahl2019}.

The same method was adopted to reprocess the LOSS data of SN~2016gkg (originally published by \citealt{Bersten2018}), except that no subtraction procedure was applied to SN~2016gkg; the calibration source was also chosen from the Pan-STARRS1 catalogue. Photometry of SN~2016gkg at two epochs was also obtained with the 3.6\,m Devasthal optical  telescope (DOT) using the 4K$\times$4K CCD Imager. SN~2016gkg was the first SN detected by the 3.6\,m DOT during its initial commissioning phases. For the data obtained from the 3.6\,m DOT, the \citet{Landolt1992} photometric standard fields PG 0918, PG 1633, and PG 1657 were observed on 2021 Feb. 07 along with the SN field in the $UBVRI$ bands under good photometric conditions. These three Landolt fields have standard stars with a $V$-band magnitude range of 12.27--15.26\,mag and a $B-V$ colour range of $-$0.27 to +1.13\,mag. The SN fields observed in 2021 were used for template subtraction to remove the host-galaxy contributions from the source images. Template subtraction was performed with standard procedures by matching the full width at half-maximum intensity (FWHM) and flux values of respective images. The optical photometric data reduction and calibration were made with a standard process discussed by \citet[][]{Kumar2021} and Python scripts hosted on \textsc{RedPipe} \citep[][]{Singh2021}. The average atmospheric extinction values in the $U$, $B$, $V$, $R$, and $I$ bands for the Devasthal site were adopted from \citet{2022JApA...43...27K}. The recalibrated KAIT data of SN~2016gkg, along with those observed at later epochs using the 4K$\times$4K CCD Imager mounted at the axial port of the 3.6\,m DOT were utilised for the construction of bolometric light curves as described in the following sections.

A single optical spectrum of SN~2016iyc was obtained on 2016 Dec. 23 with the Kast double spectrograph mounted on the 3\,m Shane telescope at Lick Observatory.  The 2700\,s exposure was taken at or near the parallactic angle to minimise slit losses caused by atmospheric dispersion \citep[][]{1982PASP...94..715F}. The observations were conducted with a $2''$-wide slit, 600/4310 grism on the blue side, and 300/7500 grating on the red side.  This instrument configuration has a combined wavelength range of $\sim 3500$--10,400\,\AA and spectral resolving power of $R \approx 800$.  Data reduction followed standard techniques for CCD processing and spectrum extraction \citep[][]{2012MNRAS.425.1789S} utilising IRAF routines and custom Python and IDL codes.  Low-order polynomial fits to comparison-lamp spectra were used to calibrate the wavelength scale, and small adjustments derived from night-sky lines in the target frames were applied.  Observations of appropriate spectrophotometric standard stars were used to flux calibrate the spectrum.

\section{Photometric Properties}
\label{sec:Photometric_ch4}
In this section, we discuss the photometric properties of SN~2016iyc, including colour evolution, extinction, bolometric light curves, and various blackbody parameters. Most of the analyses in this Chapter have been performed with respect to the phase of $V$-band maximum brightness.  The photometric data of SN~2016iyc lack dense coverage near peak brightness; thus, to find the phase of $V$-band maximum, we used the $V$-band light curve of SN~2013df as a template having a rising timescale similar to that of SN~2016iyc (Figure~\ref{fig:V_max}). We fit a fourth-order polynomial to the template light curve and find the date of $V$ maximum to be MJD $57752.7 \pm 0.2$. The left panel of Figure~\ref{fig:BVRIC_LC} shows the $BVRI$- and $C$-band light curves of SN~2016iyc. The characteristic extended shock-breakout (hereafter, extended-SBO) feature typically observed in SNe~IIb is seen in all of the bands. Multiple mechanisms and/or ejecta/progenitor properties have been theorised to explain such enhancement in the luminosity before the primary peak, including an increase in the progenitor radius up to a few 100\,R$_{\odot}$ \citep[e.g.][]{Nomoto1993,Podsiadlowski1993,Woosley1994}; an interaction with CSM similar to the case of Type IIn SNe \citep[][]{Schlegel1990}; in a close-binary system, the interaction with the companion \citep[][]{Kasen2010,Moriya2015}; and sometimes enhanced $^{56}$Ni mixing into the outer ejecta \citep[e.g.,][]{Arnett1989}. The right-hand panel of Figure~\ref{fig:BVRIC_LC} shows the $BVRI$ and $C$ light curves along with the late-time upper limits in each band. These upper limits are very useful in constraining the upper limit on $M_{\rm Ni}$.

\begin{figure*}
\centering
    \includegraphics[width=.49\columnwidth]{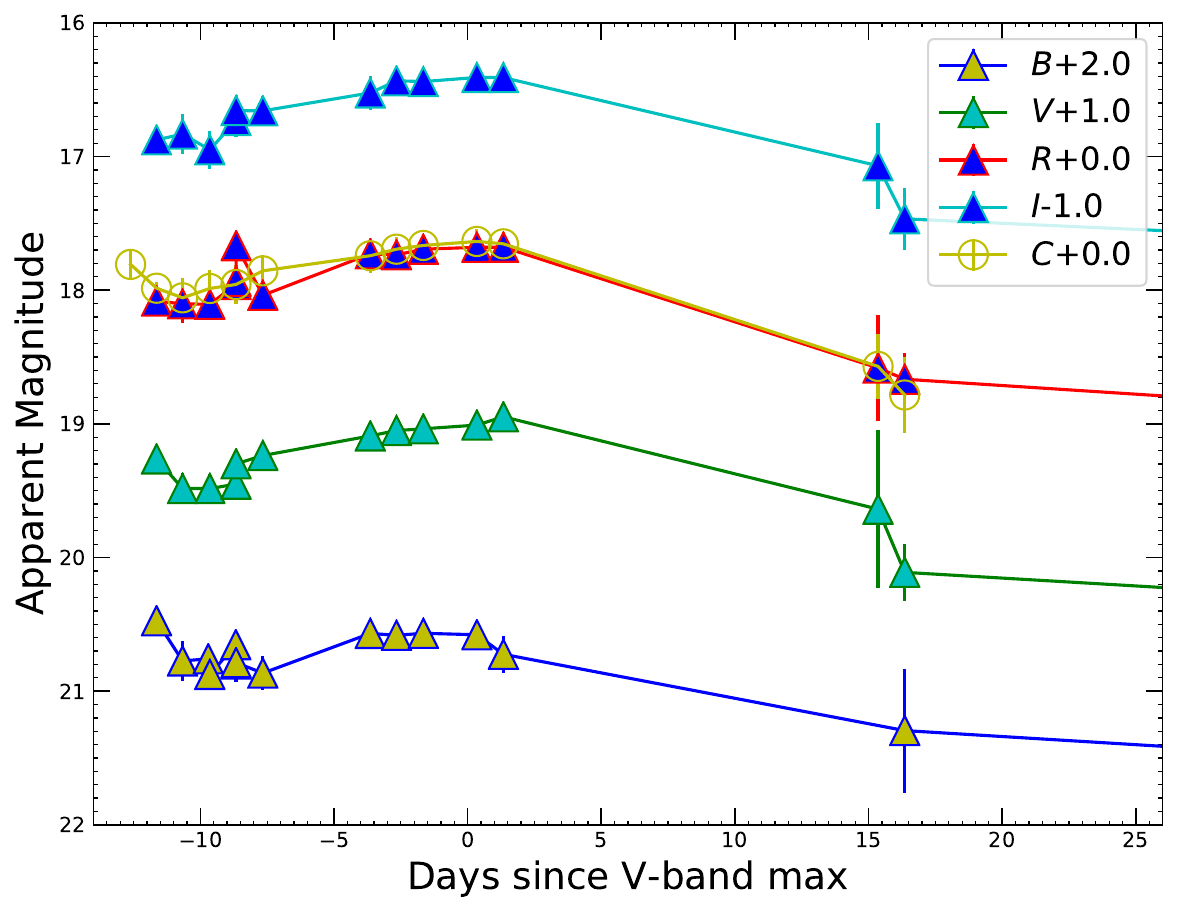}
    \includegraphics[width=.49\columnwidth]{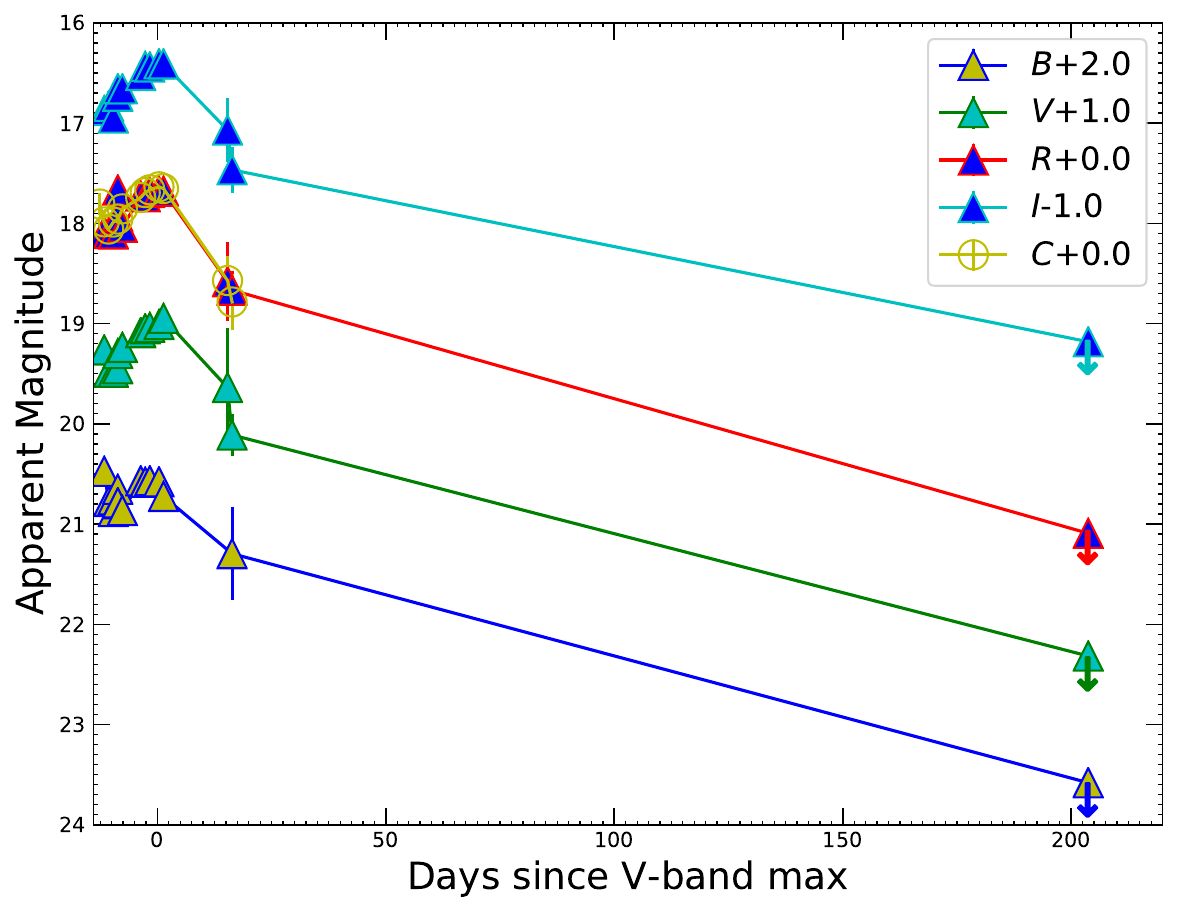}
   \caption{ {\em Left:}  The $BVRI$- and $C$-band light curves of SN~2016iyc, obtained with KAIT. The generic extended-SBO feature of SNe~IIb is visible in each band. {\em Right:} The $BVRI$ and $C$ light curves along with the upper limits in each band using the Las Cumbres Observatory global telescope network. The upper limits in the last epoch are extremely useful for setting an upper limit on  $M_{\rm Ni}$.}
    \label{fig:BVRIC_LC}
\end{figure*}

\subsection{Distance estimation of SN~2016iyc}

\begin{figure}	
    \includegraphics[width=\columnwidth]{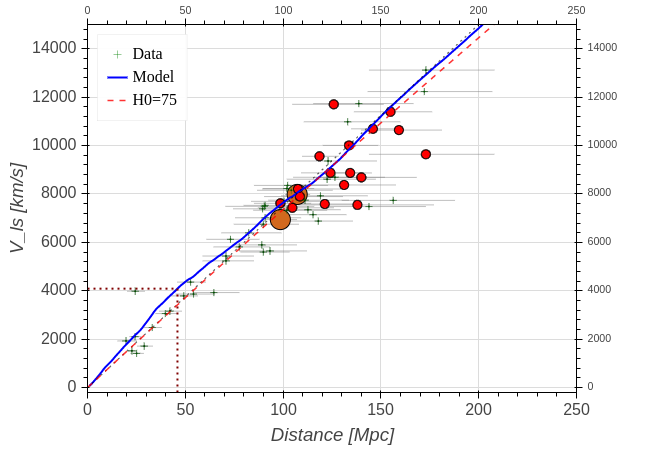}
    \caption{ Distance estimation for the nearby ($z$ = 0.012685) SN~2016iyc using the method described by \citet[][]{Kourkchi2020} with \citet[][]{Graziani2019} models. The distance value estimated with this method is $\sim 10$\,Mpc ($\sim 20$\%) nearer than that reported using the published redshift value \citep[][]{Planck2016} for SN~2016iyc.}
    \label{fig:distance}
\end{figure}

Distance determinations from redshifts ($z$) are severely biased for nearby SNe because of the peculiar motions of nearby galaxies that are comparable to the Hubble flow. So, the redshift-based distance estimates can be used only for SNe having $z$\,$>$\,0.1. Hence, the redshift-based distance for SN~2016iyc, published by \citet[][]{Planck2016}, could be spurious. SN~2016iyc being nearby ($z$ = 0.012685), we cross-verified the redshift-based distance estimate (56.6837\,Mpc as mentioned by \citealt[][]{Planck2016}) with an advanced tool (Figure~\ref{fig:distance}) recently featured by \citet[][]{Kourkchi2020}\footnote{http://edd.ifa.hawaii.edu/CF3calculator/}, known as the Distance-Velocity ($D$--$V$) Calculator. Corresponding to a heliocentric velocity $V_{\rm h} \approx 3804$\,km\,s$^{-1}$, the observed velocity ($V_{\rm ls}$) at the location of SN~2016iyc is found to be $\sim 4089$\,km\,s$^{-1}$ by utilising Eq.~5 of \citet[][]{Kourkchi2020}. Corresponding to $V_{\rm ls} = 4089$\,km\,s$^{-1}$, the $D$--$V$ Calculator gives a distance of $\sim 46$\,Mpc, which is $\sim 20$\% less than  \citet[][]{Planck2016}. The distance modulus for SN~2016iyc corresponding to this distance is 33.31\,mag and is adopted for all further analyses in this Chapter. The distances for the SNe used as a comparison sample are well established in the literature and are used as such for the estimation of their respective bolometric luminosities. The distance of each SN in the comparison sample, along with the corresponding distance modulus is presented in Table~\ref{tab:comparison_Sample}.

\begin{landscape}
\begin{table*}
  \caption {The adopted total extinction values, distances, and corresponding distance moduli of a subset of SNe considered here.}
%  in the present analysis along with those estimated for SN~2016iyc.}
\label{tab:comparison_Sample}
\begin{center}
{\scriptsize
\begin{tabular}{ccccccccccccc}
\hline\hline
	SN name    &	$E(B-V)_{\rm tot}$	&	Adopted distance  & Distance modulus & $V_{\rm max}$ & log\,$(L_{BVRI})_p$ \\
               &	(mag)	&	(Mpc) & (mag) & (mag) & (erg\,s$^{-1}$) \\
\hline
SN~1987A  	 	&	0.16 \citep[][]{Bose2021}	&	0.05 \citep[][]{Bose2021} & 18.44 & $-15.52 \pm 0.02$ & 42.555 $\pm$ 0.004\\\\

SN~1993J  	 	&	0.18 \citep[][]{Richmond1996}	&	3.68 \citep[][]{Bose2021} & 27.82 & $-16.97 \pm 0.03$ & 42.01 $\pm$ 0.01\\\\

SN~2003bg  	 	&	0.02 \citep[][]{Mazzali2009}	&	24 \citep[][]{Mazzali2009}   & 31.90 & -17.8$\pm$0.2 & 42.31 $\pm$ 0.03\\\\

SN~2008ax  	    &	0.3 \citep[][]{Tsvetkov2009}	&	 9.6 \citep[][]{Pastorello2008}   & 29.92 & $-16.35 \pm 0.05$   & 42.07  $\pm$ 0.03\\\\

SN~2011dh  	 	&	0.035 \citep[][]{Sahu2013}	&	 8.4 \citep[][]{Sahu2013}   & 29.62 & $-17.06 \pm 0.02$ & 41.99 $\pm$ 0.03\\\\

SN~2011fu 	 	&	0.22 \citep[][]{Kumar2013}	&	 77.0 \citep[][]{Kumar2013}   & 34.46 & $-17.51 \pm 0.03$ & 42.426 $\pm$ 0.006\\\\

SN~2011hs  	 	&	0.17 \citep[][]{Bufano2014}	&	 23.44$^{*}$     & 31.85 & $-16.03 \pm 0.03$ & 41.74 $\pm$ 0.02\\\\

SN~2013df  	 	&	0.098 \citep[][]{Morales2015}	&	 16.6 \citep[][]{Van2014}   & 31.1 & $-16.47 \pm 0.05$ & 41.87 $\pm$ 0.01\\\\

SN~2016gkg 	&	0.017 \citep[][]{Bersten2018}	&	 26.4 \citep[][]{Kilpatrick2017}   & 32.11 & $-17.03 \pm 0.05$ & 41.98 $\pm$ 0.02\\\\

SN~2016iyc 	&	0.137 	&	 46.0   &  33.31 & $-15.32 \pm 0.05$ & 41.44 $\pm$ 0.01\\

\hline\hline
\end{tabular}}
\end{center}
{The sources for the $BVRI$ light curves for SNe in the comparison sample are as follows. SN~1987A, \citet[][]{Menzies1987} and \citet[][]{Makino1987}; SN~1993J, \citet[][]{Zhang2004}; SN~2003bg, \citet[][]{Hamuy2009}; SN~2008ax, \citet[][]{Pastorello2008}; SN~2011dh, \citet[][]{Sahu2013}; SN~2011fu, \citet[][]{Kumar2013}; SN~2011hs, \citet[][]{Bufano2014}; SN~2013df, \citet[][]{Morales2015}. Adopted distances have been used to calculate the distance moduli. The total extinction correction and distance moduli for all the SNe in the comparison sample have been considered while calculating the bolometric light curves. $^*$For SN~2011hs, the distance modulus is 31.85\,mag \citep[][]{Bufano2014}, which is used to back-calculate a distance of 23.44\,Mpc.}
\end{table*}
\end{landscape}

\subsection{Colour evolution and extinction correction}
For SN~2016iyc, we corrected for the Milky Way (MW) extinction using NED, following \citet[][]{Schlafly2011}. In the direction of SN~2016iyc, $E(B-V)_{\rm MW} = 0.067$\,mag, so the MW extinction corrections for the $B$, $V$, $R$, and $I$ bands are 0.278, 0.207, 0.155, and 0.099\,mag, respectively.

\begin{figure}	
    \includegraphics[width=\columnwidth]{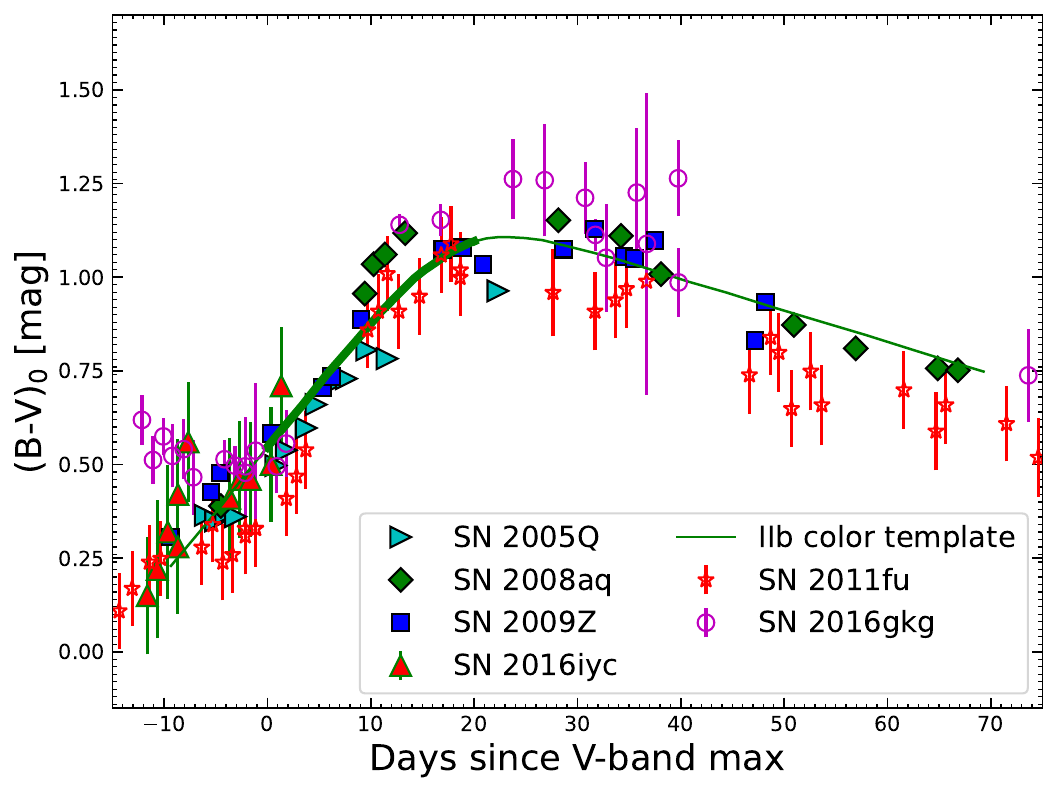}
    \caption{ The total-extinction-corrected $(B-V)_{0}$ colour curves of SN~2016iyc, SN~2011fu, and SN~2016gkg, plotted along with other SNe~IIb. The data for SN~2005Q, SN~2008aq, and SN~2009Z were taken from \citet[][]{Stritzinger2018}, with these three SNe analysed to have negligible host-galaxy extinction. The green curve shows the template $(B-V)_{0}$ curve for SNe~IIb having negligible host-galaxy extinction. The thick portion of the template curve shows the 0 to +20\,d period that should be considered when determining the colour excess for the reasons mentioned by \citet[][]{Stritzinger2018}.}
    \label{fig:color_curve}
\end{figure}

Only one spectrum of SN~2016iyc is available, and it does not exhibit a clear Na~I~D absorption line produced by gas in the host galaxy, suggesting that there is negligible host-galaxy extinction. However, neglecting host-galaxy extinction based on only the absence of obvious Na~I~D could be spurious. 
In a Lick/KAIT data-release paper for various stripped-envelope SNe, \citet[][]{Zheng2022} performed a comprehensive analysis to determine the host-galaxy contamination and found $E(B-V)_{\rm host} = 0.07$\,mag for SN~2016iyc. We also performed a simple analysis to put an upper limit on the host-galaxy extinction. Five early epochs were selected, and the spectral energy distribution (SED) was fitted with blackbody curves by assuming different $E(B-V)_{\rm host}$ values (Figure~\ref{fig:extinction_trial}). We found that going beyond 0.07\,mag of host-galaxy extinction results in blackbody temperature exceeding 11,200\,K. Such high temperatures are generally not seen in SNe~IIb. Following \citet[][]{Ben2015}, the early-time blackbody temperatures associated with SN~1993J, SN~2011dh, and SN~2013df are 8200\,K, 8200\,K, and 7470\,K, respectively. There have been only a few cases where the early blackbody temperature exceeds 11,000\,K; one such example is SN~2001ig \citet[][]{Ben2015}, but this SN may have come from a compact WR binary progenitor system\citep[][]{Ryder2004}.

Based on the above analyses and the results of \citet[][]{Zheng2022}, we adopt a host-galaxy extinction of 0.07\,mag throughout this Chapter. Thus, a total (Milky Way + host-galaxy) extinction of $E(B-V)_{\rm tot} = 0.137$\,mag is adopted for SN~2016iyc. Figure~\ref{fig:color_curve} shows the comparison of total extinction corrected $(B-V)_{0}$ colour of SN~2016iyc with other similar SNe.

\subsection{Bolometric light curves}
\label{subsec3.3}
Before computing the bolometric light curves, the absolute $V$-band light curve of SN~2016iyc is compared with a few other similar SNe~IIb. The left panel of Figure~\ref{fig:abs_bol} shows that SN~2016iyc lies toward the fainter end of the distribution.

Furthermore, to obtain the quasi-bolometric light curve, we make use of the {\tt SUPERBOL} code \citep{Nicholl2018}. The extinction-corrected $B$, $V$, $R$, and $I$ data are provided as input to {\tt SUPERBOL}. The light curve in each filter is then mapped to a common set of times through the processes of interpolation and extrapolation. Thereafter, {\tt SUPERBOL} fits blackbody curves to the SED at each epoch,  up to the observed wavelength range (4000--9000\,\AA), to give the quasi-bolometric light curve by performing trapezoidal integration.

The right-hand panel of Figure~\ref{fig:abs_bol} shows the comparison of the quasi-bolometric light curve of SN~2016iyc with  other well-studied SNe~IIb as listed in Table~\ref{tab:comparison_Sample}. The peak quasi-bolometric luminosity (log\,$(L_{BVRI})_p$) of each SN has also been calculated by fitting a third-order polynomial to the quasi-bolometric light curve. As indicated by the right-hand panel of Figure~\ref{fig:abs_bol}, SN~2016iyc lies toward the fainter limit of SNe~IIb in the comparison sample. It is also worth mentioning that the low-luminosity SNe with low $^{56}$Ni yields are thought to arise from progenitors having masses near the threshold mass for producing a CCSN \citep[][]{Smartt2009b}.

Furthermore, the bolometric luminosity light curve of SN~2016iyc is also produced after considering the additional blackbody corrections to the observed $BVRI$ quasi-bolometric light curve by fitting a single black body to observed fluxes at a particular epoch and integrating the fluxes trapezoidally for a wavelength range of 100--25,000\,\AA\ using {\tt SUPERBOL}. Figure \ref{fig:bb_fits} shows the blackbody fits to the SED of SN~2016iyc. The top panel of Figure~\ref{fig:BB_param_SN2016iyc} shows the resulting quasi-bolometric and bolometric light curves of SN~2016iyc.

\begin{figure*}
\centering
    \includegraphics[height=6cm,width=0.49\columnwidth]{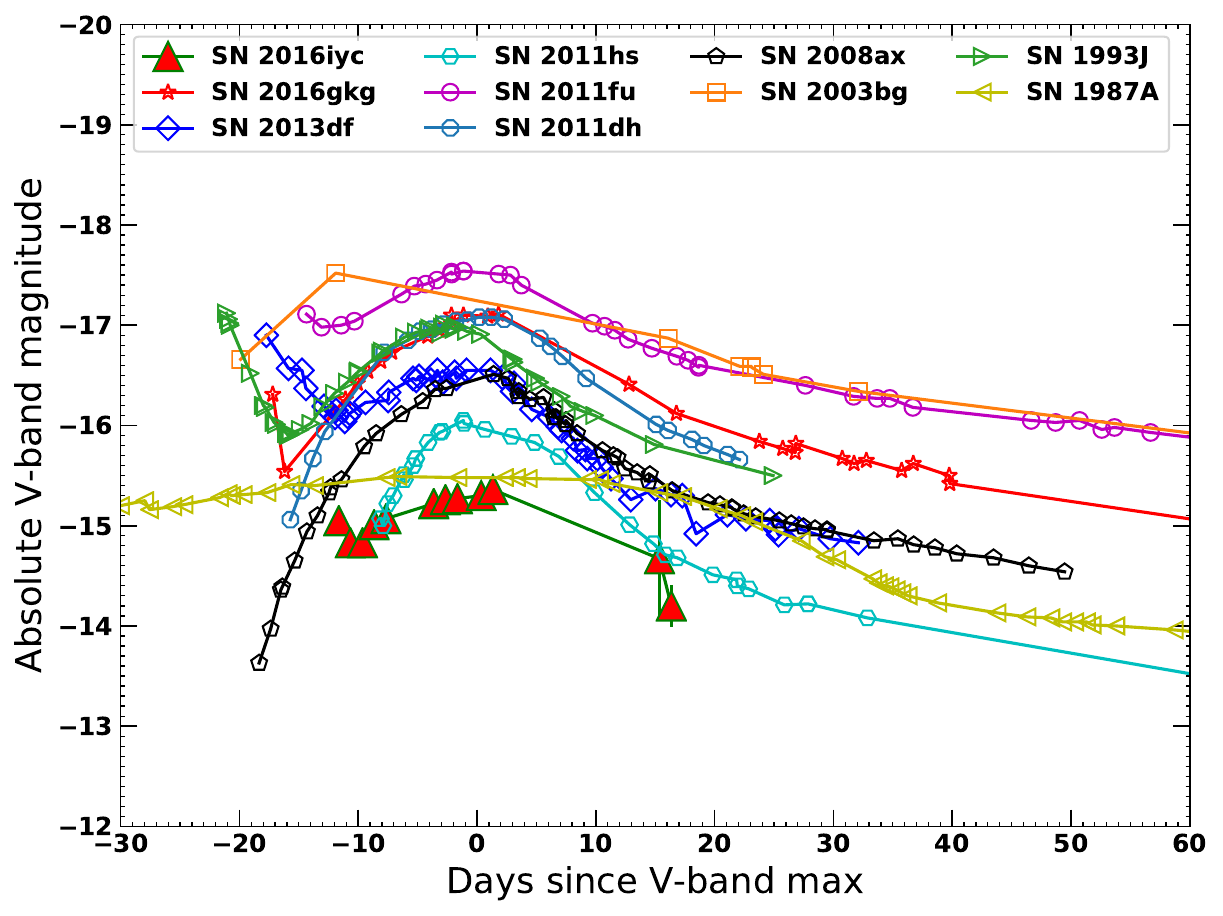}
    \includegraphics[height=6cm,width=0.49\columnwidth]{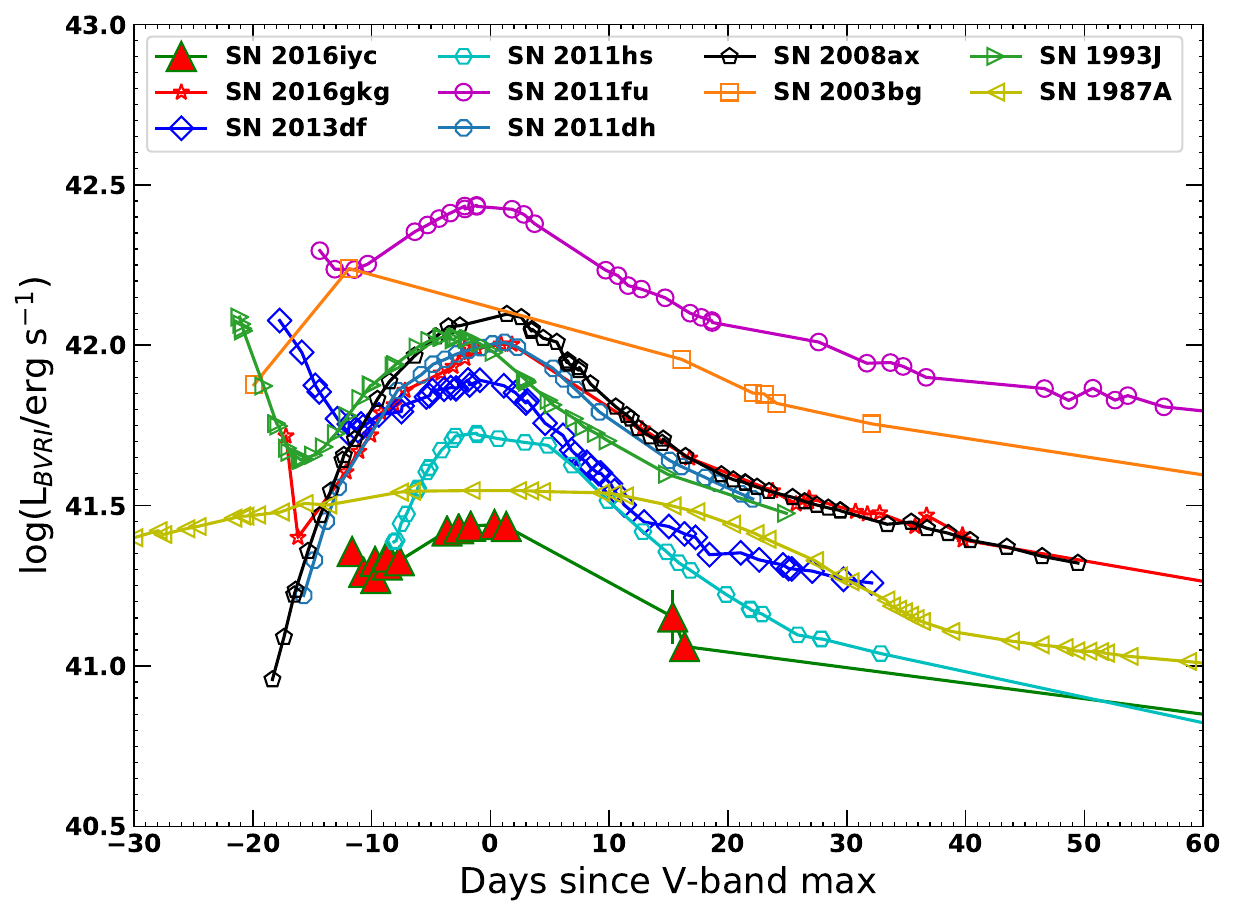}
   \caption{ {\em Left:} Comparison of the absolute $V$-band light curves of SN~2016iyc and SN~2016gkg with other well-studied SNe~IIb ( the peculiar Type II SN~1987A, included in the sample because of its low luminosity). {\em Right:} Comparison of the quasi-bolometric light curves of SN~2016iyc and SN~2016gkg with other well-studied SNe~IIb obtained by integrating the fluxes over the $BVRI$ bands. SN~2016iyc lies toward the faint end of SNe~IIb. The total extinction correction and distance moduli for all the SNe in the comparison sample have been taken into account while calculating these light curves.}
    \label{fig:abs_bol}
\end{figure*}

\begin{figure}
\includegraphics[width=\columnwidth]{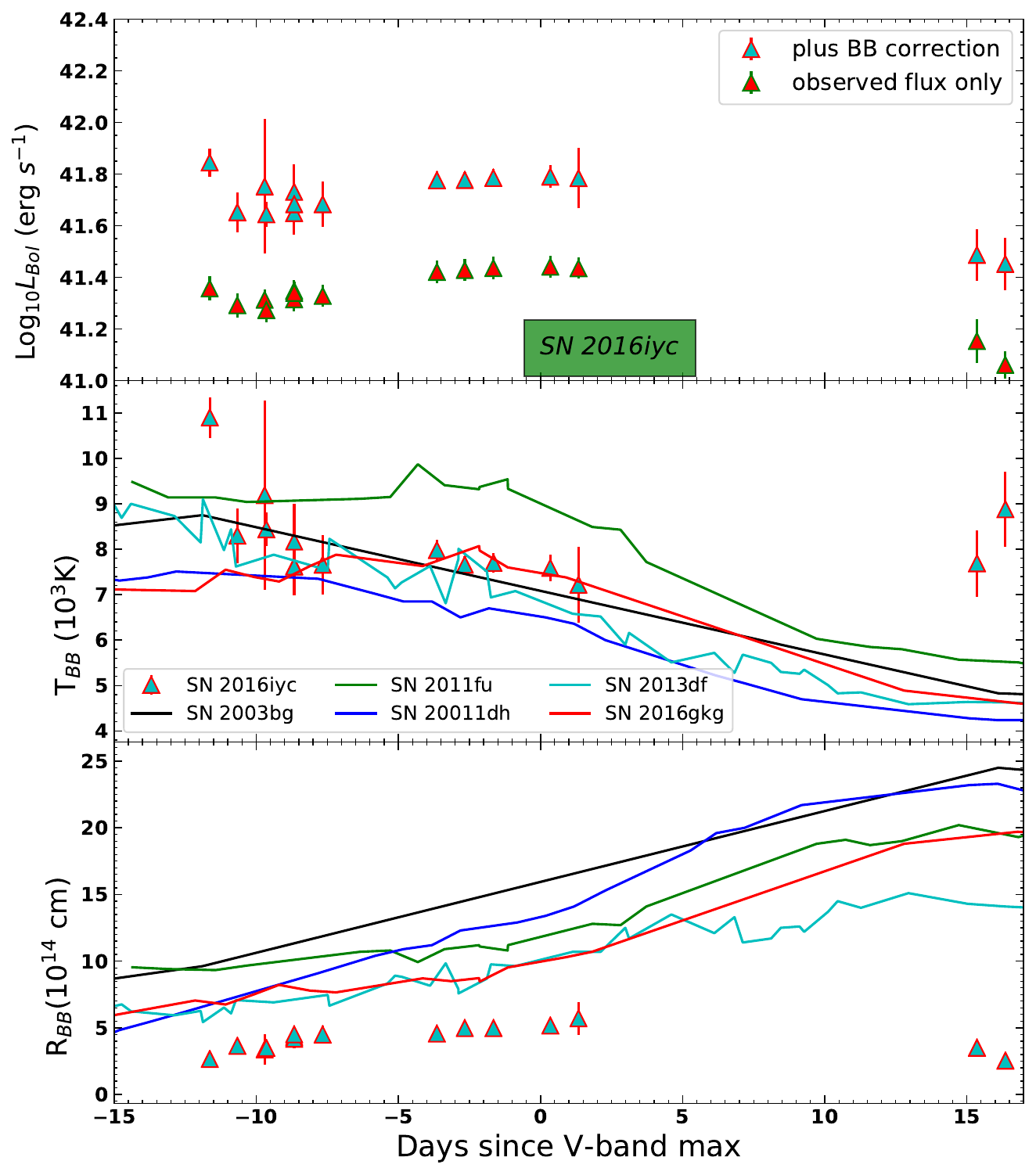}
\caption{The top panel shows the bolometric and quasi-bolometric light curves of SN~2016iyc. The luminosity corresponding to the late upper limits on $BVRI$ is shown later. The second and third panels from the top display the blackbody temperature and radius evolutions of SN~2016iyc, respectively.}
\label{fig:BB_param_SN2016iyc}
\end{figure}

\subsection{Temperature and radius evolution}

From {\tt SUPERBOL}, the blackbody temperature ($T_{\rm BB}$) and radius ($R_{\rm BB}$) evolution of SN~2016iyc are also obtained. During the initial phases, the photospheric temperature is high, reaching $\sim 10,900$\,K at $-10.63$\,d. Furthermore, the SN seems to evolve very rapidly; its temperature quickly drops to $\sim 7600$\,K in only a few days around $-7.7$\,d and then remains nearly constant (Figure~\ref{fig:BB_param_SN2016iyc}, second panel from top). Along with SN~2016iyc, the temperature evolutions of a few more similar SNe~IIb  are also shown in this panel. The blackbody temperature of SN~2016iyc seems to follow the typical temperature evolution as seen in SNe~IIb.  

A conventional evolution in radius is also seen. Initially, at an epoch of $-10.63$\,d, the blackbody radius is $2.64 \times 10^{14}$\,cm. Thereafter, the SN expands and its radius increases, reaching a maximum radius of $\sim$\,$5.8 \times 10^{14}$\,cm, beyond which the photosphere seems to recede into the SN ejecta (Figure~\ref{fig:BB_param_SN2016iyc}, third panel from top). Along with SN~2016iyc, the blackbody radius evolutions of a few more similar SNe~IIb are also shown. SN~2016iyc seems to exhibit anomalous behaviour, with its blackbody radii at various epochs being the smallest among other similar SNe~IIb. This result can be attributed to the low ejecta velocity of SN~2016iyc.

\section{Spectral Studies of SN~2016iyc}
\label{sec:Spectral_ch4}

In this section, we identify the signatures of various lines by modelling the only available spectrum of SN~2016iyc using {\tt SYN++} \citep[][]{Branch2007, Thomas2011}. We discuss various spectral features of SN~2016iyc, and the spectrum is also compared with other similar SNe. 

\subsection{Spectral modelling}
A single optical spectrum of SN~2016iyc was obtained on 2016 Dec. 23 with the Kast double spectrograph \citep[][]{Miller&Stone1993} mounted on the 3\,m Shane telescope at Lick Observatory. Figure~\ref{fig:syn++} shows the spectral modelling of it, corresponding to a phase of $-6.6$\,d. The individual lines corresponding to various elements and ions are also indicated for better identification of the features. The profiles of H$\alpha$ near 6563\,\AA, He\,I near 5876\,\AA, and Ca\,II~H\&K are very nicely reproduced by {\tt SYN++} modelling. A very strong H$\alpha$ feature near 6563\,\AA\ classifies SN~2016iyc as an SN~IIb. The observed velocities obtained using H$\alpha$, He\,I, and Fe\,II features in the spectrum are $\sim 10,000$\,km\,s$^{-1}$, $\sim 6700$\,km\,s$^{-1}$, and $\sim 6100$\,km\,s$^{-1}$ (respectively), while the respective velocities of these lines from {\tt SYN++} modelling are 10,100\,km\,s$^{-1}$, 6800\,km\,s$^{-1}$, and 6100\,km\,s$^{-1}$, very close to the observed ones. The parameterisation velocity and photospheric velocity used to produce the {\tt SYN++} model are 6000\,km\,s$^{-1}$ and 6100\,km\,s$^{-1}$, respectively. Also, a photospheric temperature of 6300\,K is employed to produce the model spectrum.

\begin{figure}
\centering
\includegraphics[width=\columnwidth]{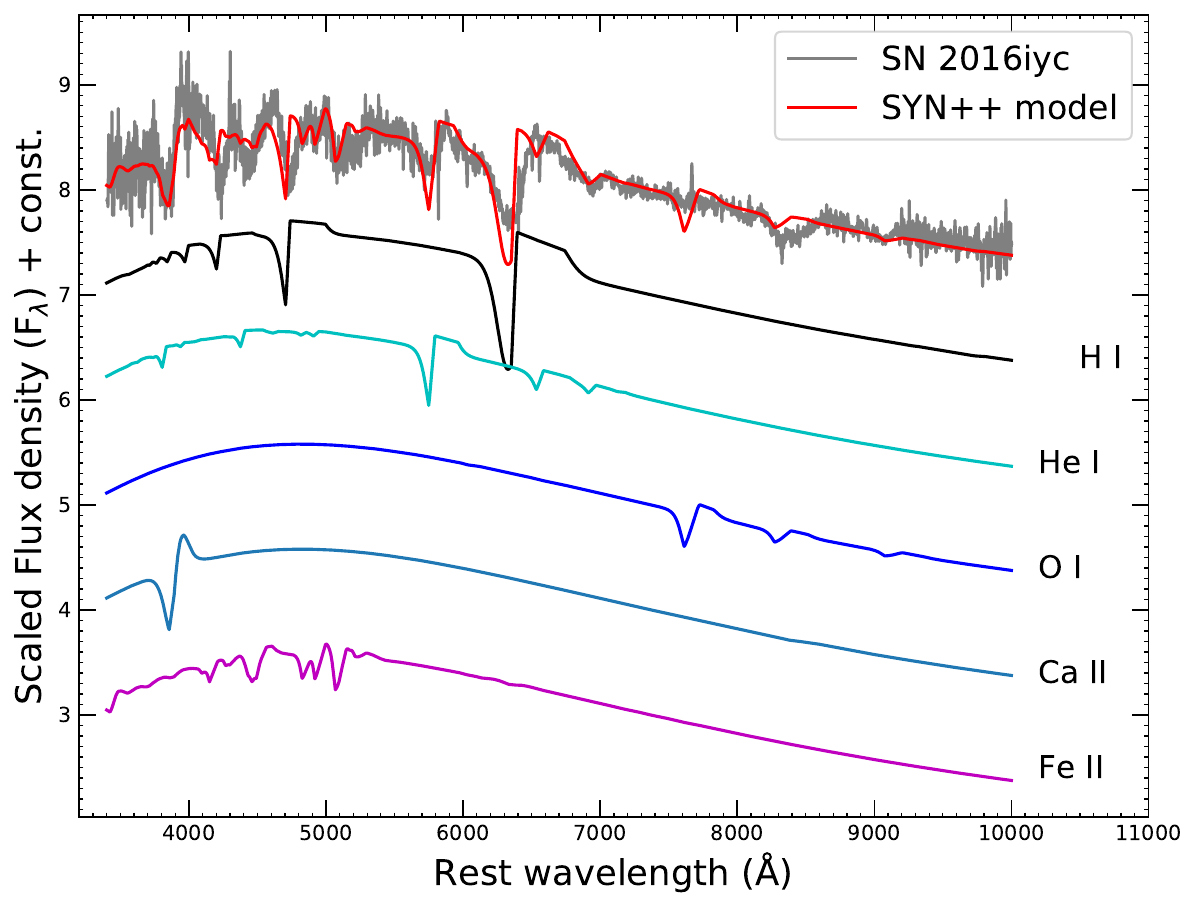}
\caption{{\tt SYN++} modelling of the spectrum of SN~2016iyc at a phase of $-6.6$\,d. The effects of various elements present in the SN ejecta and contributing to the spectrum are also displayed individually.}
\label{fig:syn++}
\end{figure}

\begin{figure}
\includegraphics[height=20cm,width=\columnwidth]{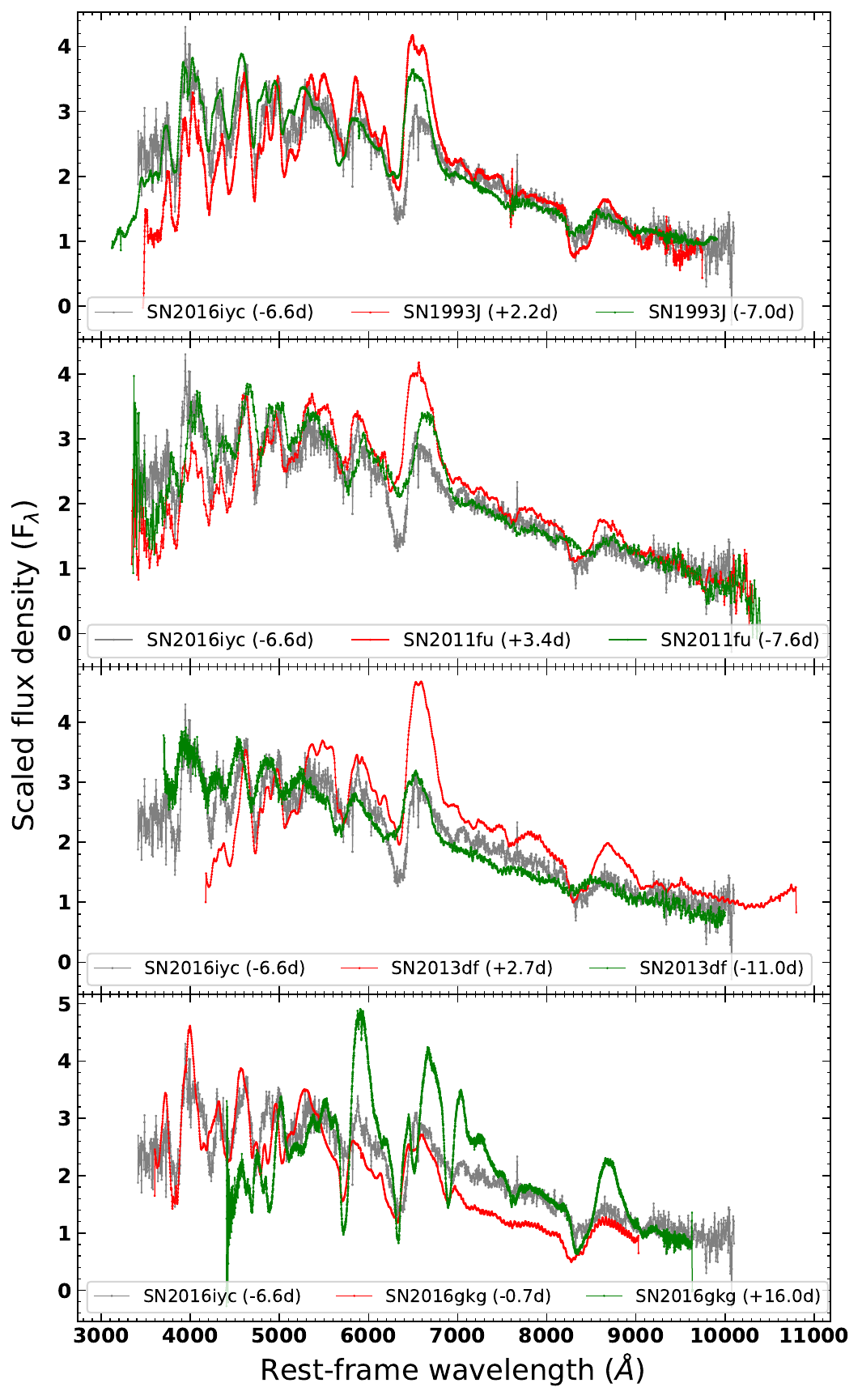}
\caption{Comparison of the $-6.6$\,d spectrum of SN~2016iyc with spectra of other well-studied SNe~IIb, including SN~1993J, SN~2011fu, SN~2013df, and SN~2016gkg. }
\label{fig:spec_comparison}
\end{figure}

\subsection{Spectral comparison}
Figure~\ref{fig:spec_comparison} shows a comparison of the normalised spectrum of SN~2016iyc with other well-studied SNe~IIb. The top plot displays the comparison with the spectra of SN~1993J at +2.3\,d and $-7.0$\,d; we see that the spectral features of SN~2016iyc closely resemble those of SN~1993J spectra. In the second panel from the top, the spectrum of SN~2016iyc is compared with the spectra of SN~2011fu at +3.4\,d and $-7.0$\,d; the match is close, except for the H$\alpha$ feature where the spectra of SN~2011fu are slightly off. The third panel from the top shows the spectral comparison of SN~2016iyc with spectra of SN~2013df at epochs of +2.7\,d and $-11.0$\,d, revealing a good match with the $-11.0$\,d spectrum. The progenitor of SN~2013df is also thought to be arising from the lower-mass end. In the bottom panel, the spectrum of SN~2016iyc is compared with the spectra of SN~2016gkg at epochs of $-0.7$\,d and +16\,d. The $-0.7$\,d spectrum of SN~2016gkg resembles the spectrum of SN~2016iyc toward the bluer side, while features in the redder part of the spectrum are slightly off. The +16\,d spectrum of SN~2016gkg does not show a good resemblance with the spectrum of SN~2016iyc. From Figure~\ref{fig:spec_comparison}, we conclude that the spectrum of SN~2016iyc shows a close resemblance with the spectra of other well-studied SNe~IIb, thereby providing good evidence for the classification of SN~2016iyc as an SN~IIb.

\section{Possible Progenitor Modelling and the Results of Synthetic Explosions for SN~2016iyc}
%as possible progenitor}
\label{sec:mesa_snec_ch4}
\begin{figure*}
\centering
    \includegraphics[height=7.0cm,width=.49\columnwidth]{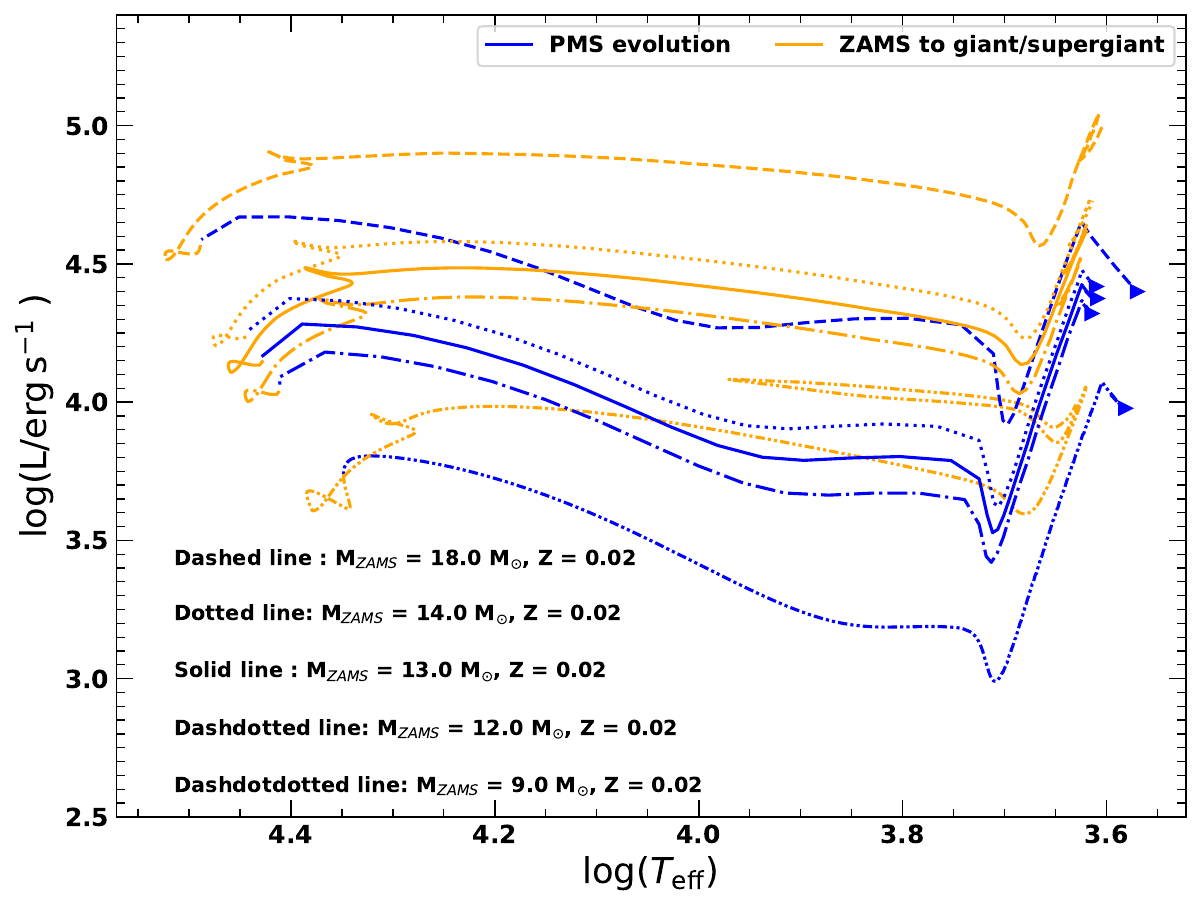}
    \includegraphics[height=7.0cm,width=.49\columnwidth]{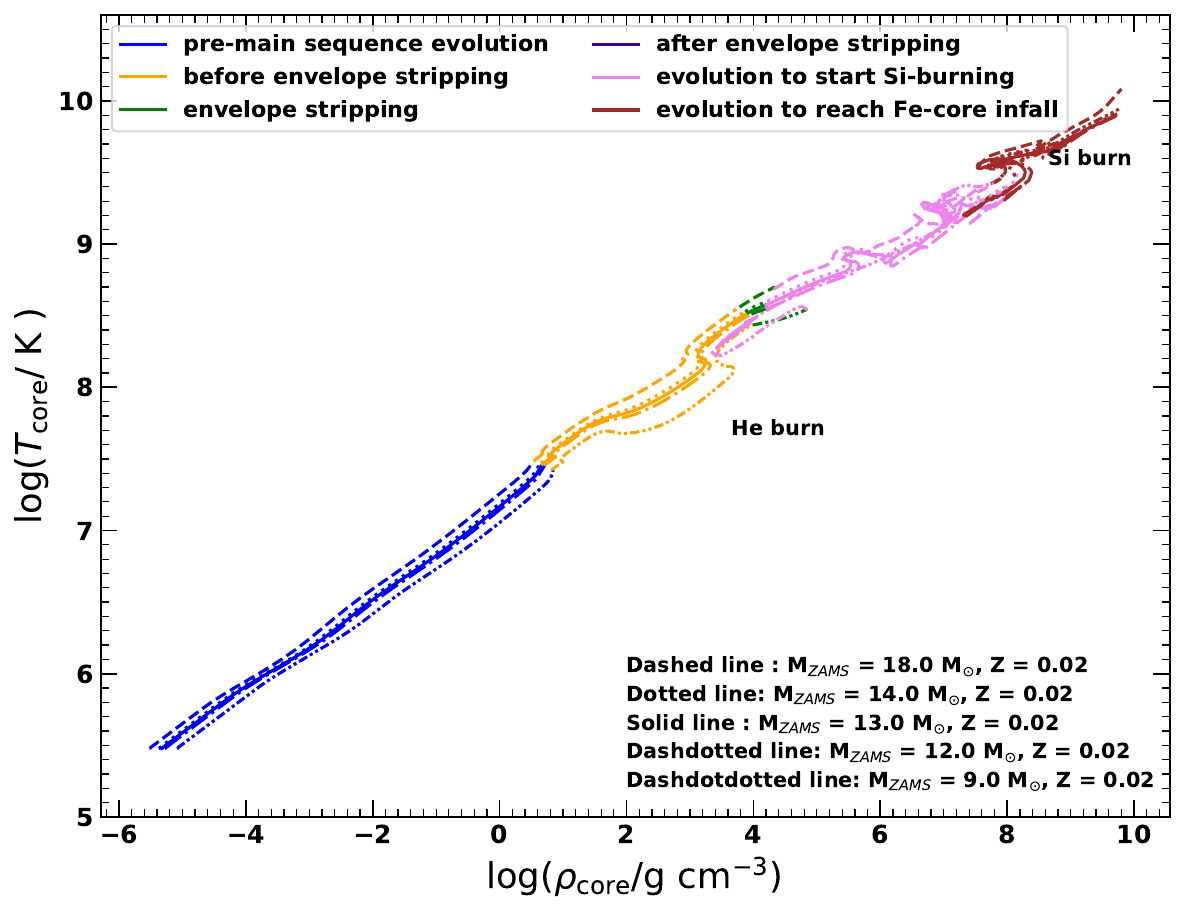}
   \caption{ {\em Left:} The evolution of 9.0\,M$_{\odot}$, 12\,M$_{\odot}$, 13\,M$_{\odot}$, 14\,M$_{\odot}$, and 18.0\,M$_{\odot}$ ZAMS progenitors with Z = 0.02 on the HR diagram. The models begin evolution on the pre-main sequence (blue curve), then reach the main sequence and evolve until they become giants/supergiants with being ready to strip their outer envelopes (orange curve). {\em Right:} The variation of core-temperature with a core-density as the models evolve through various phases on the HR diagram. The core-He and core-Si burning phases are marked. The onset of Fe-core infall in the models is marked by the core temperatures and core densities reaching above $10^{10}$\,K and $10^{10}$\,g\,cm$^{-3}$, respectively.}
    \label{fig:HR_Rho_T}
\end{figure*}

To constrain the physical properties of the possible progenitor of SN~2016iyc, we attempted several progenitor models. Following the available literature, SN~2016iyc lies near the faint limit (see Table~\ref{tab:comparison_Sample}), with $M_{\rm ej}$ also close to the lowest limit (Table~\ref{tab:mejecta_comparison}). As mentioned earlier, low-luminosity SNe with low $^{56}$Ni production are thought to arise from progenitors having masses near the threshold mass for producing CCSNe \citep[][]{Smartt2009}. With low $M_{\rm ej}$ among typical SNe~IIb and having intrinsically low luminosity, we started with the nearly lowest possible ZAMS progenitor mass of 9\,M$_{\odot}$ for a Type IIb SN. Starting from the pre-main sequence, the model could be evolved up to the onset of core collapse. But the 9\,M$_{\odot}$ model at the pre-SN phase in our simulation is very compact, having a radius of only 0.14\,R$_{\odot}$. Such a compact progenitor cannot generate the generic extended-SBO feature of typical SNe~IIb. Furthermore, no direct observational signatures have been found for an SN~IIb arising from a progenitor having ZAMS mass $\leq 11$\,M$_{\odot}$, so we do not make any further attempt to model progenitors having masses $\leq 11$\,M$_{\odot}$. Thus, we select models having ZAMS masses of 12, 13, and 14\,M$_{\odot}$, and evolve them up to the onset of core collapse. Such models originating from the lower limits of progenitor mass systems lack sufficiently strong winds to suffer much stripping; thus, the models are artificially stripped to mimic the effect of a binary companion. A brief description of our models is provided below.

\begin{table}
\caption {Ejecta masses of various SNe~IIb and SN~2016iyc.}
%  from available literature.}
\label{tab:mejecta_comparison}
\begin{center}
{\scriptsize
\begin{tabular}{ccccccccccccc}
\hline\hline
	SN name    &	$M_{\rm ej}$	&	source  \\
\hline

SN~1993J  	 	&	1.9--3.5 	& \citet[][]{Young1995} \\\\

SN~2003bg  	 	&	4	        & \citet[][]{Mazzali2009} \\\\

SN~2008ax  	    &	2--5	        & \citet[][]{Taubenberger2011}   \\\\

SN~2011dh  	 	&	1.8--2.5  	& \citet[][]{Bersten2012} \\\\

SN~2011fu 	 	&	3.5      	& \citet[][]{Morales2015} \\\\

SN~2011hs  	 	&	1.8     	& \citet[][]{Bufano2014}   \\\\

SN~2013df  	 	&	0.8--1.4 	& \citet[][]{Morales2014} \\\\

SN~2016gkg 	    &	3.4     	& \citet[][]{Bersten2018}	 \\\\

SN~2016iyc 	    &	1.2     	& \citet[][]{Zheng2022}   \\

\hline\hline
\end{tabular}}
\end{center}
\end{table}

We first evolve the nonrotating 9, 12, 13, and 14\,M$_{\odot}$ ZAMS stars until the onset of core collapse, using the 1-dimensional stellar evolution code {\tt MESA} \citep[][] {Paxton2011,Paxton2013,Paxton2015,Paxton2018}.  

For the 9\,M$_{\odot}$ model, $\alpha_{\rm MLT} = 2.0$ is used throughout the evolution, except for the phase when the model evolves to reach the beginning of core-Si burning (i.e., in the {\tt inlist\_to\_si\_burn} file), where $\alpha_{\rm MLT} = 0.01$ is used. At this phase, the evolution of the models is extremely sensitive to this $\alpha_{\rm MLT}$, since even a slight change (say, 0.02) results in the failure of the beginning of core-Si burning. Although $\alpha_{\rm MLT} = 0.01$ seems to be very low, this is required for the successful evolution of considered models through the last phases of their evolution. Furthermore, as mentioned by \citet[][]{Joyce2018}, $\alpha_{\rm MLT}$ is neither a physical constant nor a computational one; it is rather a free parameter, so its value must be determined on an individual basis in each stellar evolution code. Thus, as $\alpha_{\rm MLT} = 0.01$ is helpful for evolving the models beyond the beginning of core-Si burning, it is acceptable.
For the 12, 13, and 14\,M$_{\odot}$ models, $\alpha_{\rm MLT} = 3.0$ is used throughout the evolution.

Convection is modelled using the mixing theory of \citet[][]{Henyey1965}, adopting the Ledoux criterion. Semiconvection is modelled following \citet[][]{Langer1985} with an efficiency parameter of $\alpha_{\mathrm{sc}} = 0.01$. For the thermohaline mixing, we follow \citet[][]{Kippenhahn1980}, and set the efficiency parameter as $\alpha_{\mathrm{th}} = 2.0$. We model the convective overshoot with the diffusive approach of \citet[][]{Herwig2000}, with $f= 0.001$ and $f_0 = 0.007$ for all the convective cores and shells. We use the ``Dutch" scheme for the stellar wind, with a scaling factor of 1.0. The ``Dutch" wind scheme in MESA combines results from several papers. Specifically, when $T_{\mathrm{eff}} > 10^4$\,K and the surface mass fraction of H is greater than 0.4, the results of \citet[][]{Vink2001} are used, and when $T_{\mathrm{eff}} > 10^4$\,K and the surface mass fraction of H is less than 0.4, the results of \citet[][]{Nugis2000} are used. In the case when $T_{\mathrm{eff}} < 10^4$\,K, the \citet[][]{dejager1988} wind scheme is used.

SNe~IIb have been considered to originate from massive stars which have retained a significant amount of their H-envelope. We have adopted a mass-loss rate of $\dot{M} \gtrsim 10^{-4}$\,M$_{\odot}\,\mathrm{yr}^{-1}$ to artificially strip the models until the final $M_{\rm H}$ reaches in range of (0.013--0.055)\,M$_{\odot}$. Such extensive mass loss rates are supported by the studies performed by \citet[][]{Ouchi2017}, where they mention that the binary scenario for the progenitors of SNe~IIb leads to such high mass-loss rates. Furthermore, \citet[][]{VanLoon2005}  have also reported extensive mass-loss rates reaching $\dot{M} = 10^{-4}$\,M$_{\odot}\,\mathrm{yr}^{-1}$, solely by a single stellar wind.

\begin{landscape}
\begin{figure*}
\centering
    \includegraphics[height=8.0cm,width=8.5cm,angle=0]{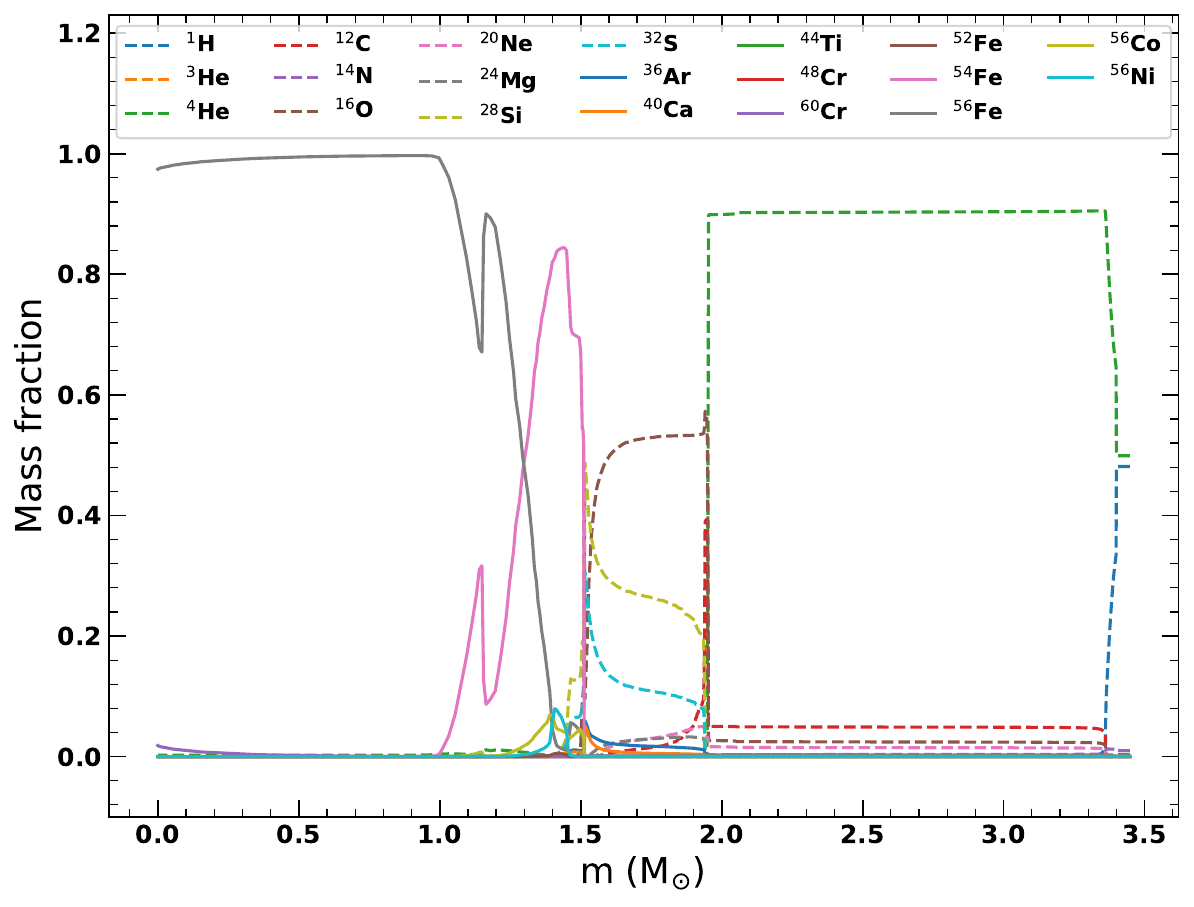}
    \includegraphics[height=8.0cm,width=8.5cm,angle=0]{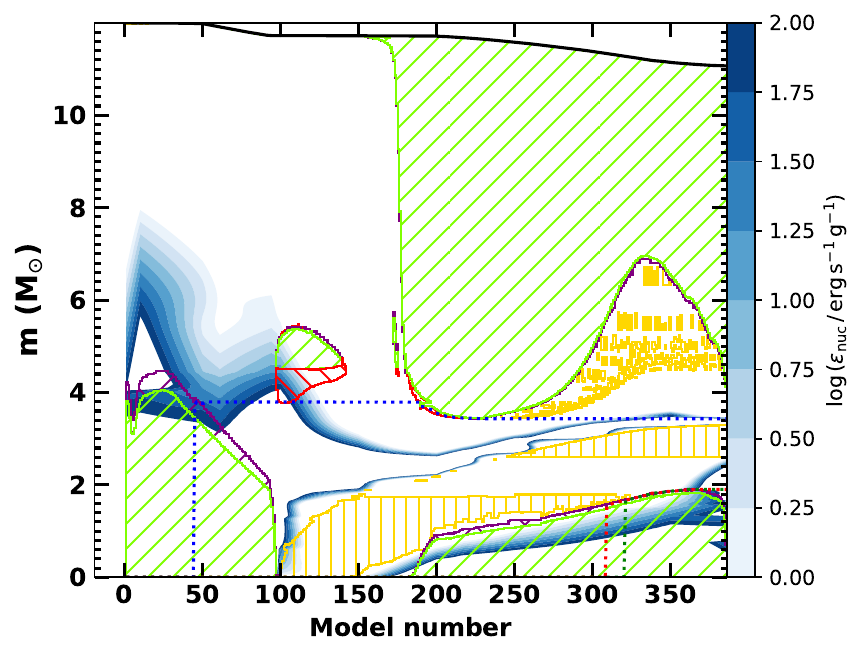}
   \caption {{\em Left:} The mass fractions of a few key elements when the 12\,M$_{\odot}$ ZAMS progenitor model with Z = 0.02 has just reached the stage of Fe-core infall. Notice the very high mass fraction of $^{56}$Fe in the core compared to other species. {\em Right:}  The Kippenhahn diagram of the same model for a period from the beginning of main-sequence evolution to the stage when the model is ready to be stripped.}
    \label{fig:mass_Kipp}
\end{figure*}
\end{landscape}

\begin{figure*}
\centering
    \includegraphics[height=8.0cm,width=.49\columnwidth]{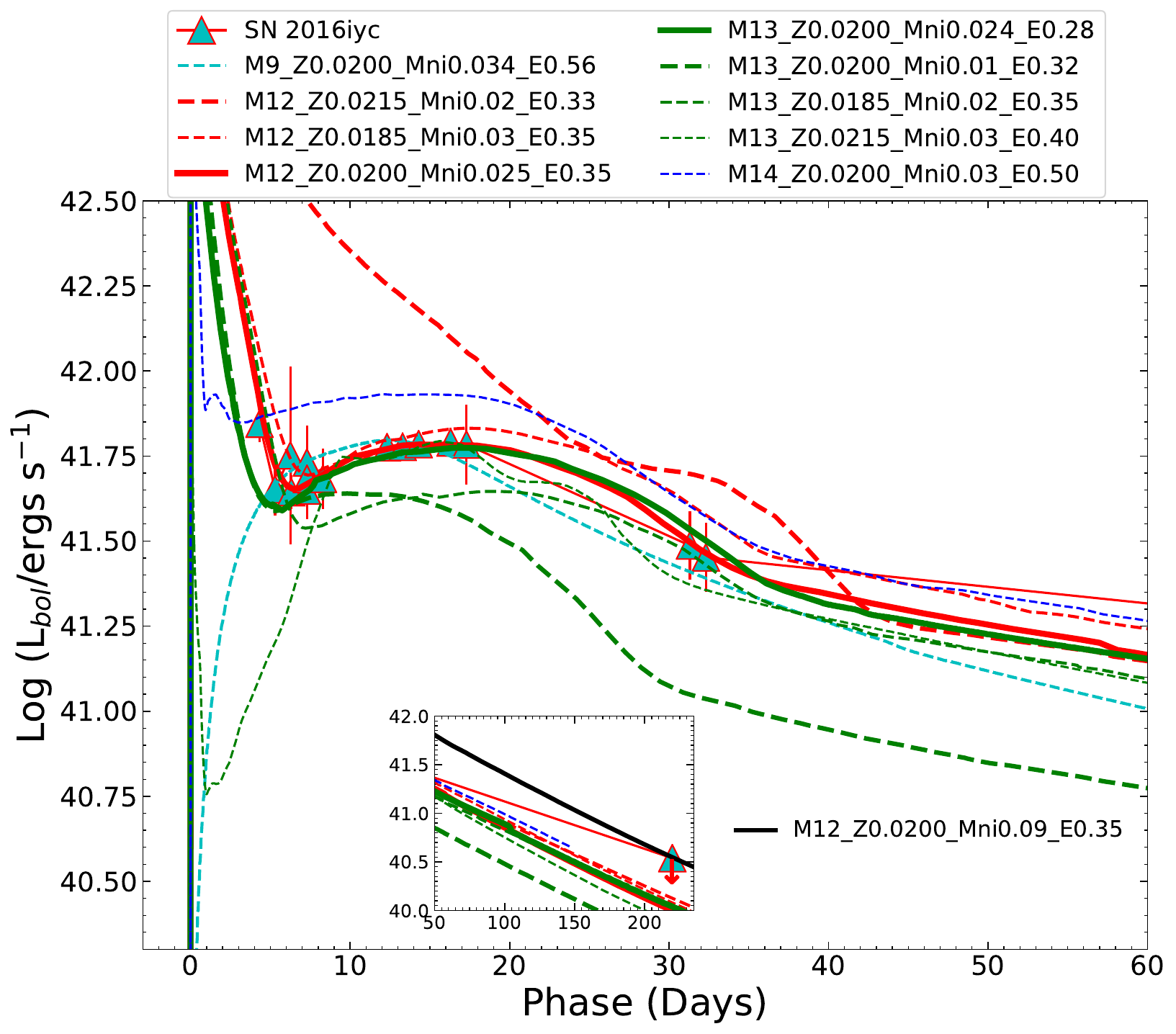}
    \includegraphics[height=8.0cm,width=.49\columnwidth]{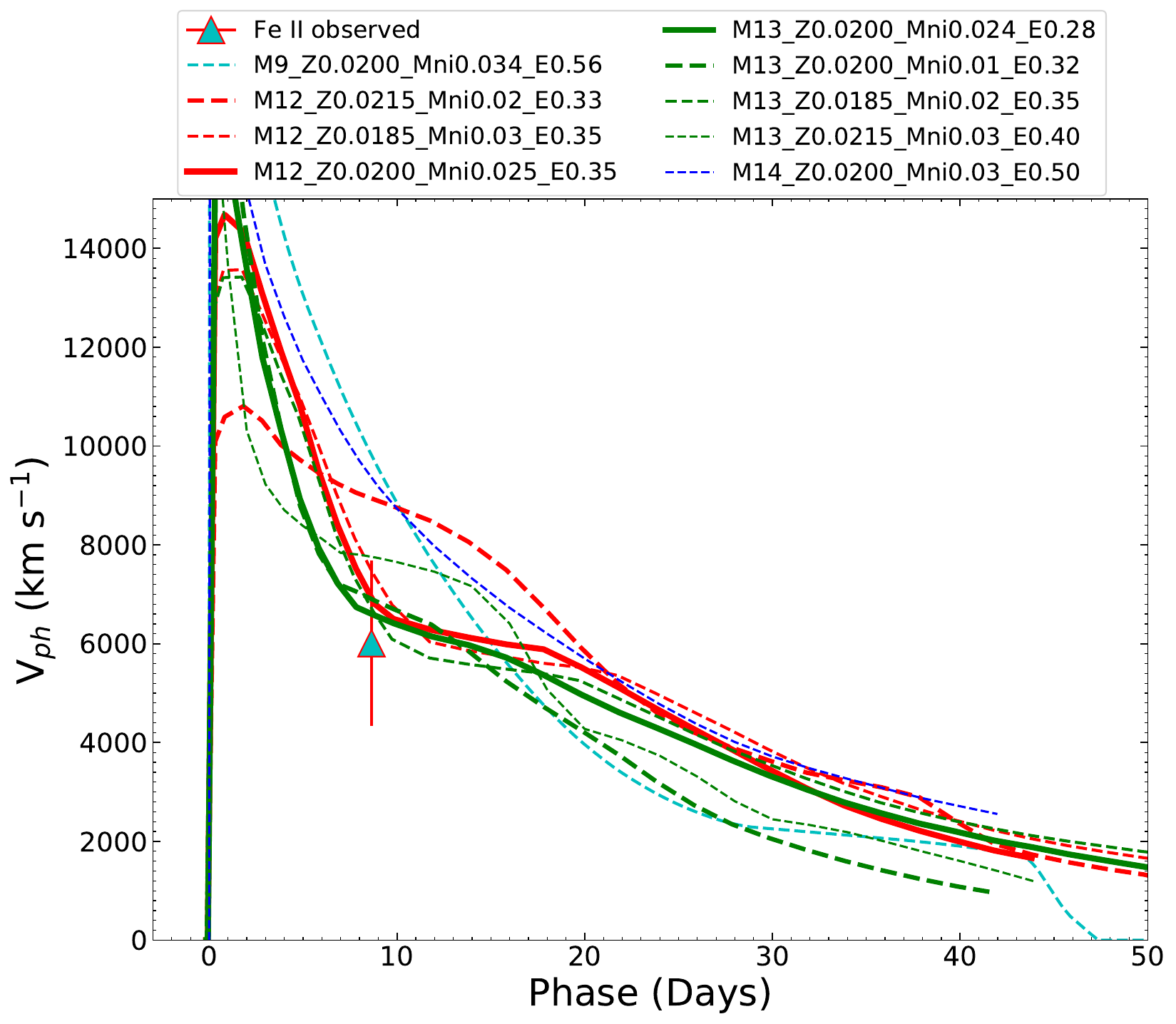}
   \caption {The results of the synthetic explosions produced using {\tt STELLA/SNEC} by assuming  9\,M$_{\odot}$, 12\,M$_{\odot}$, 13\,M$_{\odot}$, and  14\,M$_{\odot}$ ZAMS stars as the possible progenitors for SN~2016iyc. {\em Left:} The bolometric luminosity light curves corresponding to different models having different metallicities, explosion energies, and nickel masses compared with the observed bolometric light curve of SN~2016iyc. {\em Right:} The corresponding photospheric velocity evolution comparison. In both the panels, the ``Phase'' is (approximate) days since the explosion. Following \citet[][]{Zheng2022}, the adopted phase of explosion is $-4.64^{+0.67}_{-0.76}$ since first detection that corresponds to MJD $57736.47^{+0.67}_{-0.76}$. The velocities produced by the models are well within the error bar of the observed photospheric velocities indicated by the Fe\,II line velocity.}
    \label{fig:hydrodynamical_lc}
\end{figure*}

Once the models have stripped off up to the specified limit of the H-envelope, we switch off the artificial mass loss and further evolve the models until the onset of core collapse. Corresponding to various ZAMS masses, the amount of remaining H varies. Massive progenitors with a similar rate of stripping as less-massive progenitors will retain a larger amount of H. In our simulations, the specified limit on H mass depends primarily on (a) the model's ability to evolve up to the stage of core collapse by retaining the specified amount of H (also, there is a limit on stripping), and (b) the radius of the pre-SN progenitor. If we need a compact progenitor, the amount of retained H is less, and if pre-SN progenitors are required to have extended envelopes, the amount of retained H is more.

The evolution of the models using {\tt MESA} takes place in various steps. The models start to evolve on the pre-main sequence and reach the main sequence. The arrival of the models on the main sequence is marked when the ratio of the luminosity due to nuclear reactions and the total luminosity of the models is 0.8. Later, the models further evolve on the main sequence, becoming giants or supergiants. As a next step, artificial stripping of the models is performed, after which they are allowed to settle down. Once the stripping of the models reaches the specified H-envelope mass limit and the models have settled down, they further evolve until the ignition of Si~burning in their core. Once the Si~burning has started in the core, the models begin to develop an inert iron core, which is responsible for their cores to collapse.

The evolution of such models having ZAMS masses of 9, 12, 13, and 14\,M$_{\odot}$ with metallicity Z = 0.0200 on the Hertzsprung-Russell (HR) diagram is illustrated in the left panel of Figure~\ref{fig:HR_Rho_T}. We simulated a total of 9 models covering progenitor masses of 9--14\,M$_{\odot}$ and also covering subsolar to supersolar metallicity wherever necessary. The pre-explosion properties using {\tt MESA} and explosion properties using {\tt STELLA/SNEC} are listed in Table~\ref{tab:MESA_MODELS}. The models have been so named that they include the pieces of information of ZAMS mass, metallicity, $M_{\rm Ni}$, and $E_{\rm exp}$. Thus, the model {\tt M9.0\_Z0.0200\_Mni0.034\_E0.56} has a ZAMS mass of 9\,M$_{\odot}$, Z = 0.0200, $M_{\rm Ni} = 0.034$\,M$_{\odot}$, and $E_{\rm exp} = 0.56\times10^{51}$\,erg.

The right-hand panel of Figure~\ref{fig:HR_Rho_T} shows the variation of core temperature ($T_{\rm core}$) with core density ($\rho_{\rm core}$) as the models evolve through various phases on the HR diagram. The core-He and core-Si burning phases are marked. The onset of core collapse is marked by $T_{\rm core}$ and $\rho_{\rm core}$ reaching above $\sim10^{10}$\,K and $10^{10}$\,g\,cm$^{-3}$, respectively. The left panel of Figure~\ref{fig:mass_Kipp} shows the mass fractions of various species present when the 12\,M$_{\odot}$ model (with Z = 0.0200) has achieved Fe-core infall. The core comprises $^{56}$Fe with negligible fractions of other species. Significant fractions of heavier metals are seen toward the surface of the pre-explosion progenitor. The right-hand panel of Figure~\ref{fig:mass_Kipp} shows the Kippenhahn diagram for the 12\,M$_{\odot}$ model (with Z = 0.0200) for a period from the beginning of main-sequence evolution to the stage when the model is ready to begin envelope stripping. In this figure, the convective regions are marked by the hatchings with the logarithm of the specific nuclear energy generation rate ($\epsilon_{\rm nuc}$) inside the stellar interiors marked with blue colours. The dark-yellow regions indicate the stellar interior where the thermohaline mixing is going on.

\begin{landscape}
\begin{table*}
\caption{{\tt MESA} model and {\tt STELLA/SNEC} explosion parameters of various models for SN~2016iyc.}
\label{tab:MESA_MODELS}
\begin{center}
{\scriptsize
\begin{tabular}{ccccccccccccc} 
\hline\hline
Model Name	& $M_{\rm ZAMS}$	& Z  & $M_{\mathrm{H}}^{a}$ & $R_{\mathrm{0}}^{b}$ & $f_{ov}^{c}$ &	$M_{\mathrm{f}}^{d}$	& $M_\mathrm{ci}^{e}$	&	$M_\mathrm{cf}^{f}$ & $M_{\mathrm{ej}}^{g}$	&	$M_{\mathrm{Ni}}^{h}$ &	$E_{\mathrm{exp}}^{i}$ 	\\
	&	(M$_{\odot}$) &	&	(M$_{\odot}$)	 &  (R$_{\odot}$) &  & (M$_{\odot}$)	&	(M$_{\odot}$) 	&	(M$_{\odot}$)  & (M$_{\odot}$) & (M$_{\odot}$) & 	($10^{51}$\,erg) 	\\
\hline
\hline

M9.0\_Z0.0200\_Mni0.034\_E0.56     &	9.0  	&	0.0200  &  0.013   & 0.14  & 0.007  &    2.17  & 1.4 & 1.4 &  0.77 & 0.034 &  0.56		\\

M12.0\_Z0.0215\_Mni0.02\_E0.33     &	12.0  	&	0.0215  &  0.035   & 596  & 0.007  &    3.96  & 1.54 & 1.54 & 2.42 & 0.02 &  0.33		\\

M12.0\_Z0.0185\_Mni0.03\_E0.35     &	12.0  	&	0.0185  &  0.055   & 315 & 0.007  &    3.49  & 1.46 & 1.46 & 2.03 & 0.03 &  0.35		\\

M12.0\_Z0.0200\_Mni0.025\_E0.35     &	12.0  	&	0.0200  &  0.05    & 300 & 0.007  &    3.45  & 1.52 &  1.52 & 1.93 & 0.025 &  0.35		\\

M12.0\_Z0.0200\_Mni0.09\_E0.35     &	12.0  	&	0.0200  &  0.05    & 300 & 0.007  &    3.45  & 1.52 & 1.52 & 1.93 & 0.09 &  0.35		\\

M13.0\_Z0.0200\_Mni0.024\_E0.28     &	13.0  	&	0.0200  &  0.04    & 204 & 0.007  &    3.79  & 1.64 & 1.90 & 1.88 & 0.024 &  0.28 		\\

M13.0\_Z0.0200\_Mni0.01\_E0.32     &	13.0  	&	0.0200  &  0.04    & 204 & 0.007  &    3.79  & 1.64 & 1.64 & 2.15 & 0.01 &  0.32 		\\

M13.0\_Z0.0185\_Mni0.02\_E0.35     &	13.0  	&	0.0185  &  0.06    & 318 & 0.007  &    3.92  & 1.53 & 1.56 & 2.36 & 0.02 &  0.35		\\

M13.0\_Z0.0215\_Mni0.03\_E0.40     &	13.0  	&	0.0215  &  0.015   & 10  & 0.007  &    3.81  & 1.61 & 1.62 &  2.19 & 0.03 &  0.40		\\

M14.0\_Z0.0200\_Mni0.03\_E0.50     &	14.0  	&	0.0200  &  0.03    & 55  & 0.007  &    4.23  & 1.54 & 1.54 & 2.69 & 0.03 &  0.50		\\

\hline\hline
\end{tabular}}
\end{center}
%\par
{$^a$Amount of hydrogen retained after stripping.
$^b$pre-SN progenitor radius.
$^c$Overshoot parameter.
$^d$Final mass of pre-SN model.
$^e$Initial mass of the central remnant.
$^f$Final mass of the central remnant.
$^g$Ejecta mass.
$^h$Nickel mass.
$^i$Explosion energy.}\\

\end{table*} 
\end{landscape}

Using the progenitor models on the verge of core-collapse obtained through {\tt MESA}, we carried out radiation hydrodynamic calculations to simulate the synthetic explosions. For this purpose, we used {\tt STELLA}\citep[][]{Blinnikov1998, Blinnikov2000, Blinnikov2006} and {\tt SNEC} \citep[][]{Morozova2015}. {\tt STELLA} solves the radiative transfer equations in the intensity momentum approximation in each frequency bin, while {\tt SNEC} is a 1-dimensional Lagrangian hydrodynamic code that solves the radiation energy transport equations in the flux-limited diffusion approximation. {\tt STELLA} and {\tt SNEC}, both generate the bolometric light curve and the photospheric velocity evolution of the SN, along with a few other observed parameters. The radioactive decay of $^{56}$Ni to $^{56}$Co is considered to be one of the prominent mechanisms for powering the primary peak of SNe~IIb. Both of these codes incorporate this model by default. Thus in this section, we model the entire bolometric light curve of the SN~2016iyc assuming this powering mechanism. Here, we provide the setup of the explosions to incorporate the Ni--Co decay model. The setups for simulating the synthetic explosion using {\tt SNEC} and {\tt STELLA} closely follow \citet[][]{Ouchi2019} and the settings used in Chapter~\ref{Ch:3}, respectively. Here, we briefly summarise the important parameters and modifications.

We simulate the synthetic explosions of the 9\,M$_{\odot}$ model using {\tt SNEC}. First, the innermost 1.4\,M$_{\odot}$ is excised before the explosion, assuming that the model collapses to form NSs. The number of grid cells is set to be 1000 so that the light curves and photospheric velocities of the SNe from synthetic explosions are well converged in the time domain of interest.

For the M9.0\_Z0.0200\_Mni0.034\_E0.56 model, the synthetic explosion is carried out using {\tt SNEC}. The explosion is simulated as a {\tt Thermal\_Bomb} by adding $0.56 \times 10^{51}$\,erg of energy in the inner 0.1\,M$_{\odot}$ of the model for a duration of 0.1\,s. {\tt SNEC} does not include a nuclear-reaction network, so the amount of $^{56}$Ni is set by hand. A total of 0.034\,M$_{\odot}$ of $^{56}$Ni is distributed from the inner boundary up to the mass coordinate $m(r) = 2.0$\,M$_{\odot}$.

For the models having ZAMS masses of 12, 13, and 14\,M$_{\odot}$, we used {\tt STELLA} to simulate the synthetic explosions. The pre-SN model masses from 12\,M$_{\odot}$ models lie in the range of (3.45--3.96)\,M$_{\odot}$, while from 13\,M$_{\odot}$ models, the pre-SN model masses lie in the range (3.79--3.81)\,M$_{\odot}$. Furthermore, the 14\,M$_{\odot}$ model has a pre-SN mass of 4.23\,M$_{\odot}$ and is thus prone to produce a much higher $M_{\rm ej}$ than required for SN~2016iyc. For producing the synthetic explosions, the hydrodynamic simulations are performed using {\tt Thermal\_Bomb}-type explosion. Various explosion parameters, including the ejecta masses, synthesised nickel masses, and explosion energies corresponding to different models, are presented in Table~\ref{tab:MESA_MODELS}

The results of the hydrodynamic simulations are shown in Figure~\ref{fig:hydrodynamical_lc}. The left panel of Figure~\ref{fig:hydrodynamical_lc} shows the comparison of the {\tt SNEC}- and {\tt STELLA}-calculated bolometric light curves with the observed bolometric light curve (see Sec.~\ref{subsec3.3} for details on bolometric light curves) produced by fitting black bodies to the SEDs and integrating the fluxes over the wavelength range of 100--25,000\,\AA, while the right-hand panel shows the comparison of the corresponding photospheric velocities with the photospheric velocity obtained using the only available spectrum indicated by the Fe~II line velocity. The M9\,M\_Z0.0200\_Mni\_0.034\_E0.56 model could match the observed stretch factor and the peak of the bolometric light curve, but it fails to reproduce the early extended SBO feature. The failure in producing the generic extended-SBO feature could be associated to the compactness of the pre-SN model having a radius of only 0.14\,R$_{\odot}$.

Moreover, all of the remaining models generate the generic extended-SBO feature. Still, only the M12.0\_M\_Z0.0200\_Mni0.025\_E0.35 could generate the extended-SBO and overall bolometric light curve that could match with the actual bolometric light curve of SN~2016iyc. Another model that could nearly match the SN~2016iyc bolometric light curve is M13.0\_M\_Z0.0200\_Mni0.024\_E0.28. The remaining models deviate largely from the observed bolometric light curve of SN~2016iyc (left panel, Figure~\ref{fig:hydrodynamical_lc}). From the right-hand panel of Figure~\ref{fig:hydrodynamical_lc}, we find that the photospheric velocity evolution generated by the models, M13.0\_M\_Z0.0200\_Mni0.01\_E0.25 and M12.0\_M\_Z0.0200\_Mni0.025\_E0.35 pass closely to the observed line velocity from Fe~II which is a good indicator of observed photospheric velocity.

We also have an upper limit on the bolometric luminosity of SN~2016iyc at a phase nearly 220\,d since the explosion. In order to match the upper limit of the luminosity at that epoch, the model M12.0\_M\_Z0.0200\_Mni0.025\_E0.35 requires $M_{\rm Ni} = 0.09$\,M$_{\odot}$ (please take a look at the inset plot in the left panel of Figure~\ref{fig:hydrodynamical_lc}; M12.0\_M\_Z0.0200\_Mni0.09\_E0.35 is the corresponding model), serving as an upper limit on the synthesised nickel in SN~2016iyc.

\citet[][]{Anderson2019} has estimated a median value of 0.102\,M$_{\odot}$ for the nickel mass by considering 27 SNe~IIb. Moreover, \citet[][]{Afsariardchi2021} have also estimated the Ni mass for eight SNe~IIb and found that except for SN\,1996cb ($M_{\rm Ni} = 0.04\pm0.01$\,M$_{\odot}$) and SN\,2016gkg ($M_{\rm Ni} = 0.09\pm0.02$\,M$_{\odot}$), each SN~IIb has higher $M_{\rm Ni}$ than 0.09\,M$_{\odot}$. These comparisons show that SN~2016iyc definitely suffered low nickel production. Thus, the 1-dimensional stellar evolution of various models, along with the hydrodynamic simulations of their explosions, suggest that a ZAMS progenitor having mass in the range 12--13\,M$_{\odot}$ with $M_{\rm ej}$ in the range (1.89--1.93)\,M$_{\odot}$, $M_{\rm Ni} < 0.09$\,M$_{\odot}$, and $E_{\rm exp}$ = (0.28--0.35) $\times 10^{51}$\,erg could be the possible progenitor of SN~2016iyc.

\begin{figure*}
\centering
    \includegraphics[height=8.0cm,width=.49\columnwidth]{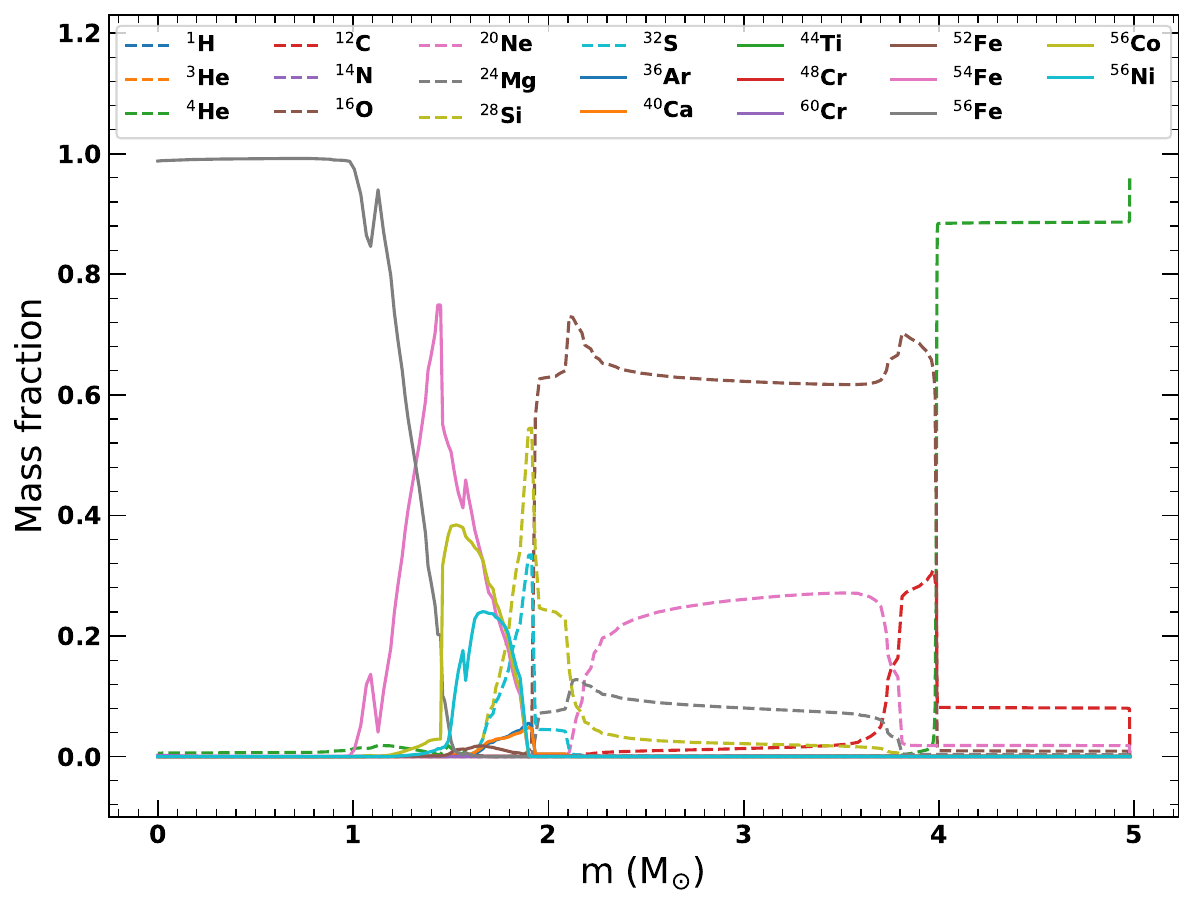}
    \includegraphics[height=8.0cm,width=.49\columnwidth]{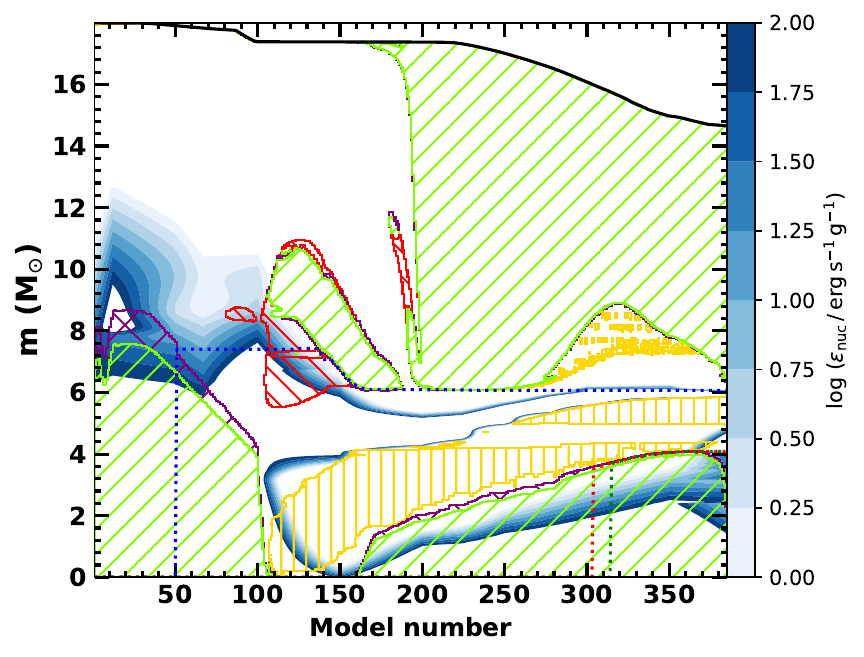}
   \caption{ {\em Left:} The mass fractions of a few key elements when the 18\,M$_{\odot}$ ZAMS progenitor model with Z = 0.02 has just reached the stage of Fe-core infall. The mass fraction of $^{56}$Fe in the centre is much higher compared to other species. {\em Right:} The Kippenhahn diagram of the model for a period from the beginning of main-sequence evolution to the stage when the model is ready to be stripped.}
    \label{fig:mass_Kipp_2}
\end{figure*}

\begin{figure*}
\centering
    \includegraphics[height=8.0cm,width=.49\columnwidth]{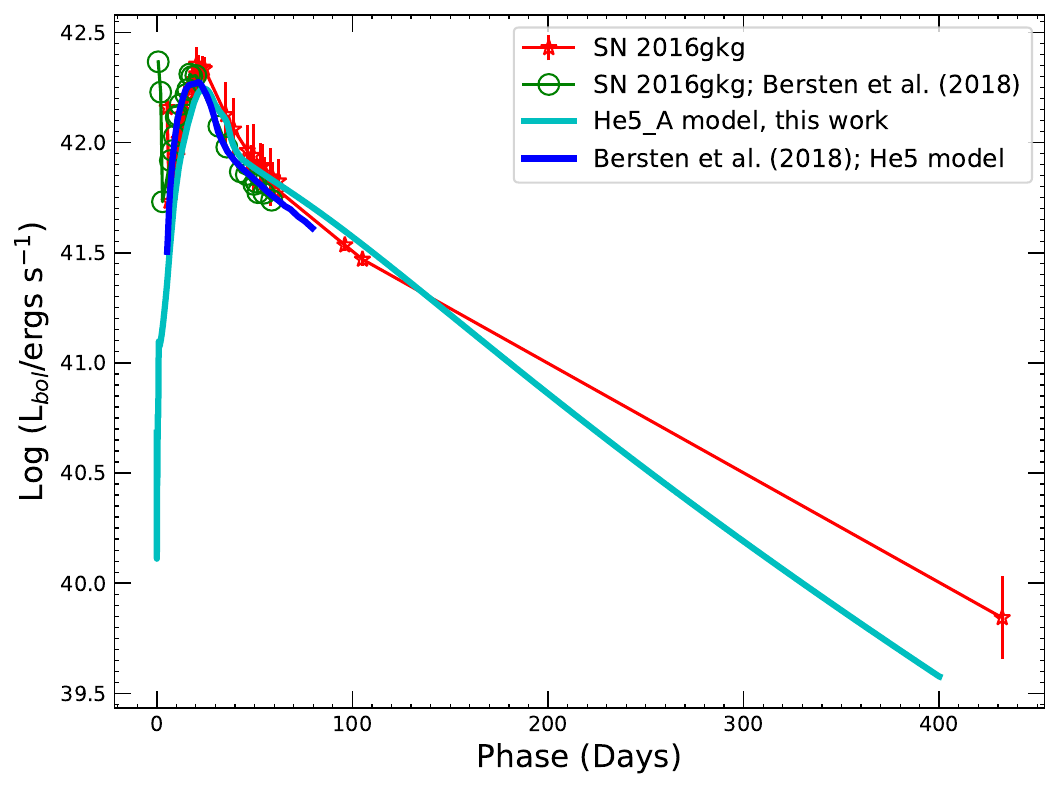}
    \includegraphics[height=8.0cm,width=.49\columnwidth]{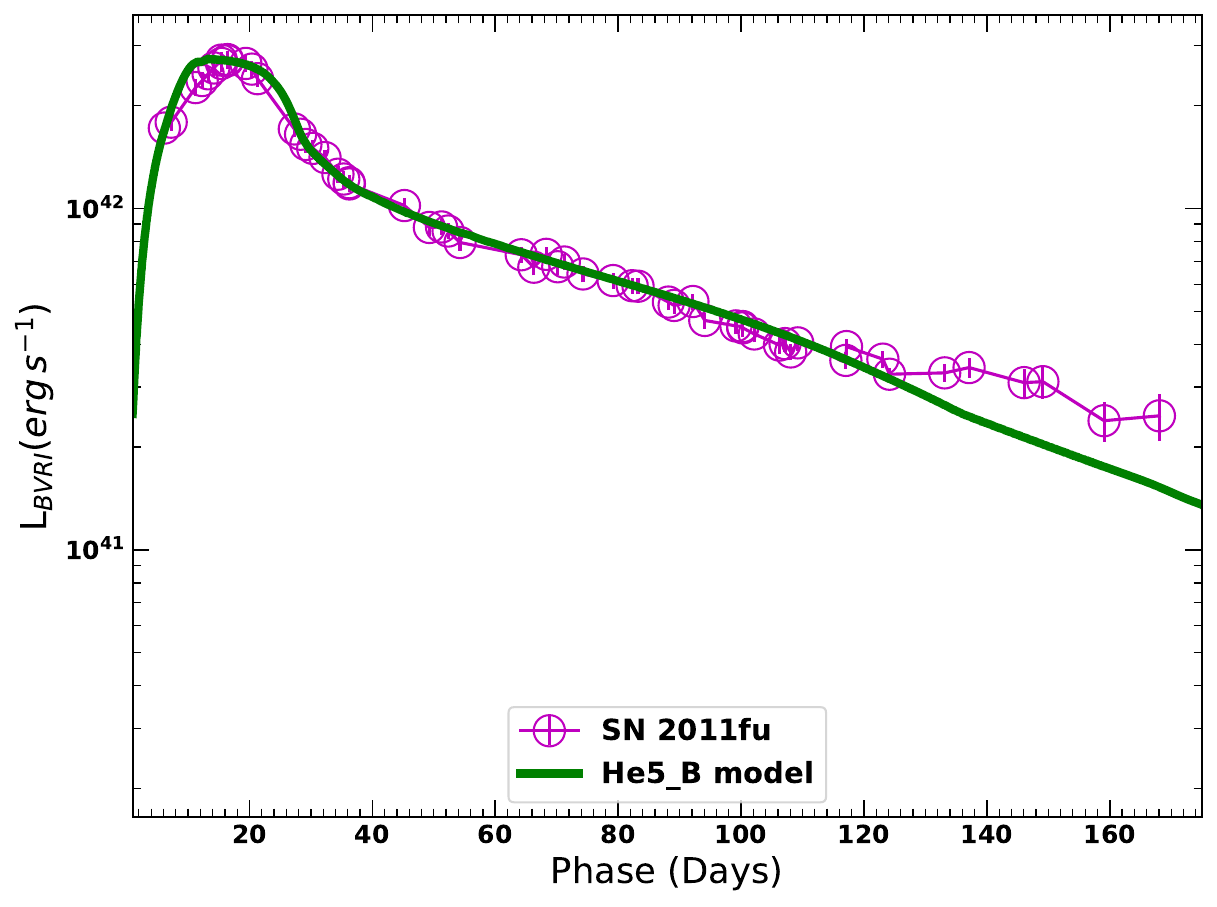}
   \caption{ The results of 1-dimensional stellar evolution of models using {\tt MESA} and their synthetic explosion using {\tt SNEC} for SN~2016gkg and SN~2011fu. {\em Left:} Comparison of the quasi-bolometric light curve of SN~2016gkg with that obtained using {\tt SNEC} by taking into account the $^{56}$Ni and $^{56}$Co decay model and keeping the parameters close to those of \citet[][]{Bersten2018}. {\em Right:} Result of a similar analysis for SN~2011fu.}
    \label{fig:snec_2}
\end{figure*}

Recent studies suggest the masses of possible progenitors of Type IIb CCSNe to be usually higher than 9\,M$_{\odot}$, lying in the range 10--18\,M$_{\odot}$ \citep[][]{Van2013, Folatelli2014, Smartt2015}. However, there has been no direct observational evidence of an SN~IIb arising from a ZAMS progenitor of $\lesssim 12$\,M$_{\odot}$. The present analysis indicates that SN~2016iyc arises from the lower-mass end of the SN~IIb progenitor channel. As part of our study, we also performed the 1-dimensional stellar evolutions of the possible progenitors of SN~2016gkg and SN~2011fu, and we simulated their hydrodynamic explosions in the next section to cover the range of faintest (SN~2016iyc), intermediate (SN~2016gkg), and highest (SN~2011fu) luminosity SNe in the comparison sample. 

\section{Stellar Modelling and Synthetic Explosions for SN 2016gkg and SN 2011fu}
% of the possible progenitors of SN~2016gkg and SN~2011fu}
\label{sec:SN2016gkg_model_ch4}

In this section, we perform hydrodynamic simulations of explosions from the possible progenitors of an intermediate-luminosity SN~2016gkg and the most-luminous SN~2011fu in the comparison sample to cover the higher end of the progenitor masses of SNe~IIb. After modelling their progenitors using {\tt MESA}, we simulate the synthetic explosions using {\tt SNEC} and match the {\tt SNEC} produced bolometric light curves with the observed ones. 

To construct the bolometric light curve of SN~2016gkg, we used the recalibrated $BVRI$ KAIT data along with the data from the 3.6\,m DOT at two epochs and incorporated {\tt SUPERBOL}. The photometric data of SN~2016gkg in this work are presented in Table~\ref{tab:optical_observations_2016gkg1} and Table~\ref{tab:optical_observations_2016gkg2}. Previously, \citet[][]{Bersten2018} also used KAIT data calibrated from an older KAIT reduction pipeline. Figure~\ref{fig:SN2016gkg_comparison} shows the comparison between the KAIT data used by \citet[][]{Bersten2018} and the recalibrated KAIT data. To construct the bolometric light curve of SN~2011fu, we make use of {\tt SUPERBOL} as we did in earlier sections by incorporating the $BVRI$ data from \citet[][]{Kumar2013}. 

To model the possible progenitor of SN~2016gkg, we closely follow the HE5 model from \citet[][]{Bersten2018}. Also, \citet[][]{Morales2015} suggests a nearly similar model for the possible progenitor of SN~2011fu. An 18\,M$_{\odot}$ ZAMS progenitor mass is employed for both SNe. The modelling and explosion parameters are listed in Table~\ref{tab:MESA_MODELS_11fu_n_16gkg}. Starting from the ZAMS, the model is evolved up to the stage where the core starts to collapse. The evolution of the model on the HR diagram is shown in the left panel of Figure~\ref{fig:HR_Rho_T}. Various physical processes during the evolution on the HR diagram have been indicated. Also, the right-hand panel displays the variation of $T_{\rm core}$ with $\rho_{\rm core}$. It is indicated that during the last evolutionary phases, the core density and temperatures have reached over $10^{10}$\,g\,cm$^{-3}$ and $10^{10}$\,K, respectively. Such high core density and temperatures mark the onset of core collapse. The left panel of Figure~\ref{fig:mass_Kipp_2} shows the mass fractions of various elements at the stage when the model has just reached the stage of Fe-core infall. As another piece of evidence for the onset of core collapse, we can see that the core is mainly composed of inert $^{56}$Fe. The right-hand panel of Figure~\ref{fig:mass_Kipp_2} shows the Kippenhahn diagram of the model for a period from the beginning of main-sequence evolution to the stage when the model is ready to be stripped.

Models He5\_A and He5\_B are used for SN~2016gkg and SN~2011fu, respectively. Although the parameters, including the ZAMS mass, metallicity, rotation, and overshoot parameter, are the same for these two models, different explosion parameters are employed using {\tt SNEC} to simulate the synthetic explosions.

The left panel of Figure~\ref{fig:snec_2} illustrates the comparison of our hydrodynamic simulation of synthetic explosions for SN~2016gkg with the results of \citet[][]{Bersten2018}. The difference between the bolometric light curve from \citet[][]{Bersten2018} and calculated using KAIT revised photometry are within the error bars. Our model could explain the bolometric light curve of SN~2016iyc very well. Furthermore, the right-hand panel of Figure~\ref{fig:snec_2} shows the comparison of the {\tt SNEC}-calculated bolometric light curve with the observed quasi-bolometric light curve of SN~2011fu. The 1-dimensional stellar modelling of possible progenitors using {\tt MESA} along with their hydrodynamic simulation of explosions using {\tt SNEC} explain the observed light curves of SN~2016gkg and SN~2011fu very well. Now, we have performed the stellar modelling of the possible progenitors and the hydrodynamic explosions of SN~2016iyc, SN~2016gkg, and SN~2011fu to cover a range of faintest (SN~2016iyc), intermediate (SN~2016gkg), and highest (SN~2011fu) luminosity SNe in the comparison sample.

\section{Discussion on Findings}
\label{sec:Discussions_ch4}

Detailed photometric and spectroscopic analyses of the low-luminosity Type IIb SN~2016iyc are performed in this work. The extinction-corrected data of SN~2016iyc are used to construct the quasi-bolometric and bolometric light curves using {\tt SUPERBOL}. Comparisons of the absolute $V$-band and quasi-bolometric light curves of SN~2016iyc with other well-studied SNe~IIb indicate that SN~2016iyc lies toward the faint limit of this subclass. Low-luminosity SNe~IIb with low $^{56}$Ni production are thought to arise from progenitors having masses near the threshold mass for producing a CCSN.

%The blackbody radius evolution indicates that SN~2016iyc has the smallest radius at any epoch. This anomalous behaviour could be attributed to its low ejecta velocity.
Our study indicates that among the comparison sample in this work, SN~2016iyc has the smallest blackbody radius at any given epoch. This anomalous behaviour could be attributed to its low ejecta velocity.

Based on the low $M_{\rm ej}$ and the lowest intrinsic brightness among SNe in the comparison sample, 9--14\,M$_{\odot}$ ZAMS progenitors are modelled as the possible progenitor of SN~2016iyc using {\tt MESA}. The results of synthetic explosions simulated using {\tt STELLA} and {\tt SNEC} are in good agreement with the observed ones.  

The 1-dimensional stellar modelling of the possible progenitor using {\tt MESA} and simulations of hydrodynamic explosions using {\tt SNEC}/{\tt STELLA} indicate that SN~2016iyc originated from a (12--13)\,M$_{\odot}$ ZAMS progenitor, near the lower end of progenitor masses for SNe~IIb. The models show a range of parameters for SN~2016iyc, including $M_{\rm ej} =$ (1.89--1.93)\,M$_{\odot}$ and $E_{\rm exp} =$ (0.28--0.35) $\times 10^{51}$\,erg. We also put an upper limit of 0.09\,M$_{\odot}$ on the amount of nickel synthesised by the SN. The pre-SN radius of the progenitor of SN~2016iyc is (240--300)\,R$_{\odot}$.

Stellar evolution of the possible progenitors and hydrodynamic simulations of synthetic explosions of SN~2016gkg and SN~2011fu have also been performed to compare the intermediate- and high-luminosity ends among well-studied SNe~IIb using {\tt MESA} and {\tt SNEC}. The results of stellar modelling and synthetic explosions for SN~2016iyc, SN~2016gkg, and SN~2011fu exhibit a diverse range of mass of the possible progenitors for SNe~IIb.

\section{Summary and Conclusions}
\label{sec:Conclusions_ch4}

We performed detailed photometric and spectroscopic analyses of SN~2016iyc, a Type IIb SN discovered by LOSS/KAIT. The observed photometric properties of SN~2016iyc were unique in many ways: low luminosity, low ejecta mass, and small blackbody radius.  Attempts to model the possible progenitor were made using the 1-dimensional hydrodynamic code {\tt MESA}. As a part of the present work,  hydrodynamic modelling of the synthetic explosion of an intermediate-luminosity Type IIb SN~2016gkg using recalibrated KAIT data and late-time data from the 3.6\,m DOT, along with an optically very luminous Type IIb SN~2011fu, were also performed. The main results based on the present analysis are as follows.

\begin{enumerate}

\item{Based on the low value of $M_{\rm ej}$, ZAMS stars having masses of 9--14\,M$_{\odot}$ were adopted to model the possible progenitor of SN~2016iyc using {\tt MESA}. The results of synthetic explosions simulated using {\tt SNEC} and {\tt STELLA} were in good agreement with observed properties for ZAMS progenitor masses of 12--13\,M$_{\odot}$ having a pre-SN radius of (240--300)\,R$_{\odot}$. Thus, SN~2016iyc likely had a progenitor arising from the lower end of the progenitor mass channel of an SN~IIb.}\\

\item{We concluded that the overall detailed hydrodynamic simulations of the explosions from various models showed a range of parameters for SN~2016iyc, including an $M_{\rm ej}$ of (1.89--1.93)\,M$_{\odot}$, an $E_{\rm exp}$ of (0.28--0.35) $\times 10^{51}$\,erg, and an upper limit of  $< 0.09$\,M$_{\odot}$ on the amount of nickel synthesised by SN~2016iyc.}\\   
 
\item{Finally, 1-dimensional stellar evolution models of possible progenitors and the hydrodynamic explosions of SN~2016gkg and SN~2011fu were also performed to compare intermediate- and high-luminosity examples among well-studied SNe~IIb. The results for SN~2016iyc, SN~2016gkg, and SN~2011fu showed a diverse range of mass [(12.0--18.0)\,M$_{\odot}$] for the possible progenitors of SNe~IIb considered in this work. Discovery of more such events through survey projects in the near future should provide additional data with which to establish the lower mass limits of such explosions.}
\end{enumerate}

 %SN2016iyc 
\chapter{\sc Investigating the evolutions of Rotating, Population III star of 25\,M$_{\odot}$ and the strength of Resulting Supernovae}
\label{Ch:5} 
\blfootnote{ The results of this Chapter are published in: \textbf{{Aryan}, A.,} {Pandey}, S. B., {Gupta}, Rahul et al., 2023, {\textit{MNRAS:Letters}, {\textbf{521}}, L17-L23}.}
\ifpdf
    \graphicspath{{Chapter5/Chapter5Figs/PNG/}{Chapter5/Chapter5Figs/PDF/}{Chapter5/Chapter5Figs/}}
\else
    \graphicspath{{Chapter5/Chapter5Figs/EPS/}{Chapter5/Chapter5Figs/}}
\fi

\section{Pop III Stars and their Impact in the Early Universe}
\label{sec:intro}

Having comprehensively discussed the temporal and spectral properties of H-deficient Type Ib SNe in Chapter~\ref{Ch:3} followed by the detailed photometric and spectroscopic analysis of a Type IIb SN retaining an intermediate amount of H-envelope in Chapter~\ref{Ch:4}, this Chapter discusses H-rich and H-deficient CCSNe together that originate from progenitors each starting with a mass of 25\,M$_{\odot}$ at ZAMS and zero metallicity. The first generations of stars formed out of uncontaminated matter, initially comprised only of the first two stable elements, hydrogen and helium, from the periodic table, are considered population III (Pop III) stars. 

Due to the insufficiency of coolants in the primordial gas, it is hypothesised that the Pop III stars were massive intrinsically \citep[][]{1983MNRAS.205..705S,1997ApJ...474....1T,1999ApJ...527L...5B,  2001ApJ...548...19N, 2002Sci...295...93A, 2007ApJ...661...10B, 2010MNRAS.401L...5S, 2015MNRAS.448..568H}. However, there have been multiple studies to apprehend the possibility of the existence of Pop III stars having low masses. In recent simulations, it has been found that the formation of pristine, metal-free stars at low to intermediate masses could potentially be due to the fragmented accretion disks around massive Pop III protostars \citep[][]{2009Sci...325..601T,2010MNRAS.403...45S,2011Sci...334.1250H,2011Sci...331.1040C,2012MNRAS.424..399G,2014ApJ...781...60H, 2015MNRAS.448..568H,2016MNRAS.462.1307S,  2018MNRAS.479..667R, 2020MNRAS.494.1871W}. 
On the-low mass Pop III stars, \citet[][]{2016ApJ...826....9I} have found that the Pop III stars having masses $<$\,0.8\,M$_{\odot}$ would have longer lifetimes as compared to the cosmic time; therefore such low-mass stars could linger around to be detected in our Milky Way itself. 

Further, Pop III stars were responsible for the enrichment of the early universe by spreading metals heavier than He through violent supernova (SN) explosions or possibly through sporadic mass losses due to vigorous stellar winds \citep[][]{2000MNRAS.319..539F,2001ApJ...557..126A}. 
The study by \citet[][]{2018MNRAS.475.4378C} shows that a CCSN from a Pop III star could cause the minihalo to undergo internal-enrichment. This causes the metallicity to be -5 $\lesssim$\,[Fe/H]\,$\lesssim$-3 in the recollapsing region. Thus, internal-enrichment caused by a CCSN from a Pop III star can explain the stars which are extremely metal-poor.
In relatively recent work, the authors of \citet[][]{2020MNRAS.491.4387K} have estimated the dose of heavy elements introduced by massive Pop III stars. In doing so, they considered the amount of heavy elements synthesised only from PISNe or CCSNe explosions of massive Pop III stars. They found that the heavy elements introduced by Pop III stars are usually much more than those from galaxies found in the low-density regions. Besides the above-mentioned studies involving Pop III stars, there have also been investigations to recognise the influence of Pop III stars on cosmic reionization \citep[e.g.,][]{1997ApJ...476..458H,2000ApJ...528L..65T,2001PhR...349..125B,2001MNRAS.324..381C, 2013RPPh...76k2901B}, and dust formation \citep[][]{2001MNRAS.325..726T}.

There have been multiple studies to understand the evolution of the Pop III stars. \citet[][]{2003A&A...399..617M} and \citet[][]{2008A&A...489..685E} have studied the evolution of Pop III stars by assuming solid-body and differential rotation, respectively. \citet[][]{2010ApJ...724..341H} have discussed the nucleosynthesis and evolutions of non-rotating Pop III stars. They also generated the light curves of the resulting transients from the non-rotating models corresponding to different explosion energies. \citet[][]{2012A&A...542A.113Y} have discussed the evolution of massive Pop III stars having masses in the range of [10 -- 1000]\,M$_{\odot}$ and have investigated the consequences of including rotation and magnetic fields. Due to the chemically homogeneous evolution, the rapidly rotating high-mass stars could result in a class of energetic transients, including Type Ib/c SNe, GRBs, HNe, and PISNe. In their work, \citet[][]{2012A&A...542A.113Y} have prepared a phase diagram in the plane of mass and rotational velocity at ZAMS and discussed the culminating fates of Pop III stars. The authors of \citet[][]{2018ApJS..234...41W} have investigated the evolutions of non-rotating Pop III stars in the mass range of [1 -- 1000]\,M$_{\odot}$ considering no mass loss and have also discussed chances of the observability  of an individual Pop III star. In recent work, \citet[][]{2021MNRAS.501.2745M} have studied a grid of Pop III models having masses in the range of [1.7 -- 120]\,M$_{\odot}$ and explored the effect of changing the initial rotational velocity from 0 to 40 per cent of critical rotational velocity.

Taking such studies one step further, in this work, we study the entire evolution (from ZAMS up to the stage of the onset of core collapse) of a 25\,M$_{\odot}$ Pop III star and investigate the effect of rotation on the final fates. 
Following the phase diagram in \citet[][]{2012A&A...542A.113Y}, the resulting supernovae (SNe) from a 25\,M$_{\odot}$ ZAMS star will either be weak Type II or Type Ib/c depending upon the initial rotations. For the first time in this work, we have evolved the rotating and non-rotating Pop III models together up to the onset of core collapse. We further performed the hydrodynamic simulations of their synthetic explosions, showing the light curves of resulting transients.

We have divided this Chapter into four sections. After providing a brief introduction of literature in Section~\ref{sec:intro}, we discuss the numerical setups and physical properties of the models in Section~\ref{sec:mesa_ch5}. The numerical setups to simulate the hydrodynamic explosion of models are discussed in Section~\ref{sec:snec_ch5}. Finally, the major outcomes from the entire evolutions of the models, along with their synthetic explosions, are discussed in Section~\ref{sec:results and discussion}. In this Section, we also provide the implications and discussions of the simulation results presented in the underlying work.

\section{Stellar Evolution Using {\tt MESA}}
\label{sec:mesa_ch5}
\subsection{Numerical setups}
We employ one of the state-of-the-art and 1-dimensional stellar evolution codes, {\tt MESA} to perform the stellar evolutions of 25\,M$_{\odot}$ ZAMS stars with zero metallicity (Z = 0.00) and different initial rotations. In this simulation work, we have utilised the {\tt MESA} version r22.05.1 \citep[][]{Paxton2011, Paxton2013, Paxton2015, Paxton2018}. We start with a non-rotating, zero metallicity model and increase the angular rotational velocity ($\Omega$) in units of 0.2 times the critical angular rotational velocity ($\Omega_{\rm crit}$) up to $\Omega$ / $\Omega_{\rm crit}$ $=$ 0.8. Following \citet[][]{Paxton2013}, the critical angular rotational velocity is expressed as $\Omega_{\rm crit}^2 = (1-L/L_{\rm edd})GM/R^3$, where $L_{\rm edd}$ is the Eddington luminosity. As specified in the default {\tt MESA} setups, when the ratio of the luminosity from nuclear processes and the overall luminosity of the model at a particular stage reaches 0.4 (set by {\tt Lnuc\_div\_L\_zams\_limit} = 0.4 in {\tt MESA}), the model is assumed to have reached ZAMS. Adopting the Ledoux criterion, convection is modelled utilising the mixing theory of \citet[][]{Henyey1965}. We have adopted a mixing-length-theory parameter ($\alpha_{\rm MLT}$) $=$ 2.0 in the whole analysis. Following \citet[][]{Langer1985}, semi-convection is modelled by setting an efficiency parameter of $\alpha_{\mathrm{sc}}$ = 0.01.  To model the thermohaline mixing, we follow \citet[][]{Kippenhahn1980} by fixing the efficiency parameter to $\alpha_{\mathrm{th}}$ = 2.0. To model the convective overshooting, we use the diffusive approach as presented in \citet[][]{Herwig2000} and set $f = 0.004$ and $f_{0} = 0.001$ for all convective cores and shells. The stellar winds from rotating Pop III stars are also theorised to make contributions to the enrichment of the early universe with heavy metals, so to model the stellar winds of the Pop III stars in our study, we use the 'Dutch' wind scheme \citep[][]{2009A&A...497..255G} and set the scaling factor ($\eta$) to 0.5. This wind scheme incorporates the outcomes from multiple works for different situations, with m$_{\rm H}$ representing the surface mass fraction of hydrogen, (a) when the effective temperature, T$_{\rm eff} > 10^4$\,K along with m$_{\rm H}$ being greater than 0.4, the outcomes of \citet[][]{Vink2001} are used; (b) when T$_{\rm eff} > 10^4$\,K combined with the m$_{\rm H}$ being lesser than 0.4, the results of \citet[][]{Nugis2000} are used; and finally (c) the wind scheme presented in \citet[][]{dejager1988} is used in {\tt MESA} for the condition with T$_{\rm eff} < 10^4$\,K. 

Starting from the pre-ZAMS, the 1-dimensional stellar evolutions of the Pop III models are performed till they reach the stage of the onset of the Iron core collapse. The onset of core collapse is marked when the infall velocity of the Iron core exceeds the specified Iron core infall velocity limit of 100\,km\,s$^{-1}$ (set by {\tt fe\_core\_infall\_limit = 1d7} in {\tt MESA}). In this work, the models have been named in a way that they contain the pieces of information, including the ZAMS mass, metallicity, and rotation ($\Omega$ / $\Omega_{\rm crit}$ indicated by ``Rot"). For example, the model M25\_Z0.00\_Rot0.0 indicates a 25\,M$_{\odot}$ ZAMS star with Z = 0.00, and $\Omega$ / $\Omega_{\rm crit}$ = 0.0. For comparison purposes, we have also performed the evolution of a solar metallicity (Z = 0.02) model with the same ZAMS mass of 25\,M$_{\odot}$.

\begin{figure}	
    \includegraphics[width=\columnwidth]{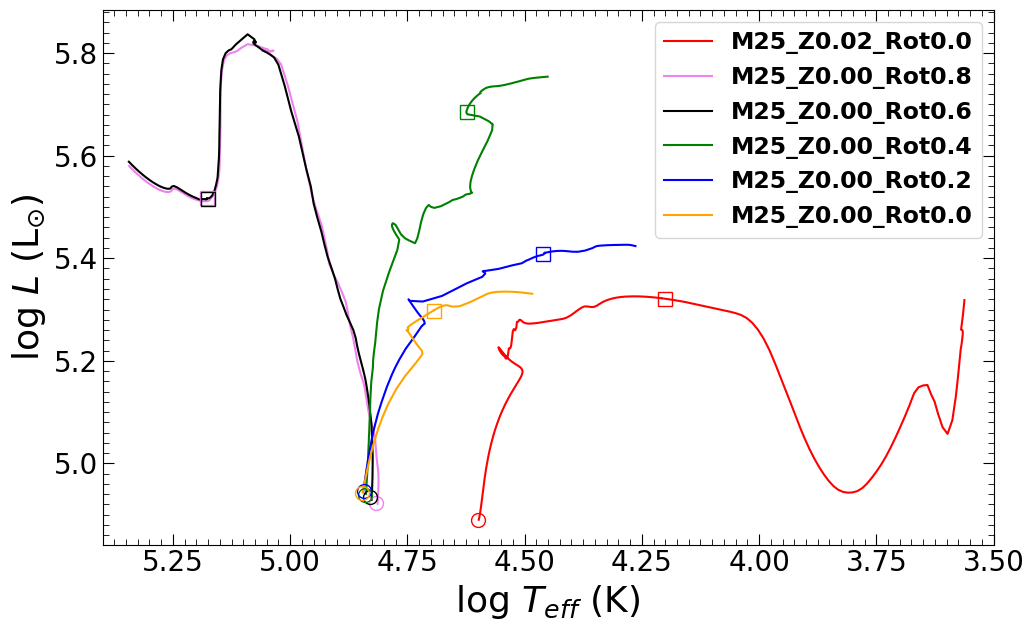}
    \caption{Evolution of the Pop III models having different rotations on the HR diagram. The arrivals of the models on ZAMS are shown by hollow circles, while the core-He exhaustion stages of the models have been marked by hollow squares. The solar metallicity (Z = 0.02) model evolutionary track has also been shown for comparison.}
    \label{fig:HR_Diagram}
\end{figure}

\begin{figure}	
    \includegraphics[width=\columnwidth]{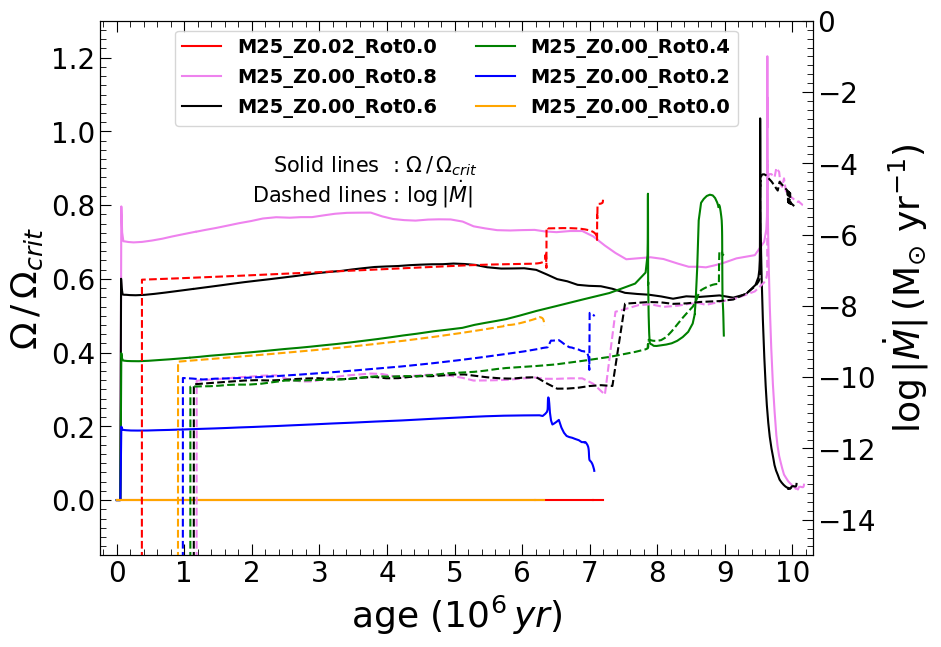}
    \caption{ The evolution of angular rotational velocity ($\Omega$) in units of critical angular rotational velocity ($\Omega_{\rm crit}$) along with corresponding mass loss rate (log $|\dot{M}|$) evolution.}
    \label{fig:Omega_plot}
\end{figure}

\subsection{Physical properties of the models}
%\subsubsection{Evolution on HR diagram}
Figure~\ref{fig:HR_Diagram} displays the evolutions of Pop III models with 25\,M$_{\odot}$ ZAMS mass each, on the Hertzsprung–Russell (HR) diagram along with a similar ZAMS mass model having solar metallicity (Z = 0.02) for comparison purposes.  Compared to a solar metallicity model, the Pop III models reach the ZAMS at higher effective temperatures ($T_{\rm eff}$) but show nearly similar ZAMS luminosities. Thus, the Pop III models are bluer than the solar metallicity model.
Among Pop III models, at ZAMS, the models with higher initial angular rotational velocities possess lower luminosities and lower effective temperatures; a well-known effect of rotation as mentioned in \citet[][]{2008A&A...489..685E}. It is also evident from this figure that beyond ZAMS, the models with higher initial angular rotational velocities possess higher luminosities and higher effective temperatures as well for most of their evolutionary paths. Similar results were also obtained in the case of \citet[][]{2012A&A...542A.113Y}.

\begin{figure}
\includegraphics[width=\columnwidth,angle=0]{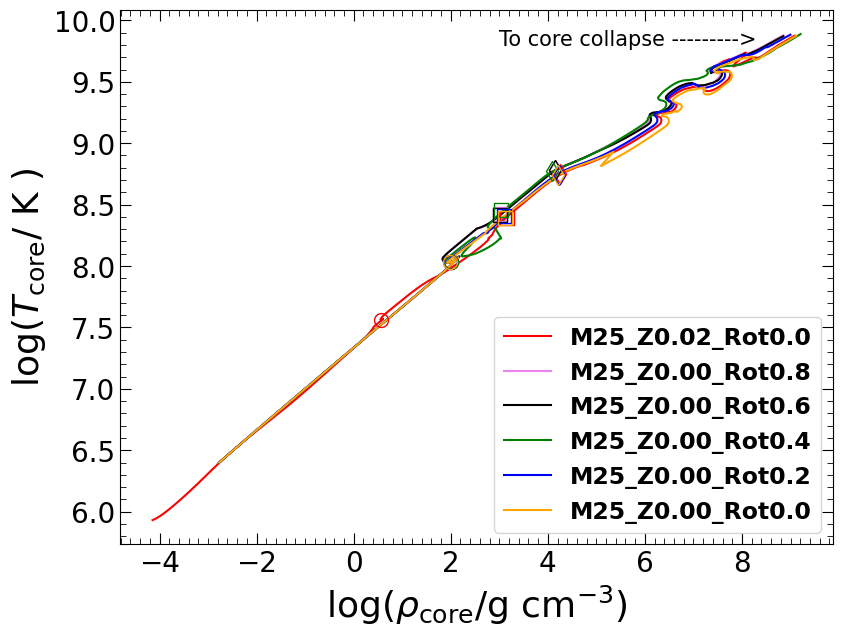}
   \caption{The variations of core-temperature ($T_{\rm core}$) vs core-density ($\rho_{\rm core}$) curves throughout the course of evolution of the models on the HR Diagram. The arrival on the ZAMS, exhaustion of core-He burning, and exhaustion of core-C burning have been marked by hollow circles, squares, and diamonds, respectively.}
    \label{fig:density_temp}
\end{figure}
%\subsubsection{Rotations and Mass losses}
Figure~\ref{fig:Omega_plot} shows the variations of $\Omega$ / $\Omega_{\rm crit}$ and corresponding mass loss rates (log $|\dot{M}|$) as the models evolve up to the stage of the onset of core collapse. Initially, the models touch the specified $\Omega$ / $\Omega_{\rm crit}$ values and then settle to new $\Omega$ / $\Omega_{\rm crit}$ values as they evolve further. This shows that a perturbation imposed on an equilibrium model experiences a transient response before the system settles into a new equilibrium configuration. During the last evolutionary stages, the rapidly rotating models (with $\Omega$ / $\Omega_{\rm crit}$ = 0.6 and 0.8) show chaotic rotations exceeding the critical rotational velocities which are responsible for the dynamic events to occur as indicated by corresponding heavy mass loss rates during these phases. Our rapidly rotating massive Pop III stars dredge up a large amount of CNO elements up to the surface during the core-He burning stage. It dramatically increases the surface metallicity, which eventually strongly boosts the radiative mass loss through a mechanism similar to that discussed in \citet[][]{2007A&A...461..571H}. Thus, the rapidly rotating models are significantly stripped compared to the slow- or non-rotating models. We have also shown the Kippenhahn diagram for one slow-rotating model (M25\_Z0.00\_Rot0.2) and one rapidly rotating model (M25\_Z0.00\_Rot0.8) in Figure~\ref{fig:kippenhahn}.
A few more important physical properties, including radii and effective temperatures ($T_{\rm eff}$) at various stages, are listed in Table~\ref{tab:MESA_MODELS5}.

%\subsubsection{Pre-SN radii and the evolution of $\rho_{\rm core}$ vs $T_{\rm core}$}

The overall variations of the core-density ($\rho_{\rm core}$) vs core-temperature ($T_{\rm core}$) curves for the entire evolutions of the models up to the onset of core collapse are shown in Figure~\ref{fig:density_temp}. The arrival of the models on ZAMS, the exhaustion of core-He burning phases, and the exhaustion of core-C burning stages have been indicated by the hollow circles, squares, and diamonds, respectively. Compared to a solar metallicity model, the Pop III models ignite the H-burning in their respective cores at higher $\rho_{\rm core}$ and $T_{\rm core}$, which is due to the lack of CNO elements in Pop III stars needed to ignite the CNO cycle \citep[][]{2008A&A...489..685E}. During the last evolutionary stages, all the models have exceeded the core temperatures of $\sim$10$^{9.9}$\,K; the perfect condition for the cores to collapse under their own gravity. Thus, the models have now reached the stage of the onset of core collapse.

\begin{figure*}
\centering
    \includegraphics[height=6.5cm,width=0.49\columnwidth,angle=0]{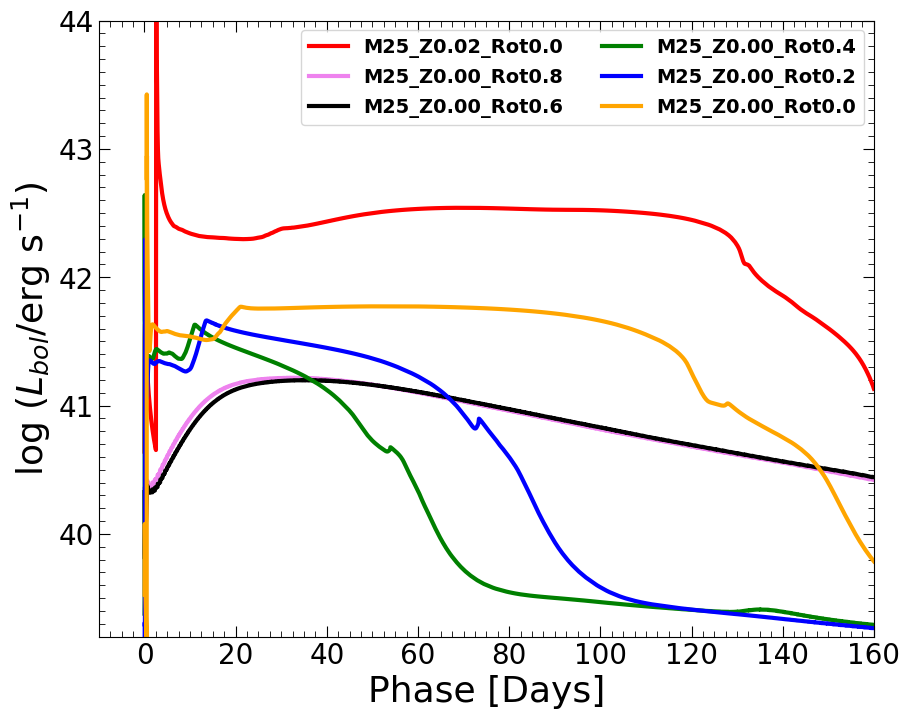}
    \includegraphics[height=6.5cm,width=0.49\columnwidth,angle=0]{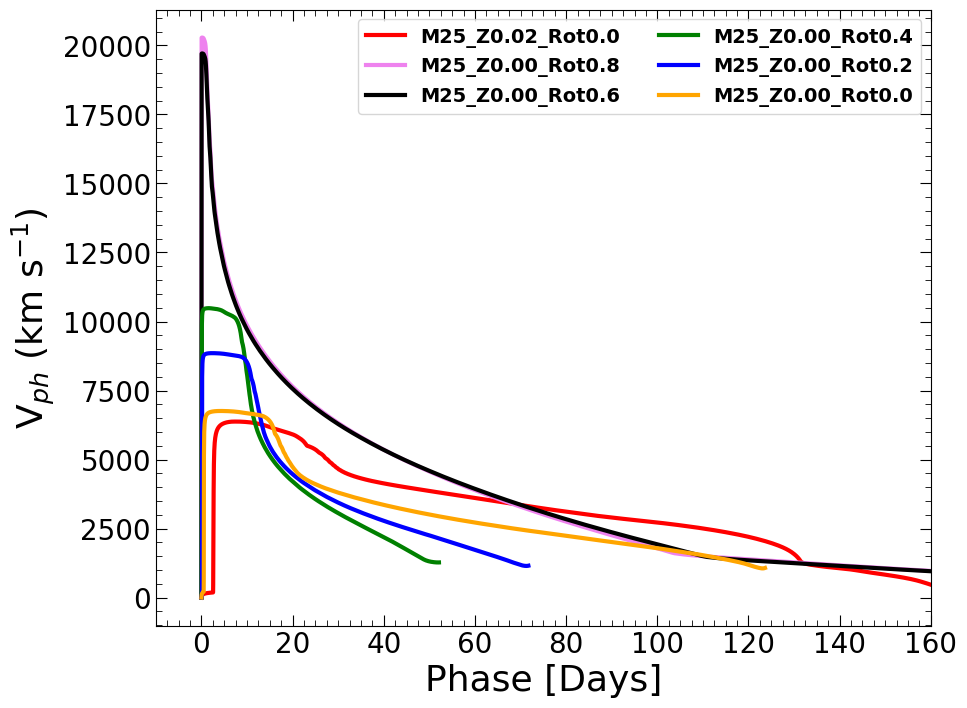}
   \caption {{\em Left:} The bolometric luminosity light curves resulting from the synthetic explosions of models using {\tt SNEC}. {\em Right:}  Corresponding photospheric velocity evolutions. Results of the non-rotating solar metallicity model are also shown for comparison.}
    \label{fig:blc_vel}
\end{figure*}

\begin{landscape}
\begin{table*}
\caption{The ZAMS and pre-SN properties of the Pop III models using {\tt MESA} along with the {\tt SNEC} explosion parameters. }
\label{tab:MESA_MODELS5}
\begin{center}
{\scriptsize
\begin{tabular}{ccccccccccccc}
\hline \hline

 & &  ZAMS & & \hspace{1.3cm}\vline&  &  &  pre-SN &\hspace{1.3cm}\vline & &  Explosion\\
\hline
Model Name	& $M_{\rm ZAMS}^{a}$	& $T_{\mathrm{eff}}$  & $R_{\mathrm{ZAMS}}^{b}$ & $L_{\rm ZAMS}^{c}$  &	$M_{\rm pre-SN}^{d}$	& $T_{\mathrm{eff}}$  & $R_{\mathrm{pre-SN}}^{e}$ & $L_{\rm pre-SN}^{f}$ 	&	$M_{\mathrm{c}}^{g}$ & $M_{\mathrm{ej}}^{h}$ & $M_{\mathrm{Ni}}^{i}$ &	$E_{\mathrm{exp}}^{j}$ 	\\
	&	(M$_{\odot}$) & K	&	(R$_{\odot}$)	 &  (L$_{\odot}$) &      (M$_{\odot}$) & K	&	(R$_{\odot}$)	 &  (L$_{\odot}$) & (M$_{\odot}$) & (M$_{\odot}$) & (M$_{\odot}$) &	($10^{51}$\,erg)\\ 	
\hline
\hline

M25\_Z0.00\_Rot0.0     &	25.0  	&	70069  &  2.01   & 4.94 & 24.99  &    10319  & 195 & 5.58  &  1.70 & 23.29 &  0.001 & 1.0	\\

M25\_Z0.00\_Rot0.2     &	25.0  	&	69877  &  2.02   & 4.94  & 24.99  &    17216  & 57 & 5.42 & 2.00 & 22.99 &  0.001 & 1.0		\\

M25\_Z0.00\_Rot0.4     &	25.0  	&	68882  &  2.07   & 4.94 & 24.96  &    29784  & 32 & 5.87 & 1.80 & 23.16 &  0.001	& 1.0	\\

M25\_Z0.00\_Rot0.6     &	25.0  	&	67069  &  2.16    & 4.93 & 11.94  &    140668  & 1.5 &  5.90 & 2.10 & 9.84 &  0.05 & 1.0	\\

M25\_Z0.00\_Rot0.8     &	25.0  	&	65107  &  2.26    & 4.91 & 11.79 &    175858  & 0.6 & 5.53 & 2.10 & 9.69 &  0.05	& 1.0	\\

M25\_Z0.02\_Rot0.0     &	25.0  	&	3962  &  5.91    & 4.88 & 22.64  &    3623  & 1219 & 5.36 & 1.90 & 22.74 &  0.001 & 1.0 		\\

\hline\hline
\end{tabular}}
\end{center}
%\par
{$^a$Mass at ZAMS.
$^b$Progenitor radius at ZAMS,
$^c$Luminosity at ZAMS,
$^d$Final mass of pre-SN model,
$^e$pre-SN phase radius,
$^f$pre-SN phase luminosity,
$^g$Mass of the central remnant in simulation,
$^h$Ejecta mass,
$^i$Amount of synthesised nickel used in the explosion,
$^j$Explosion energy.}\\

\end{table*} 
\end{landscape}

\section{Synthetic Explosions Using {\tt SNEC}}
\label{sec:snec_ch5}
Once the models have reached the stage of core collapse marked by the infall velocity exceeding the specified Iron-core infall velocity, the outputs of {\tt MESA} in appropriate forms are provided as input to {\tt SNEC} \citep[][]{Morozova2015}. {\tt SNEC} is a 1-dimensional Lagrangian hydrodynamic code that simulates the synthetic explosions of the stellar models at the stage of the onset of their core collapse. {\tt SNEC} solves the radiation energy transport equations within the flux-limited diffusion approximation to simulate the explosions.

Further, in this work, to simulate the synthetic explosions of the models which have already arrived at the stage of the onset of core collapse, we closely follow the setups of \citet[][]{Ouchi2019} along with \citet[][] {Aryan2021, 2022MNRAS.517.1750A} for {\tt SNEC}. However, the major changes are summarised here. 
First, for each model, the innermost mass $M_{\rm c}$ representing the mass of the central remnant is excised before the explosion by assuming that the model will finally collapse to form an NS. The central remnant mass is decided by the final mass of the Iron-core when the model has reached the stage of the onset of core collapse. Further, a set of 800 grid cells are used to simulate the synthetic explosion of the model. With 800 grid cells, the light curves and photospheric velocities of the resulting SN from simulations are very well converged in the interested domains of time. The explosion of each model is simulated as {\tt thermal bomb} by adding $E_{\rm exp}$ amount of energy for a duration of 0.1\,s in the inner 0.1\,M$_{\odot}$ section of the model. As discussed in \citet[][]{Morozova2015}, {\tt SNEC} lacks a nuclear-reaction network; thus, the synthesised amount of nickel ($^{56}$Ni) in an SN is decided and fixed by the individual user. For each model, an amount of $^{56}$Ni specified by corresponding $M_{\rm Ni}$ in Table~\ref{tab:MESA_MODELS5}, is distributed between the excised central remnant mass ($M_{\rm c}$) cut and the chosen mass coordinate which is close to the outer surface of the selected model. For models with $\Omega$ / $\Omega_{\rm crit}$\,$\leq$\,0.4, the amount of $^{56}$Ni is set to 0.001\,M$_{\odot}$ while the remaining models with heavy rotations and suffering significant mass losses, the amount of $^{56}$Ni is set to 0.05\,M$_{\odot}$. Choosing a slightly greater amount of $M_{\rm Ni}$ for stripped models is followed by 
\citet[][]{Afsariardchi2021}. The ejecta mass ($M_{\rm ej}$) for each CCSN is estimated by finding the difference between the pre-SN mass ($M_{\rm Pre-SN}$) and $M_{\rm c}$. The detailed explosion parameters are listed in Table~\ref{tab:MESA_MODELS5}.

Finally, the $UBVRI$ -bands light curves generated through synthetic explosions are shown in Figure~\ref{fig:lc}. As shown in this figure, the slow-rotating models (i.e., models M25\_Z0.00\_Rot0.0, M25\_Z0.00\_Rot0.2, M25\_Z0.00\_Rot0.4, and the solar metallicity model M25\_Z0.02\_Rot0.0) retaining a significant amount of their outer H-envelope result into Type II CCSNe. In contrast, the rapidly rotating models result in CCSNe Type Ib/c. These results are also complementing the results as predicted in the phase diagram of \citet[][]{2012A&A...542A.113Y} (in Figure 12). 
In Figure~\ref{fig:lc}, a few important simulation results of the H-rich Pop III CCSNe (i.e. models with $\Omega$ / $\Omega_{\rm crit}$\,$\leq$\,0.4) are also displayed; First, the peak magnitudes of the shock breakout (SBO) features from these models are much fainter than a typical solar metallicity H-rich CCSN;  second, the absolute magnitudes of the plateau of the H-rich Pop III CCSNe are at least 1.5 magnitudes fainter than the solar metallicity H-rich CCSN; thus the H-rich Pop III CCSNe are pretty faint within the considered limits of $E_{\rm exp}$ and $M_{\rm Ni}$ in this study. The effect of bolometric light curves becoming less luminous as metallicity decreases has been explored in \citet[][]{2009ApJ...703.2205K} and \citet[][]{Paxton2018}. The primary cause of this behaviour is associated with the smaller pre-SN radius and less total mass loss as an effect of lower metallicity. However, in their work, they have not calculated the light curves corresponding to Z = 0.00; and the third, surprisingly, although the pre-SN radius of the non-rotating H-rich Pop III model is much smaller than a non-rotating solar metallicity H-rich model, the earlier model shows almost a similar plateau duration. The non-rotating solar model has a larger pre-SN radius compared to the M25\_Z0.00\_Rot0.0 model, but the latter has a more massive H-envelope. From Figure~\ref{fig:mass_fraction}, the M25\_Z0.02\_Rot0.0 model has an H-envelope starting from a mass coordinate, m(M$_{\odot}$)$\sim$8\,M$_{\odot}$ while the M25\_Z0.00\_Rot0.0 has a more-massive H-envelope starting from m(M$_{\odot}$)$\sim$5\,M$_{\odot}$. Thus, the presence of extra hydrogen could be responsible for the increased plateau duration in the non-rotating Pop III model. Finally, the rapidly rotating H-less Pop III models result into much fainter Type Ib/c SNe. These explosions are fainter because of the less explosion energy (and an M$_{\rm Ni}$ of 0.05\,M$_{\odot}$) considered in our study. With higher explosion energies and more nickel production, they might result in more luminous SNe or HNe \citep[][]{2013ARA&A..51..457N}. 
{\tt SNEC} could also produce the bolometric luminosity light curves and the corresponding photospheric velocity evolutions for all the models as shown in the left and the right panels of Figure~\ref{fig:blc_vel}, respectively. In the left panel, the bolometric light curves of the Pop III CCSNe display a similar behaviour as earlier in the case of $UBVRI$ -bands light curves comparison with the solar metallicity model. In the right panel, as expected, the stripped models display higher photospheric velocities compared to the H-rich models.

\section{Results and Discussion}
\label{sec:results and discussion}
In this work, we have performed the 1-dimensional stellar evolutions of Pop III models up to the stage of the onset core collapse and then simulated their synthetic explosions. Utilising the 1-dimensional simulations performed in this work, we summarise our findings below:
\begin{enumerate}

    \item The peak absolute magnitudes of the SBO features of Pop III CCSNe are much smaller than that of a CCSN, resulting from a solar Type model with similar ZAMS mass.
    
    \item The H-rich CCSNe from Pop III models are fainter than the H-rich SN resulting from a solar metallicity model. The plateau magnitudes of Pop III star H-rich CCSNe are at least 1.5 magnitudes fainter than the latter. In the earlier epochs, the stripped CCSNe from Pop III models are much fainter than SNe resulting from H-rich Pop III models.
    
    \item One of the most intriguing results from our simulations is that although the pre-SN radius of a non-rotating H-rich Pop III model is much smaller than a non-rotating H-rich solar Type model, both models show nearly similar plateau durations. One of the reasons for the increased plateau duration, despite a relatively smaller pre-SN radius in non-rotating Pop III CCSNe, could be associated with the increased amount of hydrogen mass.
    
    \item Among the discussed Pop III models, SN resulting from the non-rotating H-rich model is the brightest. It has a nearly constant absolute magnitude of around -16.5 mag in the V-band for the plateau phase. This would correspond to an apparent magnitude of $\sim$35.5\,mag at a redshift of $z$ = 10 (using the cosmology of a Hubble constant, H$_{0}$ = 73, $\Omega_{\rm M}$ = 0.3, and $\Omega_{\rm vac}$ = 0.7). Currently, no ground- or space-based observatory can go this faint to detect a Pop III CCSN resulting from an individual star; however, with the major advancement in observational technologies having large diameters could possibly detect such events in the near future.
    
    \item Thus, through our work, we find that within the considered limits of explosion energies and nickel masses, these transient events are very faint, making it difficult for them to be detected at high redshifts.
\end{enumerate}
 %PopIII SNe
\chapter{\sc Evolution of a non-rotating 100\,M$_{\odot}$ ZAMS star resulting in to Type IIP-like supernovae}
\label{Ch:6} 
\blfootnote{ The results of this Chapter are published in: \textbf{{Aryan}, A.,} {Pandey}, S. B., {Gupta}, Rahul et al., 2023, {\textit{JApA}, {\textbf{43}}, 2}.}
\ifpdf
    \graphicspath{{Chapter6/Chapter6Figs/PNG/}{Chapter6/Chapter6Figs/PDF/}{Chapter6/Chapter6Figs/}}
\else
    \graphicspath{{Chapter6/Chapter6Figs/EPS/}{Chapter6/Chapter6Figs/}}
\fi

\section{Final Fates of Very Massive Stars}
\label{Intro_ch6}

Chapters~\ref{Ch:3}, \ref{Ch:4}, and \ref{Ch:5} discuss the properties of canonical CCSNe arising from progenitors having ZAMS masses of 25\,M$_{\odot}$ or less. However, in this Chapter, we present the pure simulation-based results of the stellar evolution of a 100\,M$_{\odot}$ ZAMS progenitor star and discuss the resulting transients.

In this work, we take a 100\,M$_{\odot}$ ZAMS star with metallicity, Z = 0.01, and evolve it up to the stage of core-collapse, and finally, the synthetic explosions are simulated.
Depending on the mass loss rate, rotation, and even on metallicity, such a heavy progenitor could result in various transients, including PISN, PPISN, Type IIp, Type IIn, Type Ibn, and Type Icn SNe also \citep[][]{Sukhbold2016,Woosley2017}.  The numerical simulations of such heavy progenitors are always a challenging task. With the mass loss and rotation included, the simulations are further complicated. The Type IIp SNe progenitors suffer minimal  mass loss (as they retain almost all of their H-envelope) and possess (very) low rotation. Thus, to keep things simple yet close to reality, we try to model a massive ZAMS progenitor that can result in a Type IIP-like SN. One very recent study by \citet[][]{Nicholl2020} indicates a total (SN ejecta + CSM) mass likely exceeding 50--100\,M$_{\odot}$ for a Type II superluminous SN~2016aps. Attempts are underway to simulate such massive progenitors with much lower metallicities close to zero and even models with zero metallicity to understand the internal structure and resulting transients in the future. Although a few studies have been performed in this direction by (among others) \citet[][]{Ohkubo2006}, such studies are going to be extremely useful in understanding the final fates of massive primordial stars. The primary reason to choose a 100\,M$_{\odot}$ ZAMS progenitor is that the mass ranges, including such heavy progenitors, are least explored. Only a handful of studies \citep[eg.][]{Sukhbold2016,Woosley2017} have been performed which discuss the evolution and final fates of such massive zero-age main-sequence progenitors. This study is a preliminary step to enhance our understanding of CCSNe originating from very high mass progenitors. Further, as mentioned above, a recent study by \citet[][]{Nicholl2020} indicates a total (SN ejecta + CSM) mass exceeding 50--100\,M$_{\odot}$, so our study is also going to be important to look into the progenitors of such events. Moreover, in the coming TMT\footnote{https://www.tmt.org/}, ELT\footnote{https://elt.eso.org/} and other large telescopes, such massive stars are likely to be discovered; thus, it will always be an interesting task to compare the simulations with the actual observations. All the analyses performed in this work make use of publicly available tools. A very nice review has been provided (among others) in \citet[][]{Aryan2021b} depicting the usefulness of publicly available tools and how these tools can be a boon to the transient community.

This Chapter has been divided into six sections. The basic assumptions and methods of modelling the 100\,M$_{\odot}$ progenitor have been discussed in section~\ref{Models_ch6}, thereafter the properties and parameters to synthetically simulate the explosions are discussed in section~\ref{Explosion_ch6}. The results of synthetic explosions have been discussed in section~\ref{Results_ch6}.  Later, we discuss the major outcomes of our studies in section~\ref{Discussion_ch6} and provide our concluding remarks in section~\ref{Conclusion_ch6}.

\section{1-dimensional stellar evolution of the 100\,M$_{\odot}$ ZAMS progenitor using {\tt MESA}} 
\label{Models_ch6}

\begin{figure*}
  \includegraphics[width=\columnwidth]{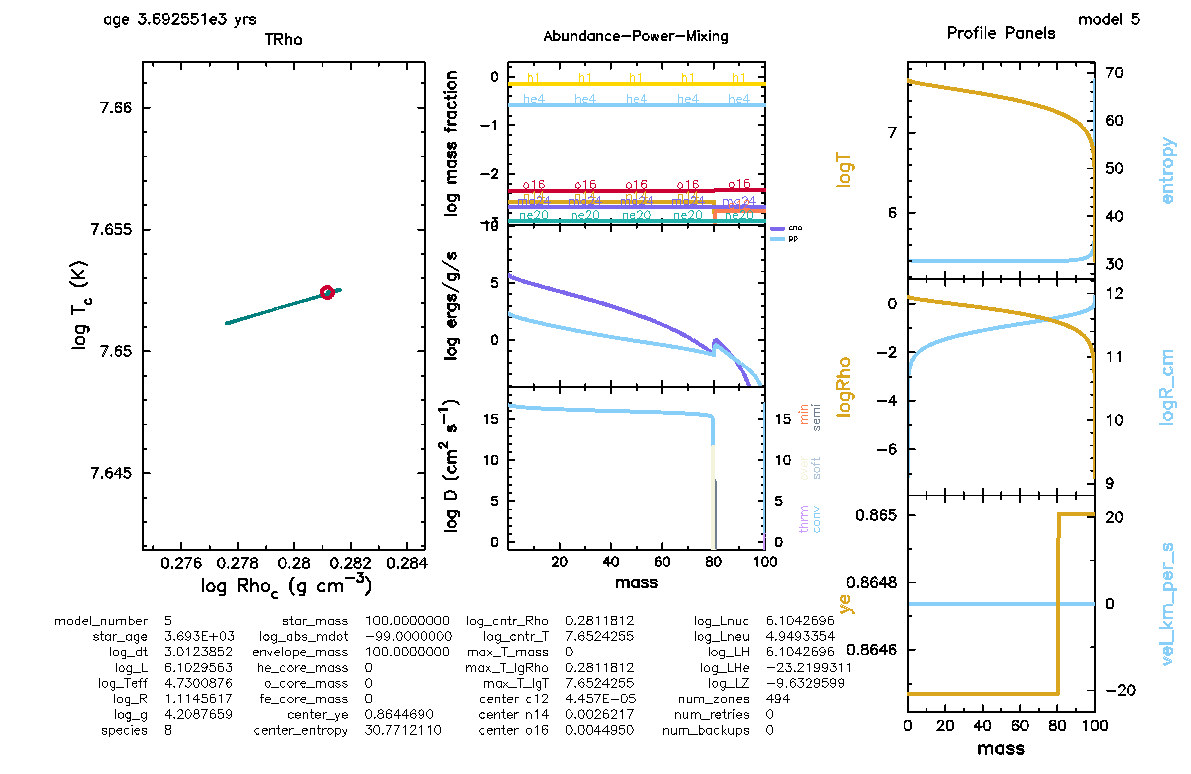}
  \caption{The physical condition and chemical engineering of our model when it has just arrived on the ZAMS. This figure is one of the many figures generated by {\tt MESA} while the evolution of the models through various evolutionary phases.}
  \label{fig:mesa1}
\end{figure*}

In this work, we have taken a 100 M$_{\odot}$ ZAMS progenitor and evolved it up to the onset of core-collapse using {\tt MESA} (version r11701). For this purpose, we used the default {\tt example\_make\_pre\_ccsn} directory available in {\tt MESA}. Most of the settings closely follow \citet[][]{Aryan2021} and  \citet[][]{Pandey2021} with only a few minor deviations. We do not consider rotation and assume an initial metallicity of Z = 0.01. Convection is modelled using the mixing theory of \citet[][]{Henyey1965} by adopting the Ledoux criterion. We set the mixing-length parameters to $\alpha$ = 3.5 in the region where the mass fraction of hydrogen exceeds 0.5 and set it to 1.5 in the other regions. Semi-convection is modelled following \citet[][]{Langer1985} with an efficiency parameter of $\alpha_{\mathrm{sc}}$ = 0.01. For the thermohaline mixing, we follow \citet[][]{Kippenhahn1980}, and set the efficiency parameter as $\alpha_{\mathrm{th}}$ = 2.0. We model the convective overshooting with the diffusive approach of \citet[][]{Herwig2000}, with $f$= 0.004 and $f_0$ = 0.001 for all the convective core and shells. We do not consider the presence of any strong stellar winds and hence consider no progenitor moss loss before the core collapse. 

\begin{figure*}
  \includegraphics[width=\columnwidth]{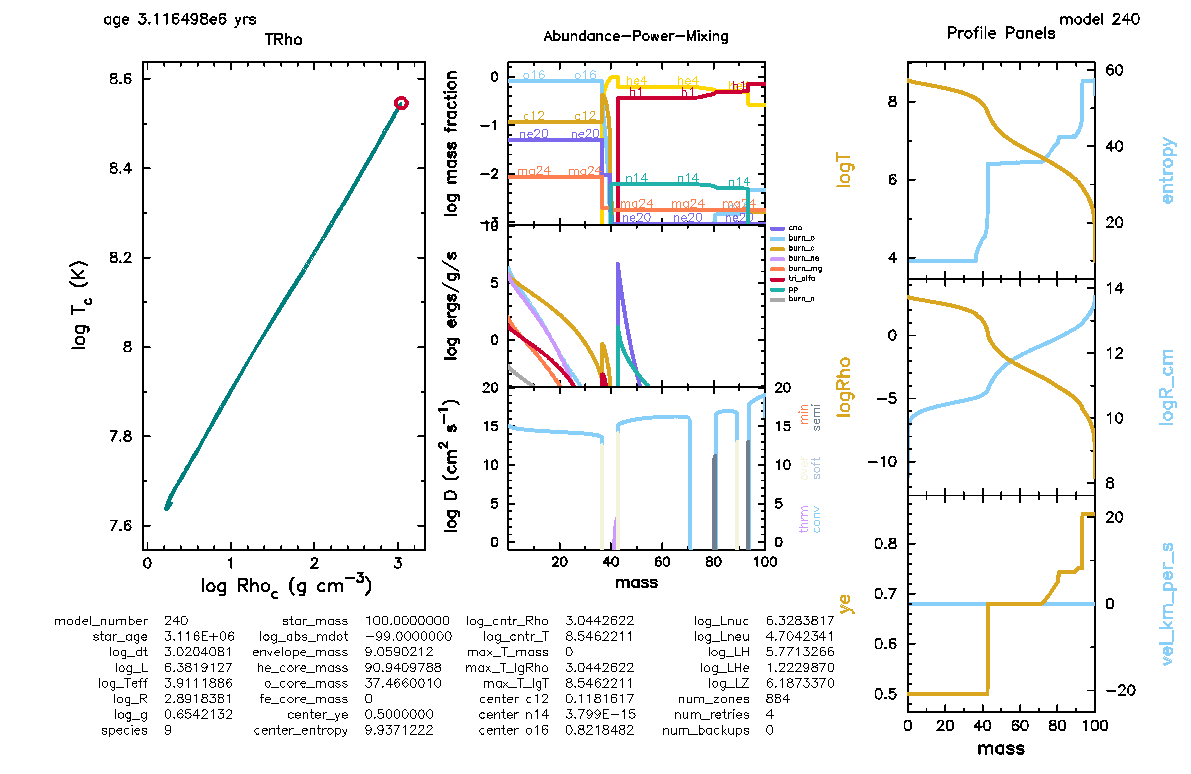}
  \caption{Same as fig.~\ref{fig:mesa1}, but at the stage when the model has evolved through the main sequence and has become red supergiant.}
  \label{fig:mesa2}
\end{figure*}

\begin{figure*}[!t]
  \includegraphics[width=\columnwidth]{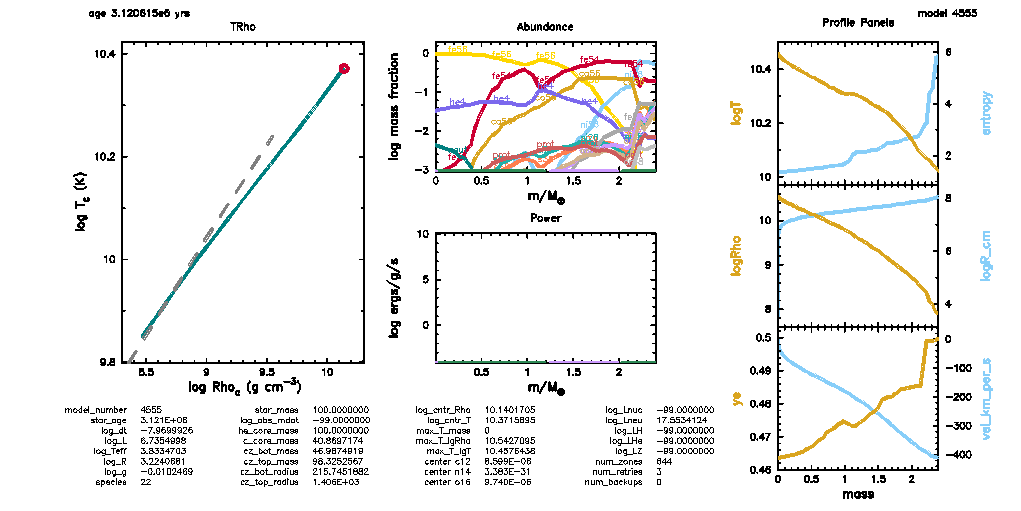}
  \caption{A glimpse of the physical and chemical structure of our model star just when the Iron-core infall has been achieved. We could see that the core-temperature and core-density have reached in excess of $10^{10}$\,K and $10^{10}$\,g\,cm$^{-3}$ respectively, which are supposed to be the perfect conditions for the core to collapse.}
  \label{fig:mesa3}
\end{figure*}

Further, we briefly explain the evolution of our models through the various stages. Landing on the ZAMS, our model evolves through the main-sequence,  becomes supergiant, and finally approaches its death by developing an inert Fe-core in its centre. Figure~\ref{fig:mesa1} displays the physical condition and chemical engineering of our model when it has just landed on the main sequence. The arrival of our model on the main sequence is marked by the control parameter {\tt Lnuc\_div\_L\_zams\_limit}, which checks the ratio of luminosity due to nuclear reactions and total luminosity at ZAMS. Once this ratio is $\sim$\,0.8, the onset of the main-sequence phase is marked. The leftmost panel of figure~\ref{fig:mesa1} shows the variation of the core-temperature (log\,T$_{c}$)  and core density (log\,Rho$_{c}$) as the model evolves in time. On the beginning of main-sequence, the log\,T$_{c}$ is $\sim$\,7.65\,K and log\,Rho$_{c}$ is $\sim$\,0.281\,g\,cm$^{-3}$. The top plot in the middle panel of figure~\ref{fig:mesa1} displays the mass fraction of various elements present in the star. We can see that as the model has just arrived on the main-sequence, the fractions of H and He are much larger than any other heavy metals depending on the metallicity of our model. The middle plot in the middle panel of figure~\ref{fig:mesa1} shows the specific luminosity corresponding to the various processes inside the star. It can be seen that only two processes have kicked in as our model star is just beginning its journey through the main-sequence. The bottom plot in the middle panel shows the variation of the kinematic viscosity inside our model star. Now the rightmost panel of figure~\ref{fig:mesa1} shows various physical properties of our model star. Here, the topmost plot shows the variation of temperature and entropy with the mass coordinate, the middle plot shows the variation of density (logRho) and radius (logR\_cm) of the model star with the mass coordinate, and finally, the bottom plot shows the variation of baryon fraction (ye) and convectional velocity (vel\_km\_per\_s) inside the star.        

The figure~\ref{fig:mesa2} shows similar properties as in figure~\ref{fig:mesa1}, but at a stage when our model star has evolved through main-sequence and has become a supergiant. The leftmost panel in figure~\ref{fig:mesa2} shows that now the core temperature and density have risen considerably. Here, the top plot in the middle panel of figure~\ref{fig:mesa2} indicates the presence of heavier elements near the core while H and He are present in the envelope towards the higher mass coordinates. The middle plot in the middle panel of figure~\ref{fig:mesa2} shows that multiple energy generation processes have kicked in, while the bottom plot shows the further variation of kinematic viscosity inside our model star. Now, the rightmost panel of figure~\ref{fig:mesa2} depicts the variation of various physical parameters. From the radius plot, we can see that the model star has swollen up to $\sim$\,10$^{14}$\,cm, which is roughly around 1400\,R$_{\odot}$ while from the temperature plot, we can see that the temperature (logT plot) near the surface is $\sim$\,10000\,K. Such radius and temperature clearly indicate that our model is currently in the supergiant phase. 

Finally, figure~\ref{fig:mesa3} shows similar properties but at a stage when the core of our model star is about to collapse. The leftmost panel of figure~\ref{fig:mesa3} shows the log\,T$_{c}$ and log\,Rho$_{c}$ have reached in excess of 10$^{10}$\,K and 10$^{10}$\,gm\,cm$^{-3}$ respectively. Such core temperatures and core densities are thought to be the perfect conditions for the core to undergo collapse. Thus, the core of our model star is about to collapse. The mass fraction and specific luminosity plots in the middle panel of figure~\ref{fig:mesa3} also depict perfect conditions for the core to collapse. In the mass fraction plot, a very high fraction of $^{56}$Fe is seen, which indicates that the model star has developed an inert Fe-core, also in the specific luminosity plot, the absence of any energy generation process is seen. The plots in the rightmost panels of figure~\ref{fig:mesa3} show the physical properties of the inner--core regions. All these plots indicate the arrival of the stage of core collapse.

Post core-collapse, our model is evolved from a few seconds after the central explosion triggered by core-collapse to a time just before the outgoing shock reaches the stellar surface as discussed in \citet[][]{Paxton2018}. For this purpose, the default directory {\tt example\_ccsn\_IIp} from {\tt MESA}  is used.

\begin{figure*}
\centering
    \includegraphics[height=7.0cm,width=0.49\columnwidth,angle=0]{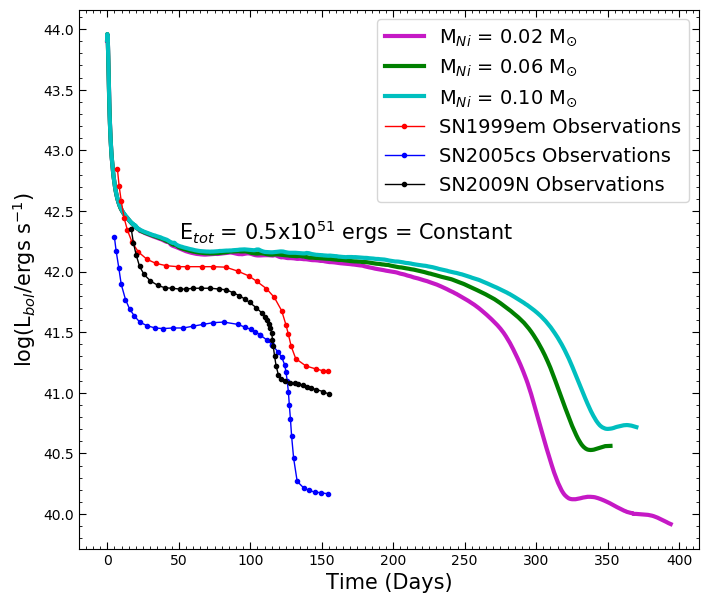}
    \includegraphics[height=7.0cm,width=0.49\columnwidth,angle=0]{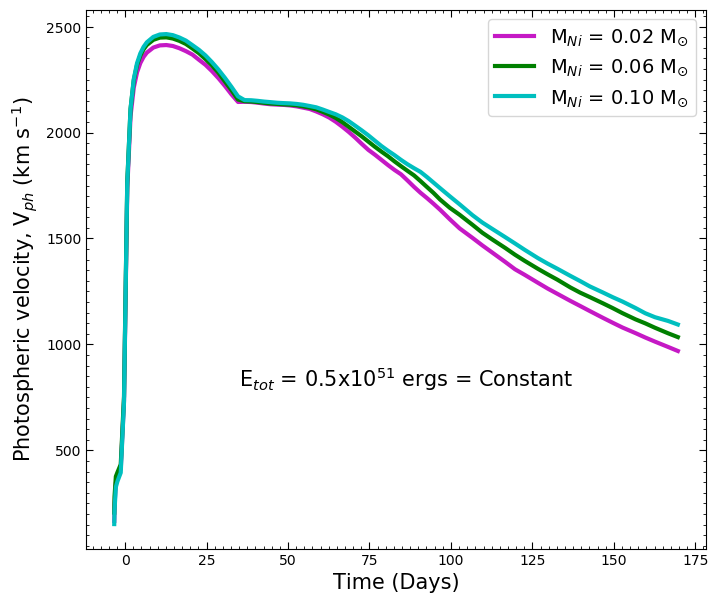}
   \caption{ Results of the synthetic explosions simulated by {\tt STELLA}, for comparison of the plateau phase, the bolometric light curves of three normal Type IIP SNe, namely SN1999em, SN2005cs, and SN2009N have also been plotted. The explosion energy has been kept fixed at 0.5$\times$10$^{51}$\,erg and the nickel mass is varied to 0.10\,M$_{\odot}$, 0.06\,M$_{\odot}$, and 0.02\,M$_{\odot}$. {\em Left:} The behaviour of the bolometric light curve by the variation of nickel mass with the explosion energy kept constant has been displayed. The duration of the plateau increases with the increase in nickel mass. {\em Right:} The effect on the photospheric velocity when the nickel mass is varied, and the explosion energy is kept constant. Minimal changes are seen in the photospheric velocities. }
    \label{fig:stella1}
\end{figure*}
\begin{figure*}
\centering
    \includegraphics[height=7.0cm,width=0.49\columnwidth,angle=0]{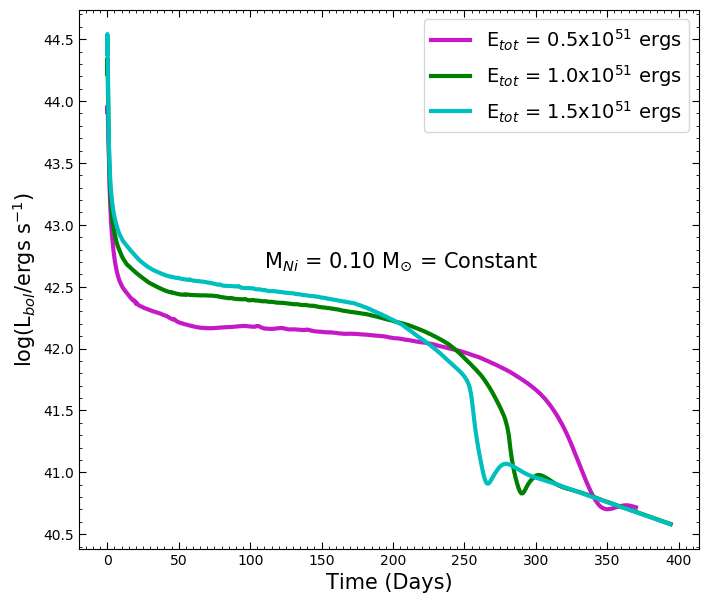}
    \includegraphics[height=7.0cm,width=0.49\columnwidth,angle=0]{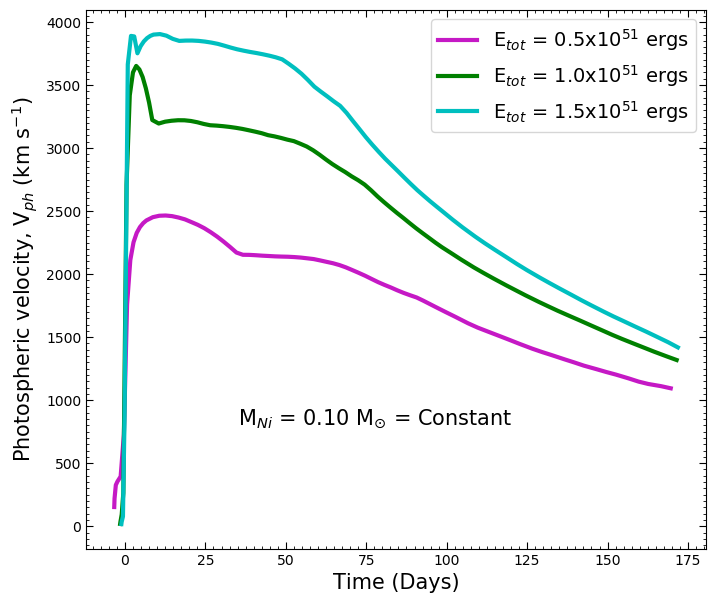}
    \caption{ Results of the synthetic explosions simulated by {\tt STELLA}. The nickel mass has been kept fixed to 0.10\,M$_{\odot}$ and the explosion energy is varied to 0.5$\times$10$^{51}$\,erg, 1.0$\times$10$^{51}$\,erg and, 1.5$\times$10$^{51}$\,erg. {\em Left:} The behaviour of the bolometric light curve by the variation of explosion energy with the nickel mass kept constant has been displayed. The overall plateau luminosity increases with an increase in explosion energy, while the plateau duration decreases with an increase in explosion energy. {\em Right:} The effect on the photospheric velocity when the explosion energy is varied, and the nickel mass is kept constant. The photospheric velocities increase with the increase in explosion energy.}
    \label{fig:stella2}
\end{figure*}

\section{Synthetic Explosions Using {\tt STELLA}} \label{finding}
\label{Explosion_ch6}

The output of {\tt MESA} when the shock has reached just before the stellar surface, is fed as input to {\tt STELLA}, which is a hydrodynamic code used to simulate the energetic CCSN explosions by evolving the model through shock-breakout. It solves the radiative transfer equations in the intensity momentum approximation in each frequency bin. In {\tt MESA}, {\tt STELLA} is  run using 40 frequency groups, which is enough to produce spectral energy distribution, but  insufficient to produce spectra. The opacity is computed based on over 153,000 spectral lines from \citet{Kuruz1995} and \citet{Verner1996}. To calculate the overall opacity, the code uses 16 species, which include H, He, C, N, O, Ne, Na, Mg, Al, Si, S, Ar, Ca, a sum of stable Fe and radioactive $^{56}$Co, stable Ni and radioactive $^{56}$Ni. The expansion opacity formalism from \citet{Eastman1993} is used to compute the line opacity by considering the effect of high-velocity gradients. The photo-ionization, electron scattering, and free-free absorption are also the sources of opacity that are included. Local thermodynamic equilibrium (LTE) is assumed in the plasma so that the Boltzmann-Saha distribution for ionization and level populations can be used. No nuclear networks except radioactive decay of $^{56}$Ni to $^{56}$Co which further decays to $^{56}$Fe, are used. The energy from the radioactive decay of $^{56}$Ni and $^{56}$Co is deposited into positrons and gamma-photons. This energy is treated in a one-group transport approximation as suggested by \citet{Swartz1995}.

In {\tt STELLA}, a Lagrangian co-moving grid is used to solve the 1-dimensional equations for mass, momentum, and total energy. The artificial viscosity comprising of the standard von Neumann artificial viscous pressure is used for stabilising solution \citep{Von1950} and a so-called cold artificial viscosity is used to smear shocks \citep{Blinnikov1996}. So, {\tt STELLA} enables one to properly compute the shock propagation along the ejecta and the shock breakout event. The coupled equations of radiation hydrodynamics are solved through an implicit high-order predictor–corrector procedure based on the methods of \citet{Brayton1972} (more details in \citealt{Blinnikov1996}).

Our {\tt STELLA} calculations include the synthetic explosions of a 100 M$_{\odot}$ ZAMS progenitor by assuming a set of explosion energies (E$_{tot}$) and nickel masses (M$_{Ni}$). The radioactive decay of $^{56}$Ni and $^{56}$Co powering mechanism has been employed in {\tt STELLA}.  We explored the effect on the bolometric light curve and photospheric velocity by changing one variable out of (E$_{tot}$) and (M$_{Ni}$) and keeping the other constant. In the first case, the explosion energy is kept constant to 0.5$\times$10$^{51}$\,erg while the nickel mass is changed to 0.10\,M$_{\odot}$, 0.06\,M$_{\odot}$, and 0.02\,M$_{\odot}$. The masses of the central remnants for these cases were $\sim$41\,M$_{\odot}$. For the second case, the nickel mass is kept constant to 0.10\,M$_{\odot}$ and the explosion energy is varied to 0.5$\times$10$^{51}$\,erg, 1.0$\times$10$^{51}$\,erg and, 1.5$\times$10$^{51}$\,erg. The masses of the central remnants were different corresponding to different explosion energies. Corresponding to the explosion energies of 0.5$\times$10$^{51}$\,erg, 1.0$\times$10$^{51}$\,erg and, 1.5$\times$10$^{51}$\,erg, the masses of the central remnants were $\sim$\,41\,M$_{\odot}$, 38\,M$_{\odot}$ and 31\,M$_{\odot}$ respectively. 

\section{Result of Synthetic Explosions}
\label{Results_ch6}

We evolved a 100 M$_{\odot}$ ZAMS progenitor up to the onset of core-collapse using {\tt MESA}, and the synthetic explosions of the model are simulated using {\tt STELLA}. The effects of explosion energy and nickel mass variations on the bolometric luminosity and photospheric velocity are explored. The left and right panels of figure~\ref{fig:stella1} show the effect of variation of nickel mass by keeping the explosion energy fixed on the bolometric light curve and photospheric velocity evolutions, respectively. First, an increased duration of the plateau is noticed in comparison to normal Type IIP SNe. Such an increased plateau duration is attributed to the huge, inflated H-envelope of the model star. For comparison purposes, we have also plotted three normal Type IIP SNe, namely, SN~1999em, SN~2005cs, and SN~2009N (the light curves have been taken from \citealt[][]{Paxton2019}) in the left panel of figure~\ref{fig:stella1}. From modelling their bolometric light curves, \citet[][]{Paxton2019} found that their progenitors have masses in the range 13\,M$_{\odot}$ to 19\,M$_{\odot}$. These SNe have a much lower extent of H-envelope than our 100\,M$_{\odot}$, so they have a much lower plateau phase. Thus, the H-envelope of an SN progenitor plays a very important role in determining the length of the plateau phase. 

Further, it is noticed that with the increase in the nickel mass, the duration of the plateau is also increased (left panel of figure~\ref{fig:stella1}). Due to the increased amount of nickel, there is an increase in the energy deposition in the expanding ejecta by the gamma-rays produced by the radioactive decay of $^{56}$Ni and $^{56}$Co, which prolongs the plateau duration. Since the explosion energy is kept fixed, the ejecta move with nearly similar velocity for each case of different nickel mass; thus, minimal variation is seen on the photospheric velocities ( right panel of figure~\ref{fig:stella1}).

Figure~\ref{fig:stella2} shows the effect of variation of explosion energy by keeping nickel mass constant on the bolometric luminosity and photospheric velocity evolutions. In the left panel of figure~\ref{fig:stella2}, we can see two prominent effects when the nickel mass is kept fixed and the explosion energy is varied; first, the overall increase in the luminosity of the plateau phase and second, the decrease in the plateau duration with an increase in explosion energy. Due to the increase in the explosion energy, the supernova is intrinsically bright and powerful, which is why the intrinsic brightness during the plateau phase increases with the increase in explosion energy. Also, higher explosion energy implies an increase in the kinetic energy of ejecta; as a result, the SN expands and cools faster. Thus the reionizing front that is responsible for the plateau is extinguished early, and the duration of the plateau decreases. With the increase in the explosion energy, the kinetic energy of ejecta also increases, resulting in faster-moving ejecta, which is clearly evident in the right panel of figure~\ref{fig:stella2}.

\section{Discussion of the Present Work}
\label{Discussion_ch6}

In this work, a 100 M$_{\odot}$ ZAMS progenitor model has been evolved from ZAMS up to the stage where it can undergo core-collapse with the help of publicly available tool {\tt MESA}. The output of {\tt MESA} is fed as input to {\tt STELLA}, which simulates the synthetic explosions. Based on the mass loss rate, rotation and metallicity, such a heavy progenitor could result in various transients, including PISN, PPISN, Type IIp, Type IIn, Type Ibn and Type Icn also. Still, for our case, we explored the possible outcomes when such a massive ZAMS progenitor undergoes a Type IIP like CCSNe by displaying a huge plateau in the light curves. Such heavy stars under prevailing physical circumstances might lead towards exploding as superluminous SNe, but analysis of such types of explosions is beyond the scope of the present thesis. In the following points, we briefly summarise our present work : 

1) A 100 M$_{\odot}$ ZAMS progenitor model has  a much larger H-envelope compared to the typical progenitors of Type IIP SNe, due to which an extended duration of the plateau is seen.

2) The effect of variation of nickel mass is explored. On increasing the nickel mass while keeping the explosion energy constant, the duration of the plateau in the light curve is increased. This behaviour due to the increase in nickel mass is attributed to the increase in the energy deposition in the expanding ejecta by the gamma-rays that are produced by the radioactive decay of $^{56}$Ni and $^{56}$Co. Also, minor changes are seen in the photospheric velocity evolution. This behaviour of photospheric velocity is attributed to the constant explosion energy. Due to the constant energy of the explosion, the kinetic energy of the ejecta having different nickel masses is nearly similar, resulting in nearly similar photospheric velocities.     

3) The effect of variation of explosion energy is also explored. On increasing the explosion energy and keeping the nickel mass constant, the overall plateau luminosity increases but the duration of the plateau decreases. Such behaviour is attributed to the supernova becoming intrinsically bright and powerful with the increase in explosion energy. Due to the supernova becoming intrinsically bright, the luminosity of the plateau phase is increased. Also, with the increase in the explosion energy, the supernova expands and cools faster. Because of this, the reionizing front that is responsible for the creation of the plateau gets extinguished faster, and the plateau duration decreases.

\section{Conclusion}
\label{Conclusion_ch6}

We could simulate synthetic CCSN explosions from a 100 M$_{\odot}$ ZAMS progenitor. As discussed earlier, based on the mass loss rate, rotation and metallicity, such a heavy progenitor could result in various transients, including PISN, PPISN, Type IIp, Type IIn, Type Ibn and Type Icn also. In this work, we explored the outcomes of the synthetic explosions from a 100 M$_{\odot}$ ZAMS progenitor resulting in Type IIP-like SNe. The presence of a very large H-envelope before the supernova explosion resulted in an extremely large plateau for the SNe. The effects of the variation of explosion energy and nickel mass on the bolometric luminosity and photospheric velocity were also explored.    %100 solar mass SN 
\chapter{\sc Summary and future prospects}\label{Ch:8}  

\ifpdf
    \graphicspath{{Chapter8/Chapter8Figs/PNG/}{Chapter8/Chapter8Figs/PDF/}{Chapter8/Chapter8Figs/}}
\else
    \graphicspath{{Chapter8/Chapter8Figs/EPS/}{Chapter8/Chapter8Figs/}}
\fi

\ifpdf
    \graphicspath{{Chapter8/Chapter8Figs/JPG/}{Chapter8/Chapter8Figs/PDF/}{Chapter8/Chapter8Figs/}}
\else
    \graphicspath{{Chapter8/Chapter8Figs/EPS/}{Chapter8/Chapter8Figs/}}
\fi

%\newpage
%\thispagestyle{empty}
%\mbox{}

\section{Summary}

Knowledge about possible progenitors of CCSNe is the most fundamental aspect towards understanding the physics behind these catastrophic transients. The possible progenitors of Type Ib/IIb SNe are primarily connected to the absence/presence of a small amount of hydrogen at the pre-explosion stages. The amount of hydrogen and its depletion before the explosion is also connected to the assumption of the single vs binary channel of the evolution of the progenitor system. The present understanding of the single vs binary channels is highly biased because of the incomplete sample of such observed SNe in the nearby universe. The most appropriate way to shed light upon the possible progenitors of underlying CCSNe will be the detection of the progenitors in high-resolution pre-explosion images. But, obtaining pre-explosion images of the progenitors is highly uncertain because one can not predict the temporal and spatial position of the occurrence of an SN. Owing to these uncertainties, only a handful of pre-explosion detections are there in archival images.

It is also relevant to mention here that the uncertainty in the estimation of the amount of hydrogen retained by the CCSNe progenitors at their pre-explosion stages is directly connected to observational uncertainties in the estimation of extinction and distance of the well-known sample set of available Type Ib and Type  IIb SNe. The pre-explosion mass-loss history of the underlying progenitor also plays an important role on the amount of hydrogen retained. Correct estimation of the amount of hydrogen at the pre-explosion stage affects not only the peak luminosity but also the stretch of the light curve and spectral parameters that are directly related to underlying physical mechanisms leading to these energetic explosions.

To address above mentioned yet least explored but important problems shedding light on connections between Type Ib and Type IIb SNe and their possible channels of progenitors, we adopted an amalgamation of optical observations (both photometric and spectroscopic) and 1-dimensional hydrodynamical simulations of a sample of Ib/IIb SNe. We also extended our studies to understand the nature of explosions of Pop III stars and massive stars up to 100\,M$_{\odot}$ under similar physical conditions to shed some light on such energetic transients expected to be discovered using future missions.

Within the context of the research work in this thesis, we attempted to unveil various properties of CCSNe, including the possible progenitors, powering mechanisms and the surrounding environments. We utilised photometric and spectroscopic data of several Type Ib/IIb SNe acquired using multiple national and international telescopes to meet the goals specified in the proposed synopsis. As a large part of the work, we also utilised multiple analysis tools, including {\tt MINIM}, {\tt MOSFiT}, {\tt MESA}, {\tt STELLA}, {\tt SNEC}, etc., for the research works conducted under this thesis. With the combination of photometric and spectroscopic analysis coupled with the simulations involving 1-dimensional stellar evolutions of possible progenitors and the hydrodynamic simulations of their synthetic explosions, we attempted to explore the specified goals in the proposed synopsis for several CCSNe.

To achieve some of the goals mentioned above, the first project consisted of thorough investigations of the photometric and spectroscopic properties of two stripped, H-deficient Type Ib SNe, namely, SN~2015ap and SN~2016bau. The SN~2015ap proved to have intermediate luminosity among the subset of similar types of SNe, while SN~2016bau showed to be highly extinguished due to large host galaxy extinction. The photometric and spectroscopic analysis, in conjunction with the hydrodynamic simulations of synthetic explosions, indicated a 12\,M$_{\odot}$ ZAMS star as the possible progenitor for SN~2015ap. At the same time, SN~2016bau proved to be originated from a slightly less massive progenitor. To understand aspects like the role of hydrogen mass and the evolution of light curves and spectral features providing direct evidence to establish overlap between Type Ib and Type IIb SNe, we investigated the photometric and spectroscopic behaviour of a partially stripped Type IIb SN~2016iyc. A detailed study by \citet[][]{2019ApJ...885..130S} explored the impact of single- and binary-progenitor channels for Type IIb CCSNe and found that at solar metallicity, the single- or binary-channels contributed to nearly the same number of Type IIb CCSNe. Further, the binary-progenitor channel dominates at lower metallicities. Recent studies by \citet[][]{2022MNRAS.511..691G} also investigated the possible progenitor channels for Type IIb and the amount of hydrogen retained by them. The authors found that the post-interaction mass-loss rate played an important role in the amount of hydrogen retained in the envelope by the Type IIb CCSNe progenitors at the time of the explosion. For SN~2016iyc, we attempted several single-progenitor models having ZAMS mass in the range of (9--14)\,M$_{\odot}$. Our analysis could put important constraints on the ZAMS mass of the possible progenitor for SN~2016iyc. Additionally, the amount of hydrogen retained by SN\,2016iyc was found to be somewhere between that of Type Ib and hydrogen-rich Type II SNe.  Besides, under this project, we also modelled the likely progenitors of two other Type IIb SNe, namely SN~2016gkg and SN~2011fu. By doing so, we modelled three Type IIb SNe spanning a range of luminosity and found that the Type IIb SNe exhibit a diverse range of progenitor masses. 
Further, utilising the learnings from progenitor modelling and the observational properties of Type Ib/IIb SNe, we investigated the important aspects of the evolutions of the very first-generation stars having a range of initial rotations. In this work, we performed the 1-dimensional stellar evolutions of 25\,M$_{\odot}$ with different initial rotations and no-metal content, starting from ZAMS up to the stage of the onset of core collapse and investigated their terminating fates. Owing to their initial rotations, the resulting transients fell under weak Type II and Type Ib/c SNe. Finally, we also extended our studies to perform the stellar evolution of a 100\,M$_{\odot}$ ZAMS star and simulated the synthetic explosions by providing a range of explosion energy and amount of nickel synthesised. Depending upon mass loss rates, initial rotations and the specified metallicity, such a massive star can terminate its life as various classes of transients, including PISN, PPISN, Type IIP-like, Type IIn-like and Type Icn-like SNe too. We investigated its behaviour when the 100\,M$_{\odot}$ would die as Type IIP-like SNe. These studies have been able to achieve the goals mentioned in the synopsis to a large extent.

Further, we provide a comprehensive summary of each project carried out as part and partial of the thesis:
\\

$\bullet$~Chapter~\ref{Ch:3} provides the details of the photometric and spectroscopic analysis of two Type Ib SNe, SN~2015ap and SN~2016bau. The photometric comparison of these two SNe with a set of similar SNe revealed that SN~2015ap had an intermediate luminosity while SN~2016bau was heavily extinguished due to extensive host galaxy contamination. We utilised the extinction-corrected multi-band data of these SNe to calculate their bolometric/quasi-bolometric luminosity light curves. From our analysis, we found that SN~2015ap synthesised an amount of 0.14$\pm$0.02\,M$_{\odot}$ of $^{56}$Ni mass while for SN~2016bau, the amount of $^{56}$Ni synthesised was 0.055$\pm$0.006\,M$_{\odot}$.   
Determining the amount of $^{56}$Ni generated in a CCSN is crucial to understand the progenitor scenarios and explosion processes associated with different CCSNe, as it heavily relies on both the explosion characteristics and the core structure of the underlying progenitor. As a consequence, the measurement of the amount of synthesised nickel is very important to comprehend the progenitors of CCSNe and their explosion mechanisms \citep[][]{2019MNRAS.483.3607S}. Additionally, the estimated ejecta masses for SN~2015ap and SN~2016bau were 2.2$\pm$0.6\,M$_{\odot}$ and 1.6$\pm$0.3\,M$_{\odot}$, respectively. We also studied the evolutions of a number of blackbody parameters, including the radius, temperature and velocity for both the SNe.

Further, the spectra of SN~2015ap and SN~2016bau at various epochs displayed strong He~I features near 5876\,\AA\, confirming their Type Ib nature. The spectra of these two SNe were also compared with other similar SNe. The spectra of both these SNe at various epochs evolved similarly to other SNe of similar types.  We also modelled their spectra at various epochs using {\tt SYN++} and found that the strong He~I features were produced nicely, further confirming the Type Ib nature of these two SNe. Matching (after appropriate flux-scaling due to the phase differences) the 100\,d spectrum for various ZAMS mass progenitors from \citet[][]{Jerkstrand2015} with the +98.75\,d spectrum of SN~2015ap and the +121\,d spectrum of SN~2016bau, we found that these two SNe had originated from progenitors having ZAMS masses close to 12\,M$_{\odot}$.  

After performing spectroscopic analysis and modelling, the semi-analytical light curve fittings were also performed using {\tt MINIM} to fit the quasi-bolometric light curves of SN~2015ap and SN~2016bau assuming different powering mechanisms for the light curves of these two SNe. The Ni-Co radioactive decay model is considered the prominent powering mechanism for Type Ib SNe, so we utilised the RD model to fit the quasi-bolometric light curves of SN~2015ap and SN~2016bau. We found that for both these SNe, the semi-analytical RD model fails to fit the early and late parts of their corresponding quasi-bolometric light curves. This issue was later attributed to the assumption of the constant ejecta density profile and also to the constant opacity assumed in {\tt MINIM} code. In earlier studies, the light curves of a few Type Ib SNe were also found to be powered by the spin-down of a magnetar model. However, utilising the spin-down of a magnetar model using {\tt MINIM} as the powering mechanisms for SN~2015ap and SN~2016bau resulted in unphysical parameters. For SN~2015ap, we obtained a very slow initial rotation of the magnetar with an initial period, $P_{i}\approx40\,ms$ and a low magnetar rotational energy, $E_{p}\approx10^{49}$\,erg. Similarly, for SN~2016bau, we obtained a further slow $P_{i}\approx70\,ms$ and an order less $E_{p}\approx10^{48}$\,erg. The unphysical parameters obtained for the MAG model discarded the possibilities of SN~2015ap and SN~2016bau powered by the magnetar spin-down.

Utilising the observational parameters obtained through the photometric and spectroscopic studies, we tried to model the possible progenitors of SN~2015ap and SN~2016bau using {\tt MESA}. To obtain a significantly stripped pre-SN model in {\tt MESA}, we were required to artificially strip the models by specifying a mass loss rate $\geq$\,10$^{-4}$\,M$_{\odot}$\,yr$^{-1}$. A hybrid mass loss mechanism has been proposed to account for such high mass loss rates for Type Ib SNe progenitor in \citep[][]{2022arXiv220905283S} where the mass losses due to the binary interaction and stellar winds are considered. The synthetic explosions of the pre-SN models were also simulated using {\tt STELLA}  and {\tt SNEC}. The comparisons of the {\tt STELLA/SNEC} produced bolometric light curves and velocity evolutions with actual observations indicated a 12\,M$_{\odot}$ ZAMS progenitor for SN~2015ap while SN~2016bau proved to originate from a progenitor with slightly less ZAMS mass. 
\\

$\bullet$~ Type IIb SNe are thought to be the interconnecting link between H-rich and H-deficient CCSNe. Thus, after discussing the properties of two stripped, H-deficient Type Ib SNe in Chapter~\ref{Ch:3}, we performed a detailed photometric and spectroscopic analysis of a partially stripped Type IIb SN~2016iyc in Chapter~\ref{Ch:4}. The total extinction corrected $BVRI$-bands light curves were utilised to compute the bolometric/quasi-bolometric light curve. The comparisons of the absolute $V$-band light curve and the quasi-bolometric light curve with other similar SNe indicated that SN~2016iyc lied towards the faint limit of Type IIb SNe. Additionally, SN~2016iyc also suffered from low nickel production. The faint SNe suffering low nickel production are thought to result from the explosions of low-mass ZAMS stars.

For SN~2016iyc, the blackbody parameters, including the temperature and radius evolutions, were also studied and compared with other similar SNe. The blackbody temperature showed a conventional evolution like similar SNe. Further, the comparison of the blackbody radius evolution indicated an anomalous behaviour as the blackbody radius of SN~2016iyc at each epoch was smallest among other SNe Type IIb. This behaviour in the blackbody radius evolution can be associated with low ejecta velocity.

Further, we identified the spectral features present in the only available spectrum of SN~2016iyc utilising {\tt SYN++}. The spectrum displayed prominent features of H$_{\alpha}$ near 6563\,\AA\, and He~I near 5876\,\AA\, which strongly supported the Type IIb classification of SN~2016iyc. The line velocities associated with the H$_{\alpha}$, He~I, and Fe~II in {\tt SYN++} were $\sim$10,100, $\sim$6800, and $\sim$6100\,km\,s$^{-1}$, respectively. These model-produced velocities were very close to the observed velocities. The photospheric velocity, photospheric temperature, and the parameterisation velocity employed to produce the model spectrum using {\tt SYN ++} were 6100\,km\,s$^{-1}$,  6300\,K  and 6000\,km\,s$^{-1}$, respectively. The comparison of the only available spectrum of SN~2016iyc showed a close resemblance with the spectra of other well-studied Type IIb SNe.

Based on the low ejecta mass and low nickel mass from literature along with SN~2016iyc lying towards the fainter limit of Type IIb SNe in the comparison sample, 9--14\,M$_{\odot}$ ZAMS progenitors were modelled as the possible progenitor using {\tt MESA} up to the stage of the onset of core collapse. Thereafter, the synthetic explosions of the models at the stage of the core collapse were simulated using {\tt STELLA} and {\tt SNEC}. The outcomes of {\tt STELLA/SNEC} were compared with the actual observations. Collectively, the 1-dimensional stellar modelling of the possible progenitor using {\tt MESA} and simulations of hydrodynamic explosions using {\tt SNEC}/{\tt STELLA} indicated that SN~2016iyc probably originated from a (12--13)\,M$_{\odot}$ ZAMS progenitor, near the lower end of progenitor masses for SNe~IIb. The pre-SN radius of the progenitor of SN~2016iyc was likely in the range of (240--300)\,R$_{\odot}$. Further, the models showed a range of parameters for SN~2016iyc, including $M_{\rm ej} =$ (1.89--1.93)\,M$_{\odot}$ and $E_{\rm exp} =$ (0.28--0.35) $\times 10^{51}$\,erg. An upper limit of 0.09\,M$_{\odot}$ on the amount of nickel synthesised by the SN was estimated. 

Finally, the stellar evolution of the possible progenitors and hydrodynamic simulations of the synthetic explosions of SN~2016gkg and SN~2011fu were performed to compare the intermediate- and high-luminosity ends among well-studied SNe~IIb using {\tt MESA} and {\tt SNEC}. The results of stellar modelling and synthetic explosions for SN~2016iyc, SN~2016gkg, and SN~2011fu exhibited a diverse range of mass of the possible progenitors for Type IIb SNe.

Our studies involving possible progenitor modelling of Type Ib (in Chapter~\ref{Ch:3}) and Type IIb SNe using {\tt MESA} indicate that the Type Ib SNe originate from significantly stripped stars retaining only a negligible amount of hydrogen, while the Type IIb SNe originate from more extended stars having radii of a few 100\,R$_{\odot}$ retaining some trace amount of hydrogen.\\ 

$\bullet$~ We presented a discussion on two H-deficient Type Ib SNe in Chapter~\ref{Ch:3}, followed by extensive details on a partially stripped Type IIb SN in Chapter~\ref{Ch:4}. Thus, in Chapter~\ref{Ch:5}, we attempted to utilise our experience to build physical models of H-rich and H-deficient CCSNe, originating from massive Pop III stars. Each model star had a mass of 25\,M$_{\odot}$ at ZAMS and different initial rotations. Pop III stars were thought to play an important role in cosmic re-ionization. These hypothetical objects were also responsible for the enrichment of the early universe by spreading heavier metals through violent SNe explosions or possibly through sporadic mass losses through stellar winds. This Chapter provided extensive details of the 1-dimensional stellar evolutions of rotating, 25\,M$_{\odot}$, Pop III stars up to the onset of core collapse utilising {\tt MESA}. The hydrodynamic simulations of the resulting CCSNe were also simulated using {\tt SNEC}.

The models with different initial rotations were evolved from ZAMS up to the onset of core collapse. Comparing their evolutions on the HR diagram, we found that the Pop III models reach the ZAMS at higher effective temperatures in comparison to the solar metallicity (i.e., Z$=$0.02) model. Among Pop III models, at ZAMS, the model with higher initial angular rotational velocities possessed lower luminosity and lower effective temperature, a well-known effect of rotation as also found by \citet[][]{2008A&A...489..685E} earlier. We also explored the evolutions of the angular rotational velocities and corresponding mass loss rates for Pop III models. We discovered that the rapidly rotating models suffered chaotic mass losses beyond their main-sequence phases once they ignited core-He burning. The extensive mass losses in the rapidly rotating models were attributed to the dramatic increase in their surface metallicities due to the deposition of a large amount of CNO elements due to strong mixing. In turn, the dramatic rise in surface metallicities of the models was responsible for the chaotic increase in corresponding mass loss rates. The evolutions of the core-temperature and core-density curves for different models were also explored. During the last evolutionary stages, all the models had exceeded the core-temperature of $\sim$\,10$^{9.9}$; the perfect condition for the cores to collapse under their own gravity.

Once the models arrived at the stage of the onset of core collapse, the outputs of {\tt MESA} were provided as input to {\tt SNEC} in appropriate forms. The hydrodynamic simulations of the synthetic explosions of the models on the verge of core-collapse were simulated using {\tt SNEC}. Depending upon the initial rotations, the resulting transients spanned a class of SNe from weak Type II and Type Ib/c. The non-rotating and slowly rotating models suffered minimal mass losses and retained a larger H-envelope at their pre-SN stage. The resulting SNe from these models displayed a plateau in their multi-band light curves resembling H-rich, Type IIP SNe. Contrary to the slow-rotating models, the rapidly rotating models suffered tremendous mass losses and stripped off all of their H-envelopes. The resulting light curves from the {\tt SNEC} simulations closely resembled Type Ib/c SNe.  Heavy metals dominated the outer layers of the rapidly rotating models at their pre-SN stages, and thus, their spectra would be dominated by the corresponding heavy metal features.

Among the Pop III models, SN resulting from the non-rotating H-rich model proved to be the brightest. It had a nearly constant absolute magnitude of around -16.5 mag in the $V$-band for the plateau phase that resulted to an apparent magnitude of $\sim$35.5\,mag at a redshift of z = 10 (using the cosmology of a Hubble constant, H$_{0}$ = 73, $\Omega_{\rm M}$ = 0.3, and $\Omega_{\rm vac}$ = 0.7). Currently, no ground- or space-based observatory can go this faint to detect a Pop III CCSN resulting from an individual star; however, with the major advancement in observational technologies having large diameters, such events could possibly be detected in the future. Based on the present research, we found that within the specified limits of the explosion energies and nickel masses, these transient events were potentially very faint, making it difficult for them to be detected at high redshifts.
\\

$\bullet$~ Chapter~\ref{Ch:3}, Chapter~\ref{Ch:4}, and Chapter~\ref{Ch:5} provided details of the properties of CCSNe arising from progenitors having ZAMS masses of 25\,M$_{\odot}$ or less. Thus, in Chapter~\ref{Ch:6}, we performed the 1-dimensional stellar evolution of a 100\,M$_{\odot}$ ZAMS star using {\tt MESA} and studied its final fates as Type IIP-like SNe. Depending upon initial rotations, mass loss, and metallicity, such a massive star could serve as a progenitor for several types of transient events, including PISNe, PPISNe, Type IIP, Type IIn, Type Ibn, Type Icn, etc. types of SNe. Starting from the ZAMS, the model star evolved to the onset of core collapse, and the synthetic explosions were later simulated using {\tt STELLA}. In this Chapter, the effects of changing the explosion energy and nickel mass over the resulting CCSNe light curves were explored. In comparison to typical Type IIP SNe progenitor at the pre-SN stage, the 100\,M$_{\odot}$ model had retained a much larger amount of hydrogen; as a result, the Type IIP SN from 100\,M$_{\odot}$ model displayed much larger plateau in its light curve.

The study investigated the impact of varying the nickel mass on the light curve and velocity evolution. By maintaining a constant explosion energy and increasing the nickel mass, it was observed that the duration of the plateau in the light curve extended. This behaviour arising due to the increase in nickel mass could be attributed to the increase in the energy deposition in the expanding ejecta by the gamma-rays produced by the radioactive decay of $^{56}$Ni and $^{56}$Co. 
Additionally, minor changes were observed in the evolution of the photospheric velocity. This behaviour of photospheric velocity could be linked to the constant energy of the explosion. Since the explosion energy remained constant, the kinetic energy of the ejecta with varying nickel masses remained nearly identical, leading to similar photospheric velocities.

Moreover, the study also investigated the impact of varying the explosion energy. By increasing the explosion energy and keeping nickel mass constant, the overall luminosity of the plateau increased. However, a corresponding decrease in the duration of the plateau was observed. This behaviour could be ascribed to the SN becoming inherently luminous and powerful with the increase in the explosion energy.  Additionally, as the explosion energy increased, the SN expanded and consequently cooled faster. As a result, the re-ionizing front responsible for the formation of the plateau in the light curves extinguished faster, consequently reducing the plateau duration.

As part of the present thesis, we explored the possible connections between Type Ib and Type IIb CCSNe originating due to the difference in the amount of hydrogen retained by their progenitor at the pre-explosion stages. However, the correct estimation of hydrogen retained by the exploding progenitors is dubious owing to the uncertainties associated with determining the extinction and distance of underlying CCSNe. We utilised an amalgamation of observational analysis and hydrodynamic modelling to constrain the physical properties of the progenitors of a set of Type Ib/IIb CCSNe. Further, utilising our learnings from these projects, we extended our analysis to study the evolution and explosions of massive Pop III stars. The research work presented here as a part of the thesis has not only addressed open issues related to the nature of progenitors of diverse sets of CCSNe but also shed light on new avenues of exploration about the properties of Pop III and very massive stars. In the following section, we provide a comprehensive overview of the potential future prospects arising from this thesis.
\\ 

\section{Future prospects}\label{sec:future_prospects}

The research works in the context of the underlying thesis have investigated various aspects of CCSNe with the help of photometric and spectroscopic data from multiple national and international telescopes. This thesis has also demonstrated the importance of various publicly and privately available analysis tools to understand the diverse nature of catastrophic CCSNe. The research works under this thesis have now opened new doors not limited to CCSNe only. Provided below are a few future prospects that originated as a result of the research in the context of the presented thesis:

$\bullet$~ {\it \bf 3-dimensional modelling of CCSNe}

Despite the remarkable progress made in CCSN observations and physics, the majority of studies conducted thus far have been significantly constrained by their dependence on spherically symmetric progenitors as the starting point. Furthermore, all simulation works featured in the current thesis have been limited to 1-dimensional models. It is essential to recognise that the final outcome of an SN explosion is intricately associated with the specific characteristics of the structure of the underlying progenitor.

To advance the field of CCSNe further, a critical and imperative next step is to investigate the impact of multidimensional progenitors before the collapse phase. Understanding the role of such complex progenitor structures will be instrumental in significantly progressing our comprehension of CCSNe. A few state-of-the-art 3-dimensional tools, including {\tt FORNAX} \citep[][]{2019ApJS..241....7S} and {\tt FLASH} \citep[][]{2000ApJS..131..273F} have begun to enhance our comprehension of these catastrophic transients significantly.

$\bullet$~ {\it \bf Simulations of possible the progenitors of long-Gamma Ray Bursts' progenitors}

There have been multiple incidences of the CCSNe association with long-duration gamma-ray bursts (LGRBS) as reported earlier in (among many others) \citet[][]{1998Natur.395..670G,1999Natur.401..453B} and recently in (among many others) \citet[][]{2021A&A...646A..50H,2022NewA...9701889K}. Such associations have indicated that a fraction of LGRBs results from the collapsar model (i.e., death of massive stars) as reported in \citet[][]{2017AdAst2017E...5C} and references therein. The spin-down of a millisecond magnetar produced in the core of a massive dying star and working as a central engine is one of the prominent powering mechanisms for the GRBs; however, a thorough understanding of the detailed nature of the underlying mechanism is still lacking. With the ability of {\tt MESA} to evolve the massive stars up to the stage of core collapse, an attempt to understand the nature of such events can be made. As a result of the collapse of a massive star, a proto-Neutron star might be born in the centre. Considering the iron-core masses as the baryonic mass of the Neutron star, a number of parameters, including the initial spin rate ($P_{NS}$), the gravitational mass ($M_{NS}$), and the average surface magnetic field ($B_{NS}$) can be estimated for the millisecond magnetar powering the GRBs. These parameters can be compared with the actual observations. Such types of studies are going to be exceptionally vital in improving the current understanding of the possible progenitors of GRBs. People have already started working in this direction. One example of such type of study is presented in \citet[][]{2023arXiv230105401S}.

$\bullet$~ {\it \bf Study of the compact binary merger event progenitor systems}

A few years ago, the detection of the first high-frequency gravitational wave (GW) signals from the merger of two black holes started a new era of astrophysics. The corresponding event GW150914 was reported in \citet[][]{2016PhRvL.116f1102A}. Thereafter, multiple merger events have been detected where the merging members are either two black holes \citep[][]{2021PhRvX..11b1053A} or two neutron stars \citep[][]{2017PhRvL.119p1101A,2020ApJ...892L...3A}. Beyond these two scenarios, mixed (one black and one neutron star) events \citep[][]{2021ApJ...915L...5A} have also been detected. With the help of {\tt MESA}, \citet[][]{2021ApJ...920L..36J} have investigated a possible progenitor system for the double neutron star merger event GW170817. In comparison to a total mass of 2.74$^{+0.04}_{-0.01}$\,M$_{\odot}$ for the GW170817, their double neutron star progenitor system had a total mass of 2.79\,M$_{\odot}$. There are only a handful of such studies available to explore the progenitor channels for GW events. Therefore, the studies to understand the progenitor channel for the GW events utilising {\tt MESA} will prove to be very crucial in the future.

$\bullet$~ {\it \bf Search for the Pop III stars or associated transients}

The research works under this thesis have also touched upon the evolution of massive Pop III stars and resulting transients. With major advancements in observational technologies starting with the era of the Thirty Meter Telescope\footnote{https://www.tmt.org/page/uselt} (TMT), Giant Magellan Telescope\footnote{https://giantmagellan.org/} (GMT) and Extremely Large Telescope\footnote{https://elt.eso.org/} (ELT), OverWhelmingly Large\footnote{https://www.eso.org/sci/facilities/eelt/owl/index.html} (OWL) Telescope in the coming future, we expect the search for the very first generations of metal-free stars and resulting transients to grow.

$\bullet$~ {\it \bf To shed light on Type IIL SNe progenitors}

Among H-rich Type II CCSNe, the Type IIL SNe display linear decay in their light curves after maximum brightness, while the Type IIP SNe show plateau in their light curves \citep[][]{Barbon1979}. The plateau in Type IIP SN light curves is attributed to the presence of a large H-envelope in the corresponding progenitor. Initially, in a Type IIP SN, the H-envelope is almost totally ionized, and only the photons from the outermost surfaces can escape. Later, as the SN cools, the ionized hydrogen recombines, and the photons from the inner layers become visible. The temperature of recombination is nearly constant, so as the SN expands and cools, the re-ionization front continues to recede more inwards, and a plateau is displayed in its light curves. On the contrary to a Type IIP SN, even after being H-rich, the Type IIL SNe lack such a plateau in their light curves. This aspect of Type IIL SNe is not very well explored, and it has been theorised that their progenitors have smaller H-envelopes \citep[][]{1971Ap&SS..10...28G,1989ApJ...342L..79Y,1993A&A...273..106B}.  With the help of {\tt MESA}, we can model the possible progenitors of Type IIL SNe and simulate their synthetic explosions using {\tt STELLA}, and {\tt SNEC} to match the simulated light curves and velocity evolutions with the observations. Such studies would help us to understand the possible progenitors of Type IIL SNe more robustly. People have already started working in this direction; however, there have been only a handful of such studies where observations and simulations complement each other. One such type of study is presented in \citet[][]{2019MNRAS.485.5120B}.
 
$\bullet$~ {\it \bf Use the combination of {\tt MESA}, {\tt STELLA} and {\tt SNEC} to model more CCSNe progenitors}

Understanding the possible progenitor of various classes of CCSNe is a challenging task. The thesis presented here tried to make a small attempt in the relevant direction utilising the photometric/spectroscopic data in combination with {\tt MESA}, {\tt STELLA}, and {\tt SNEC}. With our simulations complementing the observations, we could slightly enhance the current understanding of the progenitor system for a few stripped/partially-stripped and H-rich CCSNe. In the coming future, we plan to model the observational data for more CCSNe using {\tt MESA}, {\tt STELLA}, and {\tt SNEC} to enhance further the comprehension of physical and chemical aspects of the possible progenitors. %Summary n future
\appendix
\chapter{\sc Constraints on the progenitor masses of two Type Ib Supernovae, SN 2015ap and SN 2016bau}

\ifpdf
    \graphicspath{{Appendix/ChapterAFigs/PNG/}{Appendix/ChapterAFigs/PDF/}{Appendix/ChapterAFigs/}}
\else
    \graphicspath{{Appendix/ChapterAFigs/EPS/}{Appendix/ChapterAFigs/}}
\fi

\section{Logs of spectroscopic and photometric observations}

\begin{table*}
\caption {Spectroscopic observations of SN~2016bau.}
\label{tab:SN2016bau_spec_obs}
\begin{center}
Note: All spectra were taken with the Kast spectrograph on the 3.0\,m Shane telescope at Lick Observatory.
\smallskip
\small\addtolength{\tabcolsep}{-2pt}

%\small
\begin{tabular}{c c c c c c}
\hline \hline
UT Date	   &     MJD            \\
           &                    \\
\hline    
2016/03/16  &	57463.719       \\
2016/03/19  &	57466.919       \\
2016/04/03  &	57481.671       \\
2016/04/17  &	57495.902       \\
2016/05/02  &	57510.671       \\
2016/05/16  &	57524.860       \\
2016/05/30  &	57538.741       \\
2016/07/15  &	57584.710       \\
\hline

\end{tabular}
\end{center}
\end{table*}

\begin{table*}
\caption{Photometric observations of SN~2015ap.}
\centering
\smallskip
\begin{tabular}{c c c c c c c c}
\hline \hline
MJD  	    &  $B$              &  $V$          &  $R$          &  $I$        & Telescope               \\
         &(mag)		    & (mag)             & (mag)         & (mag)         & (mag)                     \\
\hline                           
57274.4258  & 17.35 $\pm$ 0.09 &	17.28 $\pm$  0.07   &   17.07 $\pm$ 0.05  &   16.75 $\pm$  0.06  &    KAIT  \\  
57275.3867  & 16.94 $\pm$ 0.09 &	16.83 $\pm$  0.06   &   16.67 $\pm$ 0.05  &   16.39 $\pm$  0.06  &    KAIT  \\
57276.3516  & 16.40 $\pm$ 0.17 &	16.40 $\pm$  0.11   &   16.39 $\pm$ 0.08  &   16.05 $\pm$  0.07  &    KAIT   \\
57279.4336  & 15.79 $\pm$ 0.11 &	15.84 $\pm$  0.06   &   15.64 $\pm$ 0.05  &   15.39 $\pm$  0.05  &    KAIT   \\
57283.4375  & 15.44 $\pm$ 0.01 &	15.26 $\pm$  0.01   &   15.14 $\pm$ 0.01  &   14.93 $\pm$  0.01  &    Nickel   \\
57285.5078  & 15.64 $\pm$ 0.08 &	15.44 $\pm$  0.04   &   15.20 $\pm$ 0.03  &   14.90 $\pm$  0.03  &    KAIT   \\
57286.4062  & 15.70 $\pm$ 0.04 &	15.48 $\pm$  0.03   &   15.14 $\pm$ 0.02  &   14.87 $\pm$  0.02  &    KAIT   \\ 
57287.4375  & 15.91 $\pm$ 0.05 &	15.47 $\pm$  0.03   &   15.18 $\pm$ 0.03  &   14.87 $\pm$  0.03  &    KAIT   \\
57288.4102  & 15.89 $\pm$ 0.05 &	15.51 $\pm$  0.03   &   15.17 $\pm$ 0.03  &   14.86 $\pm$  0.03  &    KAIT   \\
57290.3555  & 16.07 $\pm$ 0.07 &	15.55 $\pm$  0.05   &   15.20 $\pm$ 0.03  &   14.87 $\pm$  0.03  &    KAIT   \\
57291.3594  & 16.30 $\pm$ 0.09 &	15.58 $\pm$  0.04   &   15.19 $\pm$ 0.03  &   14.88 $\pm$  0.013  &    KAIT   \\
57297.4023  & 17.06 $\pm$ 0.20 &	16.27 $\pm$  0.07   &   15.71 $\pm$ 0.04 &   15.19 $\pm$  0.04  &    KAIT   \\
57298.3906  & 17.30 $\pm$ 0.12 &	16.20 $\pm$  0.04   &   15.75 $\pm$ 0.03  &   15.23 $\pm$  0.03  &    KAIT   \\
57299.5039  & 17.35 $\pm$ 0.15 &	16.53 $\pm$  0.05   &   15.91 $\pm$ 0.03  &   15.22 $\pm$  0.03  &    KAIT   \\
57301.4258  & 17.52 $\pm$ 0.11 &	16.48 $\pm$  0.04   &   15.98 $\pm$ 0.03  &   15.35 $\pm$  0.03  &    KAIT   \\
57302.3125  & 17.43 $\pm$ 0.01 &	16.37 $\pm$  0.01   &   15.86 $\pm$ 0.01  &   15.33 $\pm$  0.01  &    Nickel   \\
57302.3672  & 17.47 $\pm$ 0.08 &	16.52 $\pm$  0.04   &   16.05 $\pm$ 0.03  &   15.46 $\pm$  0.03  &    KAIT   \\
57304.4727  & 17.68 $\pm$ 0.12 &	16.76 $\pm$  0.07   &   16.19 $\pm$ 0.04  &   15.57 $\pm$  0.03  &    KAIT   \\
57305.3828  & 17.90 $\pm$ 0.12 &	16.76 $\pm$  0.05   &   16.27 $\pm$ 0.03  &   15.61 $\pm$  0.03  &    KAIT   \\
57306.3516  & 17.88 $\pm$ 0.18 &	16.84 $\pm$  0.06   &   16.26 $\pm$ 0.04  &   15.65 $\pm$  0.04  &    KAIT   \\
57307.3438  & 17.74 $\pm$ 0.01 &	16.72 $\pm$  0.01   &   16.18 $\pm$ 0.01  &   15.63 $\pm$  0.01  &    Nickel   \\
57307.3945  & 17.82 $\pm$ 0.13 &	16.86 $\pm$  0.04   &   16.27 $\pm$ 0.03  &   15.76 $\pm$  0.04  &    KAIT   \\
57308.3633  & 17.82 $\pm$ 0.12 &	16.83 $\pm$  0.05   &   16.34 $\pm$ 0.03  &   15.75 $\pm$  0.04  &    KAIT   \\

\hline                                   
\end{tabular}
\label{tab:optical_observations_SN2015ap1}      
\end{table*}

\begin{table*}
\caption{Photometric observations of SN~2015ap, continued...}
\centering
\smallskip
\begin{tabular}{c c c c c c c c}
\hline \hline
MJD  	    &  $B$              &  $V$          &  $R$          &  $I$        & Telescope               \\
         &(mag)		    & (mag)             & (mag)         & (mag)         & (mag)                     \\
\hline          
57309.4609  & 17.91 $\pm$ 0.16 &	16.89 $\pm$  0.08   &   16.35 $\pm$ 0.04  &   15.73 $\pm$  0.04  &    KAIT   \\
57317.3047  & 17.98 $\pm$ 0.02 &	17.10 $\pm$  0.01   &   16.64 $\pm$ 0.01  &   15.99 $\pm$  0.01  &    Nickel  \\
57317.4492  & 18.31 $\pm$ 0.24 &	17.25 $\pm$  0.08   &   16.75 $\pm$ 0.05  &   16.02 $\pm$  0.04  &    KAIT   \\
57318.3008  & 18.19 $\pm$ 0.27 &	17.30 $\pm$  0.09   &   16.72 $\pm$ 0.05  &   16.12 $\pm$  0.05  &    KAIT   \\
57319.3984  & 18.22 $\pm$ 0.20 &	17.48 $\pm$  0.12   &   16.80 $\pm$ 0.05  &   16.12 $\pm$  0.07  &    KAIT   \\
57325.4297  & 18.25 $\pm$ 0.38 &	17.14 $\pm$  0.14   &   17.15 $\pm$ 0.09  &   16.19 $\pm$  0.07  &    KAIT   \\
57326.3594  & 18.42 $\pm$ 0.29 &	17.43 $\pm$  0.07   &   17.03 $\pm$ 0.05  &   16.35 $\pm$  0.04  &    KAIT   \\
57327.4297  & 17.98 $\pm$ 0.26 &	17.40 $\pm$  0.12   &            -        &        -             &    KAIT   \\                 
57330.3906   & 18.37 $\pm$ 0.22 &	17.58 $\pm$  0.08   &   17.20 $\pm$ 0.06  &   16.46 $\pm$  0.05  &    KAIT   \\
57331.3359  & 18.63 $\pm$ 0.22 &	17.72 $\pm$  0.08   &   17.16 $\pm$ 0.04  &   16.59 $\pm$  0.05  &    KAIT   \\
57332.3047  & 18.21 $\pm$ 0.01 &	17.47 $\pm$  0.01   &   17.11 $\pm$ 0.01  &   16.47 $\pm$  0.01  &    Nickel   \\
57332.3555  & 18.31 $\pm$ 0.18 &	17.57 $\pm$  0.07   &   17.21 $\pm$ 0.04  &   16.56 $\pm$  0.04  &    KAIT   \\
57333.3477  & 18.19 $\pm$ 0.17 &	17.73 $\pm$  0.09   &   17.23 $\pm$ 0.06  &   16.56 $\pm$  0.05  &    KAIT  \\
57334.3789  & 18.35 $\pm$ 0.16 &	17.66 $\pm$  0.08   &   17.25 $\pm$ 0.06  &   16.51 $\pm$  0.05  &    KAIT   \\
57337.3281  & 18.33 $\pm$ 0.20 &	18.01 $\pm$  0.18   &   17.33 $\pm$ 0.05  &   16.85 $\pm$  0.06  &    KAIT   \\
57338.2852  & 18.62 $\pm$ 0.28 &	17.79 $\pm$  0.12   &   17.30 $\pm$ 0.05  &   16.73 $\pm$  0.05  &    KAIT   \\
57339.3125  & 18.45 $\pm$ 0.22 &	17.83 $\pm$  0.08   &   17.46 $\pm$ 0.05  &   16.75 $\pm$  0.06  &    KAIT   \\
57339.3438  & 18.28 $\pm$ 0.02 &	17.64 $\pm$  0.01   &   17.27 $\pm$ 0.02  &   16.66 $\pm$  0.01  &    Nickel   \\
57340.3594  & 18.52 $\pm$ 0.18 &	17.76 $\pm$  0.07   &   17.40 $\pm$ 0.06  &   16.73 $\pm$  0.06    &    KAIT   \\
57343.3906  & 18.31 $\pm$ 0.02 &	17.66 $\pm$  0.01   &   17.38 $\pm$ 0.02  &   16.64 $\pm$  0.01  &    Nickel   \\
57347.3164  & 18.51 $\pm$ 0.03 &	17.73 $\pm$  0.01   &   17.51 $\pm$ 0.02  &   17.02 $\pm$  0.02  &    Nickel   \\
57361.2656  & 18.55 $\pm$ 0.02 &	18.14 $\pm$  0.01   &   17.93 $\pm$ 0.02  &   17.19 $\pm$  0.01  &    Nickel   \\
57425.1641  & 19.22 $\pm$ 0.06 &	19.63 $\pm$  0.07   &            -        &        -     &    Nickel   \\
57431.1250  & 19.51 $\pm$ 0.10 &	19.48 $\pm$  0.10   &   19.05 $\pm$ 0.34  &        -     &    Nickel   \\
57444.1328  & 19.67 $\pm$ 0.09 &	20.06 $\pm$  0.10   &   19.82 $\pm$ 0.09  &   18.93 $\pm$  0.06  &    Nickel   \\

\hline                                   
\end{tabular}
\label{tab:optical_observations_SN2015ap2}      
\end{table*}

\begin{table*}
\caption {Log of {\it HST} observations of SN~2015ap.\label{tab:HST_log}}
Note: All exposures were 710\,s and there were no detections.
\begin{center}
 
\begin{tabular}{c c c c c c}
\hline \hline
UT Date	   &    Filter        \\
           &                  \\
\hline    
2017/08/16  &    	F555W     \\
2017/02/19  &    	F814W     \\
2016/04/03  &    	F555W     \\
2016/04/17  &    	F814W     \\

\hline
\end{tabular}
\end{center}
\end{table*}

\begin{landscape}
\begin{table*}
\caption{Photometric observations of SN~2016bau}
\centering
\smallskip
\begin{tabular}{c c c c c c c c}
\hline \hline
MJD  	    &  $B$              &  $V$          &  $R$    & $clear$      &  $I$        & Telescope               \\
         &(mag)		    & (mag)             & (mag)         & (mag)         & (mag)                     \\
\hline                           

57462.384   &    -              &       -           &   -              &  16.896	 $\pm$ 0.190	    &    -            \\
57463.381	& 17.196 $\pm$	 0.160	& 16.691 $\pm$	 0.131	& 16.488 $\pm$	 0.165	& 16.553 $\pm$	 0.226   	& 16.381 $\pm$	 0.124  & KAIT \\
57464.338	& 16.843 $\pm$	 0.119	& 16.408 $\pm$	 0.080	& 16.180 $\pm$	 0.102	& 16.199 $\pm$	 0.103   	& 16.079 $\pm$	 0.086  & KAIT \\
57465.324	& 16.879 $\pm$	 0.120	& 16.192 $\pm$	 0.066	& 16.027 $\pm$	 0.070	& 15.967 $\pm$	 0.131	    & 15.809 $\pm$	 0.106  & KAIT \\
57466.310	& 16.726 $\pm$	 0.134	& 16.005 $\pm$	 0.057	& 15.852 $\pm$	 0.055	& 15.838 $\pm$	 0.105	    & 15.661 $\pm$	 0.063  & KAIT \\
57467.330	& 16.787 $\pm$	 0.253	& 15.890 $\pm$	 0.103	& 15.608 $\pm$	 0.098	& 15.767 $\pm$	 0.156	    & 15.451 $\pm$	 0.128  & KAIT \\
57470.385	& 16.762 $\pm$	 0.165	& 15.741 $\pm$	 0.071	& 15.376 $\pm$	 0.074	& 15.374 $\pm$	 0.061	    & 15.165 $\pm$	 0.091  & KAIT \\
57471.397	& 16.401 $\pm$	 0.176	& 15.589 $\pm$	 0.059	& 15.278 $\pm$	 0.053	& 15.288 $\pm$	 0.062	    & 15.114 $\pm$	 0.056  & KAIT \\
57472.397	& 16.358 $\pm$	 0.140	& 15.477 $\pm$	 0.061	& 15.202 $\pm$	 0.051	& 15.254 $\pm$	 0.065      & 15.031 $\pm$	 0.055  & KAIT \\ 
57473.355	& 16.282 $\pm$	 0.114	& 15.467 $\pm$	 0.049	& 15.196 $\pm$	 0.049	& 15.206 $\pm$	 0.072	    & 14.982 $\pm$	 0.055  & KAIT \\ 
57474.383	& 16.106 $\pm$	 0.315	& 15.464 $\pm$	 0.083	& 15.107 $\pm$	 0.063	& 15.154 $\pm$	 0.075	    & 14.882 $\pm$	 0.114  & KAIT \\ 
57476.373	& 16.079 $\pm$	 0.131	& 15.398 $\pm$	 0.048	& 15.026 $\pm$	 0.049	& 15.067 $\pm$	 0.079	    & 14.806 $\pm$	 0.052  & KAIT \\ 
57477.367	& 16.274 $\pm$ 	 0.088	& 15.348 $\pm$	 0.042	& 15.048 $\pm$	 0.042	& 15.073 $\pm$	 0.046	    & 14.787 $\pm$	 0.042  & KAIT \\ 
57478.332	& 16.340 $\pm$	 0.082	& 15.376 $\pm$	 0.044	& 15.005 $\pm$	 0.042	& 15.055 $\pm$	 0.080	    & 14.744 $\pm$	 0.040  & KAIT \\
57479.395	& 16.290 $\pm$	 0.123	& 15.365 $\pm$	 0.054	& 14.989 $\pm$	 0.053	& 15.039 $\pm$	 0.089	    & 14.750 $\pm$	 0.060  & KAIT \\ 
57480.345	& 16.440 $\pm$	 0.087	& 15.422 $\pm$	 0.034	& 15.037 $\pm$	 0.038	& 15.062 $\pm$	 0.045	    & 14.753 $\pm$	 0.056  & KAIT \\
57481.338	& 16.487 $\pm$	 0.207	& 15.444 $\pm$	 0.070	& 15.102 $\pm$	 0.065	& 15.105 $\pm$	 0.073	    & 14.713 $\pm$	 0.062  & KAIT \\ 
57482.341	& 16.801 $\pm$	 0.142  & 15.527 $\pm$	 0.053	& 15.057 $\pm$	 0.045	& 15.132 $\pm$	 0.065	    & 14.784 $\pm$	 0.049  & KAIT \\
57485.358	& 17.453 $\pm$	 0.207	& 15.691 $\pm$	 0.115	& 15.237 $\pm$	 0.093	& 15.320 $\pm$	 0.090	    & 15.039 $\pm$	 0.090  & KAIT \\
57491.352	& 17.914 $\pm$	 0.243	& 16.282 $\pm$	 0.118	& 15.730 $\pm$	 0.119	& 15.744 $\pm$	 0.141	    & 15.224 $\pm$	 0.116  & KAIT \\
57492.352	& 17.765 $\pm$	 0.159	& 16.376 $\pm$	 0.103	& 15.776 $\pm$	 0.085	& 15.824 $\pm$	 0.137	    & 15.262 $\pm$	 0.073  & KAIT \\
57494.257	& 17.979 $\pm$	 0.169	& 16.484 $\pm$	 0.108	& 15.896 $\pm$	 0.111	& 15.941 $\pm$	 0.148	    & 15.392 $\pm$	 0.125  & KAIT \\
57495.299	& 18.142 $\pm$	 0.134	& 16.553 $\pm$ 	 0.069	& 15.914 $\pm$	 0.071	& 16.018 $\pm$	 0.095	    & 15.427 $\pm$	 0.064  & KAIT \\
57496.272   & 18.106 $\pm$	 0.131	& 16.603 $\pm$	 0.065	& 15.970 $\pm$	 0.063	& 16.048 $\pm$	 0.093	    & 15.445 $\pm$	 0.058  & KAIT \\
57497.268	& 18.194 $\pm$	 0.186	& 16.721 $\pm$	 0.086	& 15.966 $\pm$	 0.065	& 16.100 $\pm$	 0.084	    & 15.492 $\pm$	 0.067  & KAIT \\
57498.330   & 18.137 $\pm$	 0.145	& 16.646 $\pm$	 0.094	& 16.065 $\pm$	 0.087	& 16.129 $\pm$	 0.114	    & 15.557 $\pm$   0.079  & KAIT \\
57499.317	& 18.147 $\pm$	 0.329	& 16.675 $\pm$	 0.149	& 16.111 $\pm$	 0.141	& 16.193 $\pm$	 0.230	    & 15.555 $\pm$	 0.131  & KAIT \\
\hline                                   
\end{tabular}
\label{tab:optical_observations_2016bau1}      
\end{table*}
\end{landscape}

\begin{landscape}
\begin{table*}
\caption{Photometric observations of SN~2016bau, continued...}
\centering
\smallskip
\begin{tabular}{c c c c c c c c}
\hline \hline
MJD  	    &  $B$              &  $V$          &  $R$    & $clear$      &  $I$        & Telescope               \\
         &(mag)		    & (mag)             & (mag)         & (mag)         & (mag)                     \\
\hline                           

57502.338	& 18.204 $\pm$	 0.277	& 16.935 $\pm$	 0.170	& 16.180 $\pm$	 0.204	& 16.201 $\pm$	 0.147	    & 15.674 $\pm$	 0.095  & KAIT \\
57504.271	& 18.340 $\pm$	 0.173	& 16.893 $\pm$	 0.111	& 16.291 $\pm$	 0.114	& 16.313 $\pm$	 0.165	    & 15.738 $\pm$ 0.095  & KAIT \\
57505.327	&     -             &         -         &         -         & 16.297	  $\pm$ 0.148	    &     -           & KAIT \\
57507.265	& 18.362 $\pm$	 0.179	& 16.974 $\pm$	 0.109	& 16.372 $\pm$	 0.116	& 16.408 $\pm$	 0.184	    & 15.803 $\pm$	 0.089  & KAIT \\
57508.241	& 18.348 $\pm$	 0.248	& 16.993 $\pm$	 0.197	& 16.445 $\pm$	 0.210	& 16.408 $\pm$	 0.237	    & 15.837 $\pm$	 0.163  & KAIT \\
57509.285	& 18.414 $\pm$	 0.156	& 17.023 $\pm$	 0.110	& 16.401 $\pm$	 0.119	& 16.441 $\pm$	 0.181	    & 15.825 $\pm$	 0.098  & KAIT \\
57510.229	&	-               &      -            &	      -         & 16.458	 $\pm$ 0.172	    &     -             & KAIT \\
57511.236	& 18.498 $\pm$	 0.154	& 17.092 $\pm$	 0.121	& 16.484 $\pm$	 0.086	& 16.447 $\pm$	 0.178	    & 15.848 $\pm$	 0.087  & KAIT \\
57513.195	& 18.490 $\pm$	 0.417	& 17.150 $\pm$	 0.163	& 16.552 $\pm$	 0.134	& 16.517 $\pm$	 0.195	    & 15.929 $\pm$	 0.112  & KAIT \\
57517.272	& 18.501 $\pm$	 0.189	& 17.238 $\pm$	 0.104	& 16.619 $\pm$	 0.116	& 16.594 $\pm$	 0.221	    & 16.001 $\pm$	 0.096  & KAIT \\
57518.210	& 18.590 $\pm$	 0.151	& 17.221 $\pm$	 0.129	& 16.668 $\pm$	 0.114	& 16.591 $\pm$	 0.139	    & 15.993 $\pm$	 0.094  & KAIT \\
57519.223	& 18.514 $\pm$	 0.156	& 17.211 $\pm$	 0.094	& 16.671 $\pm$	 0.095	& 16.612 $\pm$ 	 0.127	    & 16.043 $\pm$	 0.101  & KAIT \\
57520.193	& 18.453 $\pm$	 0.167	& 17.284 $\pm$	 0.148	& 16.669 $\pm$	 0.099	& 16.651 $\pm$	 0.349	    & 16.039 $\pm$	 0.102  & KAIT \\
57521.203	& 18.651 $\pm$	 0.179	& 17.276 $\pm$	 0.126	& 16.737 $\pm$	 0.121	& 16.699 $\pm$	 0.119	    & 16.057 $\pm$	 0.093  & KAIT \\
57522.186	& 18.638 $\pm$	 0.215	& 17.359 $\pm$	 0.186	& 16.744 $\pm$	 0.190	& 16.727 $\pm$	 0.263	    & 16.074 $\pm$	 0.155  & KAIT \\
57523.259	& 18.510 $\pm$	 0.235	& 17.396 $\pm$	 0.194	& 16.697 $\pm$	 0.149	& 16.672 $\pm$	 0.322	    & 16.029 $\pm$	 0.118  & KAIT \\
57524.206	& 18.643 $\pm$	 0.147	& 17.346 $\pm$	 0.121	& 16.717 $\pm$	 0.116	& 16.753 $\pm$	 0.229	    & 16.068 $\pm$	 0.102  & KAIT \\
57525.193	& 18.675 $\pm$	 0.168	& 17.392 $\pm$	 0.090	& 16.802 $\pm$	 0.118	& 16.729 $\pm$	 0.134	    & 16.117 $\pm$	 0.110  & KAIT \\
57526.218	& 18.464 $\pm$	 0.205	& 17.358 $\pm$	 0.120	& 16.748 $\pm$	 0.113	& 16.739 $\pm$	 0.243	    & 16.097 $\pm$	 0.101  & KAIT \\
57527.200	& 18.490 $\pm$	 0.166	& 17.401 $\pm$ 	 0.092	& 16.824 $\pm$	 0.116	& 16.846 $\pm$	 0.201	    & 16.115 $\pm$  	 0.089  & KAIT \\
57531.212	& 18.761 $\pm$	 0.175	& 17.450 $\pm$	 0.096	& 16.854 $\pm$	 0.102	& 16.832 $\pm$	 0.221	    & 16.214 $\pm$	 0.114  & KAIT \\
57533.227	& 18.663 $\pm$	 0.143	& 17.541 $\pm$ 	 0.109	& 16.914 $\pm$	 0.143	& 16.848 $\pm$	 0.249	    & 16.209 $\pm$	 0.075  & KAIT \\
57534.202	& 18.583 $\pm$	 0.130	& 17.519 $\pm$	 0.085	& 16.984 $\pm$	 0.133	& 16.933 $\pm$	 0.264  	& 16.271 $\pm$	 0.102  & KAIT \\
57535.215	& 18.683 $\pm$	 0.163	& 17.558 $\pm$	 0.110	& 16.998 $\pm$	 0.112	& 16.975 $\pm$	 0.277	    & 16.269 $\pm$	 0.105  & KAIT \\

\hline                                   
\end{tabular}
\label{tab:optical_observations_2016bau2}      
\end{table*}
\end{landscape}

\begin{landscape}
\begin{table*}
\caption{Photometric observations of SN~2016bau, continued...}
\centering
\smallskip
\begin{tabular}{c c c c c c c c}
\hline \hline
MJD  	    &  $B$              &  $V$          &  $R$    & $clear$      &  $I$        & Telescope               \\
         &(mag)		    & (mag)             & (mag)         & (mag)         & (mag)                     \\
\hline                           
57536.219	& 18.693 $\pm$	 0.177	& 17.587 $\pm$	 0.101	& 16.981 $\pm$	 0.103	& 16.958 $\pm$	 0.257	    & 16.275 $\pm$	 0.076  & KAIT \\
57537.232	& 18.818 $\pm$	 0.224	& 17.562 $\pm$	 0.098	& 17.001 $\pm$	 0.093	& 17.010 $\pm$	 0.270	    & 16.257 $\pm$	 0.093  & KAIT \\
57541.231	& 18.793 $\pm$	 0.151	& 17.682 $\pm$	 0.086	& 17.137 $\pm$	 0.117	& 16.981 $\pm$	 0.263	    & 16.355 $\pm$	 0.097  & KAIT \\
57552.211	& 18.943 $\pm$	 0.150	& 17.819 $\pm$	 0.124	& 17.316 $\pm$	 0.100	& 17.231 $\pm$	 0.070	    & 16.525 $\pm$	 0.088  & KAIT \\
57572.191	& 19.171 $\pm$	 0.331	& 18.126 $\pm$ 	 0.132	& 17.587 $\pm$	 $\pm$0.117	& 17.422 $\pm$	 0.139	& 16.842 $\pm$	 0.118  & KAIT \\
57574.190	&      -                &      -                &         -                      & 17.403 $\pm$   0.091	&      -                & KAIT \\
57576.191	&      -                &      -                &         -                      & 17.544 $\pm$ 0.519	&      -                & KAIT \\
57471.326	& 16.267 $\pm$	 0.022	& 15.589 $\pm$	 0.015	& 15.274$\pm$       0.019    	& 	-	                & 15.065 $\pm$	 0.027  & KAIT \\
57484.382	& 16.823 $\pm$	 0.029	& 15.501 $\pm$	 0.034	& 15.213 $\pm$	 0.014    	& 	-	                & 14.842 $\pm$	 0.022  & KAIT \\
57498.264	& 17.822 $\pm$	 0.081	& 16.741 $\pm$	 0.057  & 16.069 $\pm$	 0.031	    &   -	                & 15.551 $\pm$	 0.035  & KAIT \\
57511.285	& 18.314 $\pm$	 0.376	& 17.054 $\pm$	 0.046	& 16.397 $\pm$	 0.025   	&   -	                & 15.843 $\pm$	 0.616  & KAIT \\
57535.248	& 18.563 $\pm$	 0.058	& 17.544 $\pm$	 0.035  & 16.948 $\pm$	 0.044    	& 	-	                & 16.289 $\pm$	 0.072  & KAIT \\
57541.241	& 18.555 $\pm$	 0.719	& 17.565 $\pm$	 0.070	& 17.014 $\pm$    0.034    	&   -	                & 16.347 $\pm$	 0.028  & KAIT \\
57546.216	& 18.816 $\pm$	 0.229	& 17.691 $\pm$	 0.070	& 17.139 $\pm$	 0.028    	&   -	                & 16.447 $\pm$	 0.099  & KAIT \\
57587.185	& 19.431 $\pm$	 0.701	& 18.453 $\pm$	 0.309	&      -                         &   -                   &       -               & KAIT      \\
\hline                                   
\end{tabular}
\label{tab:optical_observations_2016bau3}      
\end{table*}
\end{landscape}

\chapter{\sc Low-luminosity Type IIb SN~2016iyc arising from a low mass progenitor}
\ifpdf
    \graphicspath{{Appendix/ChapterAFigs/PNG/}{Appendix/ChapterAFigs/PDF/}{Appendix/ChapterAFigs/}}
\else
    \graphicspath{{Appendix/ChapterAFigs/EPS/}{Appendix/ChapterAFigs/}}
\fi

\section{Additional Figures and Tables}

\begin{figure*}
\includegraphics[width=\columnwidth]{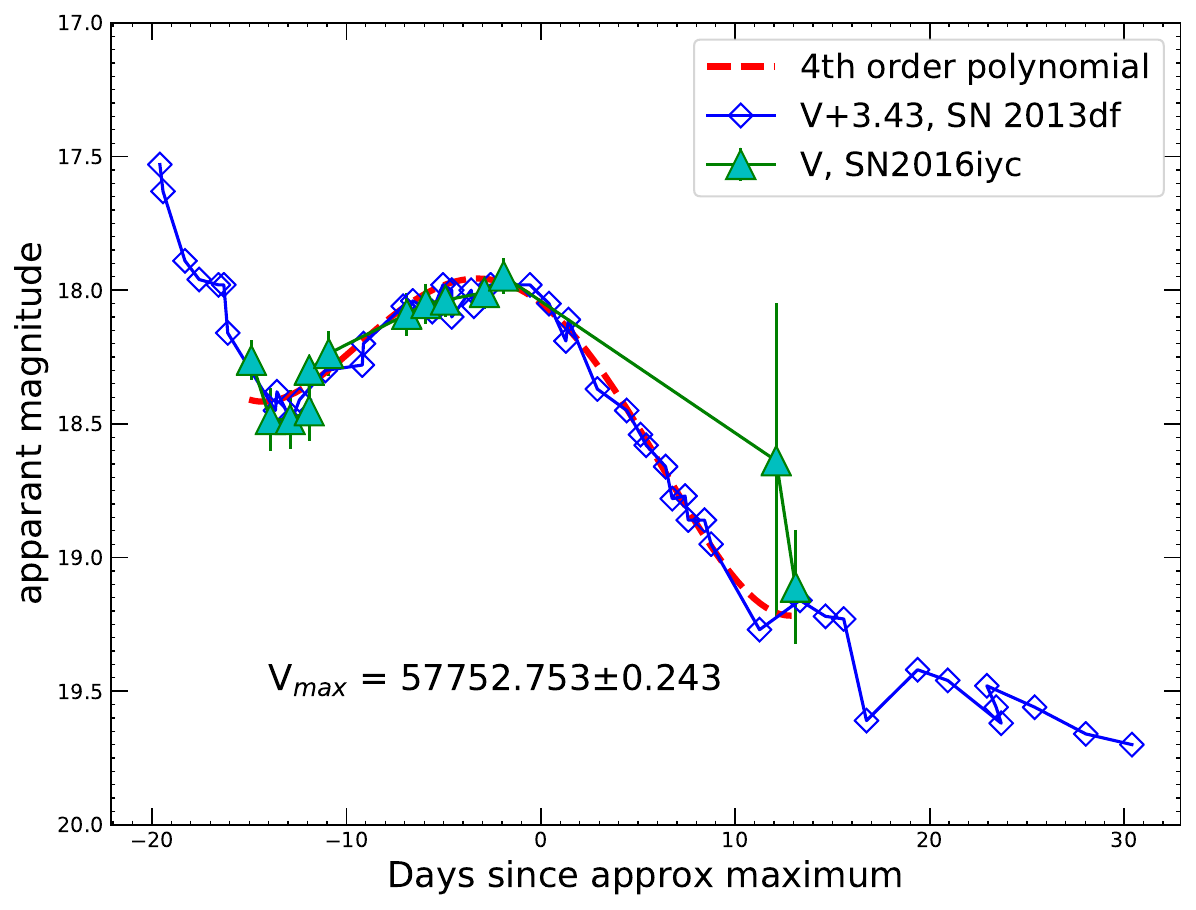}
\caption{Estimation of $V$-band maximum.}
\label{fig:V_max}
\end{figure*}

\begin{figure*}
\centering
    \includegraphics[height=6.0cm,width=0.49\columnwidth]{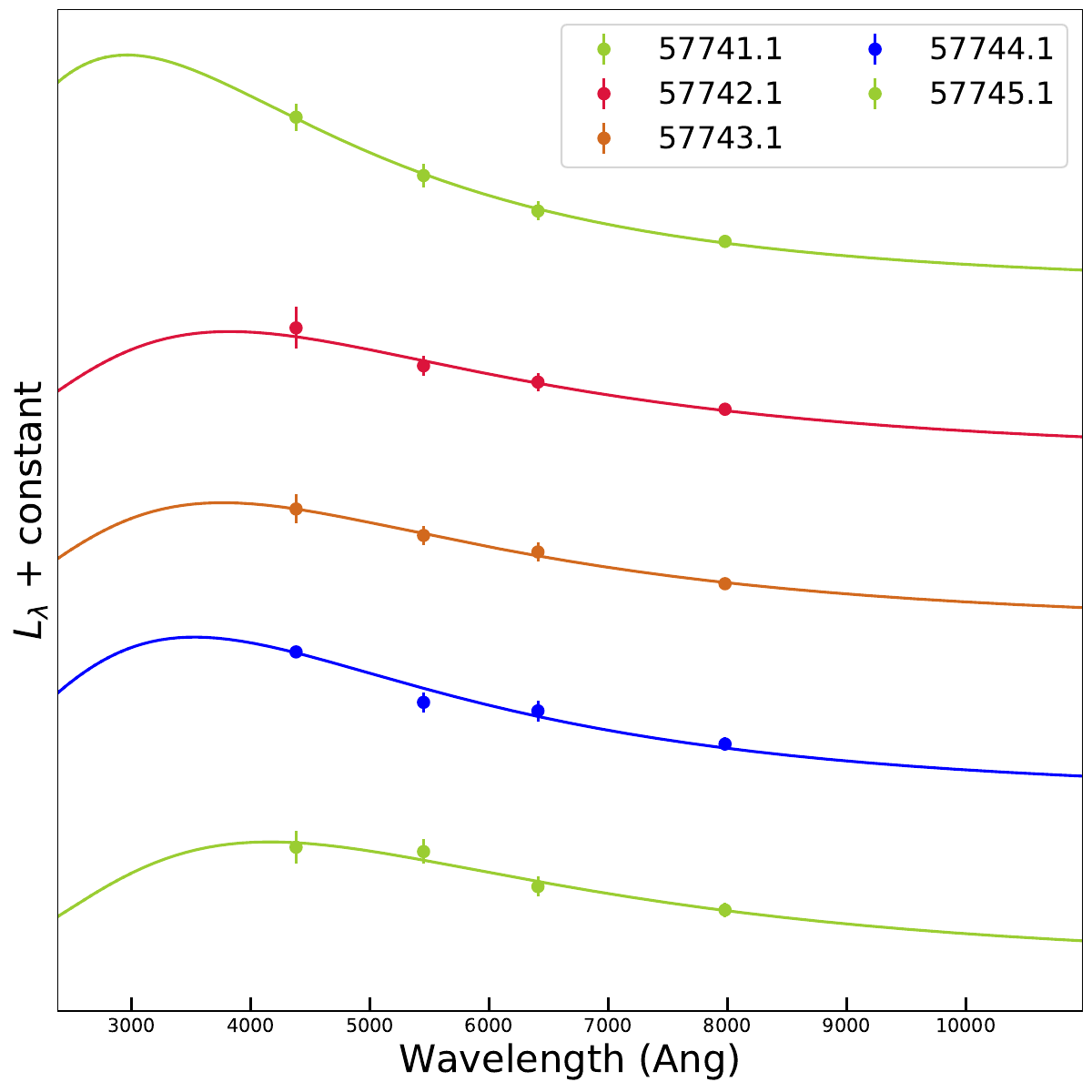}
    \includegraphics[height=6.0cm,width=0.49\columnwidth]{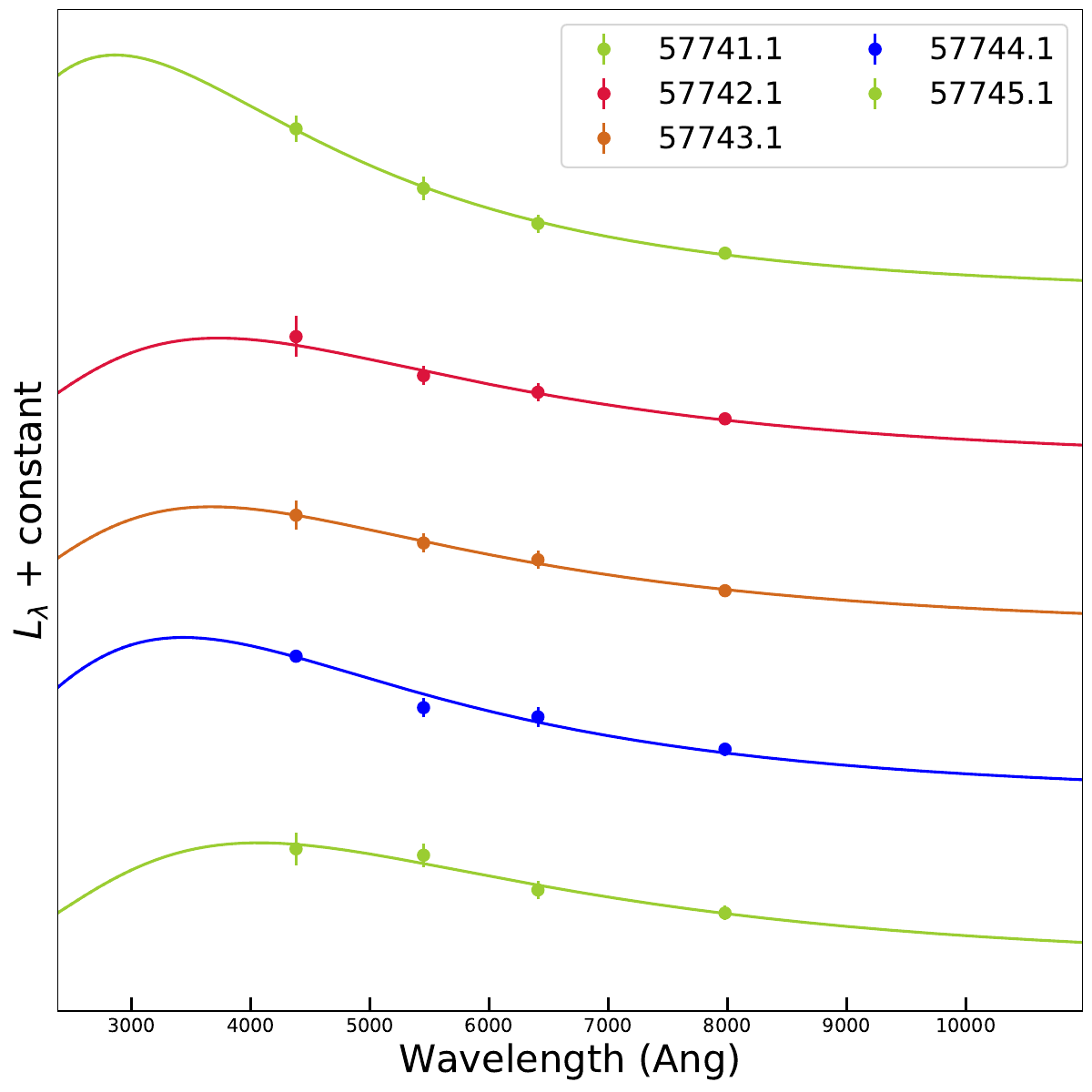}
    \includegraphics[height=6.0cm,width=0.49\columnwidth]{bb_fits_iyc_0.00_BVRI.pdf}
    \includegraphics[height=6.0cm,width=0.49\columnwidth]{bb_fits_iyc_0.02_BVRI.pdf}
    \includegraphics[height=6.0cm,width=0.49\columnwidth]{bb_fits_iyc_0.00_BVRI.pdf}
    \includegraphics[height=6.0cm,width=0.49\columnwidth]{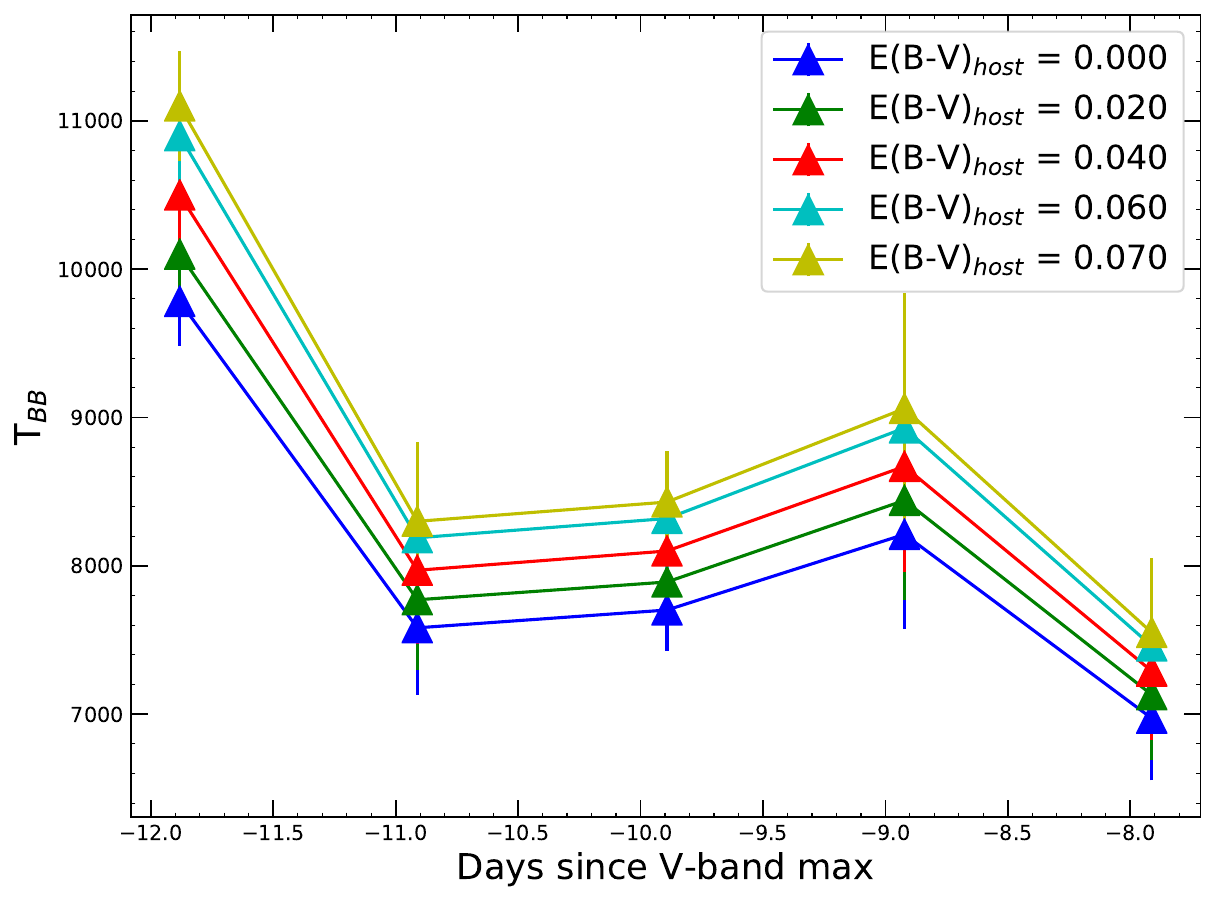}
   \caption{Fitting black-body curves to a few early epochs of SN~2016iyc by assuming different host-galaxy extinctions. The top-left and right panels show black-body fits to a few early-epoch SEDs of SN~2016iyc corresponding to host-galaxy extinctions of 0.00\,mag and 0.02\,mag, respectively. The middle-left and right panels show black-body fits to a few early-epoch SEDs of SN~2016iyc corresponding to host-galaxy extinctions of 0.04\,mag and 0.06\,mag, respectively. The bottom-left panel shows black-body fits to a few early-epoch SEDs of SN~2016iyc corresponding to a host-galaxy extinction of 0.07\,mag, while the bottom-right panel shows the variation of the black-body temperature obtained using black-body fits to the SEDs corresponding to different host-galaxy extinctions.}
    \label{fig:extinction_trial}
\end{figure*}

\begin{figure*}
\includegraphics[width=\columnwidth]{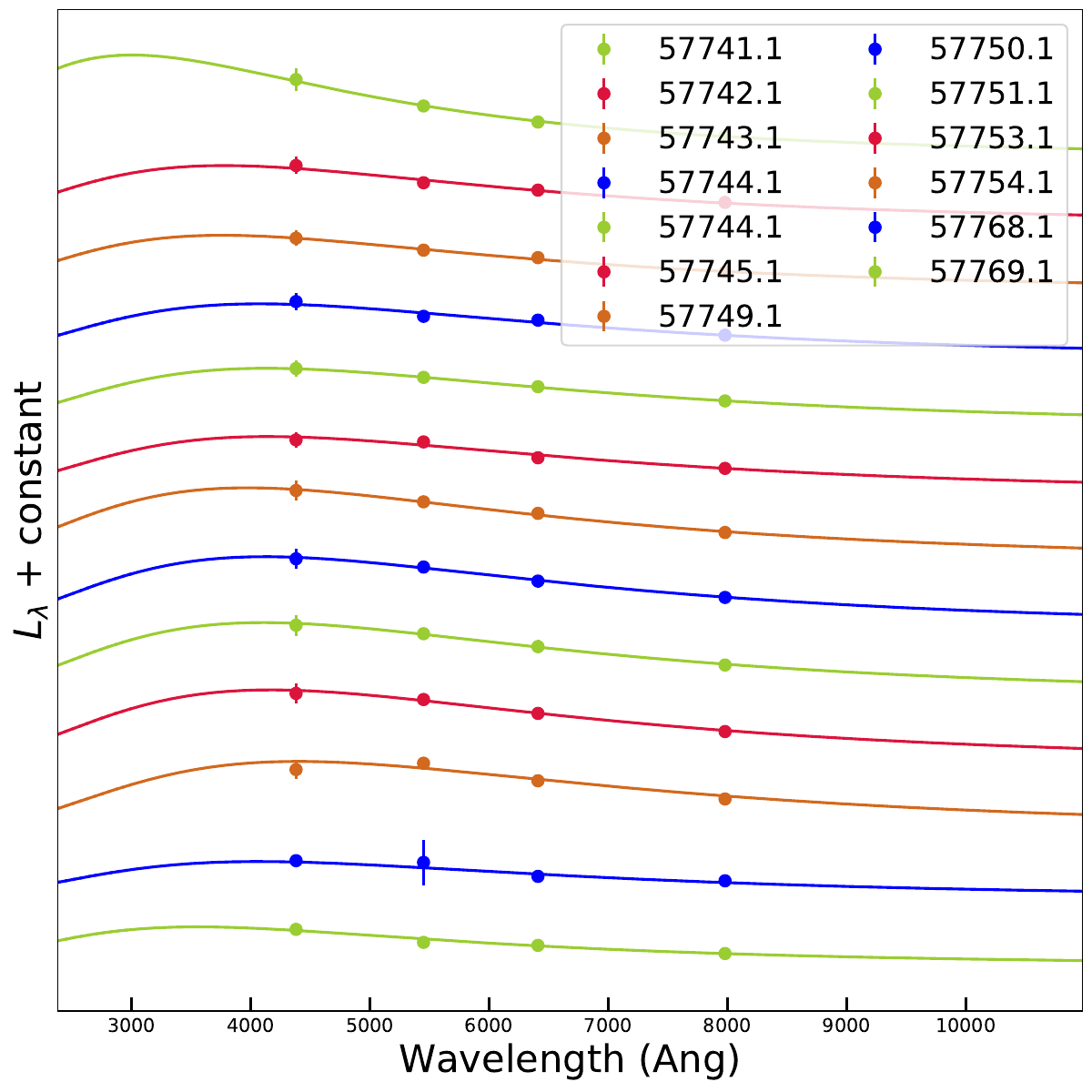}
\caption{The black-body fits to the SED of SN~2016iyc to estimate the bolometric light curve generated from {\tt SUPERBOL}.}
\label{fig:bb_fits}
\end{figure*}

\begin{figure*}
\includegraphics[width=\columnwidth]{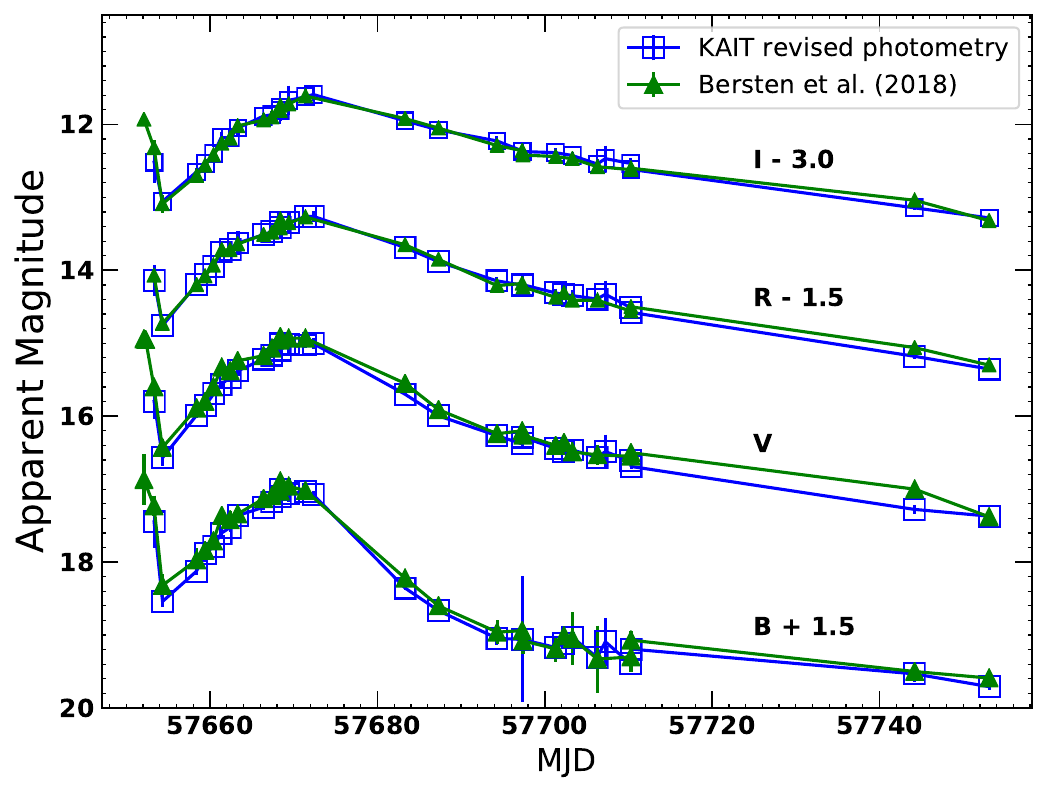}
\caption{Comparison between the KAIT revised photometry and the KAIT data used by \citet[][]{Bersten2018} for SN~2016gkg.}
\label{fig:SN2016gkg_comparison}
\end{figure*}

\begin{table*}
\caption{{\tt MESA} model and {\tt SNEC} explosion parameters of SN~2011fu and SN~2016gkg.}
\label{tab:MESA_MODELS_11fu_n_16gkg}
\begin{center}
{\scriptsize
\begin{tabular}{ccccccccccccc} 
\hline\hline
SN name &Model Name	& $M_{\rm ZAMS}$	& $Z$	 & $f_{ov}^{a}$ &	$M_{\mathrm{f}}^{b}$	 & $M_\mathrm{c}^{c}$	 & $M_{\mathrm{ej}}^{d}$	&	$M_{\mathrm{Ni}}^{e}$ &	$E_{\mathrm{exp}}^{f}$ 	\\
&	&	(M$_{\odot}$)	&	 &  & (M$_{\odot}$)	&	(M$_{\odot}$)  & (M$_{\odot}$) & (M$_{\odot}$) & 	($10^{51}$\,erg) 	\\
\hline
\hline
SN~2016gkg & He5\_A  &	18.0  	&	0.0200      & 0.01   &    5.00  & 1.6 &  3.40  & 0.087  &  1.30		\\
SN~2011fu & He5\_B  &	18.0  	&	0.0200       & 0.01   &    5.00  & 1.5 &  3.50 & 0.140  &  1.25		\\
\hline\hline
\end{tabular}}
\end{center}
%\par
{$^a$Overshoot parameter.
$^b$Final mass of the pre-SN model.
$^c$Final mass of the central remnant
$^d$Ejecta mass.
$^e$Nickel mass
$^f$Explosion energy.}\\

\end{table*}

\begin{landscape}
\begin{table*}
\caption{Photometry of SN~2016iyc}
\centering
\smallskip
\begin{tabular}{c c c c c c c c}
\hline \hline
MJD  	    &  $B$              &  $V$          &  $R$    & $C$      &  $I$        & Telescope               \\
         &(mag)		    & (mag)             & (mag)         & (mag)         & (mag)                     \\
\hline                           

57740.144	& ...	& ...  &  ...	& 17.807 $\pm$	 0.111   	& ...  & KAIT \\
57741.116	& 18.473 $\pm$	 0.075	& 18.262 $\pm$	 0.074	& 18.081 $\pm$	 0.091	& 17.985 $\pm$	 0.048   	& 17.875 $\pm$	 0.103  & KAIT \\
57742.089	& 18.774 $\pm$	 0.149	& 18.484 $\pm$	 0.118	& 18.102 $\pm$	 0.147	& 18.057 $\pm$	 0.149   	& 17.832 $\pm$	 0.148  & KAIT \\
57743.109	& 18.875 $\pm$	 0.114	& 18.484 $\pm$	 0.111	& 18.105 $\pm$	 0.117	& 17.987 $\pm$	 0.135	    & 17.949 $\pm$	 0.142  & KAIT \\
57744.087	& 18.796 $\pm$	 0.139	& 18.451 $\pm$	 0.113	& 17.957 $\pm$	 0.113	& 17.957 $\pm$	 0.143	    & 17.731 $\pm$	 0.126  & KAIT \\
57744.091	& 18.787 $\pm$	 0.054	& 18.299 $\pm$	 0.050	& 17.933 $\pm$	 0.067	& -- $\pm$	 --	    & 17.656 $\pm$	 0.119  & Nickel \\
57745.092	& 18.865 $\pm$	 0.129	& 18.237 $\pm$	 0.084	& 18.038 $\pm$	 0.106	& 17.855 $\pm$	 0.115	    & 17.658 $\pm$	 0.114  & KAIT \\
57749.111	& 18.568 $\pm$	 0.105	& 18.090 $\pm$	 0.081	& 17.728 $\pm$	 0.098	& 17.742 $\pm$	 0.131	    & 17.524 $\pm$	 0.130  & KAIT \\
57750.090	& 18.582 $\pm$	 0.100	& 18.050 $\pm$	 0.075	& 17.735 $\pm$	 0.080	& 17.693 $\pm$	 0.092	    & 17.433 $\pm$	 0.084  & KAIT \\
57751.095	& 18.567 $\pm$	 0.085	& 18.037 $\pm$	 0.063	& 17.693 $\pm$	 0.070	& 17.665 $\pm$	 0.055      & 17.439 $\pm$	 0.073  & KAIT \\ 
57751.095	& 18.567 $\pm$	 0.085	& 18.037 $\pm$	 0.063	& 17.693 $\pm$	 0.070	& 17.665 $\pm$	 0.055      & 17.439 $\pm$	 0.073  & KAIT \\
57753.099	& 18.578 $\pm$	 0.083	& 18.009 $\pm$	 0.061	& 17.679 $\pm$	 0.063	& 17.635 $\pm$	 0.091      & 17.408 $\pm$	 0.082  & KAIT \\
57754.091	& 18.724 $\pm$	 0.137	& 17.947 $\pm$	 0.066	& 17.678 $\pm$	 0.067	& 17.653 $\pm$	 0.090      & 17.409 $\pm$	 0.093  & KAIT \\
57768.107	& ...	& 18.638 $\pm$	 0.059	& 18.582 $\pm$	 0.395	& 18.570 $\pm$	 0.242      & 18.069 $\pm$	 0.321  & KAIT \\
57769.105	& 19.295 $\pm$	 0.464	& 19.111 $\pm$	 0.214	& 18.666 $\pm$	 0.194	& 18.784 $\pm$	 0.282      & 18.466 $\pm$	 0.232  & KAIT \\
57956.471	& $>$21.490 	& $>$21.336	& $>$21.542 	& ...       & $>$20.631   & Nickel\\
\hline                                   
\end{tabular}
\label{tab:optical_observations_2016iyc}      
\end{table*}
\end{landscape}

\begin{landscape}
\begin{table*}
\caption{Revised KAIT photometry of SN~2016gkg along with 3.6\,m DOT data}
\centering
\smallskip
\begin{tabular}{c c c c c c c c}
\hline \hline
MJD  	    &  $B$              &  $V$          &  $R$    & $C$      &  $I$        & Telescope               \\
         &(mag)		    & (mag)             & (mag)         & (mag)         & (mag)                     \\
\hline                           

57653.315	& 15.949 $\pm$	 0.357	& 15.799 $\pm$	 0.234	& 15.642 $\pm$	 0.214	& 15.661 $\pm$	 0.069   	& 15.518 $\pm$	 0.286  & KAIT \\
57654.322	& 17.039 $\pm$	 0.079	& 16.569 $\pm$	 0.113	& 16.257 $\pm$	 0.053	& 16.272 $\pm$	 0.030   	& 16.056 $\pm$	 0.068  & KAIT \\
57658.373	& 16.626 $\pm$	 0.054	& 15.990 $\pm$	 0.038	& 15.696 $\pm$	 0.045	& 15.745 $\pm$	 0.045   	& 15.652 $\pm$	 0.050  & KAIT \\
57659.444	& 16.383 $\pm$	 0.049	& 15.854 $\pm$	 0.041	& 15.557 $\pm$	 0.053	& 15.630 $\pm$	 0.042	    & 15.539 $\pm$	 0.063  & KAIT \\
57660.444	& 16.282 $\pm$	 0.047	& 15.690 $\pm$	 0.018	& 15.450 $\pm$	 0.022	& 15.429 $\pm$	 0.083	    & 15.401 $\pm$	 0.045  & KAIT \\
57661.316	& 16.106 $\pm$	 0.066	& 15.566 $\pm$	 0.052	& 15.249 $\pm$	 0.061	& 15.482 $\pm$	 0.071	    & 15.183 $\pm$	 0.079  & KAIT \\
57662.408	& 16.024 $\pm$	 0.060	& 15.466 $\pm$	 0.053	& 15.213 $\pm$	 0.059	& 15.238 $\pm$	 0.062	    & 15.182 $\pm$	 0.085  & KAIT \\
57663.334	& 15.861 $\pm$	 0.074	& 15.378 $\pm$	 0.068	& 15.125 $\pm$	 0.162	& 15.145 $\pm$	 0.068	    & 15.049 $\pm$	 0.109  & KAIT \\
57666.368	& 15.748 $\pm$	 0.039	& 15.217 $\pm$	 0.034	& 15.013 $\pm$	 0.045	& 14.992 $\pm$	 0.033	    & 14.887 $\pm$	 0.049  & KAIT \\
57667.373	& 15.676 $\pm$	 0.037	& 15.162 $\pm$	 0.038	& 14.969 $\pm$	 0.051	& 14.931 $\pm$	 0.025      & 14.868 $\pm$	 0.050  & KAIT \\ 
57668.362	& 15.504 $\pm$	 0.035	& 15.009 $\pm$	 0.012	& 14.838 $\pm$	 0.011	&   ...       & 14.763 $\pm$	 0.013  & KAIT \\
57668.375	& 15.605 $\pm$	 0.117	& 15.093 $\pm$	 0.060	& 14.916 $\pm$	 0.132	& 14.830 $\pm$	 0.084      & 14.805 $\pm$	 0.146  & KAIT \\
57669.369	& 15.565 $\pm$	 0.117	& 15.011 $\pm$	 0.139	& 14.845 $\pm$	 0.128	& 14.758 $\pm$	 0.180      & 14.671 $\pm$	 0.024  & KAIT \\
57671.420	& 15.536 $\pm$	 0.058	& 15.024 $\pm$	 0.043	& 14.760 $\pm$	 0.051	& 14.759 $\pm$	 0.063      & 14.621 $\pm$	 0.069  & KAIT \\
57672.330	& 15.576 $\pm$	 0.048	& 15.003 $\pm$	 0.074	& 14.760 $\pm$	 0.072	& 14.723 $\pm$	 0.054      & 14.590 $\pm$	 0.090  & KAIT \\
57683.307	& 16.856 $\pm$	 0.023	& 15.699 $\pm$	 0.016	& 15.187 $\pm$	 0.014	&    ...   	& 14.951 $\pm$	 0.014  & KAIT \\
57687.298	& 17.165 $\pm$	 0.040	& 15.995 $\pm$	 0.017	& 15.383 $\pm$	 0.016	&     ...   	& 15.076 $\pm$	 0.016  & KAIT \\
57694.279	& 17.545 $\pm$	 0.096	& 16.266 $\pm$	 0.049	& 15.645 $\pm$	 0.054	& 15.692 $\pm$	 0.053	    & 15.226 $\pm$	 0.069  & KAIT \\
57696.255	& 17.544 $\pm$	 0.021  & 16.302 $\pm$	 0.016	& 15.716 $\pm$	 0.023	&   ...	    &  ...  & 3.6\,m DOT \\
57697.300	& 17.560 $\pm$	 0.863	& 16.379 $\pm$	 0.076	& 15.703 $\pm$	 0.020	&     ...	    & 15.363 $\pm$	 0.019  & KAIT \\
57697.350	& 17.562 $\pm$	 0.106	& 16.286 $\pm$	 0.106	& 15.697 $\pm$	 0.118	& 15.815 $\pm$	 0.016	    & 15.372 $\pm$	 0.120  & KAIT \\
57701.256	& 17.669 $\pm$	 0.081	& 16.440 $\pm$	 0.052	& 15.809 $\pm$	 0.059	& 15.891 $\pm$	 0.040	    & 15.387 $\pm$	 0.059  & KAIT \\
57702.253	& 17.617 $\pm$	 0.038	& 16.487 $\pm$	 0.018	& 15.833 $\pm$	 0.018	&     ...	    &     ...  & KAIT \\
57703.289	& 17.532 $\pm$	 0.108	& 16.463 $\pm$	 0.095	& 15.847 $\pm$	 0.071	& 15.884 $\pm$	 0.055	    & 15.426 $\pm$	 0.053  & KAIT \\
57706.262	& 17.805 $\pm$	 0.156	& 16.562 $\pm$	 0.071	& 15.894 $\pm$	 0.059	& 15.965 $\pm$	 0.086      & 15.543 $\pm$	 0.071  & KAIT \\

\hline                                   
\end{tabular}
\label{tab:optical_observations_2016gkg1}      
\end{table*}
\end{landscape}

\begin{landscape}
\begin{table*}
\caption{Revised KAIT photometry of SN~2016gkg along with 3.6\,m DOT data, continued...}
\centering
\smallskip
\begin{tabular}{c c c c c c c c}
\hline \hline
MJD  	    &  $B$              &  $V$          &  $R$    & $C$      &  $I$        & Telescope               \\
         &(mag)		    & (mag)             & (mag)         & (mag)         & (mag)                     \\
\hline 
57707.237	& 17.592 $\pm$	 0.329	& 16.486 $\pm$	 0.232	& 15.829 $\pm$	 0.186	& 16.023 $\pm$	 0.037      & 15.469 $\pm$	 0.178  & KAIT \\
57710.259	& 17.893 $\pm$	 0.085	& 16.612 $\pm$	 0.055	& 16.011 $\pm$	 0.063	& 16.050 $\pm$	 0.046      & 15.535 $\pm$	 0.070  & KAIT \\
57710.312	& 16.692 $\pm$	 0.084	& 16.689 $\pm$	 0.038	& 16.077 $\pm$	 0.035	&     ...      & 15.617 $\pm$	 0.021  & KAIT \\ 
57744.149	& 18.032 $\pm$	 0.109	& 17.277 $\pm$	 0.061	& 16.678 $\pm$	 0.035	&     ...      & 16.143 $\pm$	 0.027  & KAIT \\
57753.135	& 18.208 $\pm$	 0.039	& 17.371 $\pm$	 0.089	& 16.854 $\pm$	 0.029	&     ...      & 16.282 $\pm$	 0.026  & KAIT \\
58080.168	& 22.541 $\pm$	 0.258    &  21.751 $\pm$	 0.225	& 20.821 $\pm$	 0.03	&   ...	    &  20.137 $\pm$	 0.055  & 3.6\,m DOT \\

\hline                                   
\end{tabular}
\label{tab:optical_observations_2016gkg2}      
\end{table*}
\end{landscape}

\chapter{\sc Investigating the evolutions of Rotating, Population III star of 25\,M$_{\odot}$ and the strength of Resulting Supernovae}
\ifpdf
    \graphicspath{{Appendix/ChapterAFigs/PNG/}{Appendix/ChapterAFigs/PDF/}{Appendix/ChapterAFigs/}}
\else
    \graphicspath{{Appendix/ChapterAFigs/EPS/}{Appendix/ChapterAFigs/}}
\fi

\section{Additional Figures}

\begin{landscape}
\begin{figure*}
%    \centering
    \includegraphics[height=5cm]{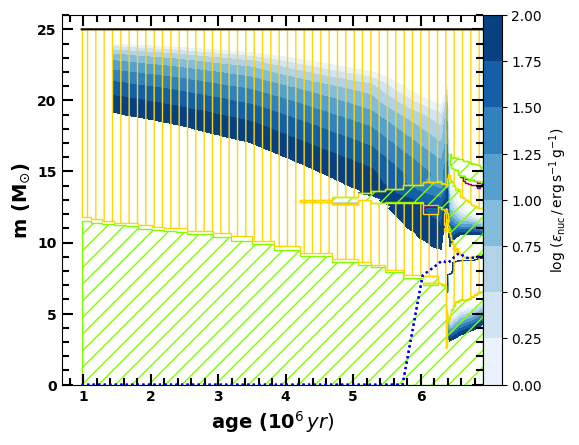}
    \includegraphics[height=5cm]{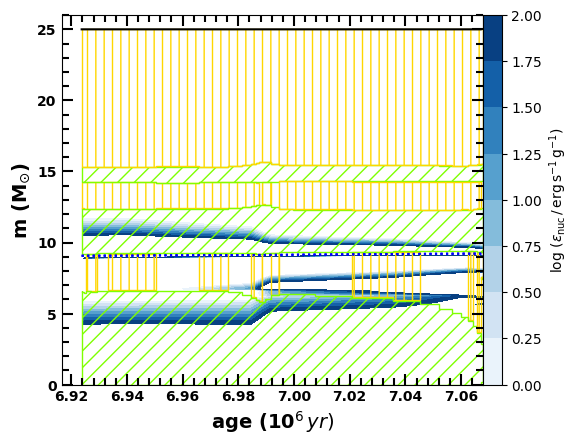}
    \includegraphics[height=5cm]{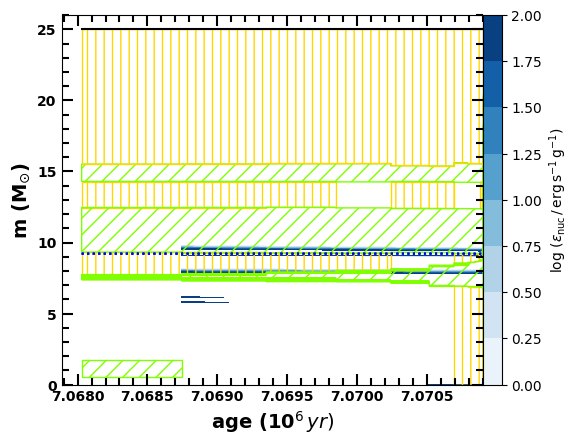}
    \includegraphics[height=5cm]{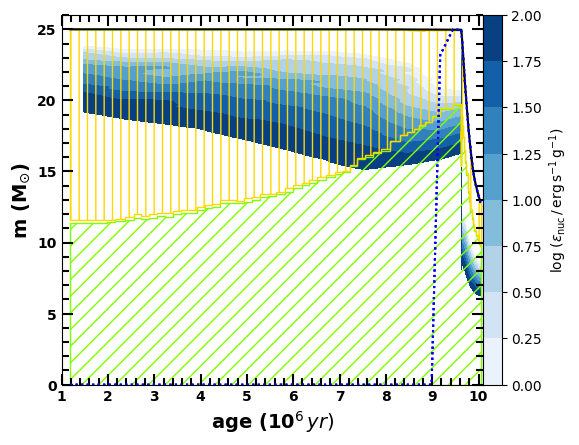}\hspace{1.45cm}
    \includegraphics[height=5cm]{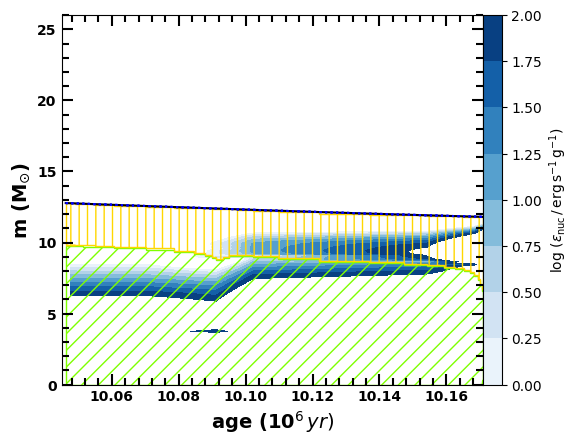}\hspace{1.45cm}
    \includegraphics[height=5cm]{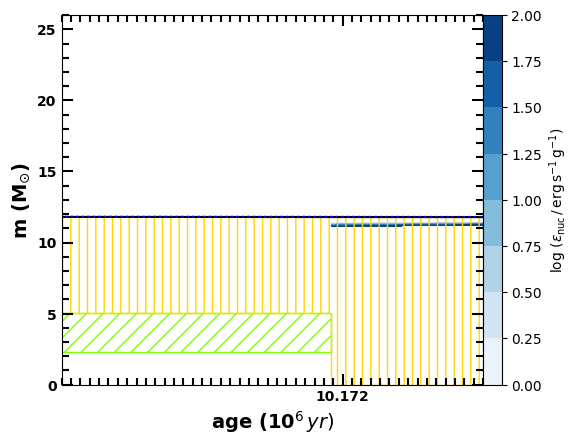}
   \caption {The kippenhahn diagrams of the models M25\_Z0.00\_Rot0.2 (top) and M25\_Z0.00\_Rot0.8 (bottom) for a period between ZAMS to close to the pre-SN stage. Here, the green hatchings indicate the convective regions and the dark-yellow regions mark the stellar interiors where the thermohaline mixing is going on. Also, the logarithm of the specific nuclear energy generation rate ($\epsilon_{\rm nuc}$) inside the stellar interiors is indicated by the blue colour gradients. The rapidly rotating model is significantly stripped.}
    \label{fig:kippenhahn}
\end{figure*}
\end{landscape}

%\begin{landscape}
\begin{figure*}
  \centering
  \includegraphics[height=9cm,width=\textwidth]{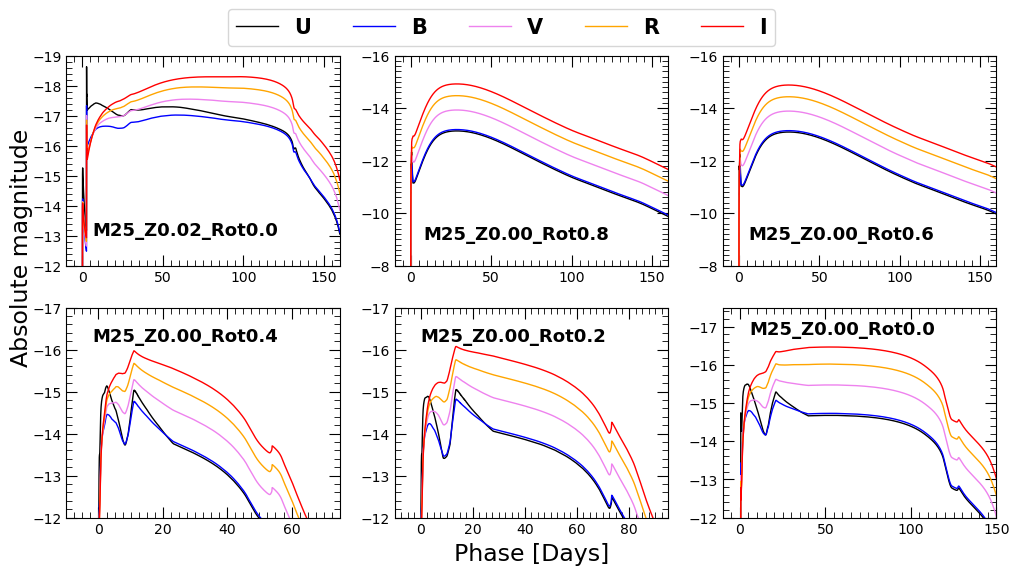}
    \caption{The U, B, V, R, and I band light curves resulting from the synthetic explosions of Pop III models using {\tt SNEC}. The non-rotating and slowly rotating models ($\Omega\,\leq\,0.4\,\Omega_{\rm crit}$) form a class of weak Type II SNe, while the rapidly rotating models ($\Omega\,\ge\,0.6\,\Omega_{\rm crit}$) result into Type Ib/c SNe within the specified limits of explosion energies and Nickel masses. The light curves resulting from the non-rotating, solar metallicity model are also shown for comparison.} 
    \label{fig:lc}
\end{figure*}
%\end{landscape}

\begin{figure*}
  \centering
  \includegraphics[height=10cm,width=\textwidth]{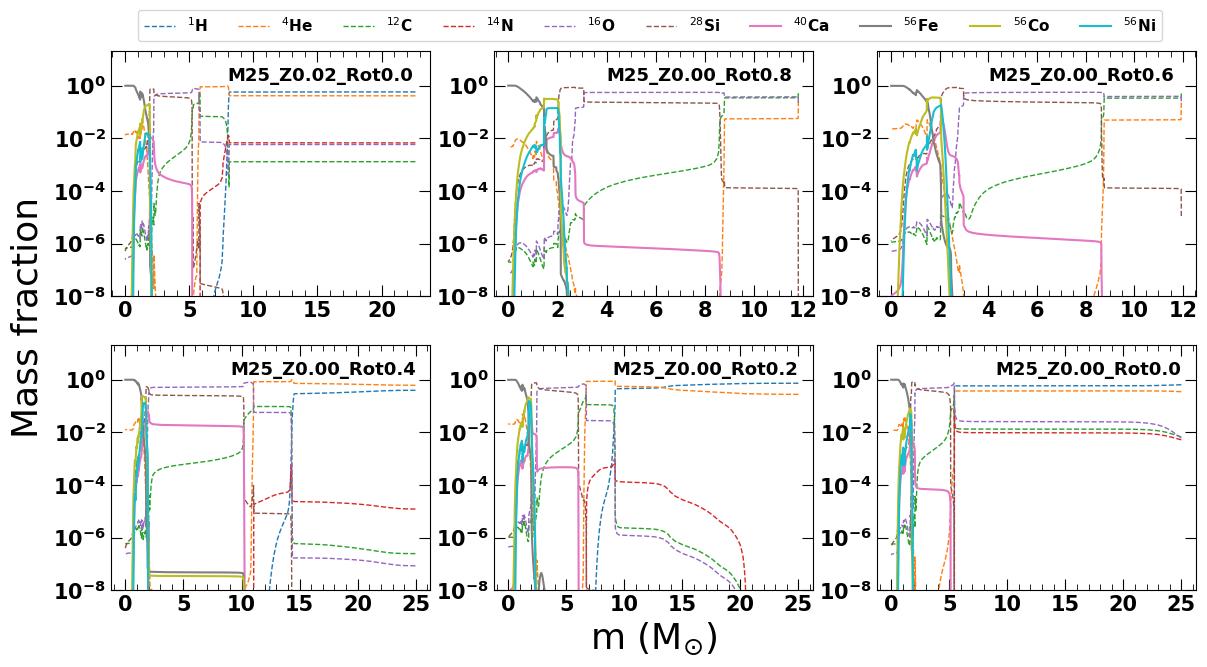}
    \caption{A combined plot showing the mass fractions of various elements for the models in this study at a stage when the models have reached the stage of the onset of core collapse.} 
    \label{fig:mass_fraction}
\end{figure*}     
\addtocounter{page}{1}
\appendix
\bibliographystyle{Classes/mnras} % Title is link 
\renewcommand{\bibname}{References} % changes the header; default: Bibliography
\bibliography{bib/ref_new}
\addcontentsline{toc}{chapter}{\sc References} %adds References to contents page
\rhead{}
\lhead{}

\end{document}